\documentclass[conference]{IEEEtran}

\usepackage{mathpartir}
\usepackage{amsmath}
\usepackage{amssymb}
\usepackage{amsthm}
\usepackage{mathtools}
\usepackage[inline]{enumitem}
\usepackage{stmaryrd}
\usepackage{hyperref}
\usepackage{color}
\usepackage{pifont}
\usepackage{subcaption}
\usepackage{xspace}
\usepackage{xstring}
\usepackage{xifthen}
\usepackage{tabularx}
\usepackage{booktabs}
\usepackage{scalerel}
\usepackage{centernot}
\usepackage{import}
\usepackage{cleveref}
\usepackage{balance}

\newtheorem{definition}{Definition}
\newtheorem{lemma}{Lemma}
\newtheorem{theorem}{Theorem}

\newcounter{ncomm}

\newlength{\sectskip}
\newcounter{lineno}
\newcommand{\zerolineno}{\setcounter{lineno}{0}}
\newcommand{\putlineno}{\stepcounter{lineno}\thelineno.}

\newcommand{\names}{\mathcal{N}}
\newcommand{\urls}{\mathcal{U}}
\newcommand{\origins}{\mathcal{O}}
\newcommand{\refs}{\mathcal{R}}
\newcommand{\vars}{\mathcal{X}}
\newcommand{\labels}{\mathcal{L}}
\newcommand{\seclabels}{\mathcal{L}^2}
\newcommand{\domains}{\mathcal{D}}
\newcommand{\ids}{\mathcal{I}}

\newcommand{\values}{\mathcal{V}}

\newcommand{\sample}[1]{#1 \leftarrow \names}
\newcommand{\domain}{\textit{dom}}

\newcommand{\subst}[2]{\{#1 \mapsto #2\}}
\newcommand{\mempty}{\{\}}
\newcommand{\mjoin}[2]{#1 \triangleleft #2}
\newcommand{\ie}{i.e.}
\newcommand{\eg}{e.g.}
\newcommand{\cf}{cf.}
\newcommand{\irule}[1]{\textsc{(#1)}}
\newcommand{\fnames}[1]{\mathit{ns}(#1)}
\newcommand{\fvars}[1]{\mathit{vars}(#1)}
\newcommand{\intval}[2]{[#1\ldots#2]}
\newcommand{\commands}[1]{\mathit{coms}(#1)}

\newcommand{\lcons}[2]{#1 :: #2}
\newcommand{\lempty}{\langle\rangle}
\newcommand{\lst}[1]{\vec{#1}}
\newcommand{\len}[1]{\StrLen{#1}[\namelen]\ifthenelse{\namelen > 1}{|\vec{#1}|}{|\vec{#1}\hspace{1pt}|}}
\newcommand{\lstexp}[1]{\langle #1 \rangle}

\newcommand{\vundef}{\bot}
\newcommand{\btrue}{\mathit{true}}
\newcommand{\bfalse}{\mathit{false}}

\newcommand{\fresh}{\textit{fresh}()}
\newcommand{\eval}[3]{\mathit{eval}_{#3}(#1, #2)}
\newcommand{\beval}[4]{\mathit{eval}_{#4}(#1, #2, #3)}

\newcommand{\tab}{\mathit{tab}}
\newcommand{\page}{\mathit{page}}

\newcommand{\hform}[2]{\mathsf{form}(\mathit{#1}, #2)}

\newcommand{\getck}[2]{\mathit{get\_ck}(#1, #2)}
\newcommand{\updck}[3]{\mathit{upd\_ck}(#1, #2, #3)}
\newcommand{\upckr}[2]{#1 \uparrow #2}

\newcommand{\load}[3]{\mathsf{load}(#1, #2, #3)}
\newcommand{\submit}[4]{\mathsf{submit}(#1, #2, #3, #4)}
\newcommand{\halt}{\mathsf{halt}}

\newcommand{\rg}[1]{@\mathit{#1}}
\newcommand{\rs}[1]{\$\mathit{#1}}
\newcommand{\param}[1]{\mathit{#1}}
\newcommand{\rf}[1]{\mathit{#1}}

\newcommand{\usr}{\mathsf{usr}}
\newcommand{\atk}{\mathsf{atk}}
\newcommand{\sid}{\iota_s}
\newcommand{\bid}{{\iota_b}}
\newcommand{\uid}{{\iota_u}}

\newcommand{\http}[1]{\mathsf{http}(#1)}
\newcommand{\https}[1]{\mathsf{https}(#1)}

\newcommand{\lauth}[3]{\sharp[ {#1} ]_{#3}^{#2}}
\newcommand{\absend}{\overline{\mathsf{req}}}
\newcommand{\abrecv}{\mathsf{res}}
\newcommand{\assend}{\overline{\mathsf{res}}}
\newcommand{\asrecv}{\mathsf{req}}
\newcommand{\bsend}[6]{\absend({#1}, {#2}, {#3}, {#4}, {#5}, {#6})}
\newcommand{\brecv}[7]{\abrecv({#1}, {#2}, {#3}, {#4}, {#5}, {#6}, {#7})}
\newcommand{\ssend}[7]{\assend({#1}, {#2}, {#3}, {#4}, {#5}, {#6}, {#7})}
\newcommand{\srecv}[6]{\asrecv({#1}, {#2}, {#3}, {#4}, {#5}, {#6})}
\newcommand{\blank}{\bullet}

\newcommand{\noorigin}{\vundef}
\newcommand{\uorigin}[1]{\text{orig}(#1)}
\newcommand{\hsts}{\Delta}
\newcommand{\ebrowser}[7]{(#1, #2, #3, #4, #5, #6)^{#7}}
\newcommand{\browser}[6]{\ebrowser{#1}{#2}{#3}{#4}{#5}{#6}{\bid}}
\newcommand{\sbrowser}[4]{B_{#4}(#1, #2, #3)}

\newcommand{\sskip}{\mathbf{skip}}
\newcommand{\incl}[2]{\mathbf{include}(#1, #2)}

\newcommand{\setdom}[3]{\mathbf{setdom}(#1, #2, #3)}
\newcommand{\ite}[3]{\mathbf{if}\ #1\ \mathbf{then}\ #2\ \mathbf{else}\ #3}
\newcommand{\cond}[2]{\mathbf{if}\ #1\ \mathbf{then}\ #2}
\newcommand{\eelse}{\mathbf{else}}

\newcommand{\para}[2]{#1 \parallel #2}
\newcommand{\xra}[1]{\xrightarrow{#1}}

\newcommand{\dom}[2]{\mathsf{dom}(#1, #2)}

\newcommand{\server}[3]{(#1, #2, #3)}
\newcommand{\shalt}{\mathbf{halt}}
\newcommand{\start}[1]{\mathbf{start}\ #1}

\newcommand{\auth}[2]{\mathbf{auth}\ #1\ \mathbf{at}\ #2}
\newcommand{\reply}[4]{\mathbf{reply}\ (#1, #2, #3)\ \mathbf{with}\ #4}
\newcommand{\ereply}[3]{\mathbf{reply}\ (#1, #2, #3)}
\newcommand{\ewith}[1]{\mathbf{with}\ #1}
\newcommand{\redr}[4]{\mathbf{redirect}\ (#1, #2, #3)\ \mathbf{with}\ #4}
\newcommand{\eredr}[3]{\mathbf{redirect}\ (#1, #2, #3)}
\newcommand{\listen}[4]{\mathit{#1}[#2](#3) \hookrightarrow #4}
\newcommand{\thread}[3]{\lceil {#1} \rfloor^{#2}_{#3}}
\newcommand{\login}[3]{\mathbf{login}\ #1, #2, #3}
\newcommand{\eerror}{\mathsf{error}}
\newcommand{\serror}{\mathsf{error}}
\newcommand{\och}[2]{\mathbf{if}\ \mathbf{originchk}(#1)\ \mathbf{then}\ #2}
\newcommand{\tokch}[3]{\mathbf{if}\ \mathbf{tokenchk}(#1, #2)\ \mathbf{then}\ #3}

\newcommand{\astep}[1]{\xra{#1}}
\newcommand{\bstep}[1]{\xra{#1}}
\newcommand{\sstep}[1]{\xra{#1}}
\newcommand{\sstepn}[2]{\sstep{#2}\!\!^{#1}\hspace{-2pt}}
\newcommand{\astepn}[2]{\astep{#2}\!\!^{#1}\hspace{-2pt}}

\newcommand{\atknow}{\mathcal{K}}

\newcommand{\atkstate}[3]{(#1, #2)\ \triangleright\ #3}

\newcommand{\proj}[2]{#1 \downarrow #2}

\newcommand{\nf}[2]{\mathit{nf}(#1, #2)}

\newcommand{\It}[1]{I(#1)}
\newcommand{\Ct}[1]{C(#1)}
\newcommand{\typelabel}[1]{\mathit{label}(#1)}
\newcommand{\cbot}{\bot_C}
\newcommand{\ctop}{\top_C}
\newcommand{\ibot}{\bot_I}
\newcommand{\itop}{\top_I}
\newcommand{\ilow}{I(\alabel)}
\newcommand{\ihigh}{\ibot}
\newcommand{\tbot}{\bot}
\newcommand{\ttop}{\top}
\newcommand{\low}{\pair{\cbot}{\itop}}
\newcommand{\unauth}{\times}

\newcommand{\isclow}[1]{#1 \ctleq C(\alabel)}
\newcommand{\ischigh}[1]{#1 \not\ctleq C(\alabel)}
\newcommand{\isilow}[1]{I(\alabel) \itleq #1}
\newcommand{\isihigh}[1]{I(\alabel) \not \itleq #1}

\newcommand{\lowest}[2]{\pair{\Ct{#1} \cmeet \Ct{#2}}{\It{#1} \ijoin \It{#2}}}

\newcommand{\attlower}[1]{\lfloor #1 \rfloor_{\alabel}}
\newcommand{\cleq}{\sqsubseteq_C}
\newcommand{\ileq}{\sqsubseteq_I}
\newcommand{\cjoin}{\sqcup_C}
\newcommand{\cmeet}{\sqcap_C}
\newcommand{\ijoin}{\sqcup_I}

\newcommand{\bigijoin}[1]{\mathop{{\bigsqcup}_I}_{#1}}

\newcommand{\imeet}{\sqcap_I}
\newcommand{\ccup}{\vee}
\newcommand{\ccap}{\wedge}
\newcommand{\icup}{\ccap}
\newcommand{\icap}{\ccup}
\newcommand{\bigccup}{\bigvee}
\newcommand{\bigccap}{\bigwedge}
\newcommand{\bigicup}{\bigccap}
\newcommand{\bigicap}{\bigccup}

\newcommand{\band}{~\mathbf{and}~}
\newcommand{\bor}{~\mathbf{or}~}
\newcommand{\uinput}[1]{\texttt{#1}}

\newcommand{\typingcontext}{C}
\newcommand{\protUrls}{\mathcal{P}}
\newcommand{\typebranch}{b}
\newcommand{\bcsrf}{\mathsf{csrf}}
\newcommand{\bhon}{\mathsf{hon}}
\newcommand{\isbcsrf}{\typebranch = \bcsrf}
\newcommand{\isbhon}{\typebranch = \bhon}
\newcommand{\alabel}{\ell_a}

\newcommand{\stype}{\tau}
\newcommand{\types}{\mathcal{T}}
\newcommand{\rtypes}{\mathcal{T}_{\refs}}

\newcommand{\pair}[2]{(#1,#2)}
\newcommand{\ctleq}{\cleq}
\newcommand{\itleq}{\ileq}
\newcommand{\tleq}{\sqsubseteq_{\alabel}}

\newcommand{\tijoin}{\tilde{\ijoin}}

\newcommand{\envu}{\Gamma_{\urls}}
\newcommand{\envv}{\Gamma_{\vars}}
\newcommand{\envrg}{\Gamma_{\refs^@}}
\newcommand{\envrs}{\Gamma_{\refs^\$}}
\newcommand{\envt}{\Gamma_{\values}}

\newcommand{\pclabel}{\texttt{pc}}

\newcommand{\seslabel}{{\ell_s}}

\newcommand{\elabelpar}[4]{{#1},{#2}\vdash^{\mathsf{se}}_{\alabel} #3 : #4}
\newcommand{\elabel}[2]{\elabelpar{\Gamma}{\seslabel}{#1}{#2}}
\newcommand{\srtypingpar}[4]{{#1},{#2}\vdash^{\mathsf{sr}}_{\alabel} #3 : #4}
\newcommand{\srtyping}[2]{\srtypingpar{\Gamma}{\seslabel}{#1}{#2}}
\newcommand{\brtypingpar}[3]{{#1} \vdash^{\mathsf{br}}_{\alabel} #2 : #3}
\newcommand{\brtyping}[2]{\brtypingpar{\Gamma}{#1}{#2}}
\newcommand{\belabelpar}[4]{#1, #2 \vdash^{\mathsf{be}}_{\alabel} #3 : #4}
\newcommand{\belabel}[2]{\belabelpar{\Gamma}{b}{#1}{#2}}
\newcommand{\ulabel}{\ell_u}

\newcommand{\sttypingpar}[2]{#1 \vdash^{\mathsf{t}}_{\alabel, \protUrls} #2}
\newcommand{\sttyping}[1]{\sttypingpar{\Gamma^0}{#1}}
\newcommand{\stypingparpctc}[7]{#1,#2,#3 \vdash^{\mathsf{c}}_{\alabel, \typingcontext} #5: #6, #7}
\newcommand{\stypingparpc}[7]{#1,#2,#3 \vdash^{\mathsf{c}}_{\alabel, (u, #4, \protUrls)} #5: #6, #7}
\newcommand{\stypingpartc}[6]{\stypingparpctc{#1}{#2}{#3}{#4}{#5}{#6}{#3}}
\newcommand{\stypingpar}[6]{\stypingparpc{#1}{#2}{#3}{#4}{#5}{#6}{#3}}
\newcommand{\stypingtc}[1]{\stypingpartc{\Gamma}{\seslabel}{\pclabel}{\typebranch}{#1}{\seslabel}}
\newcommand{\styping}[1]{\stypingpar{\Gamma}{\seslabel}{\pclabel}{\typebranch}{#1}{\seslabel}}

\newcommand{\tcre}[1]{\mathtt{cred}(#1)}
\newcommand{\tref}[1]{\mathtt{ref}(#1)}
\newcommand{\istcre}[1]{#1 = \tcre{\cdot}}

\newcommand{\ptyping}[4]{#1, #2, #3 \vdash^{\mathsf{f}}_{\alabel} #4}
\newcommand{\bstypingpar}[4]{#1,#2,#3 \vdash^{\mathsf{s}}_{\alabel, \protUrls} #4}
\newcommand{\bstyping}[2]{\bstypingpar{#1}{\pclabel}{b}{#2}}
\newcommand{\envtyping}[1]{\lambda, \alabel, #1 \vdash \diamond}

\newcommand{\prop}[1]{property \ref{#1}}

\newcommand{\timeouts}{T_O}

\newcommand{\ptodo}[2]{\todo{P{#1}: #2}}
\newcommand{\todo}[1]{\textcolor{red}{\textbf{[TODO: #1]}}}
\newcommand{\sketch}[1]{\textcolor{red}{[SKETCH: #1]}}
\renewcommand{\todo}[1]{}
\renewcommand{\sketch}[1]{#1}

\newcommand{\cremeet}{\overline{\sqcap}}
\newcommand{\crejoin}{\tilde{\sqcup}}

\newcommand{\relenv}{\Gamma}
\newcommand{\prerel}{\approx}
\newcommand{\fullrel}{\approxeq}
\newcommand{\prel}{\prerel_{\relenv}}
\newcommand{\bprel}{\prerel^{B}_{\relenv}}
\newcommand{\sprel}{\prerel^{S}_{\relenv}}
\newcommand{\rel}{\fullrel_{\relenv}}
\newcommand{\brel}{\fullrel^{B}_{\relenv}}
\newcommand{\srel}{\fullrel^{S}_{\relenv}}
\newcommand{\rbad}[1]{bad(#1)}

\newcommand{\tjoin}{\sqcup}

\newcommand{\exvdash}{\vDash}
\newcommand{\exelabelpar}[4]{{#1},{#2}\exvdash^{\mathsf{se}}_{\alabel} #3 : #4}
\newcommand{\exelabel}[2]{\exelabelpar{\Gamma}{\seslabel}{#1}{#2}}
\newcommand{\exsrtypingpar}[4]{{#1},{#2}\exvdash^{\mathsf{sr}}_{\alabel} #3 : #4}
\newcommand{\exsrtyping}[2]{\exsrtypingpar{\Gamma}{\seslabel}{#1}{#2}}
\newcommand{\exbrtypingpar}[3]{{#1} \exvdash^{\mathsf{br}}_{\alabel} #2 : #3}
\newcommand{\exbrtyping}[2]{\exbrtypingpar{\Gamma}{#1}{#2}}
\newcommand{\exbelabelpar}[4]{#1, #2 \exvdash^{\mathsf{be}}_{\alabel} #3 : #4}
\newcommand{\exbelabel}[2]{\exbelabelpar{\Gamma}{b}{#1}{#2}}
\newcommand{\exsttypingpar}[2]{#1 \exvdash^{\mathsf{t}}_{\alabel, \protUrls} #2}
\newcommand{\exsttyping}[1]{\exsttypingpar{\Gamma^0}{#1}}
\newcommand{\exstypingparpctc}[7]{#1,#2,#3 \exvdash^{\mathsf{c}}_{\alabel, \typingcontext} #5: #6, #7}
\newcommand{\exstypingparpc}[7]{#1,#2,#3 \exvdash^{\mathsf{c}}_{\alabel, (u, #4, \protUrls)} #5: #6, #7}

\newcommand{\exstypingpar}[6]{\exstypingparpc{#1}{#2}{#3}{#4}{#5}{#6}{#3}}

\newcommand{\exstyping}[1]{\exstypingpar{\Gamma}{\seslabel}{\pclabel}{\typebranch}{#1}{\seslabel}}

\newcommand{\exptyping}[4]{#1, #2, #3 \exvdash^{\mathsf{f}}_{\alabel} #4}
\newcommand{\exbstypingparu}[5]{#1,#2,#3 \exvdash^{\mathsf{s}}_{\alabel, \protUrls, #4} #5}
\newcommand{\exbstypingpar}[4]{\exbstypingparu{#1}{#2}{#3}{u}{#4}}
\newcommand{\exbstyping}[2]{\bstypingpar{#1}{\pclabel}{b}{#2}}

\newcommand{\systypingpar}[2]{#1 \exvdash_{\alabel,\usr} #2}
\newcommand{\systyping}[1]{\systypingpar{\Gamma}{#1}}
\newcommand{\servtypingpar}[2]{#1 \exvdash_{\alabel,\usr} #2}
\newcommand{\servtyping}[1]{\servtypingpar{\Gamma}{#1}}
\newcommand{\browtypingpar}[2]{#1 \exvdash_{\alabel,\usr} #2}
\newcommand{\browtyping}[1]{\browtypingpar{\Gamma}{#1}}
\newcommand{\reqtypingpar}[2]{#1 \exvdash_{\alabel,\usr} #2}
\newcommand{\reqtyping}[1]{\reqtypingpar{\Gamma}{#1}}
\newcommand{\restypingpar}[2]{#1 \exvdash_{\alabel,\usr} #2}
\newcommand{\restyping}[1]{\restypingpar{\Gamma}{#1}}

\newcommand{\trans}[1]{\overline{#1}}

\newcommand{\xRa}[1]{\xRightarrow{#1}}
\newcommand{\exstep}[1]{\xRa{#1}_{\Gamma}}
\newcommand{\exastep}[1]{\exstep{#1}}
\newcommand{\exbstep}[1]{\exstep{#1}}
\newcommand{\exsstep}[1]{\exstep{#1}}

\newcommand{\exastepn}[2]{\exastep{#2}\!\!^{#1}\hspace{-2pt}}

\newcommand{\exatkstatepar}[4]{(#1, #2)\ \triangleright_{#4}\ #3}
\newcommand{\exatkstate}[3]{\exatkstatepar{#1}{#2}{#3}{\timeouts}}
\newcommand{\exebrowser}[9]{(#1, #2, #3, #4, #5, #6)^{#7,#8,#9}}
\newcommand{\exbrowserpar}[8]{\exebrowser{#1}{#2}{#3}{#4}{#5}{#6}{\bid}{#7}{#8}}
\newcommand{\exbrowser}[6]{\exebrowser{#1}{#2}{#3}{#4}{#5}{#6}{\bid}{l}{\amode}}
\newcommand{\exhform}[3]{\hform{#1}{#2}^{#3}}
\newcommand{\exserver}[3]{\server{#1}{#2}{#3}}

\newcommand{\exupckr}[2]{#1 \uparrow' #2}
\newcommand{\exupdck}[3]{\mathit{upd\_ck'}(#1, #2, #3)}
\newcommand{\resetpc}[1]{\mathsf{reset}\ #1}

\newcommand{\batt}{\mathsf{att}}

\newcommand{\amode}{\mu}
\newcommand{\jlabel}[1]{jlabel(#1)}
\newcommand{\threads}[1]{threads(#1)}
\newcommand{\exthreadpar}[5]{\lceil {#1} \rfloor_{#2,#3}^{#4,#5}}
\newcommand{\exthread}[3]{\exthreadpar{#1}{#2}{#3}{l}{\amode}}
\newcommand{\exsrecv}[8]{\mathsf{req}({#1}, {#2}, {#3}, {#4}, {#5}, {#6})^{#7,#8}}
\newcommand{\exssend}[1]{\def\tempa{#1} \exssendcontinued}
\newcommand{\exssendcontinued}[9]{\overline{\mathsf{res}}({\tempa}, {#1}, {#2}, {#3}, {#4}, {#5}, {#6}, {#7})^{#8,#9}}
\newcommand{\exbsend}[8]{\overline{\mathsf{req}}({#1}, {#2}, {#3}, {#4}, {#5}, {#6})^{#7,#8}}
\newcommand{\exbrecv}[1]{\def\tempb{#1} \exbrecvcontinued}
\newcommand{\exbrecvcontinued}[9]{\mathsf{res}({\tempb},{#1}, {#2}, {#3}, {#4}, {#5}, {#6}, {#7})^{#8,#9}}
\newcommand{\exbeval}[5]{\mathit{eval}_{#5}(#1, #2, #3, #4)}

\newcommand{\sbad}{\mathbf{bad}}

\newcommand{\gmems}{D_{@}}
\newcommand{\smems}{D_{\$}}

\newcommand{\servers}[1]{\text{servers}(#1)}
\newcommand{\browsers}[1]{\text{browsers}(#1)}
\newcommand{\surls}[1]{\text{urls}(#1)}
\newcommand{\running}[1]{\text{running}(#1)}
\newcommand{\treft}[1]{\text{ref}_\stype(#1)}

\newcommand{\tlowestint}[1]{int_{\sqcup}(#1)}
\newcommand{\thighestint}[1]{int_{\sqcap}(#1)}
\newcommand{\tint}[1]{\text{int}(#1)}
\newcommand{\einta}[2]{{#1}@{#2}}
\newcommand{\eint}[1]{\text{sync}_I(#1)}

\newcommand{\gmem}[2]{\text{mem}_g(#1,#2)}
\newcommand{\smem}[2]{\text{mem}_s(#1,#2)}
\newcommand{\heq}[1]{=_{#1,\ihigh}}
\newcommand{\heqmpar}[1]{=_{#1,\ihigh}}
\newcommand{\heqm}{\heqmpar{\Gamma}}
\newcommand{\heqt}{=_{\Gamma,\ihigh}}
\newcommand{\heqe}{=_{\ihigh}}
\newcommand{\nexth}[1]{\mathbf{nexth}(#1)}
\newcommand{\vals}[1]{\mathit{values}(#1)}

\newcommand{\samplel}[2]{#1 \leftarrow \names_{#2}}

\setlist[itemize,1]{leftmargin=.11in}
\setlist[enumerate,1]{leftmargin=.2in}

\begin{document}

\title{Language-Based Web Session Integrity}

\author{
	\IEEEauthorblockN{
		Stefano Calzavara\IEEEauthorrefmark{1},
		Riccardo Focardi\IEEEauthorrefmark{1},
		Niklas Grimm\IEEEauthorrefmark{2}, 
		Matteo Maffei\IEEEauthorrefmark{2},
		Mauro Tempesta\IEEEauthorrefmark{2}
	}
	\IEEEauthorblockA{
		\IEEEauthorrefmark{1}Universit\`{a} Ca' Foscari Venezia\quad
		\IEEEauthorrefmark{2}TU Wien
	}
}

\maketitle

\begin{abstract}

Session management is a fundamental component of web applications: despite the apparent simplicity, correctly implementing web sessions is extremely tricky, as witnessed by the large number of existing attacks. This motivated the design of formal methods to rigorously reason about web session security which, however, are not supported at present by suitable automated verification techniques.
In this paper we introduce the first security type system that enforces session security on a core model of web applications, focusing in particular on server-side code.
We showcase the expressiveness of our type system by analyzing the session management logic of HotCRP, Moodle, and phpMyAdmin, unveiling novel security flaws that have been acknowledged by software developers.

\end{abstract}

\section{Introduction}
Since the HTTP protocol is stateless, web applications that need to keep track of state information over multiple HTTP requests have to implement custom logic for \emph{session management}.
Web sessions typically start with the submission of a login form from a web browser, where a registered user provides her access credentials to the web application.
If these  credentials are valid, the web application stores in the user's browser fresh \emph{session cookies}, which are automatically attached to all subsequent requests sent to the web application.
These cookies contain enough information to authenticate the user and to keep track of session state across requests.

Session management is essential in the modern Web, yet it is often vulnerable to a range of attacks and surprisingly hard to get right.
For instance, the theft of session cookies allows an attacker to impersonate the victim at the web application~\cite{NikiforakisMYJJ11,BugliesiCFK15,TangDK11}, while the weak integrity guarantees offered by cookies allow subtle attacks like cookie forcing, where a user is forced into an attacker-controlled session via cookie overwriting~\cite{ZhengJLDCWW15}.
Other common attacks include cross-site request forgery (CSRF)~\cite{JovanovicKK06}, where an attacker instruments the victim's browser to send forged authenticated requests to a target web application, and login CSRF, where the victim's browser is forced into the attacker's session by submitting a login form with the attacker's credentials~\cite{BarthJM08}.
We refer to a recent survey for an overview of attacks against web sessions and countermeasures~\cite{CalzavaraFST17}.

Given the complexity of session management and the range of threats to be faced on the web, a formal understanding of web session security and the design of automated verification techniques is an important research direction.
Web sessions and their desired security properties have been formally studied in several papers developing browser-side defenses for web sessions~\cite{BugliesiCFKT14,BugliesiCFK15,KhanCBGP14,CalzavaraFGM16}: while the focus on browser-side protection mechanisms is appealing to protect users of vulnerable web applications, the deployment of these solutions is limited since it is hard to design browser-side defenses that do not cause compatibility issues on existing websites and are effective enough to be integrated in commercial browsers~\cite{CalzavaraFST17}.

Thus, security-conscious developers would better rely on server-side programming practices to enforce web session security when web applications are accessed by standard browsers.
Recently, Fett et al.~\cite{FettHK19} formalized a session integrity property specific to OpenID within the  Web Infrastructure Model (WIM), an expressive web model within which proofs  are, however, manual and require a strong expertise. 

In this work, we present \emph{the first static analysis  technique for web session integrity}, focusing on sound server-side programming practices. In particular:
\begin{enumerate}
\item  we introduce a core formal model of {web systems}, representing browsers, servers, and attackers who may mediate communications between them. Attackers can also interact with honest servers to establish their own sessions and host malicious content on compromised websites.
The goal in the design of the model is to retain simplicity, to ease the presentation of the basic principles underlying our analysis technique, while being expressive enough to capture the salient aspects of session management in real-world case studies.
In this model, we formalize a generic definition  of \emph{session integrity},  inspired by prior work on browser-side security~\cite{BugliesiCFKT14}, as a semantic hyperproperty~\cite{ClarksonS10} ruling out a wide range of attacks against web sessions;

\item we design a novel type system for the verification of session integrity within our model. The type system exploits confidentiality and integrity guarantees of session data to endorse untrusted requests coming from the network and enforces appropriate browser-side invariants in the corresponding responses to guarantee session integrity;

\item we showcase the effectiveness and generality of our type system by analyzing the session management logic of HotCRP, Moodle, and phpMyAdmin.
After encoding the relevant code fragments in our formal model, we use the type system to establish a session integrity proof: failures in this process led to the discovery of critical security flaws.
We identified two vulnerabilities in HotCRP that allow an attacker to hijack accounts of authors and even reviewers, and one in phpMyAdmin, which has been assigned a CVE~\cite{phpMyAdmin-CVE-2019-12616}. 
All vulnerabilities have been reported and acknowledged by the application developers. We finally established security proofs for the fixed versions by typing.
\end{enumerate}

\section{Overview}
\label{sec:overview}

In this Section we provide a high-level overview of our approach to the verification of session integrity.
Full formal details and a complete security analysis of the HotCRP conference management system are presented in the remainder of the paper.

\subsection{Encoding PHP Code in our Calculus}

The first step of our approach consists in accessing the PHP implementation of HotCRP and carefully handcrafting a model of its authentication management mechanisms  into the core calculus we use to model web application code. While several commands are standard, our language for server-side programs includes some high-level commands abstracting functionalities that are implemented in several lines of PHP code.
The $\mathbf{login}$ command abstracts a snippet of code checking, \eg, in a database, whether the provided credentials match an existing user in the system.
Command $\mathbf{auth}$ is a \emph{security assertion} parametrized by expressions it depends on. In our encoding it abstracts code performing security-sensitive operations within the active session: here it models code handling paper submissions in HotCRP.
Command $\mathbf{start}$ takes as argument a session identifier and corresponds to the \texttt{session\_start} function of PHP, restoring variables set in the session memory during previous requests bound to that session. 

In the following we distinguish standard PHP variables from those stored in the session memory (\ie, variables in the \texttt{\$\_SESSION} array) using symbols @ and \$, respectively.
The $\mathbf{reply}$ command models the server's response in a structured way by separating the page's DOM, scripts, and cookies set via HTTP headers.

\subsection{A Core Model of HotCRP}
\label{sec:core-hotcrp}

We assume that the HotCRP installation is hosted at the domain $d_C$ and accessible via two HTTPS endpoints: \emph{login}, where users perform authentication using their access credentials, and \emph{manage}, where users can upload their papers or withdraw their submissions.
The session management logic is based on a cookie $\param{sid}$ established upon login. We now discuss the functionality of the two HTTPS endpoints; we denote the names of cookies in square brackets and the name of parameters in parentheses.
The login endpoint expects a username $\param{uid}$ and a password $\param{pwd}$ used for authentication:
\[\zerolineno\arraycolsep=1pt
\begin{array}{ll}
    \putlineno & \listen{login}{}{\param{uid}, \param{pwd}}{} \\
	\putlineno & \quad \cond{\param{uid} = \vundef \band \param{pwd} = \vundef}{} \\
	\putlineno & \quad \quad \ereply{\{\uinput{auth} \mapsto \hform{login}{\lstexp{\vundef, \vundef}}\}}{\sskip}{\mempty} \\
	\putlineno & \quad \eelse \\
	\putlineno & \quad \quad \rg{r} := \fresh;\; \login{\param{uid}}{\param{pwd}}{\rg{r}}; \\
	\putlineno & \quad \quad \start{\rg{r}}; \;\rs{user} := \param{uid}; \\
	\putlineno & \quad \quad \ereply{\{\uinput{link} \mapsto \hform{manage}{\lstexp{\vundef, \vundef, \vundef}}\}}{ \\
	\putlineno & \qquad \qquad \quad \ \ \sskip}{\{ \rf{sid} \mapsto x \})} \\
	\putlineno & \quad \quad \ \ \ewith{x = \rg{r}} \\
\end{array}
\]
If the user contacts the endpoint without providing access credentials, the endpoint replies with a page containing a login form expecting the username and password (lines 2--3). 
Otherwise, upon successful authentication via $\param{uid}$ and $\param{pwd}$, the endpoint starts a new session indexed by a fresh identifier which is stored into the variable $\rg{r}$ (lines 5--6).
For technical convenience, in the $\mathbf{login}$ command we also specify the fresh session identifier as a third parameter to bind the identity of its owner to the session.
Next, the endpoint stores the user's identity in the session variable $\rs{user}$ so that the session identifier can be used to authenticate the user in subsequent requests (line 6).
Finally, the endpoint sends a reply to the user's browser which includes a link to the submission management interface and sets a cookie $\param{sid}$ containing the session identifier stored in $\rg{r}$ (lines 7--9).
 
The submission management endpoint requires authentication, hence it expects a session cookie $\param{sid}$. It also expects three parameters: a $\param{paper}$, an $\param{action}$ (submit or withdraw) and a $\param{token}$ to protect against CSRF attacks~\cite{BarthJM08}:
\[\zerolineno\arraycolsep=-1pt
\begin{array}{ll}
    \putlineno & \listen{manage}{\rf{sid}}{\param{paper}, \param{action}, \param{token}}{} \\
	\putlineno & \quad \start{\rg{sid}}; \\
	\putlineno & \quad \cond{\rs{user} = \vundef}{} \\
	\putlineno & \quad \quad \ereply{\{\uinput{auth} \mapsto \hform{login}{\lstexp{\vundef, \vundef}}\}}{\sskip}{\mempty} \\
	\putlineno & \quad \eelse~\cond{\param{paper} = \vundef}{} \\
	\putlineno & \quad \quad \rs{utoken} = \fresh; \\
	\putlineno & \quad \quad \ereply{\{\uinput{add} \mapsto \hform{manage}{\lstexp{\vundef, \uinput{submit}, x}}, \\
	\putlineno & \qquad\qquad\quad \ \ \uinput{del} \mapsto \hform{manage}{\lstexp{\vundef, \uinput{withdraw}, x}}\}}{ \\
	\putlineno & \qquad\qquad\quad \ \sskip}{\mempty} \\
	\putlineno & \quad \quad \ \ \ewith{x = \rs{utoken}} \\
	\putlineno & \quad \eelse~\tokch{\param{token}}{\rs{utoken}}{} \\	
	\putlineno & \quad \quad \auth{\param{paper}, \param{action}}{\ell_C}; \; \ereply{\mempty}{\sskip}{\mempty} \\
\end{array}
\]
The endpoint first tries to start a session over the cookie $\param{sid}$: if it identifies a valid session, session variables from previous requests are restored (line 2).
The condition $\rs{user} = \vundef$ checks whether the session is authenticated, since the variable is only set after login: if it is not the case, the endpoint replies with a link to the login page (lines 3--4).
If the user is authenticated but does not provide any paper in her request, the endpoint replies with two forms used to submit or withdraw a paper respectively.
Such forms are protected against CSRF with a fresh token, whose value is stored in the session variable $\rs{utoken}$ (lines 5--10).
If the user is authenticated and requests an action over a given paper, the endpoint checks that the token supplied in the request matches the one stored in the user's session (line 11) and performs the requested action upon success (line 12).
This is modeled via a security assertion in the code that authorizes the requested action on the paper on behalf of the owner of the session.
The assertion has a security label $\ell_C$, intuitively meaning that authorization can be trusted unless the attacker can read or write at $\ell_C$.
Security labels have a confidentiality and an integrity component, expressing who can read and who can write.
They are typically used in the information flow literature~\cite{CalzavaraFGM16} not only to represent the security of program terms but also the attacker itself.
Here we let $\ell_C = (\https{d_C}, \https{d_C})$, meaning that authorization can be trusted unless HTTPS communication with the domain $d_C$ hosting HotCRP is compromised by the attacker.

\subsection{Session Integrity}
\label{sec:sess-int}

In this work, we are interested in \emph{session integrity}. 
Inspired by \cite{BugliesiCFKT14}, we formalize it as a relational property, comparing two different scenarios: an ideal world where the attacker does nothing and an attacked world where the attacker uses her capabilities to compromise the session.
Intuitively, session integrity requires that any authorized action occurring in the attacked world can also happen in the ideal world, unless the attacker is powerful enough to void the security assertions; this must hold for all sequences of actions of a user interacting with the session using a standard web browser.

As a counterexample to session integrity for our HotCRP model, pick an attacker hosting an HTTPS website at the domain $d_E \neq d_C$, modeled by the security label $\ell_E = (\https{d_E},\https{d_E})$.
Since $\ell_E \not\sqsupseteq \ell_C$, this attacker should not be able to interfere with authorized actions at the submission management endpoint.
However, this does not hold due to the lack of CSRF protection on the endpoint \emph{login}. In particular, pick the following sequence of user actions where \emph{evil} stands for an HTTPS endpoint at $d_E$:
\[\arraycolsep=1.4pt
\begin{array}{lcl}
  \lst{a} & = & \load{1}{\mathit{login}}{\mempty}, \\
  	& & \submit{1}{\mathit{login}}{\uinput{auth}}{\{ 1 \mapsto \usr, 2 \mapsto \uinput{pwd} \}}, \\
	& & \load{2}{\mathit{evil}}{\mempty}, \submit{1}{\mathit{login}}{\uinput{link}}{\mempty}, \\
	& & \submit{1}{\mathit{manage}}{\uinput{add}}{\{ 1 \mapsto \uinput{paper} \}}
\end{array}
\]
The user opens the login endpoint in tab $1$ and submits her username and password via the authentication form (identified by the tag \uinput{auth}). She then loads the attacker's website in tab $2$ and moves back to tab $1$ where she accesses the submission management endpoint by clicking the link obtained upon authentication. Finally, she submits a paper via the \uinput{add} form.

Session integrity is violated since the attacker can reply with a page containing a script which automatically submits the attacker's credentials to the login endpoint, authenticating the user as the attacker at HotCRP.
Thus, the last user action triggers the security assertion in the attacker's session rather than in the user's session.
Formally, this is captured by the security assertion firing the event $\lauth{\uinput{paper}, \uinput{submit}}{\usr, \atk}{\ell_C}$, modeling that the paper is submitted by the user into the attacker's session.
As such an event cannot be fired in the ideal world without the attacker, this violates session integrity.

In practice, an attacker could perform the attack against an author so that, upon uncareful submission, a paper is registered in the attacker's account, violating the paper's confidentiality.
We also discovered a more severe attack allowing an attacker to log into the victim's session, explained in \autoref{sec:case-study}.

\subsection{Security by Typing}
\label{sec:overview-typing}

Our type system allows for sound verification of session integrity and is parametric with respect to an attacker label.
In particular, typing ensures that the attacker has no way to forge authenticated events in the session of an honest user (as in a CSRF attack) or to force the user to perform actions within a session bound to the attacker's identity (\eg, due to a login CSRF).
Failures arising during type-checking often highlight in a direct way session integrity flaws. 


To ensure session integrity, we require two ingredients: first, we need to determine the identity of the sender of the request; second, we must ensure that the request is actually sent with the consent of the user, \ie, the browser is not sending the request as the attacker's deputy.
Our type system captures these aspects using two labels: a \emph{session} label and a \emph{program counter} (PC) label.
The session label models both the session's integrity (\ie, who can influence the session and its contents) and confidentiality (\ie, who can learn the session identifier used as access control token).
Since the identity associated with an authenticated event is derived from the ongoing session, the session label captures the first ingredient.
The PC label tracks who could have influenced the control flow to reach the current point of the execution. Since a CSRF attack is exactly a request of low integrity (as it is triggered by the attacker), this captures the second ingredient.
Additionally, the type system relies on a typing environment that assigns types to URLs and their parameters, to local variables and to references in the server memory.

We type-check the code twice under different assumptions.
First, we assume the scenario of an honest user regularly interacting with the page:
here we assume that all URL parameters are typed according to the
typing environment and we start with a high integrity PC label.
Second, we assume the scenario of a CSRF attack where all URL parameters have
low confidentiality and integrity (since they are controlled by the attacker) and
we start with a low integrity PC label.
In both cases, types for cookies and the server variables are taken from the typing environment since, even in a CSRF attack, cookies are taken from the cookie jar of the user's browser and the attacker has no direct access to the server memory.

We now explain on a high level why our type system fails to type-check our (vulnerable) HotCRP model.
To type the security assertion $\auth{paper, action}{\ell_C}$ in the
$\mathit{manage}$ endpoint, we need a high integrity PC label, a high
integrity session label and we require the parameters $\param{paper}$ and
$\param{action}$ to be of high integrity. While the types of the parameters are
immediately determined by the typing environment, the other two labels are influenced
by the typing derivation.

In the CSRF scenario, the security assertion is unreachable due to the presence of the token check instruction (line 11).
When typing, if we assume (in the typing environment) that $\rs{ltoken}$ is a high confidentiality reference, we can conclude that the check always fails since the parameter $\param{token}$ (controlled by the attacker) has low confidentiality, therefore we do not need to type-check the continuation.\footnote{~This reasoning is sound only when credentials (\eg, session identifiers and CSRF tokens) are unguessable fresh names. To take into account this aspect, in the type system we have special types for credentials (\cf\ \autoref{sec:types}) and we forbid subtyping for high confidentiality credentials.}

In the honest scenario, the PC label has high integrity assuming that all the preceding conditionals have high integrity guard expressions (lines 3 and 5).
The session label is set in the command $\start{\rg{sid}}$ (line 2) and depends on the type of the session identifier $\rg{sid}$. To succeed in typing, $\rg{sid}$ must have high integrity.
However, we cannot type-check the $\mathit{login}$ endpoint under this assumption: since the code does not contain any command that allows pruning the CSRF typing branch (like the token check in the $\mathit{manage}$ endpoint), the entire code must be typed with a low integrity PC label.
This prevents typing the $\mathbf{reply}$ statement  where cookie $\param{sid}$ is set (lines 7--9), since writing to a high integrity location from a low integrity context is unsound.
In practice, this failure in typing uncovers the vulnerability in our code: the integrity of the session cookie is low since an attacker can use a login CSRF attack to set a session cookie in the user's browser.


As a fix, one can protect the $\mathit{login}$ endpoint against CSRF attempts by using \emph{pre-sessions}~\cite{BarthJM08}: when the $\mathit{login}$ endpoint is visited for the first time by the browser, 
it creates a new unauthenticated session at the server-side (using a fresh cookie $pre$) and generates a token which is saved into the session and embedded into the login form. When submitting the login form, the contained token is compared to the one stored at the server-side in the pre-session and, if there is a mismatch, authentication fails: 
\[\arraycolsep=-1.5pt
\zerolineno
\begin{array}{ll}
    \putlineno & \listen{login}{\rf{pre}}{\param{uid}, \param{pwd}, \param{token}}{} \\
	\putlineno & \quad \cond{\param{uid} = \vundef \band \param{pwd} = \vundef}{} \\
	\putlineno & \quad \quad \rg{r'} := \fresh; \; \start{\rg{r'}}; \; \rs{ltoken} := \fresh; \\
	\putlineno & \quad \quad \ereply{\{\uinput{auth} \mapsto \hform{login}{\lstexp{\vundef, \vundef, x}}\}}{ \\
	\putlineno & \qquad\qquad\quad\ \sskip}{\subst{\rf{pre}}{y}} \\
	\putlineno & \quad \quad \ \ \ewith{x = \rs{ltoken}, y = \rg{r'}} \\
	\putlineno & \quad \eelse \\
	\putlineno & \quad \quad \start{\rg{pre}}; \\
	\putlineno & \quad \quad \tokch{\param{token}}{\rs{ltoken}}{} \\
	\putlineno & \quad \quad \quad \rg{r} := \fresh; \; \login{\param{uid}}{\param{pwd}}{\rg{r}}; \\
	\putlineno & \quad \quad \quad \start{\rg{r}}; \;\rs{user} := \param{uid}; \\
	\putlineno & \quad \quad \quad \ereply{\{\uinput{link} \mapsto \hform{manage}{\lstexp{\vundef, \vundef, \vundef}}\}}{ \\
	\putlineno & \qquad\qquad\qquad\ \sskip}{\{ \rf{sid} \mapsto x \}} \\
	\putlineno & \quad \quad \quad \quad  \mathbf{with}\ x = \rg{r} \\
\end{array}
\]
The session identified by $\param{pre}$ has low integrity but high confidentiality:
indeed, an attacker can cause a random $\param{pre}$ cookie to be set in the user's
browser (by forcing the browser to interact with the \emph{login} endpoint), but she has no way to learn the value of the cookie and hence cannot access the session.
We can thus assume high confidentiality for the session reference $\rs{ltoken}$ in the session identified by $\param{pre}$.

With the proposed fix, the piece of code responsible for setting the session cookie $\param{sid}$ is protected by a token check, where the parameter $\param{token}$ is compared against the high confidentiality session reference $\rs{ltoken}$ of the session identified by $\rg{pre}$ (line 9).
Similar to the token check in the \emph{manage} endpoint, this allows us to prune the CSRF typing branch and we can successfully type-check the code with a high integrity type for $\param{sid}$.
We refer the reader to \autoref{sec:typing-example} for a detailed explanation of typing the fixed $\mathit{login}$ endpoint.

The HotCRP developer acknowledged the login CSRF vulnerability and the effectiveness of the proposed fix, which is currently under development.

\section{A Formal Model of Web Systems}
We present now our model of web systems that includes the relevant ingredients for modeling attacks against session integrity and the corresponding defenses and we formally define our session integrity property.

\subsection{Expressiveness of the Model}

Our model of browsers supports cookies and a minimal client-side scripting language featuring
\begin{enumerate*}[label={\em\roman*)}]
	\item read/write access to the cookie jar and the DOM of pages;
	\item the possibility to send network requests towards arbitrary endpoints and include their contents as scripts.
\end{enumerate*}
The latter capability is used to model resource inclusion and a simplified way to perform XHR requests. 
In the model we can encode many security-sensitive aspects of cookies that are relevant for attacks involving their theft or overwriting, \ie, cookie prefixes~\cite{CookiePrefixes} and attributes \texttt{Domain} and \texttt{Secure}~\cite{Cookies}.
We also model HSTS~\cite{HSTS} which can improve the integrity guarantees of cookies set by HSTS-enabled domains.
On the server-side we include primitives used for session management and standard defenses against CSRF attacks, \eg, double submit cookies, validation of the \texttt{Origin} header and the use of CSRF tokens.

For the sake of presentation and simplicity, 
we intentionally omit some web components that are instead covered in other web models (\eg, the WIM~\cite{FettHK19}) but are not fundamental for session integrity or for modelling our case studies. In particular, we do not model document frames and cross-frame communications via the Web Messaging API, web sockets, local storage, DNS and an equational theory for cryptographic primitives.
We also exclude the \texttt{Referer} header since it conveys similar information to the \texttt{Origin} header which we already cover in our model. While we believe that our type system can be in principle extended to cover also these web components, the presentation and proof of soundness would become cumbersome, obfuscating the key aspects of our static analysis technique. 

\subsection{Syntax}

\begin{table*}[t]
	\setlength{\tabcolsep}{2pt}
	\caption{Syntax (browsers $B$ and scripts $s$ are defined in 
  \cref{sec:browser-app}).}
	\label{tab:syntax}

	\begin{subtable}[t]{\textwidth}
		\begin{tabularx}{\textwidth}{p{1.4cm}@{\hspace{2pt}}rcp{2.5cm} p{1.8cm}@{\hspace{2pt}}rcp{2.2cm} p{1.9cm}@{\hspace{2pt}}rcX}
      \multicolumn{12}{c}{{\small\textbf{Basics}}} \\
			\midrule
			Names & $n^{\ell}, i^{\ell}, j^{\ell}$ & $\in$ & $\names$ &
			References & $r$ & $\in$ & $\refs$ &
			Variables & $x$ & $\in$ & $\vars$ \\
			Identities & $\iota$ & $\in$ & $\ids \ni \usr$ &
			Domains & $d$ & $\in$ & $\domains$ &
			URLs & $u$ & $\in$ & $\urls$ \\
			Origins & $o$ & $\in$ & $\origins \supseteq O$ &
			Simple labels & $l$ & $\in$ & $\labels \supseteq L$ &
			Labels & $\ell$ & $::=$ & $(l, l)$ \\
			Types & $\stype$ & $\in$ & $\types$ &
			Numbers & $k, m$ & $\in$ & $\mathbb{N}$ &
			Primitive values & $pv$ & $::=$ & $\btrue ~|~ \bfalse ~|~ k ~|~ \ldots$ \\
			Values & $v$ & $::=$ & $pv ~|~ n ~|~ \iota ~|~ u ~|~ \vundef \in \values$ &
			Metavariables & $z$ & $\in$ & $\values \cup \vars$ &
			Forms & $f$ & $::=$ & $\mempty ~|~ f \uplus \subst{v}{\hform{u}{\lst{z}}}$ \\
			Pages & $\page$ & $::=$ & $\serror ~|~ f$ &
			Cookies & $ck$ & $::=$ & $\mempty ~|~ ck \uplus \subst{r}{z}$ &
			Memories & $M$ & $::=$ & $\mempty ~|~ M \uplus \subst{r}{v}$ \\
		\end{tabularx}
	\end{subtable}
	~\\[5pt]
	\begin{subtable}[t]{\textwidth}
		\begin{tabularx}{\textwidth}{l@{\hspace{5pt}}rcX@{\hspace{-15pt}}l@{\hspace{5pt}}rcX}
      \multicolumn{8}{c}{{\small\textbf{Servers}}} \\
			\midrule
			Expressions & $se$ & $::=$ & $x ~|~ \rg{r} ~|~ \rs{r} ~|~ v ~|~ \fresh^{\ell} ~|~ se \odot se'$ &
			Commands & $c$ & $::=$ & $\sskip ~|~ \shalt ~|~ c; c' ~|~ \rg{r} := se ~|~ \rs{r} := se$ \\
			Environments & $E$ & $::=$ & $i, \vundef ~|~ i, j$ &
      & & & $|~ \ite{se}{c}{c'} ~|~ \login{se_u}{se_{pw}}{se_{id}}$ \\ 
      Request contexts & $R$ & $::=$ & $n, u, \iota, l$ &
			& & & $|~ \start{se} ~|~ \auth{\lst{se}}{\ell}$ \\
			Databases & $D$ & $::=$ & $\mempty ~|~ D \uplus \subst{n}{M}$ &
      		& & & $|~ \tokch{e}{e'}{c}$\\
			Trust mappings & $\phi$ & $::=$ & $\mempty ~|~ \phi \uplus \subst{n}{\iota}$ &
			& & & $|~ \och{L}{c}$ \\
			Servers & $S$ & $::=$ & $(D, \phi, t)$ & 
			& & & $|~ \reply{\page}{s}{ck}{\lst{x} = \lst{se}}$ \\
			Threads & $t$ & $::=$ & $\listen{u}{\lst{r}}{\lst{x}}{c} ~|~ \thread{c}{R}{E} ~|~ \para{t}{t}$ &
			& & & $|~ \redr{u}{\lst{z}}{ck}{\lst{x}=\lst{se}}$ \\
		\end{tabularx}
	\end{subtable}
	~\\[5pt]
	\begin{subtable}[t]{.495\textwidth}
		\begin{tabularx}{\textwidth}{p{2cm}@{\hspace{2pt}}rcX}
      \multicolumn{4}{c}{{\small\textbf{User behavior}}} \\
			\midrule
    		Tab IDs & $\tab$ & $\in$ & $\mathbb{N}$ \\
    		Inputs & $p$ & $::=$ & $\mempty ~|~ p \uplus \subst{k}{v^{\stype}}$ \\
    		Actions & $a$ & $::=$ & $\halt ~|~ \load{\tab}{u}{p} ~|~ \submit{\tab}{u}{v}{p}$
		\end{tabularx}
	\end{subtable}
	~
	\begin{subtable}[b]{.495\textwidth}
		\begin{tabularx}{\textwidth}{p{3.2cm}@{\hspace{2pt}}rcX}
      \multicolumn{4}{c}{{\small\textbf{Web Systems}}} \\
			\midrule
			Attacker's Knowledge & $\atknow$ & $\subseteq$ & $\names$ \\
			Web Systems & $W$ & $::=$ & $B ~|~ S ~|~ \para{W}{W}$ \\
			Attacked Systems & $A$ & $::=$ & $\atkstate{\ell}{\atknow}{W}$
		\end{tabularx}
	\end{subtable}
\end{table*}

We write $\lst{r} = \lstexp{r_1, \ldots, r_m}$ to denote a list of elements of length $m = \len{r}$. We denote with $r_k$ the $k$-th element of $\lst{r}$ 
and we let $\lcons{r'}{\lst{r}}$ be the list obtained by prepending the element $r'$ to the list $\lst{r}$.
A map $M$ is a partial function from keys to values and we write $M(k) = v$ whenever the key $k$ is bound to the value $v$ in $M$. We let $\domain(M)$ be the domain of $M$ and $\mempty$ be the empty map.
Given two maps $M_1$ and $M_2$, we define $\mjoin{M_1}{M_2}$ as the map $M$ such that $M(k) = v$ iff either $M_2(k) = v$ or $k \notin \domain(M_2)$ and $M_1(k) = v$. We write $M_1 \uplus M_2$ to denote $\mjoin{M_1}{M_2}$ if $M_1$ and $M_2$ are disjoint.
We let $M\subst{k}{v}$ be the map obtained from $M$ by substituting the value bound to $k$ with $v$.

\subsubsection{Basics}
we let $\names$ be a set of names modeling secrets (\eg, passwords) and fresh identifiers that cannot be forged by an attacker. Names are annotated with a security label $\ell$, that we omit in the semantics since it has no semantic effect. $\refs$ is the set of references used to model cookies and memory locations, while $\vars$ is the set of variables used for parameters and server commands.
$\ids$ is the set of identities representing users: we distinguish a special identity $\usr$ representing the honest user and we assume that the other identities are under the attacker's control.

A URL $u$ is a triple $(\pi, d, v)$ where $\pi \in \{\mathsf{http}, \mathsf{https}\}$ is the protocol identifier, $d$ is the domain name, and $v$ is a value encoding the path of the accessed resource. 
We ignore the port for the sake of simplicity.
The origin of URL $u$ is the simple label $\pi(d)$.
For origins and URLs, we use $\vundef$ for a blank value.

We let $v$ range over values, \ie, names, primitive values (booleans, integers, etc.), URLs, identities and the blank value $\vundef$.
We use $z$ to range both over values and variables.

A $\page$ is either the constant $\serror$ or a map $f$ representing the DOM of the page. The $\serror$ page denotes that an error has occurred while processing a request at the server-side.
The map $f$ associates tags (\ie, strings) to links and HTML forms contained in the page. We represent them using the notation $\hform{u}{\lst{z}}$, where $u$ is the target URL and $\lst{z}$ is the list of parameters provided via the query string of a link or in the HTTP body of the request for forms.

Memories are maps from references to values. We use them in the server to hold the values of the variables during the execution,
while in the browser they are used to model the cookie jar. We stipulate that $M(r) = \vundef$ if $r \notin \domain(M)$, \ie, the access to a reference not in memory yields a blank value.

\subsubsection{Server Model}
we let $se$ range over expressions including variables, references, values, sampling of a fresh name (with label $\ell$), \eg, to generate fresh cookie values, and binary operations.
Server-side applications are represented as commands featuring standard programming constructs and special instructions for session establishment and management.
Command $\login{se_u}{se_{pw}}{se_{id}}$ models a login operation with username $se_u$ and password $se_{pw}$. The identity of the user is bound to the session identifier obtained by evaluating $se_{id}$.
Command $\start{se}$ starts a new session or restores a previous one identified by the value of the expression $se$. 
Command $\auth{\lst{se}}{\ell}$ produces an authenticated event that includes data identified by the list of expressions $\lst{se}$. The command is annotated with a label $\ell$ denoting the expected security level of the event which has a central role in the security definition presented in \autoref{sec:property}.
Commands $\tokch{x}{r}{c}$ and $\och{L}{c}$ respectively model a token check, comparing the value of a parameter $x$ against the value of the reference $r$, and an origin check, verifying whether the origin of the request occurs in the set $L$.
These checks are used as a protection mechanism against CSRF attacks.
Command $\reply{page}{s}{ck}{\lst{x} = \lst{se}}$ outputs an HTTP response containing a $\page$, a script $s$ and a sequence of \texttt{Set-Cookie} headers represented by the map $ck$. This command is a binder for $\lst{x}$ with scope $page, s, ck$, that is, the occurrences of the variables $\lst{x}$ in $page, s, ck$ are substituted with the values obtained by evaluating the corresponding expressions in $\lst{se}$.
Command $\redr{u}{\lst{z}}{ck}{\lst{x}}$ outputs a message redirect to URL $u$ with parameters $\lst{z}$ that sets the cookies in $ck$. This command is a binder for $\lst{x}$ with scope $\lst{z}, ck$.

Server code is evaluated using two memories: a global memory, freshly allocated when a connection is received, and a session memory, that is preserved across different requests.
We write $\rg{r}$ and $\rs{r}$ to denote the reference $r$ in the global memory and in the session memory respectively. 
To link an executing command to its memories, we use an environment,
which is a pair whose components identify the global memory and the session
memory ($\vundef$ when there is no active session).

The state of a server is modeled as a triple $\server{D}{\phi}{t}$ where the database $D$ is a partial map from names to memories, $\phi$ maps session identifiers (\ie, names) to the corresponding user identities, and $t$ is the parallel composition of multiple threads. Thread $\listen{u}{\lst{r}}{\lst{x}}{c}$ waits for an incoming connection to URL $u$ and runs the command $c$ when it is received. Lists $\lst{r}$ and $\lst{x}$ are respectively the list of cookies and parameters that the server expects to receive from the browser. Thread $\thread{c}{R}{E}$ denotes the execution of the command $c$ in the environment $E$ which identifies the memories of $D$ on which the command operates.
$R$ tracks information about the request that triggered the execution, including the identifier $n$ of the connection where the response by the server must be sent back, the URL of the endpoint $u$, the user $\iota$ who sent the request, and the origin of the request $l$. The user identity has no semantic import, but it is needed to spell out our security property. 

\subsubsection{User Behavior}
action $\halt$ is used when an unexpected error occurs while browsing to prevent the user from performing further actions. Action $\load{\tab}{u}{p}$ models the user entering the URL $u$ in the address bar of her browser in $tab$, where $p$ are the provided query parameters. Action $\submit{\tab}{u}{v}{p}$ models the user submitting a form or clicking on a link (identified by $v$) contained in the page at $u$ rendered in $\tab$; the parameters $p$ are the inputs provided by the user.
We represent user inputs as maps from integers to values $v^{\stype}$ annotated with their security type $\stype$. In other words, we model that the user is aware of the security import of the provided parameters, \eg, whether a certain input is a password that must be kept confidential or a public value. 

\subsubsection{Browser Model}
due to space constraints, we present the browser model in 
\cref{sec:browser-app}.
In the following we write $\sbrowser{M}{P}{\lst{a}}{\iota}$ to represent a browser without any active script or open network connection, with cookie jar $M$ and open pages $P$ which is run by the user $\iota$ performing the list of actions $\lst{a}$.

\subsubsection{Web Systems}
the state of a web system is the parallel composition of the states of browsers and servers in the system. The state of an attacked web system also includes the attacker, modeled as a pair $(\ell, \atknow)$ where the label $\ell$ defines the attacker power and $\atknow$ is her knowledge, \ie, a set of names that the attacker has learned by exploiting her capabilities.

\subsection{Labels and Threat Model}
\label{sec:label}

Let $d \in \domains$ be a domain and $\sim$ be the equivalence relation inducing the partition of $\domains$ in sets of related domains.\hspace{-0.08em}\footnote{~Two domains are related if they share the same base domain, \ie, the first upper-level domain which is not included in the public suffix list~\cite{ZhengJLDCWW15}. For instance, \texttt{www.example.com} and \texttt{atk.example.com} are related domains, while \texttt{example.co.uk} and \texttt{atk.co.uk} are not.} We define the set of simple labels $\labels$, ranged over by $l$, as the smallest set generated by the grammar:
\[
l ::= \http{d} ~|~ \https{d} ~|~ l \ccup l ~|~ l \ccap l
\]
Intuitively, simple labels represent the entities entitled to read or write a certain piece of data, inspect or modify the messages exchanged over a network connection and characterize the capabilities of an attacker.
A label $\ell$ is a pair of simple labels $(l_C, l_I)$, where $l_C$ and $l_I$ are respectively the confidentiality and integrity components of $\ell$. We let $C(\ell) = l_C$ and $I(\ell) = l_I$.
We define the confidentiality pre-order $\cleq$ as the smallest pre-order on $\labels$ closed under the following rules:
\begin{mathpar}
	\inferrule
	{i \in \{1,2\}}
	{l_i \cleq l_1 \ccup l_2}
	
	\inferrule
	{i \in \{1,2\}}
	{l_1 \ccap l_2 \cleq l_i}
	\\
	\inferrule
	{l_1 \cleq l_3 \\
	l_2 \cleq l_3}
	{l_1 \ccup l_2 \cleq l_3}
	
	\inferrule
	{l_1 \cleq l_2 \\
	 l_1 \cleq l_3}
	{l_1 \cleq l_2 \ccap l_3}
\end{mathpar}
We define the integrity pre-order $\ileq$ on simple labels such that $\forall l,l' \in \labels$ we have $l \ileq l'$ iff $l' \cleq l$, \ie, confidentiality and integrity are contra-variant. For $\cleq$ we define the operators $\cjoin$ and $\cmeet$ that respectively take the least upper bound and the greatest lower bound of two simple labels. We define analogous operators $\ijoin$ and $\imeet$ for $\ileq$. We let $\ell \sqsubseteq \ell'$ iff $C(\ell) \cleq C(\ell') \wedge I(\ell) \ileq I(\ell')$.
We also define bottom and top elements of the lattices as follows:
\begin{align*}
	\cbot &= \textstyle\bigccap_{d \in \domains} (\http{d} \wedge \https{d}) &
	\ibot &= \ctop \qquad \\
	\ctop &= \textstyle\bigccup_{d \in \domains} (\http{d} \vee \https{d}) &
	\itop &= \cbot \\
	\tbot &= \pair{\cbot}{\ibot} &
	\ttop &= \pair{\ctop}{\itop}
\end{align*}
We label URLs, user actions and cookies by means of the function $\lambda$. We label URLs with their origin, \ie, given $u = (\pi, d, v)$ we let $\lambda(u) = (\pi(d), \pi(d))$.
The label is used to:
\begin{enumerate*}
	\item characterize the capabilities required by an attacker to read and modify the contents of messages exchanged over network connections towards $u$;
	\item identify which cookies are sent to and can be set by $u$. 
\end{enumerate*}
The label of an action is the one of its URL, \ie, we let $\lambda(a) = \lambda(u)$ for $a = \load{\tab}{u}{p}$ and $a = \submit{\tab}{u}{v}{p}$.

The labelling of cookies depends on several aspects, \eg, the attributes specified by the web developer. For instance, a cookie for the domain $d$ is given the following label:
\[
	(\http{d} \ccap \https{d}, \textstyle\bigicup_{d' \sim d} (\http{d'} \icup \https{d'}))
\]
The confidentiality label models that the cookie can be sent to $d$ both over cleartext and encrypted connections, while the integrity component says that the cookie can be set by any of the related domains of $d$ over any protocol, as dictated by the lax variant of the \emph{Same Origin Policy} applied to cookies.

When the \texttt{Secure} attribute is used, the cookie is attached exclusively to HTTPS requests. However, \texttt{Secure} cookies can be set over HTTP~\cite{Cookies}, hence the integrity is unchanged.\footnote{~Although most modern browsers forbid this dangerous practice, we have decided to represent the behavior dictated by the cookie specification.}
This behavior is represented by the following label:
\[
	(\https{d}, \textstyle\bigicup_{d' \sim d} (\http{d'} \icup \https{d'}))
\]
Cookie prefixes~\cite{CookiePrefixes} are a novel proposal aimed at providing strong integrity guarantees for certain classes of cookies. In particular, compliant browsers ensure that cookies having names starting with the \texttt{\_\_Secure-} prefix are set over HTTPS and the \texttt{Secure} attribute is set.
In our label model they can be represented as follows:
\[
	(\https{d}, \textstyle\bigicup_{d' \sim d} \https{d'})
\]
The \texttt{\_\_Host-} prefix strengthens the policy enforced by \texttt{\_\_Secure-} by additionally requiring that the \texttt{Domain} attribute is not set, thus preventing related domains from setting it. This is modeled by assigning the cookie the following label:
\[
	(\https{d}, \https{d})
\]
We discuss now the impact of HSTS~\cite{HSTS} on cookie labels.
We use a set of domains $\Delta \subseteq \domains$ to represent all the domains where HSTS is enabled, which essentially corresponds to the HSTS preload list\footnote{~\url{https://hstspreload.org}} that is shipped with modern browsers.
Since HSTS prevents browsers from communicating with certain domains over HTTP, in practice it prevents network attackers from setting cookies by modifying HTTP responses coming from these domains. The label of a \texttt{Secure} cookie for domain $d$ becomes the following:
\[\small
	\begin{aligned}
		(\https{d}, \textstyle\bigicup_{\substack{d' \sim d \\ d' \notin \Delta}} \http{d'} \icup \textstyle\bigicup_{d' \sim d} \https{d'}))
	\end{aligned}
\]
The integrity label shows that the cookie can be set over HTTPS by any related domain of $d$ (as for \texttt{Secure} cookies) and over HTTP only by related domains where HSTS is not enabled.
If HSTS is activated for $d$ and all its related domains, the cookie label becomes the same as that of cookies with the \texttt{\_\_Secure-} prefix.

In the model we can also formalize attackers using labels which denote their read and write capabilities. 
Considering an attacker at label $\alabel$ and a name with label $\ell$, the name
may be learned by the attacker if $\isclow{C(\ell)}$ and may be influenced by
the attacker if $\isilow{I(\ell)}$. Here we model the following popular attackers from the web security literature:
\begin{enumerate}
	\item The web attacker hosts a malicious website on domain $d$. We assume that the attacker owns a valid certificate for $d$, thus the website is available both over HTTP and HTTPS:
	\[
		(\http{d} \ccup \https{d}, \http{d} \icap \https{d})
	\]
	\item The active network attacker can read and modify the contents of all HTTP communications:
	\[
		(\textstyle\bigccup_{d \in \domains} \http{d}, \textstyle\bigicap_{d \in \domains} \http{d})
	\]
	\item The related-domain attacker is a web attacker who hosts her website on a \emph{related domain} of a domain $d$, thus she can set (domain) cookies for $d$. Assuming (for simplicity) that the attacker controls all the related domains of $d$, we can represent her capabilities with the following label:
	\[
	\begin{array}{c}
		(\bigccup_{\substack{d' \sim d \\ d' \neq d}} (\http{d'} \ccup \https{d'}), \\
		 \ \bigicap_{\substack{d' \sim d \\ d' \neq d}} (\http{d'} \icap \https{d'}))
	\end{array}
	\]
\end{enumerate}

\subsection{Semantics}
\label{sec:semantics}

\begin{table*}[t]
	\caption{Semantics (excerpt).}
	\label{tab:semantics}
	\vspace{-5pt}
	\begin{center}
    \small\textbf{Servers}
	\end{center}
	\begin{mathpar}
		\inferrule*[lab={\footnotesize(S-Recv)}]
		{\alpha = \srecv{\bid}{n}{u}{p}{ck}{l} \\
		 R = n, u, \bid, l \\
		 \sample{i} \\\\
		 \forall k \in \intval{1}{\len{r}}.\, M(r_k) = (r_k \in \domain(ck)) ~?~ ck(r_k) : \vundef \\
		 m = \len{x} \\\\
		 \forall k \in \intval{1}{m}.\, v_k = (k \in \domain(p)) ~?~ p(k) : \vundef \\
		 \sigma = [x_1 \mapsto v_1,\ldots,x_m \mapsto v_m]}
		{\server{D}{\phi}{\listen{u}{\lst{r}}{\lst{x}}{c}}
		 \xra{\alpha}
		 \server{D \uplus \subst{i}{M}}{\phi}{\para{\thread{c\sigma}{R}{i,\vundef}}{\listen{u}{\lst{r}}{\lst{x}}{c}}}}
		
		\inferrule*[lab={\footnotesize(S-RestoreSession)}]
		{E = i, \_ \\
		 \eval{se}{D}{E} = j \\
		 j \in \domain(D)}
		{\server{D}{\phi}{\thread{\start{se}}{R}{E}}
		 \xra{\blank}
		 \server{D}{\phi}{\thread{\sskip}{R}{i, j}}}
		
		\inferrule*[lab={\footnotesize(S-NewSession)}]
		{E = i, \_ \\
		 \eval{se}{D}{E} = j \\
		 j \notin \domain(D)}
		{\server{D}{\phi}{\thread{\start{se}}{R}{E}}
		 \xra{\blank}
		 \server{D \uplus \subst{j}{\mempty}}{\phi}{\thread{\sskip}{R}{i, j}}}
		
		\inferrule*[lab={\footnotesize(S-Login)}]
		{R = n, u, \bid, l \\
		 \eval{se_{u}}{D}{E} = \sid \\\\
		 \eval{se_{pw}}{D}{E} = \rho(\sid, u) \\
		 \eval{se_{id}}{D}{E} = j}
		{\server{D}{\phi}{\thread{\login{se_{u}}{se_{pw}}{se_{id}}}{R}{E}}
		 \xra{\blank}
    	 \server{D}{\mjoin{\phi}{\subst{j}{\sid}}}{\thread{\sskip}{R}{E}}}

		\inferrule*[lab={\footnotesize(S-OChkSucc)}]
		{R = n, u, \bid, l \\
		 l \in L}
		{\server{D}{\phi}{\thread{\och{L}{c}}{R}{E}}
		 \xra{\blank}
    	 \server{D}{\phi}{\thread{c}{R}{E}}}

		\inferrule*[lab={\footnotesize(S-TChkFail)}, leftskip={8pt}]
    {\eval{e_1}{D}{E} \neq \eval{e_2}{D}{E}}
    {\server{D}{\phi}{\thread{\tokch{e_1}{e_2}{c}}{R}{E}}
		 \xra{\eerror}
    	 \server{D}{\phi}{\thread{\ereply{\serror}{\sskip}{\mempty}}{R}{E}}}
    	
    	\inferrule*[lab={\footnotesize(S-Auth)}, leftskip={8pt}]
		{R = n, u, \bid, l \\
		 j \in \domain(\phi) \\
		 \alpha = \lauth{\lst{v}}{\bid, \phi(j)}{\ell} \\\\
		 \forall k \in \intval{1}{\len{se}}.\, \eval{se_k}{D}{i, j} = v_k}
		{\server{D}{\phi}{\thread{\auth{\lst{se}}{\ell}}{R}{i,j}}
		 \xra{\alpha}
		 \server{D}{\phi}{\thread{\sskip}{R}{i,j}}}
		
		\inferrule*[lab={\footnotesize(S-Reply)}]
		{R = n, u, \bid, l \\
		 m = \len{x} = \len{se} \\
		 \forall k \in [1,m].\, \eval{se_k}{D}{E} = v_k \\\\
		 \sigma = [x_1 \mapsto v_1,\ldots,x_m \mapsto v_m] \\
		 \alpha = \ssend{n}{u}{\bot}{\lempty}{ck\sigma}{\page\sigma}{s\sigma}}
    	{\server{D}{\phi}{\thread{\reply{page}{s}{ck}{\lst{x} = \lst{se}}}{R}{E}}
    	 \xra{\alpha}
    	 \server{D}{\phi}{\thread{\shalt}{R}{E}}}
	\end{mathpar}
	\vspace{\sectskip}
	\begin{center}
    \small\textbf{Web systems}
	\end{center}
	\begin{mathpar}
		\inferrule*[lab={(A-BroSer)}]
    {W \xra{\bsend{\bid}{n}{u}{p}{ck}{l}} W' \\
      W' \xra{\srecv{\bid}{n}{u}{p}{ck}{l}} W'' \\\\
		 \atknow' = (C(\lambda(u)) \cleq C(\ell)) ~?~ (\atknow \cup \fnames{p,ck}) : \atknow}
		{\atkstate{\ell}{\atknow}{W}
		 \xra{\blank}
		 \atkstate{\ell}{\atknow'}{W''}}

		\inferrule*[lab={(A-BroAtk)}, leftskip={19pt}]
		{\alpha = \bsend{\bid}{n}{u}{p}{ck}{l} \\
		 W \xra{\alpha} W' \\\\
		 I(\ell) \ileq I(\lambda(u)) \\\\
		 \atknow' = (C(\lambda(u)) \cleq C(\ell)) ~?~ (\atknow \cup \fnames{p,ck}) : \atknow}
		{\atkstate{\ell}{\atknow}{W}
		 \xra{\alpha}
		 \atkstate{\ell}{\atknow' \cup \{n\}}{W'}}
		
		\inferrule*[lab={(A-AtkSer)}, leftskip={19pt}]
		{\sample{n} \\ \bid \neq \usr \\ \fnames{p,ck} \subseteq \atknow \\\\
		 \alpha = \srecv{\bid}{n}{u}{p}{ck}{l} \\ W \xra{\alpha} W'}
		{\atkstate{\ell}{\atknow}{W}
		 \xra{\alpha}
		 \atkstate{\ell}{\atknow \cup \{n\}}{W'}}
	\end{mathpar}
  \vspace{-20pt}
\end{table*}

We present now the most relevant rules of semantics in \ref{tab:semantics}, deferring to
\cref{sec:semantics-app}
for a complete formalization.
In the rules we use the ternary operator ``?:'' with the usual meaning: $e ~?~ e' : e''$ evaluates to $e'$ if $e$ is true, to $e''$ otherwise.

\subsubsection{Servers}
rules rely on the function $\eval{se}{D}{E}$ that evaluates the expression $se$ in the environment $E$ using the database $D$.
The formal definition is in
\cref{sec:semantics-app},
here we provide an intuitive explanation. The evaluation of $\rg{r}$ and $\rs{r}$ yields the value associated to $r$ in the global and the session memory identified by $E$, respectively.
Expression $\fresh^{\ell}$ evaluates to a fresh name sampled from $\names$ with security label $\ell$. A value evaluates to itself. Evaluation of binary operations is standard.

Rule \irule{S-Recv} models the receiving of a connection $n$ at the endpoint $u$, as indicated by the action $\srecv{\bid}{n}{u}{p}{ck}{l}$.
A new thread is spawned where command $c$ is executed after substituting all the occurrences of variables in $\lst{x}$ with the parameters $p$ received from the network. We use the value $\vundef$ for uninitialized parameters.
The environment is $i, \vundef$ where $i$ identifies a freshly allocated global memory and $\vundef$ that there is no ongoing session. The references of the global memory in $\lst{r}$ are initialized with the values in $ck$ (if provided).
In the request context we include the details about the incoming connection, including the origin $l$ of the page that produced the request (or $\vundef$, \eg, when the user opens the page in a new tab).
The thread keeps listening for other connections on the same endpoint.

The evaluation of command $\start{se}$ is modeled by rules \irule{S-RestoreSession} and \irule{S-NewSession}.
If $se$ evaluates to a name $j \in \domain(D)$, we resume a previously established session, otherwise we create a new one and allocate a new empty memory that is added to the database $D$.
We write $E = i, \_$ to denote that the second component of $E$ is immaterial. In both cases the environment is updated accordingly.

Rule \irule{S-Login} models a successful login attempt. For this purpose, we presuppose the existence of a global partial function $\rho$ mapping the pair $(\sid, u)$ to the correct password where $\sid$ is the identity of the user and $u$ is the login endpoint. 
The rule updates the trust mapping $\phi$ by associating the session identifier specified in the \textbf{login} command with the identity $\sid$.

Rules \irule{S-OChkSucc} and \irule{S-TChkFail} treat a successful origin check and a failed token check, respectively. 
In the origin check we verify that the origin of the request is in a set of whitelisted origins, while in the token check we verify that two tokens match.
In case of success we execute the continuation, otherwise we respond with an error message.
In case of a failure we produce the event $\eerror$.

Rule \irule{S-Auth} produces the authenticated event $\lauth{\lst{v}}{\bid, \sid}{\ell}$ where $\lst{v}$ is data identifying the event, \eg, $paper$ and $action$ in the HotCRP example of \autoref{sec:core-hotcrp}. The event is annotated with the identities $\bid, \sid$, representing the user running the browser and the account where the event occurred, and the label $\ell$ denoting the security level associated to the event.

Rule \irule{S-Reply} models a reply from the server over the open connection $n$ as indicated by the action $\assend$. The response contains a page $\page$, script $s$ and a map of cookies $ck$, where all occurrences of variables in $\lst{x}$ are replaced with the evaluation results of the expressions in $\lst{se}$. The third and the fourth component of $\assend$ are the redirect URL and the corresponding parameters, hence we use  $\vundef$ to denote that no redirect happens. We stipulate that the execution terminates after performing the reply as denoted by the instruction $\shalt$.

\subsubsection{Web Systems}
the semantics of web systems regulates the communications among browsers, servers and the attacker.
Rule \irule{A-BroSer} synchronizes a browser sending a request $\absend$ with the server willing to process it, as denoted by the matching action $\asrecv$.
Here the attacker does not play an active role (as denoted by action $\blank$) but she may update her knowledge with new secrets if she can read the contents of the request, modeled by the condition $C(\lambda(u)) \cleq C(\ell)$.

Rule \irule{A-BroAtk} uniformly models a communication from a browser to a server controlled by the attacker and an attacker that is actively intercepting network traffic sent by the browser. These cases are captured by the integrity check on the origin of the URL $u$. As in the previous rule, the attacker updates her knowledge if she can access the communication's contents. Additionally, she learns the network identifier needed to respond to the browser. In the trace of the system we expose the action intercepted/forged by the attacker.
Rule \irule{A-AtkSer} models an attacker opening a connection to an honest server. We require that the identity denoting the sender of the message belongs to the attacker and that the contents of the request can be produced by the attacker using her knowledge.
Sequential application of the two rules lets us model a network attacker acting as a man-in-the-middle to modify the request sent by a browser to an honest server.

\subsection{Security Definition}
\label{sec:property}

On a high level, our definition of session integrity requires that for each trace produced by the attacked web system, there exists a matching trace produced by the web system without the attacker,
which in particular implies that authenticated actions cannot be modified or forged by the attacker. 
Before formalizing this property, we introduce the notion of trace.

\begin{definition}
	The system $A$ generates the trace $\gamma = \alpha_1 \cdot \ldots \cdot \alpha_k$ iff the system can perform a sequence of steps $A \xra{\alpha_1} \ldots \xra{\alpha_k} A'$ for some $A'$ (also written as $A \astepn{*}{\gamma} A'$).
\end{definition}

Traces include attacker actions, authenticated events $\lauth{\lst{v}}{\bid, \sid}{\ell}$ and $\blank$ denoting actions without visible effects or synchronizations not involving the attacker. Given a trace $\gamma$, we write $\proj{\gamma}{(\iota, \ell)}$ for the projection containing only the authentication events of the type $\lauth{\lst{v}}{\bid, \sid}{\ell}$ with $\iota \in \{\bid, \sid\}$. A trace $\gamma$  is \emph{unattacked} if it contains only $\blank$ actions, $\eerror$ events and authenticated events, otherwise $\gamma$ is an \emph{attacked} trace.

Now we introduce the definition of session integrity.

\begin{definition}
\label{def:integrity}

A web system $W$ preserves \emph{session integrity} against the attacker ($\alabel$, $\atknow$) for the honest user $\usr$ performing the actions $\lst{a}$ if for any attacked trace $\gamma$ generated by the system $\atkstate{\alabel}{\atknow}{\para{\sbrowser{\mempty}{\mempty}{\lst{a}}{\usr}}}{W}$ there exists an unattacked trace $\gamma'$ generated by the same system such that for all labels $\ell$ we have:
\[
	I(\alabel) \not\ileq I(\ell)
    \Rightarrow \proj{\gamma}{(\usr, \ell)} = \proj{\gamma'}{(\usr, \ell)}.
  \]
\end{definition}

Intuitively, this means that the attacker can only produce authenticated events in her account or influence events produced by servers under her control. Apart from this, the attacker can only stop on-going sessions of the user but cannot intrude into them: this is captured by the existential quantification over unattacked traces that also lets us pick a prefix of any trace.

\section{Security Type System}
We now present a security type system designed for the verification of session integrity on web applications.
It consists of several typing judgments covering server programs and browser scripts. Due to space constraints, in this Section we cover only the part related to server-side code and refer to
\cref{sec:type-rules-app}
for the typing rules of browser scripts.

\subsection{Types}
\label{sec:types}

We introduce security types built upon the labels defined in \autoref{sec:label}. We construct the set of security types $\types$, ranged over by $\stype$, according to the following grammar:
\[
	\stype ::= \ell ~|~ \tcre{\ell}
\]
We also introduce the set of reference types $\rtypes = \{\tref{\stype} ~|~ \stype \in \types\}$ used for global and session references
and we define the following projections on security types:
\begin{gather*}
    \typelabel{\ell} = \ell \quad
    \typelabel{\tcre{\ell}} = \ell \quad \\
	\It{\stype} = I(\typelabel{\stype}) \quad
	\Ct{\stype} = C(\typelabel{\stype})
\end{gather*}
Security types extend the standard security lattice with the
type $\tcre{\ell}$ for credentials of label $\ell$. 
We define the pre-order $\tleq$, parametrized by the attacker label $\alabel$, with the following rules:
\begin{mathpar}
  \inferrule
  {\ell \sqsubseteq \ell'}
  {\ell \tleq \ell'}

  \inferrule
  {\isclow{\Ct{\stype} \cjoin \Ct{\stype'}} \\\\
   \isilow{\It{\stype} \imeet \It{\stype'}}}
  {\stype \tleq \stype'}
\end{mathpar}

\noindent Intuitively, security types inherit the subtyping relation for labels but this is not lifted to the credentials, \eg, treating public values as secret credentials is unsound.
However, types of low integrity and confidentiality (compared to the attacker's label) are always subtype of each other: in other words, we collapse all such types into a single one, as the attacker controls these values and is not limited by the restrictions enforced by types.

\subsection{Typing Environment}
Our typing environment $\Gamma = (\envu,\envv,\envrg,\envrs,\envt)$ is a 5-tuple and conveys the following information:
\begin{itemize}
  \item $\envu: \urls \rightarrow (\seclabels \times \lst{\types} \times \labels)$ maps URLs to labels capturing the security of the network connection, the types of the URL parameters and the integrity label of the reply;
  \item $\envv: \vars \rightarrow \types $ maps variables to types;
  \item $\envrg, \envrs: \refs \rightarrow \rtypes$ map global references and session references, respectively, to reference types;
  \item $\envt: \values \rightarrow (\seclabels \times \lst{\types} \times \labels)$ 
    maps values used as tags for forms in the DOM to the corresponding type.
    We typically require the form's type to match the one of the form's
    target URL.
\end{itemize}
Now we introduce the notion of \emph{well-formedness} which rules out inconsistent type assignments.

\begin{definition}
  \label{def:well-formed-env}
  A typing environment $\Gamma$ is \emph{well-formed} for $\lambda$ and $\alabel$
  (written $\envtyping{\Gamma}$) if the following conditions hold:
  \begin{enumerate}
    \item for all URLs $u \in \urls$ with 
      $\envu(u) = \ulabel, \lst{\stype}, l_r$ we have:
      \begin{enumerate}[ref={(\theenumi\alph*)}]
        \item \label{lbl:cons-url-lambda} $C(\ulabel) = C(\lambda(u)) \wedge I(\lambda(u)) \ileq I(\ulabel)$
        \item  for all $k \in \intval{1}{\len{\stype}}$ we have
          \begin{enumerate} [ref={(\arabic{enumi}\alph{enumii} \roman*)}]
            \item \label{lbl:cons-params-types} $\Ct{\stype_k} \cleq \Ct{\ulabel} \wedge  \It{\ulabel} \ileq
              \It{\stype_k}$ 
            \item \label{lbl:cons-params-types-ci}$\istcre{\stype_k} \wedge \isclow{\Ct{\stype_k}}
              \Rightarrow \isilow{\It{\stype_k}}$
          \end{enumerate}
      \end{enumerate}
    \item for all references $r \in \refs$ with $\envrg(r) = \stype$:
      \begin{enumerate}[ref={(\theenumi\alph*)}]
        \item \label{lbl:cons-cookie-lambda} $\Ct{\stype} \cleq C(\lambda(r))  \wedge I(\lambda(r)) \ileq \It{\stype}$ 
        \item \label{lbl:cons-cookie-url} for all $u \in \urls$, if $ C(\lambda(r)) \cleq(\lambda(u)) \wedge \isilow{I(\lambda(u))}$ then $\isclow{\Ct{\stype}}$
        \item \label{lbl:cons-conf-int} if $\isilow{I(\lambda(r))}$ and $\stype = \tcre{\cdot}$ then $\isclow{\Ct{\stype}}$
        \item \label{lbl:cons-cookie-ci} $\istcre{\stype} \wedge \isclow{\Ct{\stype}} \Rightarrow \isilow{\It{\stype}}$
      \end{enumerate}
  \end{enumerate}
\end{definition}

Conditions~\ref{lbl:cons-url-lambda} and~\ref{lbl:cons-cookie-lambda} ensure that the labels of URLs and cookies in the typing environment -- which are used for the security analysis -- are at most as strict as the labels in the function $\lambda$ introduced in \autoref{sec:label} -- which define the semantics.
For instance, a cookie $r$ with confidentiality label $C(\lambda(r)) = \http{d} \ccap \https{d}$ is attached both to HTTP and HTTPS requests to domain $d$.
It would be unsound to use a stronger label for typing, \eg, $\https{d}$, since we would miss attacks due to the cookie leakage over HTTP.
In the same spirit, we check that URLs do not contain parameters requiring stronger type guarantees than those offered by the type assigned to the URL~\ref{lbl:cons-params-types}.

Conditions~\ref{lbl:cons-params-types-ci} and~\ref{lbl:cons-cookie-ci} ensure
that low confidentiality credentials -- that can be learned and used by the
attacker -- cannot have high integrity.

Additionally, well-formedness rules out two inherently insecure type assignments for cookies.
First, if a low integrity URL can read a cookie, then the cookie must have low confidentiality since the attacker can inject a script leaking the cookies, as in a typical XSS~\ref{lbl:cons-cookie-url}.
Second, cookies that can be set over a low integrity network connection cannot be high confidentiality credentials since the attacker can set them to a value she knows~\ref{lbl:cons-conf-int}.
ci
\subsection{Intuition Behind the Typing Rules}
\label{sec:type-intuition}

The type system resembles one for standard information flow control (IFC) where we
consider explicit and implicit flows for integrity, but only explicit flows for
confidentiality: since our property of interest is web session integrity, regarding confidentiality we are only interested in preventing credentials from being leaked (since they are used for access control), while the leakage of other values does not impact our property.
The type system restricts the operations on credentials to be equality checks, hence the leak of information through implicit flows is limited to one bit: this is consistent with the way credentials are handled by real web applications.
A treatment of implicit flows for confidentiality would require a declassification mechanism to handle the bit leaked by credential checks, thus complicating our formalism without adding any tangible security guarantee.

As anticipated in~\autoref{sec:overview}, the code is type-checked twice under different assumptions: first, we consider the case of an honest user visiting the server; second, we consider a CSRF attempt where the attacker forces the user's browser to send a request to the server.
We do not consider the case of the attacker visiting the server from her own browser since we can prove that such a session is always well-typed, which is close in spirit to the opponent typability lemma employed in type systems for cryptographic protocols~\cite{FocardiM11, BackesHM14}.


To enforce our session integrity property, the type system needs to track the identity of the user owning the session and the intention of the user to perform authenticated actions. In typing, this is captured by two dedicated labels.

The \emph{session label} $\seslabel$ records the owner of the active session and is used to label references in the session memory.
The label typically equals the one of the session identifier, thus it changes when we resume or start a new session. Formally, $\seslabel \in \seclabels \cup \{\unauth\}$ where $\unauth$ denotes no active session.

The \emph{program counter label} $\pclabel \in \labels$ tracks the integrity of the control flow.
A high $\pclabel$ implies that the control flow is intended by the user. The $\pclabel$ is lowered in conditionals with a low integrity guard, as is standard in IFC type systems.
In the CSRF typing branch, the $\pclabel$ will be permanently low: we need to prune this typing branch to type-check high integrity actions.
For this purpose, we use token or origin checks: in the former, the user submits a CSRF token that is compared to a (secret) session reference or cookie, while in the latter we check whether the origin of the request is contained in a whitelist.
There are cases in which we statically know that the check will fail, allowing us to prune typing branches.

We also briefly comment on another important attack, namely  cross-site scripting (XSS): we can model XSS vulnerabilities by including a script from an attacker-controlled domain, which causes a failure in typing.
However, XSS prevention is orthogonal to the goal of our work and must be solved with alternative techniques, \eg, proper input filtering or CSP~\cite{CSP}. 

\begin{table*}
	\caption{Type system.}
	\label{tab:typesystem}
	\vspace{-5pt}
	\begin{center}
    \small\textbf{Server expressions}
	\end{center}
	{
	\begin{mathpar}

		\inferrule*[lab={(T-EName)}]{~}{\elabel{n^{\ell}}{\tcre{\ell}}}

		\inferrule*[lab={(T-EFresh)}]{~}{\elabel{\fresh^{\ell}}{\tcre{\ell}}}
    
		\inferrule*[lab={(T-EVal)}]{v \not \in \names}{\elabel{v}{\tbot}}

		\inferrule*[lab={(T-EUndef)}]{~}{\elabel{\vundef}{\stype}}

		\inferrule*[lab={(T-EVar)}]{~}{\elabel{x}{\envv(x)}}

		\inferrule*[lab={(T-EGlobRef)}]{\envrg(r) = \tref{\stype}}{\elabel{\rg{r}}{\stype}}

		\inferrule*[lab={(T-ESesRef)}]
        {\seslabel \neq \unauth \\
         \envrs(r) = \tref{\stype'} \\\\
         \ell = \lowest{\stype'}{\seslabel} \\\\
         \stype = (\stype' \neq \tcre{\cdot}) ~?~ \ell ~:~ \tcre{\ell}}
        {\elabel{\rs{r}}{\stype}} 

		\inferrule*[lab={(T-EBinOp)}]
        {\elabel{se}{\stype} \\ \elabel{se'}{\stype'} \\\\
        (\stype,\stype' \neq \tcre{\cdot} ) \vee \odot \text{ is} =}
        {\elabel{se \odot se'}{\typelabel{\stype} \sqcup \typelabel{\stype'}}} 

		\inferrule*[lab={(T-ESub)}]
		{\elabel{se}{\stype'} \\ 
		 \stype' \tleq \stype}
		{\elabel{se}{\stype}}
	\end{mathpar}
	}
	\vspace{\sectskip}
	\begin{center}
    \small\textbf{Server references}
	\end{center}
	{
	\begin{mathpar}
		\inferrule*[lab={(T-RGlobRef)}]{~}{\srtyping{\rg{r}}{\envrg(r)}}

		\inferrule*[lab={(T-RSesRef)}]
        {\seslabel \neq \unauth \\
		 \envrs(r) = \tref{\stype'} \\
         \ell = \lowest{\stype'}{\seslabel} \\\\
         \stype = (\stype' \neq \tcre{\cdot}) ~?~ \ell ~:~ \tcre{\ell}}
		{\srtyping{\rs{r}}{\tref{\stype}}}

		\inferrule*[lab={(T-RSub)}]
		{\srtyping{r}{\tref{\stype'}} \\ 
		 \stype \tleq \stype'}
		{\srtyping{r}{\tref{\stype}}}
	\end{mathpar}
	}
	\vspace{\sectskip}
	\begin{center}
    \small\textbf{Server-side commands}
	\end{center}
	{
	\begin{mathpar}
		\inferrule*[lab=(T-Skip)]
		{~}
		{\stypingtc{\sskip}}
	
		\inferrule*[lab=(T-Seq)]
    {\stypingparpctc{\Gamma}{\seslabel}{\pclabel}{\typebranch}{c}{\seslabel'}{\pclabel'} \\\\
    \stypingparpctc{\Gamma}{\seslabel'}{\pclabel'}{\typebranch}{c'}{\seslabel''}{\pclabel''}}
    {\stypingparpctc{\Gamma}{\seslabel}{\pclabel}{\typebranch}{c;c'}{\seslabel''}{\pclabel''}}
    
    	\inferrule*[lab=(T-If)]
		{\elabel{se}{\stype} \\
		 \pclabel' = \pclabel \ijoin \It{\stype} \\\\
     \stypingparpctc{\Gamma}{\seslabel}{\pclabel'}{\typebranch}{c}{\seslabel''}{\pclabel_1} \\
     \stypingparpctc{\Gamma}{\seslabel}{\pclabel'}{\typebranch}{c'}{\seslabel'''}{\pclabel_2} \\\\
     \pclabel'' = (\mathbf{reply}, \mathbf{redirect} \in \commands{c} \cup \commands{c'}) ~?~ \pclabel_1 \ijoin \pclabel_2 ~:~ \pclabel\\\\
		 \seslabel' = (\seslabel'' = \seslabel''') ~?~ \seslabel'' : \unauth}
     {\stypingparpctc{\Gamma}{\seslabel}{\pclabel}{\typebranch}{\ite{se}{c}{c'}}{\seslabel'}{\pclabel''}}

		\inferrule*[lab={(T-Login)}]
		{\elabel{se_{u}}{\stype} \\
		 \elabel{se_{pw}}{\tcre{\ell}} \\
		 \elabel{se_{sid}}{\tcre{\ell'}} \\\\
		 \Ct{\tcre{\ell}} \ctleq \Ct{\tcre{\ell'}} \\
     I(\stype) \ijoin \It{\tcre{\ell}} \ijoin \pclabel \itleq \It{\tcre{\ell'}}}
		{\stypingpartc{\Gamma}{\seslabel}{\pclabel}{\typebranch}{\login{se_{u}}{se_{pw}}{se_{sid}}}{\seslabel}}

		\inferrule*[lab={(T-Start)}]
		{\elabel{se}{\tcre{\ell}} \\\\
		 \seslabel' = (\isclow{\Ct{\tcre{\ell}}}) ~?~ \low : \ell \\\\
     \typebranch = \bhon \Rightarrow
    ((\seslabel = \unauth \vee \pclabel \ileq I(\seslabel)) \wedge
     \pclabel \ileq I(\seslabel')
   }
		{\stypingpartc{\Gamma}{\seslabel}{\pclabel}{\typebranch}{\start{se}}{\seslabel'}}

		\inferrule*[lab={(T-SetGlobal)}]
		{\srtyping{\rg{r}}{\tref{\stype}} \\
		 \elabel{se}{\stype} \\
		 \pclabel \ileq \It{\stype}}
		{\stypingtc{\rg{r} := se}}

		\inferrule*[lab={(T-SetSession)}]
		{\srtyping{\rs{r}}{\tref{\stype}} \\
		 \elabel{se}{\stype} \\
		 \pclabel \ileq \It{\stype}}
		{\stypingtc{\rs{r} := se}}

		\inferrule*[lab=(T-PruneTChk)]
		{\srtyping{r}{\tref{\tcre{\ell}}} \\
		 \elabel{x}{\stype} \\\\
     \Ct{\stype} \neq \Ct{\tcre{\ell}} \\
		 \ischigh{\Ct{\tcre{\ell}}}\\
		 \typebranch=\bcsrf}
		{\styping{\tokch{x}{r}{c}}}

		\inferrule*[lab=(T-TChk)]
		{\srtyping{r}{\tref{\tcre{\ell}}} \\
		 \elabel{x}{\tcre{\ell}} \\\\
		 \stypingpartc{\Gamma}{\seslabel}{\pclabel}{\typebranch}{c}{\seslabel'}} 
		{\stypingpartc{\Gamma}{\seslabel}{\pclabel}{\typebranch}{\tokch{x}{r}{c}}{\seslabel'}} 

		\inferrule*[lab=(T-PruneOChk)]
		{\forall l \in L. \isihigh{l} \\
		 u \in \protUrls \\ 
		 \typebranch=\bcsrf}
		{\styping{\och{L}{c}}}

		\inferrule*[lab=(T-OChk)]
		{\stypingpartc{\Gamma}{\seslabel}{\pclabel}{\typebranch}{c}{\seslabel'}} 
		{\stypingpartc{\Gamma}{\seslabel}{\pclabel}{\typebranch}{\och{L}{c}}{\seslabel'}}
		
		\inferrule*[lab=(T-Auth)]
		{\seslabel \neq \unauth \\
         \forall k \in \intval{1}{\len{se}}. \, \elabel{se_k}{\stype_k} \\\\
		 \Big(\isilow{\bigijoin{1 \leq k \leq \len{se}} \It{\stype_k} \ijoin \pclabel \ijoin I(\seslabel)}\Big) 
         \Rightarrow \isilow{I(\ell)}}
		{\stypingtc{\auth{\lst{se}}{\ell}}}

		\inferrule*[lab={(T-Reply)}]
		{\envu(u) = \ulabel, \vec{\stype}, l_r \\ 
		 \pclabel' = \pclabel \ijoin l_r \\
		 \envv' = x_1 \colon \stype_1, \ldots, x_{\len{se}} \colon \stype_{\len{se}} \\
		 \Gamma' = (\envu,\envv',\envrg,\envrs,\envt) \\
		 \forall k \in \intval{1}{\len{se}}. \, \elabel{se_k}{\stype_k} \wedge \Ct{\stype_k} \ctleq C(\ulabel) \\
		 \forall r \in \domain(ck).\, \srtyping{r}{\tref{\stype_r}} \wedge \elabelpar{\Gamma'}{\seslabel}{ck(r)}{\stype_r}  \wedge \pclabel' \ileq \It{\stype_r} \\
		 \bstypingpar{\Gamma'}{b}{\pclabel'}{s} \\
		 \isbcsrf \Rightarrow \forall x \in vars(s). \, \isclow{\Ct{\envv'(x)}} \\
		 \isbhon \Rightarrow \pclabel \ileq l_r \wedge 
      \left(\page = \serror \vee \forall v \in \domain(\page).\, \ptyping{\Gamma'}{v}{\pclabel'}{\page(v)}\right) \\ 
		 \isilow{I(\ulabel)} \Rightarrow  \forall k \in \intval{1}{\len{se}}. \, \isclow{\Ct{\stype_k}}}
		{\styping{\reply{\page}{s}{ck}{\lst{x} = \lst{se}}}}

    \inferrule*[lab={(T-Redir)}]
		{\envu(u) = \ulabel, \vec{\stype}, l_r \\ 
		 \envv' = x_1 \colon \stype_1, \ldots, x_{\len{se}} \colon \stype_{\len{se}} \\
		 \Gamma' = (\envu,\envv',\envrg,\envrs,\envt) \\
		 \forall k \in \intval{1}{\len{se}}. \, \elabel{se_k}{\stype_k} \wedge \Ct{\stype_k} \ctleq C(\ulabel) \\
		 \forall r \in \domain(ck).\, \srtyping{r}{\tref{\stype_r}} \wedge \elabelpar{\Gamma'}{\seslabel}{ck(r)}{\stype_r}  \wedge \pclabel \ileq \It{\stype_r} \\
		 \isilow{I(\ulabel)} \Rightarrow  \forall k \in \intval{1}{\len{se}}. \, \isclow{\Ct{\stype_k}} \\
     \isbcsrf \Rightarrow \forall x \in vars(\vec{z}). \, \isclow{\Ct{\envv'(x)}} \\
     u' \not \in \protUrls \\
		 \envu(u') = \ulabel', \vec{\stype'}, l_r' \\
     l_r = l_r' \\
     \isihigh{I(\ulabel)} \\
		 \isbhon \Rightarrow \left( 
		   \pclabel \itleq I(\ulabel) \wedge
		   m = \len{z} = \len{\stype'} \wedge
       \forall k \in \intval{1}{m}.\, \elabelpar{\Gamma'}{\seslabel}{z_k}{\stype_k'} \wedge \stype_k' \tleq \stype_k
   \right)}
    {\styping{\redr{u'}{\lst{z}}{ck}{\lst{x} = \lst{se}}}}
	\end{mathpar}
	}
\end{table*}

\begin{table*}
	\ContinuedFloat
	\caption{Type system (continued).}
	\begin{center}
    \small\textbf{Forms}
	\end{center}
	{
	\begin{mathpar}
    	\inferrule*[lab=(T-Form)]
		{\envt(v) = \envu(u) = \ulabel, \vec{\stype}, l_r \\
      \isihigh{I(\ulabel)} \\
		 \pclabel \itleq I(\ulabel)\\
		 m = \len{z} = \len{\stype} \\
		 \forall k \in \intval{1}{m}.\, \elabel{z_k}{\stype_k'} \wedge
		  \stype_k' \tleq \stype_k}
		{\ptyping{\Gamma}{v}{\pclabel}{\hform{u}{\lst{z}}}}
	\end{mathpar}
	}
	\vspace{\sectskip}
	\begin{center}
    \small\textbf{Server threads}
	\end{center}
	{
    \begin{mathpar}
		\inferrule*[lab={(T-Parallel)}]
		{\sttyping{t} \\ \sttyping{t'}}
		{\sttyping{\para{t}{t'}}}
	
		\inferrule*[lab={(T-Recv)}]
		{\envtyping{\Gamma^0} \\
		 \envu^0(u) = \ulabel, \lst{\stype}, l_r \\
		 m = \len{\stype} = \len{x} \\\\
		 \forall k \in \intval{1}{\len{r}}.\, \Ct{\envrg^0(r_k)} \cleq \Ct{\ulabel} 
		    \wedge \It{\ulabel} \ileq \It{\envrg^0(r_k)} \\
		 \envv = x_1 \colon \stype_1, \ldots, x_m \colon \stype_m \\
		 \stypingpar{(\envu^0,\envv,\envrg^0,\envrs^0,\envt^0)}{\unauth}{I(\ulabel) }{\bhon}{c}{\_} \\\\
		 \envv' = x_1 \colon \low, \ldots, x_m \colon \low \\
		 \stypingpar{(\envu^0,\envv',\envrg^0,\envrs^0,\envt^0)}{\unauth}{\itop}{\bcsrf}{c}{\_}}
		{\sttyping{\listen{u}{\lst{r}}{\lst{x}}{c}}}
	\end{mathpar}
	}	
\end{table*}

\subsection{Explanation of the Typing Rules}
\label{sec:type-explanation}

%


\subsubsection{Server Expressions}
typing of server expressions is ruled by the judgement $\elabel{se}{\stype}$, meaning that the expression $se$ has type $\stype$ in the typing environment $\Gamma$ within the session $\seslabel$. 
Names have type $\tcre{\ell}$ where $\ell$ is the label provided as an annotation \irule{T-EName, T-EFresh}. Values different from names are constants of type $\tbot$, \ie, they have low confidentiality and high integrity \irule{T-EVal}. 
Rule \irule{T-EUndef} gives any type to the undefined value $\vundef$. This is needed since the initial memory and empty parameters contain this value and have to be well-typed.
Types for variables and references in the global memory are read from the corresponding environments \irule{T-EVar,T-EGlobRef}.
For session references we combine the information stored in the environment with the session label $\seslabel$, which essentially acts as an upper bound on the types of references \irule{T-ESesRef}.
In a honest session, $\seslabel$ can have high confidentiality, thus the session memory can be used to store secrets.
In the attacker session, instead, the types of all session references are lowered and can never store secrets.
Typing fails if no session is active, \ie, $\seslabel = \unauth$. 
The computed type for a reference is a credential type if and only if it is so in the environment.
Binary operations are given the join of the labels of the two operands \irule{T-EBinOp}. However, on credentials we allow only equality checks to limit leaks through implicit flows. 
Note that by projecting the types to their labels we perform a declassification and hence the result of a binary operation can never be a high confidentiality credential.
Finally, \irule{T-ESub} lets us use subtyping on expressions.

\subsubsection{Server References}
typing of server references is ruled by the judgment
$\srtyping{r}{\tref{\stype}}$ meaning that the reference $r$ has
type $\tref{\stype}$ in the typing environment $\Gamma$ within the session
$\seslabel$. 
This judgement is used to derive the type of a reference we write into, in contrast to the typing of expressions which covers the typing of references from which we read. 
While \irule{T-RGlobRef} just looks up the type
of the global reference in the typing environment, in \irule{T-RSesRef} we have 
analogous conditions to \irule{T-ESesRef} for session references.
Subtyping for reference types is contra-variant to subtyping for security types
\irule{T-RSub}.

\subsubsection{Server Commands}
the judgement $\stypingparpc{\Gamma}{\seslabel}{\pclabel}{\typebranch}{c}{\seslabel'}{\pclabel'}$ states that the command $c$ (bound to the endpoint at URL $u$) can be typed against the attacker $\alabel$ in the typing branch $\typebranch \in \{\bhon,\bcsrf\}$ using typing environment $\Gamma$, session label $\seslabel$ and program counter label $\pclabel$. $\protUrls$ contains all URLs that rely on an origin check to prevent CSRF attacks.
After the execution of $c$, the session label and the PC label are respectively updated to $\seslabel'$ and $\pclabel'$.
We let $\typingcontext = (u, \typebranch, \protUrls)$ if the individual components of the tuple are not used in a rule.
The branch $\typebranch$ tracks whether we are typing the scenario of an honest request ($b = \bhon$) or the CSRF case ($b = \bcsrf$).

Rule \irule{T-Skip} does nothing, while \irule{T-Seq} types the second command with the session label and the PC label obtained by typing the first command.

Rule \irule{T-Login} verifies that the password and the session identifier are both credentials and that the latter is at least as confidential as the former, since the identifier can be used for authentication in place of the password.
Finally, we check that the integrity of username, password and $\pclabel$ are at least as high as the integrity of the session identifier to prevent an unauthorized party from influencing the identity associated to the session. 

Rule \irule{T-Start} updates the session label used for typing the following commands.
First we check that the session identifier $se$ has a credential type: if it has low confidentiality, we update the session label to $\low$ (since the attacker can access the session), otherwise we use the label $\ell$ in the type of $se$.
Furthermore, we ensure that in the honest typing branch high integrity sessions can not be started or ended (by starting a new session) in a low integrity context (\ie, in a conditional with low integrity guard), since this can potentially influence the value of high integrity references of the session memory in the continuation. For the CSRF typing branch this is not required, since due to its low PC label it can never write to high integrity references.

Rules \irule{T-SetGlobal} and \irule{T-SetSession} ensure that no explicit flow
violates the confidentiality or integrity policies, where for integrity we
also consider the PC label. 

Rule \irule{T-If} lowers the PC based on the integrity label of the guard expression of the conditional and uses it to type-check the two branches.
If one of the branches contains a $\mathbf{reply}$ or a $\mathbf{redirect}$ command, then reaching
the continuation depends on the taken branch, thus we use the join of
the PC labels returned in the two branches to type-check the continuation; 
otherwise, we use the original PC label.
If typing the two branches yields two different session labels, we use the session label $\unauth$ in the continuation to signal that the session state cannot be statically predicted and thus no session operation should be allowed. 

Rule \irule{T-Auth} ensures that the attacker cannot affect any component leading to an authenticated event (PC label, session label or any expression in $\lst{se}$) unless the event is annotated with a low integrity label.
Since authenticated events are bound to sessions, we require $\seslabel \neq \unauth$. 

Rules \irule{T-PruneTChk} and \irule{T-TChk} handle CSRF token checks.
In \irule{T-PruneTChk} we statically know that the check fails since the reference where the token is stored has a high confidentiality credential type and the parameter providing the token is a low confidentiality value, hence we do not type-check the continuation $c$.
This reasoning is sound since credentials are unguessable fresh names and we disallow subtyping for high confidentiality credentials, \ie, public values cannot be treated as secret credentials.
This rule is used only in the CSRF typing branch.
Rule \irule{T-TChk} covers the case where the check may succeed and we simply type-check the continuation $c$. We do not change the PC label since a failure in the check produces an $\serror$ page which causes the user to stop browsing.

Similarly, rules \irule{T-PruneOChk} and \irule{T-OChk} cover origin checks.
We can prune the CSRF typing branch if the URL we are typing is protected ($u \in \protUrls$) and all whitelisted origins have high integrity, since the origin of a CSRF attack to a protected URL has always low integrity.

Rule \irule{T-Reply} combines the PC label with the expected integrity label of the response $l_r$ for the current URL to compute $\pclabel'$ which is used to type the response.
In the honest typing branch, we require $\pclabel' = l_r$, which establishes an invariant used when typing an $\mathbf{include}$ command in a browser script, where we require that the running script and the included script can be typed with the same $\pclabel$ (\cf\ rule \irule{T-BInclude} in 
\cref{sec:type-rules-app}).
Using the typing environment $\Gamma'$ which contains types for the variables embedded in the response, we check the following properties:

\begin{itemize}
	\item secrets are not disclosed over a network connection which cannot guarantee their confidentiality;
	\item the types of the values assigned to cookies are consistent with those in the typing environment (where the PC label is taken into account for the integrity component);
	\item the script in the response is well-typed 
    (rules in \cref{sec:type-rules-app});
	\item secrets are not disclosed to a script in the CSRF typing branch since it might be included by an attacker's script;
	\item in the honest typing branch, we check that the returned page is
    either the $\serror$ page or all its forms are well-typed according to rule
    \irule{T-Form}. We do not perform this check in the CSRF branch since
    a CSRF attack is either triggered by a script inclusion or through a
    redirect. 
    In the first case the attacker cannot access the DOM, which in a real
    browser is enforced by the Same Origin Policy.
    In the second case, well-formed user behavior (\cf\ \autoref{def:user-actions})
    ensures that the user will not interact with the DOM in this scenario;

	\item no high confidentiality data is included in replies over a low integrity network connection, since the attacker could inject scripts to leak secrets embedded in the response.
\end{itemize}
Rule \irule{T-Redir} performs mostly the same checks as \irule{T-Reply}. Instead of typing script and DOM, we perform checks on the URL similar to the typing of forms, as discussed below. 
Additionally, we require that the target URL is not relying on an origin check
for CSRF protection ($u' \not \in \protUrls$), as the redirect would allow for a circumvention of that protection.
Finally, we also require that the expected integrity label for the response for
the current URL and the target URL are the same.

\subsubsection{Forms}
the judgement $\ptyping{\Gamma}{v}{\pclabel}{f}$ says that a form $f$ identified by the name $v$ is well-typed in the environment $\Gamma$ under the label $\pclabel$.
Our rule for typing forms \irule{T-Form} first checks that the type of the form
name matches the type of the target URL.
This is needed since for well-formed user behavior (\cf\ \autoref{def:user-actions})
we assume that the user relies on the name
of a form to ensure that her inputs are compliant with the expected types.
We require that only links to high integrity URLs are included and 
with $\pclabel \itleq I(\ulabel)$ we check that the
thread running with program counter label $\pclabel$ is allowed to trigger
requests to $u$. In this way we can carry
over the $\pclabel$ from one thread where the form has been created to 
the one receiving the request since we type-check the honest branch with 
$\pclabel = I(\ulabel)$.
Finally, we check that the types of form values comply with the expected type for the corresponding URL parameters, taking the PC into account for implicit integrity flows.

\subsubsection{Server Threads}
the judgement $\sttyping{t}$ says that the thread $t$ is well-typed in the environment $\Gamma^0$ against the attacker $\alabel$ and $\protUrls$ is the set of URLs protected against CSRF attacks via origin checking. 

Rule \irule{T-Parallel} states that the parallel composition of two threads is well-typed if both are well-typed. 
Rules for typing running threads (\ie, $t=\thread{c}{R}{E}$) are in 
\cref{sec:type-rules-app}, 
since they are needed only for proofs. 

Rule \irule{T-Recv} checks that the environment is well-formed and that the network connection type $\ulabel$ is strong enough to guarantee the types of the cookies, akin to what is done for parameters in \autoref{def:well-formed-env}.
Then we type-check the command twice with $\seslabel = \unauth$, since no session is initially active.
In the first branch we let $\typebranch = \bhon$: parameters are typed according
    to the type of $u$ in $\envu^0$ which is reflected in the environment $\envv$. 
As the honest user initiated the request, we let $\pclabel = I(\ulabel)$, \ie, we use the integrity label of the network connection as \pclabel. This allows us to 
    import information about the program counter from another (well-typed)
    server thread or browser script that injected the form into the DOM or 
    directly triggered the request.
In the second branch we let $\typebranch = \bcsrf$: parameters are chosen by the attacker, hence they have type $\low$ in $\envv'$. As the attacker initiated the request, we let $\pclabel = \itop$.

\subsection{Formal Results}
\label{sec:formal-res}

We introduce the notion of \emph{navigation flow}, which identifies a sequence of navigations among different pages occurring in a certain tab and triggered by the user's interaction with the elements of the DOM of rendered pages.
Essentially, a navigation flow is a list of user actions consisting of a $\mathsf{load}$ on a certain tab followed by all actions of type $\mathsf{submit}$ in that tab (modeling clicks on links and submissions of forms) up to the next $\mathsf{load}$ (if any).
A formal definition is presented in
\cref{sec:formal-app}.

Next we introduce the notion of \emph{well-formedness} to constrain the
interactions of an honest user with a web system.


\begin{definition}
  \label{def:user-actions}
  The list of user actions $\lst{a}$ is \emph{well-formed} for the honest user $\usr$
  in a  web system $W$ with respect to a typing environment $\Gamma^0$ and an
  attacker $\alabel$ iff
	\begin{enumerate}
		\item \label{lab:wf1} for all actions $a'$ in $\lst{a}$ we have:
		\begin{itemize}
			\item if $a' = \load{tab}{u}{p}$, $\envu(u) = \ulabel, \lst{\stype}, l_r$ then for all $k \in \domain(p)$ we have $p(k) = v^{\stype'} \Rightarrow \stype' \tleq \stype_k$;
			\item if $a' = \submit{tab}{u}{v'}{p}$, $\envt(v') = \ulabel, \lst{\stype}, l_r$ then for all $k \in \domain(p)$ we have if $p(k) = v^{\stype'}$ then $\stype' \tleq \stype_k$.
        If $\isilow{\lambda(u)}$ we additionally have $\stype' \tleq \ell_a$.
		\end{itemize}
  \item \label{lab:wf2} $\atkstate{\alabel}{\atknow_0}{\para{\sbrowser{\mempty}{\mempty}{\lst{a}}{\usr}}{W}} \astepn{*}{\gamma} \atkstate{\alabel}{\atknow'}{\para{\sbrowser{M}{P}{\lempty}{\usr}}{W'}}$ for some $\atknow', W', M, P$ where $\gamma$ is an unattacked trace, not containing the event $\serror$;
    \item \label{lab:wf3} for every navigation flow $\lst{a}\,'$ in $\lst{a}$,
      we have that $\isilow{I(\lambda(a_j'))}$ implies $
      \isilow{I(\lambda(a_k')})$ for all $j < k \leq \len{a'}$.
	\end{enumerate}
\end{definition}

Condition~\ref{lab:wf1} prevents the user from deliberately leaking secrets by enforcing that the expected parameter types are respected.
While the URL in a $\mathsf{load}$ event is the target URL and we can directly check its type, in a $\mathsf{submit}$ action it refers to the page containing the form: intuitively, this models a user who knows which page she is actively visiting with a $\mathsf{load}$ and which page she is currently on when performing a $\mathsf{submit}$.
However, we do not expect the user to inspect the target URL of a form.
Instead, we expect the user to identify a form by its displayed name (the parameter $v'$ in $\mathsf{submit}$) and input only data matching the type associated to that form name. For instance, in a form named ``public comment'', we require that the user enters only public data.
Typing hence has to enforces that all forms the user interacts with are named correctly. Otherwise, an attacker could abuse a mismatch of form name and target URL in order to steal confidential data.
For this reason we also require that the user never provides secrets to a form embedded in a page of low integrity.

Condition~\ref{lab:wf2} lets us consider only honest runs in which the 
browser terminates regularly without producing errors. Concretely, this rules out interactions 
that deliberately trigger an error at the server-side, \eg, the user loads
a page expecting a CSRF token without providing this token, or 
executions that do not terminate due to infinite loops, \eg, where a script recursively includes itself.

Condition~\ref{lab:wf3} requires that the user does not navigate a trusted
website reached by interacting with an untrusted page.
Essentially, this rules out phishing attempts where the attacker
influences the content shown to the user in the trusted website. 

Our security theorem predicates over \emph{fresh clusters}, \ie, systems composed of multiple servers where no command is running or has been run in the past.

\begin{definition}
  \label{def:cluster}
	A server $S$ is \emph{fresh} if $S = \server{\mempty}{\mempty}{t}$ where $t$ is the parallel composition of threads of the type $\listen{u}{\lst{r}}{\lst{x}}{c}$. A system $W$ is a \emph{fresh cluster} if it is the parallel composition of fresh servers.
\end{definition}


We now present the main technical result, namely that well-typed clusters preserve the session integrity property from \autoref{def:integrity} for all well-formed interactions of the honest user with the system, provided that her passwords are confidential.

\begin{theorem}
  Let $W$ be a fresh cluster, ($\alabel$, $\atknow$) an attacker, $\Gamma^0$ a
  typing environment, $\protUrls$ a set of protected URLs against CSRF via origin checking
  and let $\lst{a}$ be a list of
  well-formed user actions for $\usr$ in $W$ with respect to $\Gamma^0$ and
  $\alabel$. 
  Assume that for all $u$ with $\rho(\usr,u) = n^\ell$ we have
  $\ischigh{C(\ell)}$ and for all $n^\ell \in \atknow$ we have  
  $\isclow{C(\ell)}$.
  Then $W$ preserves \emph{session integrity} against $\alabel$ with knowledge $\atknow$
  for the honest user $\usr$ performing the list of actions $\lst{a}$
  if $\sttyping{t}$ for all servers $S=\server{\mempty}{\mempty}{t}$ in
  $W$.
\end{theorem}

The proof builds upon a simulation relation connecting a run of the system with the attacker with a corresponding run of the system without the attacker in which the honest user behaves in the same way and high integrity authenticated events are equal in the two runs.
The full security proof can be found in 
\cref{sec:proof}.

\section{Case Study}
\label{sec:case-study}

Now we resume the analysis of HotCRP, started in \autoref{sec:overview} where we described the login CSRF and proposed a fix, and describe the remaining session integrity problems we discovered by typing its model in our core calculus.
The encodings of Moodle and phpMyAdmin, including the description of the new vulnerability, are provided in 
\cref{sec:examples-app}.

\subsection{Methodology}
We type-check the HotCRP model of \autoref{sec:overview} against different attackers, including the web-, related-domain-, and network attacker.
Two scenarios motivate the importance of the related-domain attacker in our case study. 
First, many conferences using HotCRP deploy the system on a subdomain of the university organizing the event, \eg, CSF 2020: any user who can host contents on a subdomain of the university can act as the attacker.
Second, anybody can host a conference on a subdomain of \url{hotcrp.com} or access the administrative panel of \url{test.hotcrp.com}: by exploiting a stored XSS vulnerability (now fixed) in the admin panel, it was possible to show on the homepage of the conference a message containing JavaScript code that tampers with cookies to implement the attacks below.


Failures in type-checking highlight code portions that we analyze manually, as they likely suffer from session integrity flaws.
Once a problem is identified, we implement a patch in our HotCRP model and try to type-check it again; this iterative process stops when we manage to establish a security proof by typing, as shown in~\autoref{sec:typing-example}.

%
\subsection{Cookie Integrity Attacks}

Our fix against login CSRF does not ensure the integrity of session cookies against network and related-domain attackers: the former can compromise cookie integrity by forging HTTP traffic, while the latter can set cookies for the target website by using the \texttt{Domain} attribute.
Attackers can thus perform \emph{cookie forcing} to set the their session cookies  in the victim's browser, achieving the same outcome of a login CSRF.

Even worse, the lack of cookie integrity combined with a logical vulnerability on HotCRP code enables a \emph{session fixation} attack, where the attacker manages to force a known cookie into the browser of the victim before she authenticates which is used by HotCRP to identify the victim's session after login.
With the known cookie, the attacker can then access the victim's session to steal submitted papers, send fake reviews, or deanonymize reviewers.
HotCRP tries to prevent session fixation by checking during login whether the provided session cookie (if any) identifies a session where no variable is set: in such a case, the value of the cookie is changed to an unpredictable random string.
However, some session variables are not properly unset during logout, thus the above check can be voided by an attacker with an account on the target website that obtains a valid cookie by authenticating and logging out.\footnote{~To simplify the presentation, this complex behavior is not encoded in the example in~\autoref{sec:overview}. However, the possibility to perform cookie forcing, which is modeled in our example, is a prerequisite for session fixation and is detected by the type system.}
At this point, the attacker can inject this cookie into the victim's browser to perform the attack.

Both attacks are captured in typing as follows: although we have a certain liberty in the choice of our initial environment, no possible type for $\param{sid}$ leads to a successful type derivation since $\param{sid}$ must have a credential type.
As the attacker can set the cookie, it must have low integrity by well-formedness of the typing environment (\autoref{def:well-formed-env}). Since the attacker can write (low confidentiality) values of her knowledge into $\param{sid}$, it may not be a credential of high confidentiality, again by \autoref{def:well-formed-env}.
Hence we must assume that $\param{sid}$ is a credential of low confidentiality and integrity. However, since the user's password has high confidentiality, typing fails in the \emph{login} endpoint (on line 9) when applying rule \irule{T-Login}.


A possible solution against these threats relies on the adoption of \emph{cookie prefixes} (\cf\ \autoref{sec:label}) which provide high integrity guarantees against network and related-domain attackers.
This protection cannot be applied by default in HotCRP due to backward compatibility reasons, \ie, \url{hotcrp.com} relies on cookies shared across multiple domains to link different conferences under the same account.
However, the developer has fixed the bug causing the session fixation vulnerability and we have discussed with him the option to offer cookie prefixes as an opt-in security mechanism during the setup of HotCRP.

\subsection{Typing Example}
\label{sec:typing-example}

Now we show how to type-check the fixed \emph{login} endpoint (from \autoref{sec:overview-typing}) on domain $d_C$ against an attacker controlling a related-domain $d_E \sim d_C$, assuming that the session cookie is secured with the \texttt{\_\_Host-} prefix. 
We let the attacker label $\alabel = 
(\http{d_E} \ccup \https{d_E}, \http{d_E} \icap \https{d_E})$, and let 
$\ell_C = \pair{\https{d_C}}{\https{d_C}}$, $\ell_{LH} =
\pair{\cbot}{\https{d_C}}$, $\ell_{HL} = \pair{\https{d_C}}{\itop}$.
We then consider a minimal environment $\Gamma$ sufficient to type the \emph{login} endpoint, where:
\[\small
\begin{aligned}
  \envu &= \{\mathit{login} \mapsto (\ell_C, (\ell_{LH},\tcre{\ell_C},\tcre{\ell_{HL}}), \https{d_C}), \\
  &\quad\ \ \mathit{manage} \mapsto (\ell_C, (\ell_C,\ell_{LH},\tcre{\ell_{HL}}), \https{d_C})\} \\
  \envrg &= \{r \mapsto \tcre{\ell_C}, r' \mapsto \tcre{\ell_{HL}}, \\
  &\quad \ \ \rf{sid} \mapsto \tcre{\ell_C}, \rf{pre} \mapsto \tcre{\ell_{HL}} \} \\
  \envrs &= \{ \rf{user} \mapsto \ell_{LH}, \rf{ltoken} \mapsto \tcre{\ell_{HL}} \} \\
	\envt &= \{\uinput{auth} \mapsto \envu(\mathit{login}),
		\uinput{link} \mapsto \envu(\mathit{manage})\}
\end{aligned}
\]
We type-check the code under two different assumptions in \irule{T-Recv}.
Our goal is to prune the CSRF typing branch before the security critical part and type it only in the honest setting.

We start with the honest typing branch.
When typing the conditional (line 2) in rule \irule{T-If}, we do not lower $\pclabel$ since the integrity label of the guard and $\pclabel$ is $\https{d_C}$. 
In the $\mathbf{then}$ branch (line 3), we have the assignment $\rg{r'} := \fresh^{\ell_{HL}}$, which types successfully according to \irule{T-SetGlobal}.\footnote{~Here we expose the annotations of \emph{fresh()} expressions (needed for typing) that we omitted from \autoref{sec:overview} for readability purposes.}
The $\mathbf{start}$ statement with the freshly sampled value yields a session label $\seslabel = \pair{\https{d_C}}{\itop}$.
The assignment $\rs{ltoken} := \fresh^{\ell_{HL}}$ also succeeds according to \irule{T-SetSession}.
The session label does not affect the type of the reference $\rs{ltoken}$ in this case.
For the $\mathbf{reply}$ (lines 4--6) we successfully check that the URL is well-formed and may be produced with the current $\pclabel$ \irule{T-Form}, that the empty script is well-typed, and that $y = \rg{r'}$ may be assigned to the cookie $pre$ \irule{T-Reply}.
In the $\mathbf{else}$ branch of the conditional, we start a session over the cookie $\rg{pre}$ (line 8), leading to a session label $\seslabel = \pair{\https{d_C}}{\itop}$ \irule{T-Start}.
The conditions in \irule{T-TChk} are fulfilled for the $\mathbf{tokenchk}$ command (line 9) and we continue typing without any additional effect.
Since we still have $\pclabel = \https{d_C}$, the assignment $\rg{r} := \fresh^{\ell_C}$ type-checks (line 10).
As the password is of the same type as the reference $\rg{r}$ containing the session secret, the $\mathbf{login}$ also type-checks successfully \irule{T-Login}. 
The $\mathbf{start}$ statement over a credential of type $\tcre{\ell_C}$ gives us the session label $\seslabel = \ell_C$ (line 11).
For the $\mathbf{reply}$ (lines 12--14), we check that we may include the form with the current $\pclabel$ and that it is well formed (trivial since it contains only $\vundef$), that the empty script is well-typed and that we may assign the value of $\rg{r}$ to the cookie $\param{sid}$ \irule{T-Reply}.

The $\mathbf{then}$ branch of the CSRF case types similarly to the honest case, since all references used in it and the cookie $\param{pre}$ have integrity label $\itop$.
Additionally, in the CSRF branch, we do not type the DOM \irule{T-Reply}.
In the $\mathbf{else}$ branch we start a session (line 8) with label $\seslabel= \pair{\https{d_C}}{\itop}$ \irule{T-Start}.
When performing the $\mathbf{tokenchk}$ (line 9), we can apply rule \irule{T-PruneTChk}, since $\elabel{\rs{ltoken}}{\tcre{\ell_{HL}}}$ and $\elabel{token}{\alabel}$ cannot be given the same confidentiality label.
Hence, we do not have to type-check the continuation.

\section{Related Work}
Formal foundations for web security have been proposed in a seminal paper~\cite{AkhaweBLMS10}, using a model of the web infrastructure expressed in the Alloy model-checker to find violations of expected web security goals.
Since then, many other papers explored formal methods in web security: a recent survey~\cite{BugliesiCF17} covers different research lines. We discuss here the papers which are closest to our work.

In the context of web sessions, \cite{BugliesiCFK15} employed \emph{reactive non-interference}~\cite{BohannonPSWZ09} to formalize and prove strong confidentiality properties for session cookies protected with the \texttt{HttpOnly} and \texttt{Secure} attributes, a necessary condition for any reasonable notion of session integrity.
A variant of reactive non-interference was also proposed in~\cite{KhanCBGP14} to formalize an integrity property of web sessions which rules out CSRF attacks and malicious script inclusions.
The paper also introduced a browser-side enforcement mechanism based on \emph{secure multi-execution}~\cite{DevrieseP10}.
A more general definition of web session integrity, which we adapted in the present paper, was introduced in~\cite{BugliesiCFKT14} to capture additional attacks, like password theft and session fixation.
The paper also studied a provably sound browser-based enforcement mechanism based on runtime monitoring.
Finally, \cite{CalzavaraFGM16} proposed the adoption of \emph{micro-policies}~\cite{AmorimDGHPST15} in web browsers to prevent a number of attacks against web sessions and presented Michrome, a Google Chrome extension implementing the approach. 
None of these papers, however, considered the problem of enforcing a formal notion of session integrity by analyzing web application code, since they only focused on browser-side defenses.

Formal methods found successful applications to web session security through the analysis of \emph{web protocols}, which are the building blocks of web sessions when single sign-on services are available.
Bounded model-checking was employed in~\cite{ArmandoCCCT08} and~\cite{ArmandoCCCPS13} to analyze the security of existing single sign-on protocols, exposing real-world attacks against web authentication.
WebSpi is a ProVerif library designed to model browser-server interactions, which was used to analyze existing implementations of single sign-on based on OAuth 2.0~\cite{BansalBDM14} and web-based cloud providers~\cite{BansalBDM13}.

Web protocols for single sign-on have also been manually analyzed in the expressive Web Infrastructure Model (WIM): for instance, \cite{FettKS16} focused on OAuth 2.0, ~\cite{FettKS17} considered OpenID Connect, and~\cite{FettHK19} analyzed the OpenID Financial-grade API.
While the WIM is certainly more expressive than our core model, proofs are at present manual and require a strong human expertise.
In terms of security properties, \cite{FettHK19} considers a session integrity property expressed as a trace property that is specific to the OpenID protocol flow and the resources accessed thereby, while our definition of session integrity is generic and formulated as a hyperproperty.

Server-side programming languages with formal security guarantees have been proposed in several research papers.
Examples include SELinks~\cite{CorcoranSH09}, UrFlow~\cite{Chlipala10}, SeLINQ~\cite{SchoepeHS14} and JSLINQ~\cite{BalliuLSS16}.
All these languages have the ability to enforce information flow control in multi-tier web applications, potentially including a browser, a server and a database. 
Information flow control is an effective mechanism to enforce session integrity, yet these papers do not discuss how to achieve web session security; rather, they propose new languages and abstractions for developing web applications.
To the best of our knowledge, there is no published work on the formal security analysis of server-side programming languages, though the development of accurate semantics for such languages~\cite{FilarettiM14} is undoubtedly a valuable starting point for this kind of research.

\section{Conclusion}
We introduced a type system for sound verification of session integrity for web applications encoded in a core model of the web, and used it to assess the security of the session management logic of HotCRP, Moodle, and phpMyAdmin.
During this process we unveiled novel critical vulnerabilities that we responsibly disclosed to the applications' developers, validating by typing the security of the fixed versions.

We are currently developing a type-checker to fully automate the analysis, which we intend to make available as open source.
Providing type annotations is typically straightforward, as they depend on the web application specification and are easily derivable from it (\eg, cookie labels are derived from their attributes) and typing derivations are mostly deterministic, with a few exceptions (\eg, subtyping) that however follow recurrent patterns (\eg, subtyping is used in assignments to upgrade the value type to the reference type). 

Furthermore, while in this work we focused on a concise web model to better illustrate the foundational aspects of our analysis technique, it would be interesting to extend the type system to cover richer web models, \eg, the WIM model~\cite{FettHK19}, as well as additional web security properties.
We also plan to automate the verification process for PHP code, \eg, by developing an automated translation from real world code into our calculus.
Finally, we would like to formalize our theory in a proof assistant.


\section*{Acknowledgments}

This work has been partially supported by the the European Research Council (ERC) under the European Union's Horizon 2020 research (grant agreement 771527-BROWSEC); by the Austrian Science Fund (FWF) through the project PROFET (grant agreement P31621); by the Austrian Research Promotion Agency (FFG) through the Bridge-1 project PR4DLT (grant agreement 13808694) and the COMET K1 SBA.

\balance
\bibliographystyle{IEEEtranS}
\bibliography{biblio}

\begin{thebibliography}{10}
\providecommand{\url}[1]{#1}
\csname url@samestyle\endcsname
\providecommand{\newblock}{\relax}
\providecommand{\bibinfo}[2]{#2}
\providecommand{\BIBentrySTDinterwordspacing}{\spaceskip=0pt\relax}
\providecommand{\BIBentryALTinterwordstretchfactor}{4}
\providecommand{\BIBentryALTinterwordspacing}{\spaceskip=\fontdimen2\font plus
\BIBentryALTinterwordstretchfactor\fontdimen3\font minus
  \fontdimen4\font\relax}
\providecommand{\BIBforeignlanguage}[2]{{%
\expandafter\ifx\csname l@#1\endcsname\relax
\typeout{** WARNING: IEEEtranS.bst: No hyphenation pattern has been}%
\typeout{** loaded for the language `#1'. Using the pattern for}%
\typeout{** the default language instead.}%
\else
\language=\csname l@#1\endcsname
\fi
#2}}
\providecommand{\BIBdecl}{\relax}
\BIBdecl

\bibitem{AkhaweBLMS10}
D.~Akhawe, A.~Barth, P.~E. Lam, J.~C. Mitchell, and D.~Song, ``{Towards a
  Formal Foundation of Web Security},'' in \emph{Proceedings of the 23rd {IEEE}
  Computer Security Foundations Symposium, {CSF} 2010}, 2010, pp. 290--304.

\bibitem{ArmandoCCCPS13}
A.~Armando, R.~Carbone, L.~Compagna, J.~Cu{\'{e}}llar, G.~Pellegrino, and
  A.~Sorniotti, ``{An Authentication Flaw in Browser-Based Single Sign-On
  Protocols: Impact and remediations},'' \emph{Computers {\&} Security},
  vol.~33, pp. 41--58, 2013.

\bibitem{ArmandoCCCT08}
A.~Armando, R.~Carbone, L.~Compagna, J.~Cu{\'{e}}llar, and M.~L. Tobarra,
  ``{Formal Analysis of SAML 2.0 Web Browser Single Sign-On: Breaking the
  SAML-Based Single Sign-On for Google Apps},'' in \emph{Proceedings of the 6th
  {ACM} Workshop on Formal Methods in Security Engineering, {FMSE} 2008}, 2008,
  pp. 1--10.

\bibitem{BackesHM14}
M.~Backes, C.~Hri\c{t}cu, and M.~Maffei, ``{Union, Intersection and Refinement
  Types and Reasoning About Type Disjointness for Secure Protocol
  Implementations},'' \emph{Journal of Computer Security}, vol.~22, pp.
  301--353, 2014.

\bibitem{BalliuLSS16}
M.~Balliu, B.~Liebe, D.~Schoepe, and A.~Sabelfeld, ``{JSLINQ: Building Secure
  Applications across Tiers},'' in \emph{Proceedings of the 6th {ACM}
  Conference on Data and Application Security and Privacy, {CODASPY} 2016},
  2016, pp. 307--318.

\bibitem{BansalBDM13}
C.~Bansal, K.~Bhargavan, A.~Delignat{-}Lavaud, and S.~Maffeis, ``{Keys to the
  Cloud: Formal Analysis and Concrete Attacks on Encrypted Web Storage},'' in
  \emph{Proceedings of the 2nd International Conference on Principles of
  Security and Trust, {POST} 2013}, 2013, pp. 126--146.

\bibitem{BansalBDM14}
------, ``{Discovering Concrete Attacks on Website Authorization by Formal
  Analysis},'' \emph{Journal of Computer Security}, vol.~22, no.~4, pp.
  601--657, 2014.

\bibitem{Cookies}
A.~Barth, ``Http state management mechanism,'' 2011, available at
  \url{https://tools.ietf.org/html/rfc6265}.

\bibitem{BarthJM08}
A.~Barth, C.~Jackson, and J.~C. Mitchell, ``{Robust Defenses for Cross-Site
  Request Forgery},'' in \emph{Proceedings of the 15th {ACM} Conference on
  Computer and Communications Security, {CCS} 2008}, 2008, pp. 75--88.

\bibitem{BohannonPSWZ09}
A.~Bohannon, B.~C. Pierce, V.~Sj{\"{o}}berg, S.~Weirich, and S.~Zdancewic,
  ``{Reactive Noninterference},'' in \emph{Proceedings of the 16th {ACM}
  Conference on Computer and Communications Security, {CCS} 2009}, 2009, pp.
  79--90.

\bibitem{BugliesiCF17}
M.~Bugliesi, S.~Calzavara, and R.~Focardi, ``{Formal methods for web
  security},'' \emph{Journal of Logic and Algebraic Programming}, vol.~87, pp.
  110--126, 2017.

\bibitem{BugliesiCFK15}
M.~Bugliesi, S.~Calzavara, R.~Focardi, and W.~Khan, ``{CookiExt: Patching the
  Browser Against Session Hijacking Attacks},'' \emph{Journal of Computer
  Security}, vol.~23, no.~4, pp. 509--537, 2015.

\bibitem{BugliesiCFKT14}
M.~Bugliesi, S.~Calzavara, R.~Focardi, W.~Khan, and M.~Tempesta, ``{Provably
  Sound Browser-Based Enforcement of Web Session Integrity},'' in
  \emph{Proceedings of the 27th {IEEE} Computer Security Foundations Symposium,
  {CSF} 2014}, 2014, pp. 366--380.

\bibitem{CalzavaraFGM16}
S.~Calzavara, R.~Focardi, N.~Grimm, and M.~Maffei, ``{Micro-Policies for Web
  Session Security},'' in \emph{Proceedings of the 29th {IEEE} Computer
  Security Foundations Symposium, {CSF} 2016}, 2016, pp. 179--193.

\bibitem{techreport}
S.~Calzavara, R.~Focardi, N.~Grimm, M.~Maffei, and M.~Tempesta,
  ``{Language-Based Web Session Integrity},''
  \url{https://arxiv.org/abs/2001.10405}, 2020.

\bibitem{CalzavaraFST17}
S.~Calzavara, R.~Focardi, M.~Squarcina, and M.~Tempesta, ``{Surviving the Web:
  A Journey into Web Session Security},'' \emph{ACM Computing Surveys},
  vol.~50, no.~1, pp. 13:1--13:34, 2017.

\bibitem{Chlipala10}
A.~Chlipala, ``{Static Checking of Dynamically-Varying Security Policies in
  Database-Backed Applications},'' in \emph{Proceedings of the 9th {USENIX}
  Symposium on Operating Systems Design and Implementation, {OSDI} 2010}, 2010,
  pp. 105--118.

\bibitem{ClarksonS10}
M.~R. Clarkson and F.~B. Schneider, ``{Hyperproperties},'' \emph{Journal of
  Computer Security}, vol.~18, no.~6, pp. 1157--1210, September 2010.

\bibitem{CorcoranSH09}
B.~J. Corcoran, N.~Swamy, and M.~W. Hicks, ``{Cross-Tier, Label-Based Security
  Enforcement for Web Applications},'' in \emph{Proceedings of the {ACM}
  {SIGMOD} International Conference on Management of Data, {SIGMOD} 2009},
  2009, pp. 269--282.

\bibitem{AmorimDGHPST15}
A.~A. de~Amorim, M.~D{\'{e}}n{\`{e}}s, N.~Giannarakis, C.~Hritcu, B.~C. Pierce,
  A.~Spector{-}Zabusky, and A.~Tolmach, ``{Micro-Policies: Formally Verified,
  Tag-Based Security Monitors},'' in \emph{Proceedings of the 36th {IEEE}
  Symposium on Security and Privacy, {S\&P} 2015}, 2015, pp. 813--830.

\bibitem{DevrieseP10}
D.~Devriese and F.~Piessens, ``{Noninterference through Secure
  Multi-execution},'' in \emph{Proceedings of the 31st {IEEE} Symposium on
  Security and Privacy, S{\&}P 2010}, 2010, pp. 109--124.

\bibitem{FettHK19}
D.~Fett, P.~Hosseyni, and R.~K{\"{u}}sters, ``{An Extensive Formal Security
  Analysis of the OpenID Financial-Grade API},'' in \emph{Proceedings of the
  40th {IEEE} Symposium on Security and Privacy, {S\&P} 2019}, 2019, pp.
  453--471.

\bibitem{FettKS16}
D.~Fett, R.~K{\"{u}}sters, and G.~Schmitz, ``{A Comprehensive Formal Security
  Analysis of OAuth 2.0},'' in \emph{Proceedings of the 23rd {ACM} Conference
  on Computer and Communications Security, {CCS} 2016}, 2016, pp. 1204--1215.

\bibitem{FettKS17}
------, ``{The Web SSO Standard OpenID Connect: In-depth Formal Security
  Analysis and Security Guidelines},'' in \emph{Proceedings of the 30th {IEEE}
  Computer Security Foundations Symposium, {CSF} 2017}, 2017, pp. 189--202.

\bibitem{FilarettiM14}
D.~Filaretti and S.~Maffeis, ``{An Executable Formal Semantics of PHP},'' in
  \emph{Proceedings of the 28th European Conference in Object-Oriented
  Programming, {ECOOP} 2014}, 2014, pp. 567--592.

\bibitem{FocardiM11}
R.~Focardi and M.~Maffei, \emph{Types for Security Protocols}.\hskip 1em plus
  0.5em minus 0.4em\relax IOS Press, 2011, pp. 143--181.

\bibitem{HSTS}
J.~Hodges, C.~Jackson, and A.~Barth, ``Http strict transport security (hsts),''
  2012, available at \url{https://tools.ietf.org/html/rfc6797}.

\bibitem{JovanovicKK06}
N.~Jovanovic, E.~Kirda, and C.~Kruegel, ``{Preventing Cross Site Request
  Forgery Attacks},'' in \emph{Proceedings of the 2nd International Conference
  on Security and Privacy in Communication Networks, SecureComm 2006}, 2006,
  pp. 1--10.

\bibitem{KhanCBGP14}
W.~Khan, S.~Calzavara, M.~Bugliesi, W.~D. Groef, and F.~Piessens, ``{Client
  Side Web Session Integrity as a Non-interference Property},'' in
  \emph{Proceedings of the 10th International Conference on Information Systems
  Security, {ICISS} 2014}, 2014, pp. 89--108.

\bibitem{phpMyAdmin-CVE-2018-10188}
\BIBentryALTinterwordspacing
{MITRE}, ``{CVE-2018-10188},'' April 2018. [Online]. Available:
  \url{https://www.cvedetails.com/cve/CVE-2018-10188/}
\BIBentrySTDinterwordspacing

\bibitem{Moodle-CVE-2018-16854}
\BIBentryALTinterwordspacing
------, ``{CVE-2018-16854},'' November 2018. [Online]. Available:
  \url{https://www.cvedetails.com/cve/CVE-2018-16854/}
\BIBentrySTDinterwordspacing

\bibitem{phpMyAdmin-CVE-2018-19969}
\BIBentryALTinterwordspacing
------, ``{CVE-2018-19969},'' December 2018. [Online]. Available:
  \url{https://www.cvedetails.com/cve/CVE-2018-19969/}
\BIBentrySTDinterwordspacing

\bibitem{phpMyAdmin-CVE-2019-12616}
\BIBentryALTinterwordspacing
------, ``{CVE-2019-12616},'' June 2019. [Online]. Available:
  \url{https://www.cvedetails.com/cve/CVE-2019-12616/}
\BIBentrySTDinterwordspacing

\bibitem{Moodle}
\BIBentryALTinterwordspacing
{Moodle HQ}, ``{Moodle Learning Platform}.'' [Online]. Available:
  \url{https://moodle.org}
\BIBentrySTDinterwordspacing

\bibitem{NikiforakisMYJJ11}
N.~Nikiforakis, W.~Meert, Y.~Younan, M.~Johns, and W.~Joosen, ``{SessionShield:
  Lightweight Protection against Session Hijacking},'' in \emph{Proceedings of
  the 3rd International Symposium on Engineering Secure Software and Systems,
  ESSoS 2011}, 2011, pp. 87--100.

\bibitem{phpMyAdmin}
\BIBentryALTinterwordspacing
{phpMyAdmin Development Team}, ``{phpMyAdmin Database Administration
  Software}.'' [Online]. Available: \url{https://www.phpmyadmin.net}
\BIBentrySTDinterwordspacing

\bibitem{SchoepeHS14}
D.~Schoepe, D.~Hedin, and A.~Sabelfeld, ``{SeLINQ: Tracking Information Across
  Application-Database Boundaries},'' in \emph{Proceedings of the 19th {ACM}
  {SIGPLAN} International Conference on Functional Programming, ICFP 2014},
  2014, pp. 25--38.

\bibitem{TangDK11}
S.~Tang, N.~Dautenhahn, and S.~T. King, ``{Fortifying Web-Based Applications
  Automatically},'' in \emph{Proceedings of the 18th {ACM} Conference on
  Computer and Communications Security, {CCS} 2011}, 2011, pp. 615--626.

\bibitem{CSP}
\BIBentryALTinterwordspacing
{W3C}, ``{Content Security Policy Level 2},'' December 2016. [Online].
  Available: \url{https://www.w3.org/TR/CSP2/}
\BIBentrySTDinterwordspacing

\bibitem{CookiePrefixes}
\BIBentryALTinterwordspacing
M.~West, ``{Cookie Prefixes}.'' [Online]. Available:
  \url{https://tools.ietf.org/html/draft-west-cookie-prefixes-05}
\BIBentrySTDinterwordspacing

\bibitem{ZhengJLDCWW15}
X.~Zheng, J.~Jiang, J.~Liang, H.~Duan, S.~Chen, T.~Wan, and N.~Weaver,
  ``{Cookies Lack Integrity: Real-World Implications},'' in \emph{Proceedings
  of the 24th {USENIX} Security Symposium, {USENIX} Security 2015}, 2015, pp.
  707--721.

\end{thebibliography}

\newpage
\appendix
\label{sec:appendix}

\subsection{Browser Model}
\label{sec:browser-app}

The syntax of the scripting language supported in our browser model is given in \autoref{tab:syntax-browser-app}.
We let $be$ range over expressions including references (for cookies), values, DOM elements, and binary operations defined over expressions, \eg, arithmetic and logical operations.
In particular, expression $\dom{be}{be'}$ extracts a value from the DOM of the page where the script is running: the expression $be$ identifies the tag of the form in the page, while $be'$ specifies the parameter of interest in the form. For simplicity, we stipulate that $\dom{be}{be'}$ selects the URL of the form if $be'$ evaluates to 0.

Command $\sskip$ does nothing, while $s;s'$ denotes the standard command concatenation. Command $r := be$ assigns to reference $r$ the value obtained by evaluating the expression $be$.
Command $\incl{u}{\lst{be}}$ retrieves the script located at URL $u$ providing $\lst{be}$ as parameters: we use this construct to model both contents inclusion and a simplified version of XHR requests which is not subject to SOP restrictions which are applied by real browsers. Command $\setdom{be'}{u}{\lst{be}}$ substitutes a form in a page, where $be'$ is the tag of the form to be replaced, $u$ and $\lst{be}$ are respectively the URL and the parameters of the new form.

The state of a browser is $\browser{N}{M}{P}{T}{Q}{\lst{a}}$ where $\bid$ is the identity of the user who wants to perform the list of actions $\lst{a}$.
The network store $N$ maps connection identifiers to triples $(\tab, u, l)$ where $\tab$ identifies the tab that initiated the connection, $u$ is the contacted endpoint and $l$ is the origin that has been sent in the \texttt{Origin} header of the request and it is needed to correctly handle the header during redirects.
$M$ is the cookie jar of the browser, which is modeled as a map from references to values. $P$ maps tab identifiers to pairs $(u, \page)$ representing the URL and the contents of the web page and $T$ tracks running scripts: if $T = \subst{\tab}{s}$, script $s$ is running on the page contained in $\tab$. Finally, $Q$ is a queue (of maximum size 1) of browser requests that is needed to handle redirects in our model.

Finally, we presuppose the existence of the set of domains $\Delta \subseteq \domains$ containing all domains where HSTS is enabled, which essentially models the HSTS preload list\footnote{~\url{https://hstspreload.org}} that is shipped with modern browsers.

\begin{table}[t]
	\caption{Syntax of browsers.}
	\label{tab:syntax-browser-app}
	\footnotesize
	\begin{tabular}{p{2cm}@{\hspace{2pt}}rcl}
		\multicolumn{4}{c}{Browsers} \\
		\midrule
		Expressions & $be$ & $::=$ & $x ~|~ r ~|~ v ~|~ \dom{be}{be'} ~|~ be \odot be'$ \\
		Scripts & $s$ & $::=$ & $\sskip ~|~ s; s' ~|~ r := be ~|~ \incl{u}{\lst{be}}$ \\
& &     & $|~ \setdom{be'}{u}{\lst{be}}$ \\
		Connections & $N$ & $::=$ & $\mempty ~|~ \subst{n}{(tab, u, l)}$ \\
		Pages & $P$ & $::=$ & $\mempty ~|~ P \uplus \subst{\tab}{(u, \page)}$ \\
		Tasks & $T$ & $::=$ & $\mempty ~|~ \subst{\tab}{s}$ \\
		Output queue & $Q$ & $::=$ & $\mempty ~|~ \{ \alpha \}$ \\
		Browsers & $B$ & $::=$ & $\ebrowser{N}{M}{P}{T}{Q}{\lst{a}}{\iota}$
	\end{tabular}
\end{table}

\subsection{More on Cookie Labels}
\label{sec:cookies-lbl-app}

Now we resume the discussion about the labelling of cookies that we started in \autoref{sec:label}.

When a cookie is set with a \texttt{Domain} attribute whose value is a domain $d$, the cookie will be attached to all requests towards $d$ and its subdomains. This behavior is modelled by the labelling
\[\small
	\begin{aligned}
		(\textstyle\bigicup_{d' \leq d} \http{d'} \ccap \https{d'}, \textstyle\bigicup_{d' \sim d} (\http{d'} \icup \https{d'}))
	\end{aligned}
\]
where $\leq$ is a preorder defined on $\domains$ such that $d \leq d'$ iff $d$ is subdomain of $d'$.

We discuss now the impact of HSTS on cookie labels: since this security policy prevents browsers from communicating with certain domains over HTTP, essentially it prevents network attackers from setting cookies by modifying HTTP responses coming from these domains. In particular, the label for a \texttt{Secure} cookie for domain $d$ becomes the following:
\[\small
	\begin{aligned}
		(\https{d}, \textstyle\bigicup_{\substack{d' \sim d \\ d' \notin \Delta}} \http{d'} \icup \textstyle\bigicup_{d' \sim d} \https{d'}))
	\end{aligned}
\]
If HSTS is enabled for $d$ and all its related domains, then the cookie label is the same as that of cookies with the \texttt{\_\_Secure-} cookie prefix, \ie:
\[\small
	\begin{aligned}
		(\https{d}, \textstyle\bigicup_{d' \sim d} \https{d'})
	\end{aligned}
\]

\subsection{Complete Semantics}
\label{sec:semantics-app}

\subsubsection{Browsers}

\begin{table*}
	\caption{Semantics of browsers.}
	\label{tab:app-browser}
	
	\begin{center}
    \small\textbf{Expressions}
	\end{center}
	\begin{mathpar}
		\footnotesize
		\inferrule*[lab={(BE-Val)}]
		{~}
		{\beval{v}{M}{f}{\ell} = v}
		
		\inferrule*[lab={(BE-BinOp)}]
		{\beval{be}{M}{f}{\ell} = v \\ \beval{be'}{M}{f}{\ell} = v'}
		{\beval{be \odot be'}{M}{f}{\ell} = v \odot v'}
		
		\inferrule*[lab={(BE-Reference)}]
		{C(\lambda(r)) \cleq C(\ell)}
		{\beval{r}{M}{f}{\ell} = M(r)}
		
		\inferrule*[lab={(BE-Dom)}]
		{\beval{be}{M}{f}{\ell} = v' \\
		 \beval{be'}{M}{f}{\ell} = v'' \\
		 \subst{v'}{\hform{u}{\lst{v}}} \in f \\
		 v'' = 0 \Rightarrow v''' = u \\
		 v'' \neq 0 \Rightarrow v''' = v_{v''}}
		{\beval{\dom{be}{be'}}{M}{f}{\ell} = v'''}
	\end{mathpar}
	~\vspace{\sectskip}
	\begin{center}
    \small\textbf{Browser}
	\end{center}
	\begin{mathpar}
		\footnotesize
		\inferrule*[lab={(B-Load)}]
		{\sample{n} \\ ck = \getck{M}{u} \\ \alpha = \bsend{\bid}{n}{u}{p}{ck}{\noorigin} \\
    (\uorigin{u} = \http{d} \Rightarrow d \not \in \hsts)}
		{\browser{\mempty}{M}{P}{\mempty}{\mempty}{\lcons{\load{\tab}{u}{p}}{\lst{a}}}
		 \xra{\blank}
		 \browser{\subst{n}{(tab, u, \noorigin)}}{M}{P}{\mempty}{\{ \alpha \}}{\lst{a}}}
		 
		\inferrule*[lab={(B-Include)}]
		{\sample{n} \\ ck = \getck{M}{u} \\
		 \subst{\tab}{(u', f)} \in P \\\\
		 \forall k \in \intval{1}{\len{be}}: p(k) = \beval{be_k}{M}{f}{\lambda(u')} \\
		 \alpha = \bsend{\bid}{n}{u}{p}{ck}{\uorigin{u'}} \\
    (\uorigin{u'} = \http{d} \Rightarrow d \not \in \hsts)}
		{\browser{\mempty}{M}{P}{\subst{\tab}{\incl{u}{\lst{be}}}}{\mempty}{\lst{a}}
		 \xra{\blank}
   \browser{\subst{n}{(tab, u,\uorigin{u'})}}{M}{P}{\subst{\tab}{\sskip}}{\{ \alpha \}}{\lst{a}}}
		
		\inferrule*[lab={(B-RecvLoad)}]
		{\alpha = \brecv{n}{u}{\bot}{\_}{ck}{\page}{s} \\\\ M' = \updck{M}{u}{ck} \\
		 \lst{a}\,' = (\page = \serror) ~?~ (\lcons{\halt}{\lst{a}}) : \lst{a}}
   {\browser{\subst{n}{(tab, u, o)}}{M}{P}{\mempty}{\mempty}{\lst{a}}
		 \xra{\alpha}
		 \browser{\mempty}{M'}{\mjoin{P}{\subst{\tab}{(u, \page)}}}{\subst{\tab}{s}}{\mempty}{\lst{a}\,'}}
	
		\inferrule*[lab={(B-RecvInclude)}]
		{\alpha = \brecv{n}{u}{\bot}{\_}{ck}{\page}{s} \\ M' = \updck{M}{u}{ck}}
    {\browser{\subst{n}{(tab, u, o)}}{M}{P}{\subst{\tab}{s'}}{\mempty}{\lst{a}}
		 \xra{\alpha}
		 \browser{\mempty}{M'}{P}{\subst{\tab}{s;s'}}{\mempty}{\lst{a}}}
		
		\inferrule*[lab={(B-Redirect)}]
		{\alpha = \brecv{n}{u}{u'}{\lst{v}}{ck}{\_}{\_} \\
		 M' = \updck{M}{u}{ck} \\ \sample{n'} \\ ck' = \getck{M'}{u'} \\
		 \forall k \in \intval{1}{\len{v}}: p(k) = v_k \\
     o' = (o = \uorigin{u}) ~?~ o : \noorigin \\
		 \alpha' = \bsend{\bid}{n'}{u'}{p}{ck'}{o'} \\
    (\uorigin{u'} = \http{d} \Rightarrow d \not \in \hsts)}
     {\browser{\subst{n}{(tab, u, o)}}{M}{P}{\mempty}{\mempty}{\lst{a}}
		 \xra{\alpha}
    \browser{\subst{n'}{(tab, u', o')}}{M'}{P}{\mempty}{\{ \alpha' \}}{\lst{a}}}
	
		\inferrule*[lab={(B-Submit)}]
		{\subst{\tab}{(u, f)} \in P \\
		 \subst{v'}{\hform{u'}{\lst{v}}} \in f \\
		 \forall k \in \intval{1}{\len{v}}.\, p'(k) = k \in \domain(p) ~?~ p(k) : v_k \\\\
		 \sample{n} \\ ck = \getck{M}{u'} \\
		 \alpha = \bsend{\bid}{n}{u'}{p'}{ck}{\uorigin{u}} \\
    (\uorigin{u'} = \http{d} \Rightarrow d \not \in \hsts)}
		{\browser{\mempty}{M}{P}{\mempty}{\mempty}{\lcons{\submit{\tab}{u}{v'}{p}}{\lst{a}}}
		 \xra{\blank}
   \browser{\subst{n}{(tab, u', \uorigin{u} )}}{M}{P}{\mempty}{\{ \alpha \}}{\lst{a}}}
		
		\inferrule*[lab={(B-Flush)}]
		{~}
		{\browser{N}{M}{P}{T}{\{ \alpha \}}{\lst{a}}
		 \xra{\alpha}
		 \browser{N}{M}{P}{T}{\mempty}{\lst{a}}}
		
		\inferrule*[lab={(B-Seq)}]
		{\browser{\mempty}{M}{P}{\subst{\tab}{s}}{\mempty}{\lst{a}}
		 \xra{\alpha}
		 \browser{\mempty}{M'}{P'}{\subst{\tab}{s'}}{\mempty}{\lst{a}}}
		{\browser{\mempty}{M}{P}{\subst{\tab}{s;s''}}{\mempty}{\lst{a}}
		 \xra{\alpha}
		 \browser{\mempty}{M'}{P'}{\subst{\tab}{s';s''}}{\mempty}{\lst{a}}}
		
		\inferrule*[lab={(B-Skip)}]
		{~}
		{\browser{\mempty}{M}{P}{\subst{\tab}{\sskip;s}}{\mempty}{\lst{a}}
		 \xra{\blank}
		 \browser{\mempty}{M}{P}{\subst{\tab}{s}}{\mempty}{\lst{a}}}
		 
		\inferrule*[lab={(B-End)}]
		{~}
		{\browser{\mempty}{M}{P}{\subst{\tab}{\sskip}}{\mempty}{\lst{a}}
		 \xra{\blank}
		 \browser{\mempty}{M}{P}{\mempty}{\mempty}{\lst{a}}}
		
		\inferrule*[lab={(B-SetReference)}]
		{\subst{\tab}{(u, f)} \in T \\
		 \ell = \lambda(u) \\
		 \beval{be}{M}{f}{\ell} = v \\
		 I(\ell) \ileq I(\lambda(r))}
		{\browser{\mempty}{M}{P}{\subst{\tab}{r := be}}{\mempty}{\lst{a}}
		 \xra{\blank}
		 \browser{\mempty}{M\subst{r}{v}}{P}{\subst{\tab}{\sskip}}{\mempty}{\lst{a}}}
	
		\inferrule*[lab={(B-SetDom)}]
		{\ell = \lambda(u') \\
		 \beval{be'}{M}{f}{\ell} = v' \\
		 \forall k \in \intval{1}{\len{be}}.\, v_k = \beval{be_k}{M}{f}{\ell}}
		{\browser{\mempty}{M}{P \uplus \subst{\tab}{(u', f)}}{\subst{\tab}{\setdom{be'}{u}{\lst{be}}}}{\mempty}{\lst{a}}
		 \xra{\blank}
		 \browser{\mempty}{M}{P \uplus \subst{\tab}{(u', f\subst{v'}{\hform{u}{\lst{v}}})}}{\subst{\tab}{\sskip}}{\mempty}{\lst{a}}}
	\end{mathpar}
\end{table*}

We present the browser semantics in \autoref{tab:app-browser} where we exclude non-deterministic behaviors by requiring that
\begin{enumerate*}[label={\em\roman*)}]
	\item at most one network connection is open at any time;
	\item the user performs an action only when there are no pending network connections and no script is running, which amounts to asking that the user waits that the current page is completely rendered.
\end{enumerate*}
This design choice is made to simplify our security proof and it has no impact the expressiveness of our model.

First we define the semantics of expressions in terms of the function $\beval{be}{M}{f}{\ell}$ that evaluates the expression $be$ in terms of the cookie jar $M$, the DOM of the webpage $f$ and the security context $\ell$.
Rule \irule{BE-Reference} models the access to the cookie jar, which is allowed only if the confidentiality level of the reference is below that of the security context. Rule \irule{BE-Dom} selects a value from the DOM of the page depending on the values of the expressions $be$ and $be'$. Rules \irule{BE-Val} and \irule{BE-BinOp} are standard.

Our semantics relies on the auxiliary functions $get\_ck$ and $upd\_ck$ to select the cookies to be attached to an outgoing request and to update the cookie jar with the cookies provided in an incoming response, respectively. Given a cookie jar $M$ and a URL $u$, we let $\getck{M}{u}$ be the map $ck$ such that $ck(r) = v$ iff $M(r) = v$ and $C(\lambda(r)) \cleq C(\lambda(u))$. Given a cookie jar $M$, a URL $u$ and a map of cookies $ck$, we let $\updck{M}{u}{ck} = \mjoin{M}{(\upckr{ck}{u})}$ where $\upckr{ck}{u}$ is the map $ck'$ such that $ck'(r) = v$ iff $ck(r) = v$ and $I(\lambda(u)) \ileq I(\lambda(r))$.

We describe now the rules of the browser semantics.
Rule \irule{B-Load} models the loading of a new page as dictated by the action $\load{\tab}{u}{p}$. The browser opens a new network connection represented by the fresh name $n$ and sends a request to the server located at $u$ providing the parameters $p$ and attaching the cookies $ck$ selected from the cookie jar, with an empty origin header, as represented by the action $\bsend{\bid}{n}{u}{p}{ck}{\noorigin}$. If the protocol of the URL $u$ is HTTP, we only allow the request if HSTS is not activated for the domain. In the connections store we associate $n$ to the triple $(tab, u, \noorigin)$.
Similarly, rule \irule{B-Include} models the embedding of a script with the $\mathbf{include}$ directive of our scripting language. Compared to \irule{B-Load}, the main differences are that
\begin{enumerate*}[label={\em\roman*)}]
	\item the list of expressions $\lst{be}$ specified in the instruction are evaluated;
	\item the request contains the origin of the page where the script is executed.
\end{enumerate*}
Notice that the execution of the script is paused until a response is received: this behavior is similar to what happens in standard browsers when embedding scripts or using synchronous XHR requests.

Rule \irule{B-RecvLoad} models the receiving of a webpage over a pending network connection, represented by the transition label $\brecv{n}{u}{\bot}{\_}{ck}{page}{s}$. As a result, the connection $n$ is closed, the cookie jar is updated with the cookies $ck$ attached to the response, the content of the tab associated to $n$ is replaced with the received page and the script $s$ is executed in that tab. In case the page $\serror$ is received, we prepend the action $\halt$ to the list of user actions: since this action is not be consumed by any of the semantic rules, this models a cautious user that interrupts the navigation when an unexpected error occurs during the navigation.
Rule \irule{B-RecvInclude} is similar to the previous rule: the main differences are that
\begin{enumerate*}[label={\em\roman*)}]
	\item the page contained in $\tab$ is left unchanged and the one sent by the server is discarded, therefore the user continues interacting with the website even when the $\serror$ page is received by the browser;
	\item the script $s$ sent by the server is prepended to the script $s'$ that is waiting to run on the page.
\end{enumerate*}
Rule \irule{B-Redirect} models the receiving of a redirect from the server to URL $u'$ with parameters $\lst{v}$, represented by the transition label $\brecv{n}{u}{u'}{\lst{v}}{ck}{\_}{\_}$. The cookie jar is updated with the cookies $ck$ set in the response and a new request to $u'$ with the appropriate cookies and parameters is prepared by the browser and added to the output queue. 
If the origin $o$ of the original request matches the origin $\uorigin{u'}$ of the new target, the origin header remains the same for the new request, otherwise it is set to $\noorigin$.
The redirect is only allowed if it respects the HSTS settings for the new target.

Rule \irule{B-Submit} models the user clicking on a link or submitting a form in the page identified by URL $u$ which is currently open in the browser at the specified tab.
For each parameter we first check if the user has inserted a value by inspecting the map $p$, otherwise we fallback to the pre-filled parameter contained in the form.
A new network connection is opened, cookies from the cookie jar are attached to the outgoing request and the HSTS settings are checked as in \irule{B-Load}. The origin of the request is the origin of the URL $u$ of the page containing the form.
Rule \irule{B-Flush} outputs on the network the request in the output queue produced by  rules \irule{B-Load}, \irule{B-Include} \irule{B-Redirect} and \irule{B-Submit}.

The remaining rules describe how scripts are processed.
Rule \irule{B-Seq} models sequencing of script commands, \irule{B-Skip} processes the $\sskip$ command and \irule{B-End} terminates the script execution.
Rule \irule{B-SetReference} models the setting of a cookie by a script, which is allowed if the integrity label of the reference is above that of the URL of the page where the script is running. Finally, rule \irule{B-SetDom} models the update of a form in the DOM of the page where the script is running.

\subsubsection{Servers}

In \autoref{tab:app-server} we give the rules of the server semantics that were not presented in \autoref{sec:semantics}.
Rule \irule{S-Seq} is used for sequencing commands, \irule{S-Skip} to evaluate $\sskip$, \irule{S-IfTrue} and \irule{S-IfFalse} for conditionals, \irule{S-OChkFail} and \irule{S-TChkSucc} cover the missing cases of origin and token check, \irule{S-SetGlobal} and \irule{S-SetSession} respectively update the value of a reference in the global memory and in the session memory.
Rule \irule{S-Redirect} models a redirect from the server to the URL $u'$ with parameters $\lst{z}$ that sets the cookies $ck$ in the user's browser. The page and script components of the action $\assend$ are respectively the empty page and the empty script, as they will be anyway discarded by the browser. As in rule \irule{S-Reply} shown in \autoref{tab:semantics}, all occurrences of variables in $\lst{x}$ contained in the response are replaced with the results of the evaluations of the corresponding expressions in $\lst{se}$ and we stipulate that the execution terminates after sending the message.
Finally, rules \irule{S-LParallel} and \irule{S-RParallel} handle the parallel composition of threads.

\begin{table*}
	\caption{Semantics of servers (remaining rules).}
	\label{tab:app-server}
	
	\begin{center}
    \small\textbf{Expressions}
	\end{center}
	\begin{mathpar}
		\scriptsize
		\inferrule*[lab={(SE-Val)}]
		{~}
		{\eval{v}{D}{E} = v}
		
		\inferrule*[lab={(SE-BinOp)}]
		{\eval{se}{D}{E} = v \\ \eval{se'}{D}{E} = v'}
		{\eval{se \odot se'}{D}{E} = v \odot v'} \\
	
		\inferrule*[lab={(SE-ReadGlobal)}]
		{~}
		{\eval{\rg{r}}{D}{i,\_} = D(i, r)}
		
		\inferrule*[lab={(SE-ReadSession)}]
		{~}
		{\eval{\rs{r}}{D}{i,j} = D(j, r)}
		
		\inferrule*[lab={(SE-Fresh)}]
		{\sample{n}}
		{\eval{\fresh}{D}{E} = n}
	\end{mathpar}
	~\vspace{\sectskip}
	\begin{center}
    \small\textbf{Server}
	\end{center}
	\begin{mathpar}
		\scriptsize
		\inferrule*[lab={(S-Seq)}]
		{\server{D}{\phi}{\thread{c}{R}{E}} \xra{\alpha} \server{D'}{\phi'}{\thread{c'}{R}{E'}}}
		{\server{D}{\phi}{\thread{c;c''}{R}{E}} \xra{\alpha} \server{D'}{\phi'}{\thread{c';c''}{R}{E'}}}
		
		\inferrule*[lab={(S-Skip)}]
		{~}
		{\server{D}{\phi}{\thread{\sskip;c}{R}{E}}
		 \xra{\blank}
		 \server{D}{\phi}{\thread{c}{R}{E}}}
		
		\inferrule*[lab={(S-IfTrue)}]
		{\eval{se}{D}{E} = \btrue}
		{\server{D}{\phi}{\thread{\ite{se}{c}{c'}}{R}{E}} \xra{\blank} \server{D}{\phi}{\thread{c}{R}{E}}}
		
		\inferrule*[lab={(S-IfFalse)}]
		{\eval{se}{D}{E} = \bfalse}
		{\server{D}{\phi}{\thread{\ite{se}{c}{c'}}{R}{E}}
		 \xra{\blank}
		 \server{D}{\phi}{\thread{c'}{R}{E}}}

		\inferrule*[lab={\footnotesize(S-OChkFail)}]
		{R = n, u, \bid, o \\
    o \not\in O}
    {\server{D}{\phi}{\thread{\och{O}{c}}{R}{E}}
		 \xra{\eerror}
    	 \server{D}{\phi}{\thread{\ereply{\serror}{\sskip}{\mempty}}{R}{E}}}

		\inferrule*[lab={\footnotesize(S-TChkSucc)}]
    {\eval{e_1}{D}{E} = \eval{e_2}{D}{E}}
    {\server{D}{\phi}{\thread{\tokch{e_1}{e_2}{c}}{R}{E}}
		 \xra{\blank}
    	 \server{D}{\phi}{\thread{c}{R}{E}}}
		
		\inferrule*[lab={(S-SetGlobal)}]
		{E = i, \_ \\ \eval{se}{D}{E} = v}
		{\server{D}{\phi}{\thread{\rg{r} := se}{R}{E}}
		 \xra{\blank}
		 \server{D\subst{i}{D(i) \subst{r}{v}}}{\phi}{\thread{\sskip}{R}{E}}}
		
		\inferrule*[lab={(S-SetSession)}]
		{\eval{se}{D}{i,j} = v}
		{\server{D}{\phi}{\thread{\rs{r} := se}{R}{i,j}}
		 \xra{\blank}
		 \server{D\subst{j}{D(j) \subst{r}{v}}}{\phi}{\thread{\sskip}{R}{i,j}}}
		
		\inferrule*[lab={\footnotesize(S-Redirect)}]
		{R = n, u, \bid, l \\
		 m = \len{x} = \len{se} \\
		 \forall k \in [1,m].\, \eval{se_k}{D}{E} = v_k \\\\
		 \sigma = [x_1 \mapsto v_1,\ldots,x_m \mapsto v_m] \\
		 \alpha = \ssend{n}{u}{u'}{\lst{z}\sigma}{ck\sigma}{\mempty}{\sskip}}
    	{\server{D}{\phi}{\thread{\redr{u'}{\lst{z}}{ck}{\lst{x} = \lst{se}}}{R}{E}}
    	 \xra{\alpha}
    	 \server{D}{\phi}{\thread{\shalt}{R}{E}}}
		
		\inferrule*[lab={(S-LParallel)}]
		{\server{D}{\phi}{t} \xra{\alpha} \server{D'}{\phi'}{t''}}
		{\server{D}{\phi}{\para{t}{t'}} \xra{\alpha} \server{D'}{\phi'}{\para{t''}{t'}}}
		
		\inferrule*[lab={(S-RParallel)}]
		{\server{D}{\phi}{t'} \xra{\alpha} \server{D'}{\phi'}{t''}}
		{\server{D}{\phi}{\para{t}{t'}} \xra{\alpha} \server{D'}{\phi'}{\para{t}{t''}}}
	\end{mathpar}
\end{table*}

\subsubsection{Web Systems}

\begin{table*}
	\caption{Semantics of web systems (remaining rules).}
	\label{tab:app-webatk}
	\begin{mathpar}
		\scriptsize
		\inferrule*[lab={(W-LParallel)}]
		{W \xra{\alpha} W'}
		{\para{W}{W''} \xra{\alpha} \para{W'}{W''}}
		
		\inferrule*[lab={(W-RParallel)}]
		{W \xra{\alpha} W'}
		{\para{W''}{W} \xra{\alpha} \para{W''}{W'}}
	
		\inferrule*[lab={(A-Nil)}]
		{W \xra{\alpha} W' \\
		 \alpha \in \{\blank, \lauth{\lst{v}}{\bid, \sid}{\ell'}\}}
		{\atkstate{\ell}{\atknow}{W}
		 \xra{\alpha}
		 \atkstate{\ell}{\atknow}{W'}}
		
		\inferrule*[lab={(A-SerBro)}]
		{W \xra{\brecv{n}{u}{u'}{\lst{v}}{ck}{\page}{s}} W' \\
		 W' \xra{\ssend{n}{u}{u'}{\lst{v}}{ck}{\page}{s}} W'' \\\\
		 \atknow' = (C(\lambda(u)) \cleq C(\ell)) ~?~ (\atknow \cup \fnames{ck, \page, s, \lst{v}}) : \atknow}
		{\atkstate{\ell}{\atknow}{W}
		 \xra{\blank}
		 \atkstate{\ell}{\atknow'}{W''}}
		
		\inferrule*[lab={(A-SerAtk)}]
		{n \in \atknow \\ 
		 \alpha = \ssend{n}{u}{u'}{\lst{v}}{ck}{\page}{s} \\\\ 
		 W \xra{\alpha} W' \\
		 \atknow' = \atknow \cup \fnames{ck, \page, s, \lst{v}}}
		{\atkstate{\ell}{\atknow}{W}
		 \xra{\alpha}
		 \atkstate{\ell}{\atknow'}{W'}}
		
		\inferrule*[lab={(A-AtkBro)}, leftskip={8pt}]
		{\alpha = \brecv{n}{u}{u'}{\lst{v}}{ck}{\page}{s} \\
		 W \xra{\alpha} W' \\\\
		 I(\ell) \ileq I(\lambda(u)) \\
		 \{ n \} \cup \fnames{ck, \page, s, \lst{v}} \subseteq \atknow}
		{\atkstate{\ell}{\atknow}{W}
		 \xra{\alpha}
		 \atkstate{\ell}{\atknow}{W'}}
		
		\inferrule*[lab={(A-Timeout)}, leftskip={8pt}]
    {W \xra{\bsend{\bid}{n}{u}{p}{ck}{o}} W' \\
      W' \centernot{\xra{\srecv{\bid}{n}{u}{p}{ck}{o}}} \\\\
		 W' \xra{\brecv{n}{u}{\bot}{\mempty}{\mempty}{\mempty}{\sskip}} W'' \\
		 \atknow' = (C(\lambda(u)) \cleq C(\ell)) ~?~ (\atknow \cup \fnames{p,ck}) : \atknow}
		{\atkstate{\ell}{\atknow}{W}
		 \xra{\blank}
		 \atkstate{\ell}{\atknow'}{W''}}
	\end{mathpar}
\end{table*}

We report in \autoref{tab:app-webatk} the rules of the web systems semantics that were not presented in the body of the paper.
Rules \irule{W-LParallel} and \irule{W-RParallel} model the parallel composition of web systems. Rule \irule{A-Nil} is applied when no synchronizations between two entities occur.

Rule \irule{A-SerBro} models an honest server providing a response to a browser over a pending connection. Here the knowledge of the attacker is extended either if she can read the messages using her network capabilities.
Rule \irule{A-SerAtk} models the reception of a response from an honest server by the attacker. We require that the attacker knows the connection identifier $n$ to prevent her from intercepting arbitrary traffic and we extend her knowledge with the contents of the message.
Rule \irule{A-AtkBro} models the attacker providing a response to a browser either using her network capabilities or a server under her control. In this case we require that the attacker is able to produce the contents of the response using her knowledge $\atknow$, which amounts to asking that all names in the response are known to the attacker.

Finally, rule \irule{A-Timeout} is used to process requests to endpoints not present in the system $W$ (\eg, attacker-controlled endpoints in a run without the attacker): in such a case, we let the browser process an empty response.

\subsection{Typing Rules for Scripts}
\label{sec:type-rules-app}

\autoref{tab:app-type-system} presents the typing rules that were not introduced in the body of the paper due to lack of space.

\begin{table*}
	\caption{Typing rules for scripts.}
	\label{tab:app-type-system}
	\begin{center}
    \small\textbf{Browser expressions and references}
	\end{center}
	\begin{mathpar}
		\scriptsize
		\inferrule*[lab={(T-BEVar)}]{~}{\belabel{x}{\envv(x)}}

		\inferrule*[lab={(T-BERef)}]{~}{\belabel{r}{\envrg(r)}}

		\inferrule*[lab={(T-BEVal)}]{v \not \in \names}{\belabel{v}{\tbot}}
		
		\inferrule*[lab={(T-BEUndef)}]{~}{\belabel{\vundef}{\stype}}

		\inferrule*[lab={(T-BEName)}]{~}{\belabel{n^\ell}{\tcre{\ell}}}

    \inferrule*[lab={(T-BEDom)}]{\typebranch \neq \bhon}{\belabel{\dom{be}{be'}}{\alabel}} 

		\inferrule*[lab={(T-BEBinOp)}]
        {\belabel{se}{\stype} \\ \belabel{se'}{\stype'} \\\\
        (\stype = \ell \wedge \stype = \ell') \vee \odot \text{ is }=}
        {\belabel{se \odot se'}{\typelabel{\stype} \sqcup \typelabel{\stype'}}} 

    \inferrule*[lab={(T-BESub)}]
      {\belabel{be}{\stype'} \\ 
       \stype' \tleq \stype}
      {\belabel{be}{\stype}}

		\inferrule*[lab={(T-BRef)}]{~}{\brtyping{r}{\envrg(r)}}

		\inferrule*[lab={(T-BRSub)}]
		{\brtyping{r}{\tref{\stype'}} \\ 
		 \stype \tleq \stype'}
		{\brtyping{r}{\tref{\stype}}}
	\end{mathpar}
	~\vspace{\sectskip}
	\begin{center}
    \small\textbf{Scripts}
	\end{center}
	\begin{mathpar}
		\scriptsize
		\inferrule*[lab={(T-BSeq)}]
		{\bstyping{\Gamma}{s} \\
		 \bstyping{\Gamma}{s'}}
		{\bstyping{\Gamma}{s;s'}}
	
		\inferrule*[lab={(T-BSkip)}]
		{~}
		{\bstyping{\Gamma}{\sskip}}

		\inferrule*[lab={(T-BAssign)}] 
		{\brtyping{r}{\tref{\stype}} \\
		 \belabel{be}{\stype} \\
		 \pclabel \ileq \It{\stype}}
		{\bstyping{\Gamma}{r := be}}

    	\inferrule*[lab={(T-BSetDom)}]  
		{\envu(u) = \ulabel, \lst{\stype}, l_r \\
		 m = \len{be} = \len{\stype} \\
		 \forall k \in \intval{1}{m}.\, \belabel{be_k}{\stype'_k} \\\\
		 (b = \bhon \Rightarrow \envt(v) = \envu(u) \wedge 
		  \pclabel \itleq I(\ulabel) \wedge
		  \forall k \in \intval{1}{m}.\, \stype'_k \tleq \stype_k) \\\\
		 (b \neq \bhon \Rightarrow 
		  \forall k \in \intval{1}{m}.\, \stype'_k \tleq \alabel)}
		{\bstyping{\Gamma}{\setdom{v}{u}{\vec{be}}}}
	
		\inferrule*[lab={(T-BInclude)}]
		{\envu(u) = \ulabel, \vec{\stype}, l_r \\
		 m = \len{be} = \len{\stype} \\
		 \forall k \in \intval{1}{m}.\, \belabel{be_k}{\stype'_k} \\\\
		 (b = \bhon \Rightarrow \isihigh{I(\ulabel)} \wedge l_r = \pclabel \wedge
		  \pclabel \itleq I(\ulabel) \wedge
		  \forall k \in \intval{1}{m}.\, \stype'_k \tleq \stype_k \wedge 
      u \not \in \protUrls 
    ) \\\\
		 (b \neq \bhon \Rightarrow 
		  \forall k \in \intval{1}{m}.\, \stype'_k \tleq \alabel)}
		{\bstyping{\Gamma}{\incl{u}{\vec{be}}}}

	\end{mathpar}
\end{table*}

\subsubsection{Browser Expressions}

Typing of browser expressions is ruled by the judgement $\belabel{be}{\stype}$, meaning that the expression $se$ has type $\stype$ in the typing environment $\Gamma$ and typing branch $\typebranch$. Rules are similar to those for server expressions, but in this case we do not carry around the session label since there are no session references. 
Rule \irule{T-BEDom} is used to type reading data from the DOM, where we conservatively forbid reading from the DOM in the honest branch and use label 
$\alabel$ otherwise, since we then know that the type of all values in the DOM is upper bounded by $\alabel$.

\subsubsection{Browser References}

Typing of references in the browser is ruled by the judgment $\brtyping{r}{\tref{\stype}}$ meaning that the reference $r$ has reference type $\tref{\stype}$ in the environment $\Gamma$. Compared to server references, the main difference is that there are no session references on the browser side.

\subsubsection{Scripts}

The typing judgment for scripts $\bstyping{\Gamma}{s}$ reads as follows: the script $s$ is well-typed in the environment $\Gamma$ under the program counter label $\pclabel$ in the typing branch $\typebranch$. 

Three straight-forward to type scripts are \irule{T-BSkip} that trivially does nothing, \irule{T-BSeq} checks both the concatenated commands and \irule{T-BAssign} handles reference assignments just like \irule{T-SetGlobal}.

In the honest branch, \irule{T-BSetDom} performs the same checks as \irule{T-Form}, namely that the script with program counter label $\pclabel$ is allowed to trigger a request to URL $u$, that the parameters of the generated form respect the type of
the URL, and that the type associated to the name of the form matches the type
of the URL.
For the attacked case, we just require that all parameters have type $\alabel$, as in the CSRF branch in rule \irule{T-Reply}.
Notice that we restrict the first expression in $\mathbf{setdom}$ to be a value, so that we can statically look up the associated type in $\envt$.

Rule \irule{T-BInclude} performs the same checks on the URL parameters as the previous rule, but additionally requires in the honest case that the integrity of the network connection is high to prevent an attacker from injecting her own script which would then be executed in the context of the original page.
Furthermore, we require that the included URL is not protected by an origin
check as otherwise an attacker could abuse this to indirectly trigger a CSRF with the expected origin.
We also require that the expected integrity label of the reply of the included URL $u$ is the same as the $\pclabel$ used to type the current script: this is needed since executing a script that was typed with a program counter label of higher integrity leads to a privilege escalation, \eg, it could write to a high integrity reference which the current script should not be allowed to do.
Including a script of lower integrity is also problematic since we type scripts 
in the same context as the DOM of the page, thus we would allow a low integrity script to write into the current (high integrity) DOM.

\subsection{Formal Results}
\label{sec:formal-app}

\begin{definition}
  \label{def:navigation-flow}
	Let $\lst{a}$ be a list of user actions containing $a_k = \load{tab}{u}{p}$. The \emph{navigation flow} initiated from $a_k$ is the list of actions $\lcons{a_k}{\nf{\lst{a} \Downarrow k}{\tab}}$ where $\lst{a} \Downarrow k$ is the list obtained from $\lst{a}$ by dropping the first $k$ elements and function $\mathit{nf}$ is defined by the following rules:
	\begin{mathpar}
		\footnotesize
		\inferrule
		{}
		{\nf{\lempty}{\tab} = \lempty}
		
		\inferrule
		{}
		{\nf{\lcons{\load{\tab}{u}{p}}{\lst{a}}}{\tab} = \lempty}
		
		\inferrule
		{a = \submit{\tab}{u}{v}{p} \\
		 \nf{\lst{a}}{\tab} = \lst{a}\,'}
		{\nf{\lcons{a}{\lst{a}}}{\tab} = \lcons{a}{\lst{a}\,'}}

		\inferrule
    {a \neq \submit{\tab}{u}{v}{p} \\  a \neq \load{\tab}{u}{p} \\
		 \nf{\lst{a}}{\tab} = \lst{a}\,'}
		{\nf{\lcons{a}{\lst{a}}}{\tab} = \lst{a}\,'}
	\end{mathpar}
\end{definition}

\subsection{Case Studies}
\label{sec:examples-app}

\newcounter{moodleloginprefix}
\newcounter{moodleloginpostfix}
\newcounter{phpmyadminloginprefix}
\newcounter{phpmyadminloginpostfix}
\newcounter{phpmyadmindropprefix}
\newcounter{phpmyadmindroppostfix}

Besides the case study on HotCRP that we have presented in the body of the paper, we have also analyzed other two popular PHP applications: phpMyAdmin~\cite{phpMyAdmin}, a software for database administration, and Moodle~\cite{Moodle}, an e-learning platform.
We discuss now the encoding of the session management logic in these applications and some session integrity vulnerabilities affecting them, either novel or taken from recent CVEs.

\subsubsection{Moodle}
we present now the $\mathit{login}$ endpoint implementing the authentication logic on Moodle. The endpoint expects the cookie $sid$ which is used to store session data and the credentials of the user, namely the username $\param{uid}$ and the password $\param{pwd}$. Its encoding in our calculus is the following:
\[\footnotesize
\begin{array}{>{\stepcounter{moodleloginprefix}{\tiny\themoodleloginprefix.~}}lcl}
    \listen{login}{\rf{sid}}{\param{uid}, \param{pwd}}{} \\
    \quad \cond{\rg{sid} = \vundef} \\
    \quad \quad \rg{sid} = \fresh; \\
    \quad \start{\rg{sid}}; \\
   	\quad \cond{\rs{uid} \neq \vundef} \\
	\quad \quad \eredr{\mathit{profile}}{\lempty}{\mempty}; \\
	\quad \eelse~\cond{\param{uid} = \vundef}{} \\
	\quad \quad \ereply{\{\uinput{auth} \mapsto \hform{login}{\lstexp{\vundef, \vundef}}\}}{\sskip}{\{\rf{sid} \mapsto x\}} \\
	\quad \quad \ \ \ewith{x = \rg{sid}}; \\
	\quad \eelse \\
	\quad \quad \rg{sid} = \fresh;
				\login{\param{uid}}{\param{pwd}}{\rg{sid}};
				\start{\rg{sid}}; \\
	\quad \quad \rs{uid} = \param{uid};
				\rs{sesskey} = \fresh; \\
	\quad \quad \redr{\mathit{profile}}{\lempty}{\{\rf{sid} \mapsto x\}}{x = \rg{sid}};
\end{array}
\]
If no cookie $sid$ has been provided, \eg, when the user visits the website for the first time, a fresh cookie is generated (lines 2--3). The session identified by $sid$ is then started (line 4): if the identifier denotes a valid session, session variables stored when processing previous requests are restored. If the user previously authenticated on the website, the session variable $\rs{uid}$ is different from the undefined value $\vundef$ and a redirect to the $\mathit{profile}$ endpoint (that here we do not model) is sent to the browser (lines 5--6). If the user is not authenticated and did not provide a pair of credentials, the server replies with a page containing the login form and a new cookie $sid$ is set into the user's browser (lines 7--9). Finally, if the user has provided valid credentials, the endpoint starts a fresh session (to prevent fixation), stores in the session memory the user's identity and a fresh value in $\rs{sesskey}$ which is used to implement CSRF protection, then redirects the user to the $\mathit{profile}$ endpoint and sets the new session identifier in the cookie $sid$ in the user's browser (lines 10--13).

Since $\mathit{login}$ does not perform any origin or token check before performing the $\mathbf{login}$ command, the endpoint is vulnerable to Login CSRF attacks, as it was the case for Moodle until November 2018~\cite{Moodle-CVE-2018-16854}. As discussed in \autoref{sec:overview-typing} for HotCRP, this problem is captured when typing since the cookie must be of low integrity since no CSRF check is performed when it is set, therefore it cannot be used to perform authenticated actions of high integrity.

The solution implemented by Moodle developers uses pre-sessions, as we proposed for HotCRP in \autoref{sec:overview}. In particular, developers decided for convenience to use the same cookie to handle both pre-sessions and sessions: this promotion of the cookie from low integrity, to handle the pre-session, to high integrity, when the session identifier is refreshed after authentication, cannot be modeled in our type system since we have a single static type environment for references, therefore type-checking would fail.
In our encoding we model the fix by using two different cookies, $pre$ and $sid$, which are set to the same value and respectively used in the pre-session and the session.
The problem can also be solved in the type system by distinguishing two different typing environments, but we leave this for future work.
\[\footnotesize
\begin{array}{>{\stepcounter{moodleloginpostfix}{\tiny\themoodleloginpostfix.~}}lcl}
    \listen{login}{\rf{pre}}{\param{uid}, \param{pwd}, \param{ltoken}}{} \\
    \quad \cond{\rg{pre} = \vundef} \\
    \quad \quad \rg{pre} = \fresh; \\
    \quad \start{\rg{pre}}; \\
   	\quad \cond{\rs{uid} \neq \vundef} \\
	\quad \quad \eredr{\mathit{profile}}{\lempty}{\mempty}; \\
	\quad \eelse~\cond{\param{uid} = \vundef}{} \\
	\quad \quad \cond{\rs{ltoken} = \vundef} \\
	\quad \quad \quad \rs{ltoken} = \fresh; \\
	\quad \quad \ereply{\{\uinput{auth} \mapsto \hform{login}{\lstexp{\vundef, \vundef, x}}\}}{\sskip}{\{\rf{pre} \mapsto y\}} \\
	\quad \quad \ \ \ewith{x = \rs{ltoken}, y = \rg{pre}}; \\\
	\quad \eelse \\
	\quad \quad \rg{ltoken} = \rs{ltoken}; \rs{ltoken} = \fresh; \\
	\quad \quad \tokch{\param{ltoken}}{\rg{ltoken}}{} \\
	\quad \quad \quad \rg{sid} = \fresh;
					  \login{\param{uid}}{\param{pwd}}{\rg{sid}};
					  \start{\rg{sid}}; \\
	\quad \quad \quad \rs{uid} = \param{uid};
					  \rs{sesskey} = \fresh; \\
	\quad \quad \quad \eredr{\mathit{profile}}{\lempty}{\{\rf{sid} \mapsto x, \rf{pre} \mapsto y\}} \\
	\quad \quad \quad \ \ \ewith{x = \rg{sid}, y = \rg{sid}};
\end{array}
\]
The main differences compared to the previous encoding are the following:
\begin{enumerate*}[label={\em\roman*)}]
	\item the endpoint now expects a third $\param{ltoken}$ which is used to implement CSRF protection (line 1);
	\item the login form is enriched with a CSRF token which is stored in the pre-session memory (lines 8--11);
	\item the token stored in the session memory is compared to the one provided by the user before performing the authentication (line 14). 
\end{enumerate*}
After applying the fix, it is possible to perform high integrity authenticated actions within session started from the cookie $sid$ since it is possible to assign it a high integrity credential type when type-checking against the web attacker.

\subsubsection{phpMyAdmin}
we show now the encoding of the session management logic for phpMyAdmin. In the following we model two HTTPS endpoints hosted on domain $d_P$: \emph{login}, where database administrators can authenticate using their access credentials, and \emph{drop}, where administrators can remove databases from the system.

We briefly discuss some implementation details of phpMyAdmin before presenting our encoding of the endpoints:
\begin{itemize}
	\item for CSRF and login CSRF protection, phpMyAdmin inspects all incoming POST requests to check whether they contain a parameter $\param{token}$ which is equal to the value stored in the (pre-)session memory;
	\item the parameters provided by the user are retrieved using the \texttt{\$\_REQUEST} array which allows to uniformly access POST and GET parameters: in our encodings we  model this behavior by using two different variables for each input of interest, \eg, $\param{g\_pwd}$ and $\param{p\_pwd}$ for the password when provided via GET or POST, respectively;
	\item a single cookie is used for pre-sessions and sessions while, as in the case of Moodle, we use two cookies $pre$ and $sid$;
	\item upon authentication, username and password are stored encrypted in two cookies: in our model we store them in the clear and use strong cookie labels to provide cookies with the confidentiality and integrity guarantees given by encryption.
\end{itemize}
We start with the encoding of the $\mathit{login}$ endpoint. As parameters it expects the username and the password, both provided via GET and POST, and the login CSRF token, while as cookies we have $\rf{pre}$ for the pre-session, $\rf{uid}$ and $\rf{pwd}$ where the credentials are stored upon authentication.
The encoding in our calculus is the following:
\[\footnotesize
\begin{array}{>{\stepcounter{phpmyadminloginprefix}{\tiny\thephpmyadminloginprefix.~}}lcl}
    \listen{login}{\rf{pre}, \rf{uid}, \rf{pwd}}{\param{g\_uid}, \param{p\_uid}, \param{g\_pwd}, \param{p\_pwd}, \param{token}}{} \\
    \quad \cond{\rg{uid} \neq \vundef \band \rg{pwd} \neq \vundef} \\
    \quad \quad \eredr{\mathit{index}}{\lempty}{\mempty}; \\
    \quad \cond{\rg{pre} = \vundef} \\
    \quad \quad \rg{pre} = \fresh; \\
    \quad \start{\rg{pre}}; \\
    \quad \cond{\param{g\_uid} = \vundef \band \param{p\_uid} = \vundef} \\
    \quad \quad \rs{token} = \fresh; \\
    \quad \quad \mathbf{reply}\ (\{\uinput{auth} \mapsto \hform{login}{\lstexp{\vundef, \vundef, \vundef, \vundef, x}}\}, \sskip, \\
    \quad \quad \ \ \{\rf{pre} \mapsto y\})\ \ewith{x = \rs{token}, y = \rg{pre}}; \\
	\quad \eelse~\cond{\param{p\_uid} \neq \vundef} \\
	\quad \quad \tokch{token}{\rs{token}}{} \\
	\quad \quad \quad \rg{sid} = \fresh;
					  \login{\param{p\_uid}}{\param{p\_pwd}}{\rg{sid}}; \\
	\quad \quad \quad \start{\rg{sid}};
					  \rs{token} = \fresh; \\
	\quad \quad \quad \mathbf{redirect}\ (\mathit{index}, \lempty, \{\rf{uid} \mapsto x, \rf{pwd} \mapsto y, \rf{pre} \mapsto z, \\
	\quad \quad \quad \ \ \rf{sid} \mapsto z\})\ \ewith{x = \param{p\_uid}, y = \param{p\_pwd}, z = \rg{sid}}; \\
	\quad \eelse \\
	\quad \quad \rg{sid} = \fresh;
					  \login{\param{uid}}{\param{pwd}}{\rg{sid}}; \\
	\quad \quad \start{\rg{sid}};
					  \rs{token} = \fresh; \\
	\quad \quad \eredr{\mathit{index}}{\lempty}{\{\rf{uid} \mapsto x, \rf{pwd} \mapsto y, \rf{pre} \mapsto z, \rf{sid} \mapsto z\}} \\
	\quad \quad \ \ \ewith{x = \param{g\_uid}, y = \param{g\_pwd}, z = \rg{sid}};
\end{array}
\]
First the endpoint checks whether the user is already authenticated by checking whether cookies $uid$ and $pwd$ are provided: in this case, the user is redirected to the $\mathit{index}$ endpoint (that here we do not model) showing all the databases available on the website (lines 2--3).
Next the session identified by cookie $pre$ is started or a fresh one is created (lines 4--6).
If the user has not sent her credentials, the page replies with a page containing the login form. This form contains a fresh CSRF token that is randomly generated for each request and stored in the session variable $\rs{token}$. The response sent by the server contains the fresh pre-session cookie generated by the server (lines 7--10).
Finally authentication is performed: a fresh session is started, a new token for CSRF protection is generated and the user is redirected to the $\mathit{index}$ endpoint. The response sets into the user's browser the cookies for session management and those containing the credentials. The only difference is that when login is performed via POST then the token checking is performed (lines 11--16), otherwise it is not (lines 17--21).

Now we present the encoding for the $\mathit{drop}$ endpoint, where we let $\ell_P = (\https{d_P}, \https{d_P})$. The endpoint expects three cookies: the session cookie $sid$ and those containing the credentials stored during the login. As parameters, it expects the name of the database to be deleted (provided either via GET and POST) and the CSRF token. The encoding in our calculus follows:
\[\footnotesize
\begin{array}{>{\stepcounter{phpmyadmindropprefix}{\tiny\thephpmyadmindropprefix.~}}lcl}
    \listen{drop}{\rf{sid}, \rf{uid}, \rf{pwd}}{\param{g\_db}, \param{p\_db}, \param{token}}{} \\
    \quad \cond{\rg{uid} = \vundef \bor \rg{pwd} = \vundef} \\
    \quad \quad \eredr{\mathit{login}}{\lempty}{\mempty}; \\
    \quad \start{\rg{sid}}; \\
    \quad \cond{\param{p\_db} \neq \vundef} \\
    \quad \quad \tokch{\param{token}}{\rs{token}}{} \\
    \quad \quad \quad \auth{\rg{uid}, \rg{pwd}, \param{p\_db}}{\ell_P}; \\
    \quad \eelse \\
    \quad \quad \auth{\rg{uid}, \rg{pwd}, \param{g\_db}}{\ell_P}; \\
    \quad \ereply{\mempty}{\sskip}{\mempty};
\end{array}
\]
First the endpoint checks where the user is authenticated by inspecting the provided cookies: if it is not the case, the  user is redirected to the $\mathit{login}$ endpoint (lines 2--3). After starting the session identified by the cookie $sid$, the endpoint drops the specified database after authenticating to the DBMS using the credentials stored in the cookies: this operation is abstractly represented using the \textbf{auth} command. Like in the $\mathit{login}$ endpoint, the CSRF token is verified when the database to be removed is provided via POST (lines 5--7) and not if sent via GET (lines 8--9).

Both endpoints are vulnerable to CSRF attacks due to the security-sensitive commands performed without any token or origin check: the $\mathbf{login}$ command in $\mathit{login}$ on line 18 and the $\mathbf{auth}$ command in $\mathit{drop}$ on line 9. Until December 2018, several sensitive endpoints of phpMyAdmin where vulnerable to CSRF vulnerabilities analogous to the one presented for the $\mathit{drop}$ endpoint~\cite{phpMyAdmin-CVE-2018-10188, phpMyAdmin-CVE-2018-19969}. The login CSRF, instead, is a novel vulnerability that we have discovered and has been recently assigned a CVE~\cite{phpMyAdmin-CVE-2019-12616}.

Type-checking captures the issue for the login CSRF vulnerability for the same reason of the other case studies, namely that the session cookie must be typed as low integrity and this prevents performing high integrity actions in the session. The standard CSRF is captured since it is not possible to apply rule \irule{T-Auth} when typing the $\mathbf{auth}$ of the $\mathit{drop}$ endpoint in the $\bcsrf$ typing branch.

The fix implemented by phpMyAdmin developers is the same for both vulnerabilities, \ie, using the \texttt{\$\_POST} array rather than the \texttt{\$\_REQUEST} array to retrieve the parameters provided by the user: this ensures that all sensitive operations are performed via POST, thus the CSRF token is always checked. To model this fix in our encoding we just get read of the input variables that represent GET parameters and remove the authenticated actions involving them.
The encoding of the $\mathit{login}$ endpoint becomes the following:
\[\footnotesize
\begin{array}{>{\stepcounter{phpmyadminloginpostfix}{\tiny\thephpmyadminloginpostfix.~}}lcl}
    \listen{login}{\rf{pre}, \rf{uid}, \rf{pwd}}{\param{p\_uid}, \param{p\_pwd}, \param{token}}{} \\
    \quad \cond{\rg{uid} \neq \vundef \band \rg{pwd} \neq \vundef} \\
    \quad \quad \eredr{\mathit{index}}{\lempty}{\mempty}; \\
    \quad \cond{\rg{pre} = \vundef} \\
    \quad \quad \rg{pre} = \fresh; \\
    \quad \start{\rg{pre}}; \\
    \quad \cond{\param{p\_uid} = \vundef} \\
    \quad \quad \rs{token} = \fresh; \\
    \quad \quad \ereply{\{\uinput{auth} \mapsto \hform{login}{\lstexp{\vundef, \vundef, x}}\}}{\sskip}{\{\rf{pre} \mapsto y\}} \\
	\quad \quad \ \ \ewith{x = \rs{token}, y = \rg{pre}}; \\
	\quad \eelse~\tokch{token}{\rs{token}}{} \\
	\quad \quad \rg{sid} = \fresh;
		  		\login{\param{p\_uid}}{\param{p\_pwd}}{\rg{sid}}; \\
	\quad \quad \start{\rg{sid}};
				\rs{token} = \fresh; \\
	\quad \quad \eredr{\mathit{index}}{\lempty}{\{\rf{uid} \mapsto x, \rf{pwd} \mapsto y, \rf{pre} \mapsto z, \rf{sid} \mapsto z\}} \\
	\quad \quad \ \ \ewith{x = \param{p\_uid}, y = \param{p\_pwd}, z = \rg{sid}};
\end{array}
\]
The encoding of the fixed $\mathit{drop}$ endpoint is the following:
\[\footnotesize
\begin{array}{>{\stepcounter{phpmyadmindroppostfix}{\tiny\thephpmyadmindroppostfix.~}}lcl}
    \listen{drop}{\rf{sid}, \rf{uid}, \rf{pwd}}{\param{p\_db}, \param{token}}{} \\
    \quad \cond{\rg{uid} = \vundef \bor \rg{pwd} = \vundef} \\
    \quad \quad \eredr{\mathit{login}}{\lempty}{\mempty}; \\
    \quad \start{\rg{sid}}; \\
    \quad \tokch{\param{token}}{\rs{token}}{} \\
    \quad \quad \auth{\rg{uid}, \rg{pwd}, \param{p\_db}}{\ell_P}; \\
    \quad \quad \ereply{\mempty}{\sskip}{\mempty};
\end{array}
\]
After applying the fix, it is possible to successfully type-check our encoding of the phpMyAdmin session management logic against the web attacker.

\onecolumn
\section{Proof}
\label{sec:proof}

In this section we present the full formal proof for the main result of the
paper. The proof consists of two major parts: 
Subject Reduction ensures that typing and other invariants are preserved during
execution of a web system.
A relational invariant ensures that the attacked system and the unattacked
system

\subsection{Outline}

In \cref{sec:proof-prelim} we introduce notation and helper functions used in
the proof.

In \cref{sec:proof-ex-sem} we present an extended version of the semantics,
containing additional annotations, as well as additional or modified typing
rules needed to type running code. We show that the semantic rules are
equivalent to the ones presented in the paper and that typing with the original
rules implies typing with the extended typing rules.

In \cref{sec:proof-sub-red} we prove the property of subject reduction for the
system: This tells us that all components of the system are well-typed and that certain invariants are preserved during the execution of a single system.

In \cref{sec:proof-rel} we introduce a relation between two websystems, that 
intuitively captures their equality on all high integrity components. We show
that an attacked websystem is always in relation with its unattacked version
and that this relation is preserved under execution.

In \cref{sec:proof-main} we combine results from the previous sections to
show our main theorem.

\subsection{Preliminaries}
\label{sec:proof-prelim}
Here we introduce some notation that will be used in the remainder of the proof

\begin{definition}[Notation]
  We define the following functions:
  \begin{itemize} 

    \item For a websystem $W$ we define $\servers{W}$ to be the set of all
      servers in $W$. 
      For a websystem with attacker $A = \atkstate{\alabel}{\atknow}{W}$ we let
      $\servers{A} = \servers{W}$
    \item For a websystem $W$ we define $\browsers{W}$ to be the set of all
      browsers in $W$
      For a websystem with attacker $A = \atkstate{\alabel}{\atknow}{W}$ we let
      $\browsers{A} = \browsers{W}$
    \item For a server $S=\exserver{D}{t}{\phi}$ we define $\surls{S}$ to be the set
      of all threads in $t$ of the form $\listen{u}{\lst{r}}{\lst{x}}{c}$.
    \item For a server $S=\exserver{D}{t}{\phi}$ we define $\running{S}$ to be the set
      of all threads in $t$ of the form $\exthread{c}{n^u}{E}$.
    \item For a thread of the form $t=\exthread{c}{n^u}{E}$ we let
      $\tint{t}=l$
    \item For a thread of the form $t=\exthread{c}{n^u}{i,j}$ and a database
      of global memories $\gmems$ we let $\gmem{\gmems}{t}=\gmems(i)$.
    \item For a thread of the form $t=\exthread{c}{n^u}{i,j}$ and a database
      of session memories $\smems$ we let $\smem{\smems}{t}=\smems(j)$.
    \item For a reference type $\stype_r = \tref{\stype}$ we let $\treft{\stype_r} = \stype$.
    \item For a command $c$ we let $\commands{c}$ be the set containing all 
      commands in $c$.
    \item For an event $\einta{\alpha}{l}$ we define
      $\eint{\einta{\alpha}{l}}=l$ as the sync integrity of the event

    \item We define a meet $\cremeet$ between a type $\stype$ and a label $\ell$
      that limits the label of $\stype$ to the label $\ell$. Formally:
      \[ \stype \cremeet \ell =
      \begin{cases} 
        \ell' \sqcap \ell & \text{ if } \stype = \ell' \\
        \tcre{\ell \sqcap \ell'} & \text{ if } \stype = \tcre{\ell'}
      \end{cases}
      \]
    \item We define a join $\crejoin$ on types that behaves like the regular
      join $\stype_1 \sqcup \stype_2$ if it is defined and 
      $\typelabel{\stype_1} \sqcup \typelabel{\stype_2}$ otherwise
    \item We define a join $\tijoin$ as $\tcre{\ell} \tijoin l := \tcre{(\Ct{\stype}, \It{\stype} \ijoin l)}$ and 
        $\ell \tijoin l := (\Ct{\stype}, \It{\stype} \ijoin l)$
    \item For a running server thread $t = \exthread{c}{E}{R}$ we let
      $\tlowestint{t} = l \ijoin \bigijoin{ l' \in \{l' ~|~ \resetpc{l'} \in c\}}$ l'
    \item For value $v^\stype$, We define 
      $\jlabel{v^\stype} = (\isclow{\Ct{\stype}}) ~:~ \pair{\cbot}{\itop} ~?~ \typelabel{\stype}$

    \item For a typing environment $\Gamma$ and two memories $M$ and $M'$, we
      write $M \heq{\Gamma} M'$ if for all $r$ with
      $\isihigh{\It{\treft{\Gamma(r)}}}$ we have $M(r) = M'(r)$
    \item For a set of name $N$, we let $\attlower{N}$ be the set same set of
      names, where all types have been lowered to $\alabel$. 
  \end{itemize}
\end{definition}

\begin{definition}[Freshness]
  \begin{itemize}
    \item A Browser $B = \browser{N}{K}{P}{T}{Q}{\lst{a}}$ is fresh if $N =
      \mempty$, $K = \mempty$, $P = \mempty$, $T = \mempty$, $Q = \mempty$.
    \item A Server $S = \server{D}{\Phi}{t}$ is fresh if 
      $D = \mempty$ and $\Phi = \mempty$. (also see \cref{def:cluster})
    \item A Websystem $W$ is fresh if all $B \in \browsers{W}$ and all $S \in
      \servers{W}$ are fresh.
  \end{itemize}
\end{definition}

\subsection{Extended Semantics and Typing Rules}
\label{sec:proof-ex-sem}
In this section we introduce additional and modified rules for the semantics
and the type system.

The most important changes are presented here:
\begin{itemize}
\item We annotate running server threads, the browser state, the DOM, and
  network requests and replies with an integrity label $l \in \labels$ and an
  attacked state $\amode \in \{\bhon, \batt\}$.
  Intuitively, $l$ is dynamically tracking which domains have influenced the
  current state of the execution, while $\amode$ is a binary flag that tells us
  whether the attacker used his capabilities to directly influence the current
  state.
\item We annotate events with an integrity label (an additional one, using the
  notation $\einta{\alpha}{l}$) . 
  This label is used to synchronize the execution of the unattacked and the
  attacked websystem in the relation:
  High integrity events have to be processed in sync, while low integrity
  events may be processed individually.
\item We introduce a new command $\resetpc{l}$ for servers to ``reset the pc''
  after a conditional. This operation has no semantic effect, it just updates
  the integrity  annotation .
\item We partition the database $D = (\gmems, \smems)$ into two different
    mappings for global and server memories.
\item We split the rule \irule{A-Timeout} into two separate rules
  \irule{A-TimeoutSend} and \irule{A-TimeoutRecv}. We therefore introduce a
  buffer in the network state that keeps track of open connections that require
  a response.
  This is required since in the relation proof, every request and response
  needs to be atomic, so that it can be matched with the corresponding request
  or response in the other system.
  For example a request to a low integrity domain, that is intercepted by the
  attacker might be processed using a timeout in the unattacked system.
\item All values $v^{\stype} \in \values$ (in the code, in the DOM, in memory
  or in requests and responses) are now annotated with a security type that
  gives us runtime information. 
  All primitive values have by default the type $\tau = \bot$ and hence can be
  given any security label $\ell$ (due to subtyping).  
  Since for names $n^{\stype} \in \names$ we have $\stype = \tcre{\ell}$ for
  some $\ell$, we cannot use subtyping if $\ischigh{\Ct{\stype}}$ or
  $\isihigh{\It{\stype}}$. We hence partition the set of names 
  $ \names = \names_0 \biguplus_{\ischigh{C(\ell)}, \vee \isihigh{I(\ell)}}
  \names_\ell$ into one set $\names_0$ of names of low confidentiality and
  integrity and one set $\names_\ell$ for each label $\ell$ with high
  confidentiality or integrity.
\end{itemize}

We define a translation $\trans{\cdot}$ function from a fresh
websystem in the original semantics to websystems in the extended semantics.

Intuitively, the translation annotates all constants with the type $\bot$
and lets the initial browser start with high integrity and in the honest mode.

\[
\begin{array}{rcll}
  \trans{\atkstate{\alabel}{\atknow}{W}} &=& 
    \exatkstatepar{\alabel}{\atknow}{\trans{W}}{\emptyset}\\
  \trans{\para{W}{W'}} &=& \para{\trans{W}}{\trans{W'}} \\
  \trans{\server{\mempty}{\mempty}{t}} &=& \exserver{(\mempty,\mempty)}{\mempty}{\trans{t}} \\
  \trans{\para{t}{t'}} &=& \para{\trans{t}}{\trans{t'}} \\
  \trans{\listen{u}{\lst{r}}{\lst{x}}{c}} &=& 
    \listen{u}{\lst{r}}{\lst{x}}{\trans{c}} \\
  \trans{\sskip} &=& \sskip \\
  \trans{c;c;} &=& \trans{c} ; \trans{c'} \\
  \trans{\rg{r} := se} &=& \rg{r} := \trans{se} \\
  \trans{\rs{r} := se} &=& \rs{r} := \trans{se} \\
  \trans{\ite{se}{c}{c'}} &=& 
    \ite{\trans{se}}{\trans{c}}{{\trans{c'}}} \\
  \trans{\login{se_u}{se_{pw}}{se_{id}}} &=& \
    \login{\trans{se_u}}{\trans{se_{pw}}}{\trans{se_{id}}} \\
  \trans{\start{se}} &=& \start{\trans{se}} \\
  \trans{\auth{\lst{se}}{\ell}} &=& \auth{\trans{\lst{se}}}{\ell} \\
  \trans{\tokch{e}{e'}{c}} &=& \tokch{\trans{e}}{\trans{e'}}{\trans{c}} \\
  \trans{\och{L}{c}} &=& \och{L}{\trans{c}} \\
  \trans{\reply{\page}{s}{ck}{\lst{x} = \lst{se}}} &=&
    \reply{\trans{\trans{\page}}}{\trans{s}}{\trans{ck}}{\lst{x} = \trans{\lst{se}}} \\
  \trans{\redr{u}{\lst{z}}{ck}{\lst{x}=\lst{se}}} &=&
    \redr{\trans{u}}{\trans{\lst{z}}}{\trans{ck}}{\lst{x}=\trans{\lst{se}}} \\
  \trans{x} &=& x \\
  \trans{\rg{r}} &=& \rg{r} \\
  \trans{\rs{r}} &=& \rs{r} \\
  \trans{\fresh^{\ell}} &=& \fresh^{\ell} \\
  \trans{se \odot se'} &=& \trans{se} \odot \trans{se'} \\
  \trans{v} &=& v^{\bot} &\text{ with }  \not \in \names\\
  \trans{n^\ell} &=& n^{\tcre{\ell}} \\
  \trans{\vundef} &=& \vundef \\
  \trans{\browser{\mempty}{\mempty}{\mempty}{\mempty}{\mempty}{\lst{a}}} &=&
  \exbrowserpar{\mempty}{\subst{r}{\vundef^{\envrg(r)}}}{\mempty}{\mempty}{\mempty}{\lst{a}}{\ibot}{\bhon}
\end{array}
\]

The extended semantics are is presented in \cref{tab:ex-browser-sem},
\cref{tab:ex-server-sem-ex}, \cref{tab:ex-server-sem-com} and \cref{tab:ex-webatk-sem}. As a convention, we use 
$\xRa{\alpha}$ for steps derived using the extended semantics and
$\xra{\alpha}$ for steps derived using the original semantics.

\begin{table*}
	\caption{Extended semantics of browsers.}
	\label{tab:ex-browser-sem}

  $\exupdck{M}{u}{ck} = \mjoin{M}{(\exupckr{ck}{u})}$ where $\exupckr{ck}{u}$
  is the map $ck'$ such that $ck'(r) = v^{\stype \tjoin \treft{\envrg(r)}}$ iff
  $ck(r) = v^\stype$ and $I(\lambda(u)) \ileq I(\lambda(r))$.

	\begin{center}
    \textbf{Expressions}
	\end{center}
	\begin{mathpar}
		\inferrule*[lab={(BE-Val)}]
		{~}
		{\beval{v^\stype}{M}{f}{\ell} = v^\stype}
		
		\inferrule*[lab={(BE-BinOp)}]
		{\beval{be}{M}{f}{\ell} = v^\stype \\ 
     \beval{be'}{M}{f}{\ell} = v'^{\stype'} \\
    }
    {\beval{be \odot be'}{M}{f}{\ell} = (v \odot v')^{\typelabel{\stype} \sqcup \typelabel{\stype'}}}

		
		\inferrule*[lab={(BE-Reference)}]
		{C(\lambda(r)) \cleq C(\ell)}
		{\beval{r}{M}{f}{\ell} = M(r)}
		
		\inferrule*[lab={(BE-Dom)}]
    {\beval{be}{M}{f}{\ell} = v^{\stype} \\
      \beval{be'}{M}{f}{\ell} = v'^{\stype'} \\
      \subst{v}{\hform{u^{\stype_u}}{\lst{v^\stype}}} \in f \\
      v' = 0 \Rightarrow v''^{\stype''} = u^{\stype_u} \\
      v' \neq 0 \Rightarrow v''^{\stype''} = v_{v'}^{\stype_{v'}}\\
   }
   {\beval{\dom{be}{be'}}{M}{f}{\ell} = v''^{\stype'' \tijoin \It{\stype} \tijoin \It{\stype'}}}
	\end{mathpar}
	~\vspace{\sectskip}
	\begin{center}
    \textbf{Browser}
	\end{center}
	\begin{mathpar}
		\inferrule*[lab={(B-Load)}]
		{\sample{n} \\
    \alpha= \exbsend{\bid}{n}{u}{p}{ck}{\noorigin}{I(\lambda(u))}{\bhon} \\
    ck = \getck{M}{u} \\
    (\uorigin{u} = \http{d} \Rightarrow d \not \in \hsts)}
    {\exbrowserpar{\mempty}{M}{P}{\mempty}{\mempty}{\lcons{\load{\tab}{u}{\sigma}}{\lst{a}}}{\ibot}{\bhon}
      \exbstep{\einta{\blank}{\ibot}}
    \exbrowserpar{\subst{n}{(tab, u, \noorigin)}}{M}{P}{\mempty}{\{\einta{\alpha}{\ibot}\}}{\lst{a}}{\lambda(u)}{\bhon}}
		 
		\inferrule*[lab={(B-Include)}]
		{\sample{n} \\ ck = \getck{M}{u} \\
      \subst{\tab}{(u', f, l' \amode')} \in P \\
    \forall k \in \intval{1}{\len{be}}: p(k) = \beval{be_k}{M}{f}{\lambda(u')} \\
    \alpha = \exbsend{\bid}{n}{u}{p}{ck}{\uorigin{u'}}{l \ijoin I(\lambda(u))}{\amode} \\
    (\uorigin{u} = \http{d} \Rightarrow d \not \in \hsts)}
		{\exbrowser{\mempty}{M}{P}{\subst{\tab}{\incl{u}{\lst{be}}}}{\mempty}{\lst{a}}
      \exbstep{\einta{\blank}{l}}
      \exbrowser{\subst{n}{(tab, u, \uorigin{u'})}}{M}{P}{\subst{\tab}{\sskip}}{\{\einta{\alpha}{l}\}}{\lst{a}}}
	
		\inferrule*[lab={(B-Submit)}]
    { \alpha = \exbsend{\bid}{n}{u'}{p}{ck}{\uorigin{u}}{l' \ijoin I(\lambda(u'))}{\amode'} \\ 
      \subst{\tab}{(u, f,l',\amode')} \in P \\
      \subst{v'}{\hform{u'^\stype}{\lst{v^\stype}}} \in f \\
      \forall k \in \intval{1}{\len{v}}.\, p(k) = k \in \domain(p) ~?~ p(k) : v_k^{\stype_k} \\
		 \sample{n} \\ ck = \getck{M}{u}}
     {\exbrowserpar{\mempty}{M}{P}{\mempty}{\mempty}{\lcons{\submit{\tab}{u}{v'}{p}}{\lst{a}}}{\ibot}{\bhon}
      \exbstep{\einta{\blank}{\ibot}}
    \exbrowserpar{\subst{n}{(tab,u',\uorigin{u})}}{M}{P}{\mempty}{\{\einta{\alpha}{\ibot}\}}{\lst{a}}{l' \ijoin I(\lambda(u))}{\amode'}}
		
		\inferrule*[lab={(B-RecvLoad)}]
    {M' = \exupdck{M}{u}{ck} \\
     \alpha = \exbrecv{\bid}{n}{u}{\vundef}{\mempty}{ck}{page}{s}{l'}{\amode'} \\
		 \lst{a}\,' = (page = \serror \wedge \bid = \usr) ~?~ (\lcons{\halt}{\lst{a}}) : \lst{a}}
     {\exbrowser{\subst{n}{(tab, u, o)}}{M}{P}{\mempty}{\mempty}{\lst{a}}
      \exbstep{\einta{\alpha}{l}}
    \exbrowserpar{\mempty}{M'}{\mjoin{P}{\subst{\tab}{(u, page,l',\amode')}}}{\subst{\tab}{s}}{\mempty}{\lst{a}\,'}{l'}{\amode'}}
	
		\inferrule*[lab={(B-RecvInclude)}]
    {M' = \exupdck{M}{u}{ck} \\ 
      \alpha = \exbrecv{\bid}{n}{u}{\vundef}{\mempty}{ck}{page}{s}{l'}{\amode'}\\
      \amode'' = (\amode = \batt \vee \amode' = \batt) ~?~ \batt ~:~ \bhon}
      {\exbrowser{\subst{n}{(tab,u,o)}}{M}{P}{\subst{\tab}{s'}}{\mempty}{\lst{a}}
      \exbstep{\einta{}{l \imeet l'}}
   \exbrowserpar{\mempty}{M'}{P}{\subst{\tab}{s;s'}}{\mempty}{\lst{a}}{l' \ijoin l}{\amode''}}

		\inferrule*[lab={(B-Redirect)}]
    {\alpha = \exbrecv{n}{u}{u'}{\lst{v}}{ck}{\vundef}{\vundef}{\vundef}{l'}{\amode'} \\
      M' = \exupdck{M}{u}{ck} \\ \sample{n'} \\ ck' = \getck{M'}{u'} \\
		 \forall k \in \intval{1}{\len{v}}: p(k) = v_k \\
     o' = (o = \uorigin{u}) ~?~ o : \noorigin \\
     \alpha' = \exbsend{\bid}{n'}{u'}{p}{ck'}{o'}{l'}{\amode'} \\
    (\uorigin{u'} = \http{d} \Rightarrow d \not \in \hsts)
  }
    {\exbrowser{\subst{n}{(tab, u, o)}}{M}{P}{T}{\mempty}{\lst{a}}
      \exbstep{\einta{\alpha}{l \imeet l'}}
    \exbrowserpar{\subst{n'}{(tab, u', o')}}{M'}{P}{T}{\{ \einta{\alpha'}{l'} \}}{\lst{a}}{l'}{\amode'}}

		\inferrule*[lab={(B-Flush)}]
    {~}
    {\exbrowser{N}{M}{P}{T}{\{\einta{\alpha}{l'}\}}{\lst{a}}
      \exbstep{\einta{\alpha}{l'}}
		 \exbrowser{N}{M}{P}{T}{\mempty}{\lst{a}}}

		\inferrule*[lab={(B-End)}]
		{~}
		{\exbrowser{\mempty}{M}{P}{\subst{\tab}{\sskip}}{\mempty}{\lst{a}}
      \exbstep{\einta{\blank}{\ibot}}
    \exbrowserpar{\mempty}{M}{P}{\mempty}{\mempty}{\lst{a}}{\ibot}{\bhon}}
		
		\inferrule*[lab={(B-Seq)}]
    {
      \exbrowser{\mempty}{M}{P}{\subst{\tab}{s}}{\mempty}{\lst{a}}
      \exbstep{\einta{\alpha}{l}}
      \exbrowserpar{\mempty}{M'}{P'}{\subst{\tab}{s'}}{\mempty}{\lst{a}}{l'}{\amode'}}
		{\exbrowser{\mempty}{M}{P}{\subst{\tab}{s;s''}}{\mempty}{\lst{a}}
      \exbstep{\einta{\alpha}{l}}
     \exbrowserpar{\mempty}{M'}{P'}{\subst{\tab}{s';s''}}{\mempty}{\lst{a}}{l'}{\amode'}}
		
		\inferrule*[lab={(B-Skip)}]
		{~}
		{\exbrowser{\mempty}{M}{P}{\subst{\tab}{\sskip;s}}{\mempty}{\lst{a}}
      \exbstep{\einta{\blank}{l}}
		 \exbrowser{\mempty}{M}{P}{\subst{\tab}{s}}{\mempty}{\lst{a}}}
		
		\inferrule*[lab={(B-SetReference)}]
    {\subst{\tab}{(u, f, l', \amode')} \in P \\
		 \ell = \lambda(u) \\
     \beval{be}{M}{f}{\ell} = v^\stype \\
		 I(\ell) \ileq I(\lambda(r))}
		{\exbrowser{\mempty}{M}{P}{\subst{\tab}{r := be}}{\mempty}{\lst{a}}
      \exbstep{\einta{\blank}{l}}
    \exbrowser{\mempty}{M\subst{r}{v^{\stype \tjoin \treft{\envrg(r)}}}}{P}{\subst{\tab}{\sskip}}{\mempty}{\lst{a}}}
	
		\inferrule*[lab={(B-SetDom)}]
    {
     \ell = \lambda(u') \\ 
     \subst{\tab}{(u', f, l', \amode')} \in P \\
     \beval{be'}{M}{f}{\ell} = v' \\
     \forall k \in \intval{1}{\len{be}}.\, v_k'^{\stype'} = \beval{be_k}{M}{f}{\ell} \wedge v_k^\stype = v_k'^{\stype' 
      \tijoin l}\\
      \amode'' = (\amode = \batt \vee \amode' = \batt) ~?~ \batt ~:~ \bhon
   }
     {\exbrowser{\mempty}{M}{P \uplus \subst{\tab}{(u', f}}{\subst{\tab}{\setdom{be'}{u}{\lst{be}}}}{\mempty}{\lst{a}} 
      \exbstep{\einta{\blank}{l}} \\\\
    \exbrowser{\mempty}{M}{P \uplus \subst{\tab}{(u', f\subst{v'}{\hform{u^{\pair{\cbot}{l}}}{\lst{v^\stype}}},l' \ijoin l, \amode'')}}{\subst{\tab}{\sskip}}{\mempty}{\lst{a}}}
	\end{mathpar}
\end{table*}

\subsubsection{Detailed explanation of changes to browser semantics}

\begin{itemize}
  \item The definition of $\exupdck{\cdot}{\cdot}{\cdot}$ is like the original
    definition of $\updck{\cdot}{\cdot}{\cdot}$, with the difference that the
    type annotations of values are joined with the type of the reference in the
    environment $\Gamma$. 
    We will show in the proof that typing then ensures that the types of values
    in a memory reference is always equal to the type for that reference in the
    environment $\Gamma$.
  \item \irule{BE-Val} simply adds the value type
  \item \irule{BE-BinOp} adds types, and assigns the join of the labels of
    the input types to the result.
  \item \irule{BE-Reference} is the same as in the original semantics.
  \item \irule{BE-Dom} adds types. The integrity label of the returned value 
    is lowered, taking into account the integrity labels of the two parameters.
  \item \irule{B-Load} adds the integrity label of the URL and the $\bhon$ flag
    to the request.
    Additionally, the request is marked as a high integrity sync action. This
    means that all load events have to be processed in sync between the
    attacked and unattacked system.
    The integrity label of the browser state the integrity label of the URL and
    the attacked mode is honest.
  \item \irule{B-Include} uses the join of the browser integrity label and the
    URL's integrity label, as well as the browser's current attack state as
    annotations on the request. 
    The event's sync integrity label is the browser's integrity label.
    The rule does not modify the browser's integrity label or attacked state.
  \item \irule{B-Submit} uses the integrity label and attacked mode from the
    DOM for the request, combined with the integrity label of the target URL
    The rule does not modify the browser's integrity label or attacked state.
  \item \irule{B-RecvLoad} receives a response to a load event, labelled with
    an integrity label and an attacked state, and uses these labels for the DOM 
    and the browser state. 
    The sync integrity of the response event is the integrity label of the
    browser. This means that event direct responses to a load or a submit
    have to be processed in sync (since they leave the browser in a high
    integrity state).
    A redirect however can lower the integrity of a browser that is awaiting a
    response to a load or submit (see below).
  \item \irule{B-RecvInclude} joins the integrity label and attacked state of
    the current browser state with the ones from the network response and uses
    them in the continuation.
    The sync integrity label of the event is the meet of the integrity label of
    the reply and the integrity label of the browser. This means that as long
    as one of the two is high, the response to the include has to be processed
    in sync.
  \item \irule{B-Redirect} uses the integrity label and attacked state of the
    incoming event for the outgoing event and the resulting browser state.
    The sync integrity label of the event is the meet of the integrity label of
    the reply and the integrity label of the browser. This means that as long
    as one of the two is high, the response to the include has to be processed
    in sync.
  \item \irule{B-Flush} sends out the event from the buffer together with its 
    sync integrity label.
  \item \irule{B-End} resets the browser's integrity label to high integrity
    and resets the attacked mode to $\bhon$.
    The sync integrity label is high, meaning that this step always has to be
    processed in sync.
  \item \irule{B-Seq} propagates the labels from the subcommand.
  \item \irule{B-Skip} propagates the browser annotations.
   The sync integrity label is the integrity label of the browser state
  \item \irule{B-SetReference} evaluates the expression and stores it in the
    memory, with the join of computed type and the type of the reference in the
    typing environment $\Gamma$. 
  \item \irule{B-SetDom} updates the DOM labelling by joining its original
    integrity label and the attacked sate with the ones of the browser state.
    The integrity label of the value stored into the DOM is lowered using the
    integrity label of the browser.
\end{itemize}

\begin{table*}
  \footnotesize
	\caption{Extended semantics of server expressions.}
	\label{tab:ex-server-sem-ex}

	\begin{center}
    \textbf{Expressions}
	\end{center}
	\begin{mathpar}
		\inferrule*[lab={(SE-Val)}]
		{~}
		{\eval{v^\stype}{D}{E} = v^\stype}
		
		\inferrule*[lab={(SE-Fresh)}]
    {\bid = \usr \Rightarrow \samplel{n^{\stype'}}{\stype}\\\\
     \bid \neq \usr \Rightarrow \samplel{n^{\stype'}}{0}}
     {\eval{\fresh^\stype}{D}{E} = n^{\stype'}}
		
		\inferrule*[lab={(SE-BinOp)}]
		{\eval{se}{D}{E} = v^\stype
    \eval{se'}{D}{E} = v'^{\stype'}
  }
  {\eval{se \odot se'}{D}{E} = (v \odot v')^{\typelabel{\stype} \sqcup \typelabel{\stype'}}} \\

	
		\inferrule*[lab={(SE-ReadGlobal)}]
		{~}
    {\eval{\rg{r}}{(\gmems,\smems)}{i,\_} = \gmems(i, r)}
		
		\inferrule*[lab={(SE-ReadSession)}]
		{~}
    {\eval{\rs{r}}{(\gmems,\smems)}{i,j} = \smems(j, r)}
	\end{mathpar}
\end{table*}

\begin{table*}
  \footnotesize
	\caption{Extended semantics of server commands.}
	\label{tab:ex-server-sem-com}
	\begin{center}
    \textbf{Server}
	\end{center}
	\begin{mathpar}
		\inferrule*[lab={(S-Seq)}]
    {\exserver{D}{\phi}{\exthread{c}{R}{E}} \exsstep{\einta{\alpha}{l}} \exserver{D'}{\phi'}{\exthreadpar{c'}{R}{E'}{l'}{\amode}}}
    {\exserver{D}{\phi}{\exthread{c;c''}{R}{E}} \exsstep{\einta{\alpha}{l}} \exserver{D'}{\phi'}{\exthreadpar{c';c''}{R}{E'}{l'}{\amode}}}
		
		\inferrule*[lab={(S-IfTrue)}]
    {c'' = (\mathbf{reply}, \mathbf{redir}, \mathbf{tokencheck}, \mathbf{origincheck} \in \commands{c}) ~?~ c ~:~  c;\resetpc{l} \\\\
      \eval{se}{D}{E} = \btrue^\stype \\
      l' = l \ijoin \It{\stype}}
      {\exserver{D}{\phi}{\exthread{\ite{se}{c}{c'}}{R}{E}} \exsstep{\einta{\blank}{l}} \exserver{D}{\phi}{\exthreadpar{c''}{R}{E}{l'}{\amode}}}

		\inferrule*[lab={(S-IfFalse)}]
    {c'' = (\mathbf{reply}, \mathbf{redir}, \mathbf{tokencheck}, \mathbf{origincheck} \in \commands{c}) ~?~ c ~:~  c;\resetpc{l} \\\\
		 \eval{se}{D}{E} = \bfalse^\stype \\
      l' = l \ijoin \It{\stype}
    }
		{\exserver{D}{\phi}{\exthread{\ite{se}{c}{c'}}{R}{E}} \\\\
      \exsstep{\einta{\blank}{l}}
     \exserver{D}{\phi}{\exthreadpar{c''}{R}{E}{l'}{\amode}}}

		\inferrule*[lab={(S-Reset)}]
		{~}
    {\exserver{D}{\phi}{\exthread{\resetpc{l'}}{R}{E}}
      \exsstep{\einta{\blank}{l'}} \\\\
     \exserver{D}{\phi}{\exthreadpar{\sskip}{R}{E}{l'}{\amode}}}

		\inferrule*[lab={(S-Skip)}]
		{~}
		{\exserver{D}{\phi}{\exthread{\sskip;c}{R}{E}}
      \exsstep{\einta{\blank}{l}}
		 \exserver{D}{\phi}{\exthread{c}{R}{E}}}

		\inferrule*[lab={(S-TCTrue)}]
    {\eval{se}{D}{E} = v^{\stype} \\
     \eval{se'}{D}{E} = v'^{\stype'} \\
      v = v'
      }
      {\exserver{D}{\phi}{\exthread{\tokch{se}{se'}{c}}{R}{E}} \exsstep{\einta{\blank}{l}} \exserver{D}{\phi}{\exthreadpar{c}{R}{E}{l}{\amode}}}
		
		\inferrule*[lab={(S-TCFalse)}]
    {\eval{se}{D}{E} = v^{\stype} \\
     \eval{se'}{D}{E} = v'^{\stype'} \\
      v \neq v'
      }
      {\exserver{D}{\phi}{\exthread{\tokch{se}{se'}{c}}{R}{E}} \\\\ 
        \exsstep{\einta{\blank}{l}} 
      \exserver{D}{\phi}{\exthreadpar{\ereply{\serror}{\sskip}{\mempty}}{R}{E}{l}{\amode}}}

		\inferrule*[lab={(S-Recv)}]
    {\alpha = \exsrecv{\bid}{n}{u}{p}{ck}{o}{l}{\amode} \\
      \sample{i} \\\\
		 \forall k \in \intval{1}{\len{r}}.\, M(r_k) = (r_k \in \domain(ck)) ~?~ ck(r_k) : \vundef \\\\
		 m = \len{x} \\ 
		 \forall k \in \intval{1}{\amode}.\, v_k = (k \in \domain(p)) ~?~ p(k) : \vundef \\\\
   \sigma = [x_1 \mapsto v_1,\ldots,x_m \mapsto v_m]}
		{\exserver{D}{\phi}{\listen{u}{\lst{r}}{\lst{x}}{c}} 
      \exsstep{\einta{\alpha}{l}} \\\\
    \exserver{D \uplus \subst{i}{M}}{\phi}{\para{\exthread{c\sigma}{(n, u, \bid, o)}{(i,\vundef)}}{\listen{u}{\lst{r}}{\lst{x}}{c}}}}
		
		\inferrule*[lab={(S-RestoreSession)}]
		{E = i, \_ \\
		 \eval{se}{D}{E} = j^\stype \\
   j \in \domain(D) }
		{\exserver{D}{\phi}{\exthread{\start{se}}{R}{E}}
      \exsstep{\einta{\blank}{l}}
      \exserver{D}{\phi}{\exthreadpar{\sskip}{R}{i, j}{l}{\amode}}}
		
		\inferrule*[lab={(S-NewSession)}]
		{E = i, \_ \\
		 \eval{se}{D}{E} = j^\stype \\\\
   j \notin \domain(D) }
		{\exserver{D}{\phi}{\exthread{\start{se}}{R}{E}}
      \exsstep{\einta{\blank}{l}}
    \exserver{D \uplus \subst{j}{\subst{r}{\vundef^{\envrs(r) \crejoin \jlabel{j}}}}}{\phi}{\exthreadpar{\sskip}{R}{i, j}{l}{\amode}}}

		\inferrule*[lab={\footnotesize(S-OChkFail)}]
		{ R = n, u, \bid, o \\
      o \not\in O}
    {\exserver{D}{\phi}{\exthread{\och{O}{c}}{R}{E}}
		 \exsstep{\blank} \\\\
    	 \exserver{D}{\phi}{\exthread{\ereply{\serror}{\sskip}{\mempty}}{R}{E}}}
		
		\inferrule*[lab={\footnotesize(S-OChkSucc)}]
		{ R = n, u, \bid, o \\
		 o \in L}
		{\exserver{D}{\phi}{\exthread{\och{L}{c}}{R}{E}}
      \exsstep{\einta{\blank}{l}} \\\\
    	 \exserver{D}{\phi}{\exthread{c}{R}{E}}}
		
		\inferrule*[lab={(S-SetGlobal)}]
    {\eval{se}{D}{E} = v^\stype \\ D = (\gmems,\smems) \\\\
      \envrg' = (\bid = \usr) ~?~ \envrg ~:~ \subst{\_}{\alabel} \\
      \stype' = \stype 
        \tijoin l
    }
    {\exserver{D}{\phi}{\exthread{\rg{r} := se}{R}{(i,j)}}
      \exsstep{\einta{\blank}{l}} \\\\
    \exserver{D\subst{i}{\gmems(i) \subst{r}{v^{\stype'}}}}{\phi}{\exthread{\sskip}{R}{(i,j)}}}
		
		\inferrule*[lab={(S-SetSession)}]
    {\eval{se}{D}{i,j} = v \\ D = (\gmems, \smems) \\\\
      \stype' = \stype \tjoin (\treft{\envrs(r)}  \crejoin \jlabel{j}) 
    }
    {\exserver{D}{\phi}{\exthread{\rs{r} := se}{R}{(i,j)}}
      \exsstep{\einta{\blank}{l}} \\\\
    \exserver{D\subst{j}{\smems(j) \subst{r}{v^{\stype'}}}}{\phi}{\exthread{\sskip}{R}{i,j} }}
		
		\inferrule*[lab={(S-Login)}]
    {\eval{se_{usr}}{D}{E} = \sid \\
      \eval{se_{pw}}{D}{E} = \rho(\sid, u) \\\\
    \eval{se_{sid}}{D}{E} = n}
    {\exserver{D}{\phi}{\exthread{\login{se_{usr}}{se_{pw}}{se_{sid}}}{R}{E}}
      \exsstep{\einta{\blank}{l}} \\\\
    	 \exserver{D}{\mjoin{\phi}{\subst{n}{\sid}}}{\exthread{\sskip}{R}{E}}}
    	
    	\inferrule*[lab={(S-Auth)}]
      {R = n, u, \bid, o \\
		 j \in \domain(\phi) \\\\
		 \forall k \in \intval{1}{\len{se}}.\, \eval{se_k}{D}{i, j} = v_k}
		{\exserver{D}{\phi}{\exthread{\auth{\lst{se}}{\ell}}{R}{i,j}}
      \exsstep{\einta{\lauth{\lst{v}}{\bid, \sid}{\ell}}{l}}
		 \exserver{D}{\phi}{\exthread{\sskip}{R}{i,j}}}
		
		\inferrule*[lab={(S-Reply)}]
    {R = n, u, \bid, o \\ 
      \envu(u) = \_, \_,  l_r \\
		 m = \len{x} = \len{se} \\
		 \forall k \in [1,m].\, \eval{se_k}{D}{E} = v_k \\\\
     \sigma = [x_1 \mapsto v_1,\ldots,x_m \mapsto v_m] \\
     \alpha = \exssend{\bid}{n}{u}{\bot}{\mempty}{ck\sigma}{page\sigma}{s\sigma}{l \ijoin l_r}{\amode} \\\\
     c' = (page = error) ~?~ \sbad ~:~ \shalt}
    	{\exserver{D}{\phi}{\exthread{\reply{page}{s}{ck}{\lst{x} = \lst{se}}}{R}{E}}
        \exsstep{\einta{\alpha}{l}}
    	 \exserver{D}{\phi}{\exthread{c'}{R}{E}}}
		
		\inferrule*[lab={(S-LParallel)}]
    {\exserver{D}{\phi}{t} \exsstep{\einta{\alpha}{l}} \exserver{D'}{\phi'}{t''}}
    {\exserver{D}{\phi}{\para{t}{t'}} \exsstep{\einta{\alpha}{l}} \exserver{D'}{\phi'}{\para{t''}{t'}}}
		
		\inferrule*[lab={\footnotesize(S-Redirect)}]
		{ R = n, u, \bid, o \\
      \envu(u) = \_, \_,  l_r \\
		 m = \len{x} = \len{se} \\
		 \forall k \in [1,m].\, \eval{se_k}{D}{E} = v_k \\\\
		 \sigma = [x_1 \mapsto v_1,\ldots,x_m \mapsto v_m] \\
   \alpha = \exssend{\bid}{n}{u}{u'}{\lst{z}\sigma}{ck\sigma}{\mempty}{\sskip}{l \ijoin l_r}{\amode} \\\\
 }
    	{\exserver{D}{\phi}{\exthread{\redr{u'}{\lst{z}}{ck}{\lst{x} = \lst{se}}}{R}{E}}
        \exsstep{\einta{\alpha}{l}}
    	 \exserver{D}{\phi}{\exthread{\shalt}{R}{E}}}
		
		\inferrule*[lab={(S-RParallel)}]
    {\exserver{D}{\phi}{t'} \exsstep{\einta{\alpha}{l}} \exserver{D'}{\phi'}{t''}}
    {\exserver{D}{\phi}{\para{t}{t'}} \exsstep{\einta{\alpha}{l}} \exserver{D'}{\phi'}{\para{t}{t''}}}
	\end{mathpar}

\end{table*}

\subsubsection{Detailed Explanation of Changes to Server Semantics}
\begin{itemize}
  \item \irule{SE-Val} also contains the type.
  \item \irule{SE-Fresh} samples names from the partition of the set of
    names indicated by the annotation.
    If the browser id is not the one of the honest user $\usr$, then we always
    sample from $\names_0$, the set of names of low confidentiality and
    integrity.
  \item \irule{SE-BinOp} is just like \irule{BE-BinOp}
  \item \irule{SE-ReadGloabl}, \irule{SE-ReadSession} look up the reference in
    the corresponding part of the database.
  \item \irule{S-Seq} just propagates the annotations
  \item \irule{S-IfTrue}, \irule{S-IfFalse} lower the integrity label, based on
    the type of the guard. In case the code for the branch does not contain any
    command that can lead to a response, a reset command is added after the
    branch, to bring the integrity label back to its original value.
  \item \irule{S-Reset} restores the integrity label to the provided value.
    The sync integrity label is the integrity label to which the reset is
    performed. This means that returning to a high integrity context from a low
    integrity context must be processed in sync.
  \item \irule{S-Skip} just propagates the annotations
  \item \irule{S-TCTrue}, \irule{S-TCFalse} just propagate the annotations.
  \item \irule{S-Recv} takes the annotations from the request and uses them for
    the newly started thread.
  \item \irule{S-RestoreSession} just propagates the annotations.
  \item \irule{S-NewSession} initializes the new memory with $\vundef$,
    annotated with the appropriate type from $\envrs$ combined with the type 
    of the session identifier.
    The integrity label is not influenced, as by an invariant the
      integrity of all session memory references and the user identity 
      is upper bounded by the integrity of the session identifier
    \item \irule{S-OChckSucc}, \irule{S-OChckFail} just propagate the annotations.
  \item \irule{S=LParallel}, \irule{R-Parallel} juts propagate the labelling of
    the events of sub threads
  \item \irule{S-SetGlobal} stores the value with its computed type, joining
    the integrity label with the thread's integrity label.
  \item \irule{S-SetSession} stores the value with the type that results from 
    joining the value's original type with the type of the reference, limited
    by the type of the session identifier.
    We will show in the proof that typing then ensures that the types of values
    in a memory reference is always equal to the type for that reference in the
    environment $\Gamma$, limited by the type of the session identifier.
  \item \irule{S-Login} just propagates the annotations 
  \item \irule{S-Auth} just propagates the annotations 
  \item \irule{S-Reply}, uses the annotations of the current thread for the
    reply, where the integrity label is joined with the expected integrity label for the reply. In case the reply is an error message, instead of going to the
    regular $\shalt$ state, the thread will go to a $\sbad$ state.
    These two states are semantically equivalent (both cannot be processed
    further) and are just used to establish an invariant in the proofs.
  \item \irule{S-Redirect} uses the annotations of the current thread for the
    reply where again the integrity label is joined with the expected integrity
    label for the reply.

\end{itemize}

\begin{table*}
  \centering
	\caption{Extended semantics of web systems with the attacker.}
	\label{tab:ex-webatk-sem}
	\begin{mathpar}
		\inferrule*[lab={(W-LParallel)}]
    {W \exastep{\einta{\alpha}{l}} W'}
    {\para{W}{W''} \exastep{\einta{\alpha}{l}} \para{W'}{W''}}
		
		\inferrule*[lab={(W-RParallel)}]
    {W \exastep{\einta{\alpha}{l}} W'}
    {\para{W''}{W} \exastep{\einta{\alpha}{l}} \para{W''}{W'}}
	
		\inferrule*[lab={(A-Nil)}]
    { \timeouts = \mempty \\
      W \exastep{\einta{\alpha}{l}} W' \\
		 \alpha \in \{\blank, \lauth{\lst{v}}{\bid, \sid}{\ell'}\}}
		{\atkstate{\alabel}{\atknow}{W}
      \exastep{\einta{\alpha}{l}}
		 \atkstate{\alabel}{\atknow}{W'}}
%
		
		\inferrule*[lab={(A-BrowserServer)}]
    { \timeouts = \mempty \\
    W \exastep{\einta{\exbsend{\bid}{n}{u}{p}{ck}{o}{l}{\amode}}{l'}} W' \\
      W' \exastep{\einta{\exsrecv{\bid}{n}{u}{p}{ck}{o}{l}{\amode}}{l}} W'' \\
    \atknow' = (C(\lambda(u)) \cleq C(\alabel)) ~?~ (\atknow \cup \attlower{\fnames{p,ck}}) : \atknow}
		{\atkstate{\alabel}{\atknow}{W}
      \exastep{\einta{\blank}{l'}}
		 \atkstate{\alabel}{\atknow'}{W''}}

		\inferrule*[lab={(A-ServerBrowser)}]
    { \timeouts = \mempty \\
      W \exastep{\einta{\exbrecv{\bid}{n}{u}{u'}{\lst{v}}{ck}{page}{s}{l}{\amode}}{l}} W' \\
      W' \exastep{\einta{\exssend{\bid}{n}{u}{u'}{\lst{v}}{ck}{page}{s}{l}{\amode}}{l'}} W'' \\
    \atknow' = (C(\lambda(u)) \cleq C(\alabel) \vee \bid \neq \usr) ~?~ (\atknow \cup \attlower{\{ n \} \cup \fnames{ck, page, s}}) : \atknow}
		{\atkstate{\alabel}{\atknow}{W}
      \exastep{\einta{\blank}{l'}}
		 \atkstate{\alabel}{\atknow'}{W''}}

		\inferrule*[lab={(A-TimeoutSend)}]
    {W \exastep{\einta{\exbsend{\bid}{n}{u}{p}{ck}{o}{l}{\amode}}{l'}} W' \\
      W' \centernot{\exastep{\einta{\exsrecv{\bid}{n}{u}{p}{ck}{o}{l}{\amode}}{l'}}} \\
      \atknow' = (C(\lambda(u)) \cleq C(\alabel)) ~?~ (\atknow \cup \attlower{\{ n \} \cup \fnames{p,ck}}) : \atknow \\
     \timeouts = \mempty \\
     \timeouts' = \{ (\bid, n, u, l, \amode) \}
   }
		{\atkstate{\alabel}{\atknow}{W}
      \exastep{\einta{\blank}{l'}}
		 \atkstate{\alabel}{\atknow'}{W'}}

		\inferrule*[lab={(A-TimeoutRecv)}]
    { \timeouts = \{ (\bid, n, u, l, \amode) \} \\
      \timeouts' = \mempty \\
      W \exastep{\einta{\exbrecv{\bid}{n}{u}{\vundef}{\mempty}{\mempty}{\mempty}{\sskip}{l}{\amode}}{l}} W' \\
    }
		{\atkstate{\alabel}{\atknow}{W}
      \exastep{\einta{\blank}{l'}}
		 \atkstate{\alabel}{\atknow'}{W'}}

		\inferrule*[lab={(A-BroAtk)}, leftskip={22pt}]
    {\timeouts = \mempty \\
      \alpha = \exbsend{\bid}{n}{u}{p}{ck}{o}{\amode}{l} \\
      W \exastep{\einta{\alpha}{l'}} W' \\
		 I(\alabel) \ileq I(\lambda(u)) \\\\
   \atknow' = (C(\lambda(u)) \cleq C(\alabel)) ~?~ (\atknow \cup \attlower{\fnames{p,ck}}) : \atknow}
		{\atkstate{\alabel}{\atknow}{W}
      \exastep{\einta{\alpha}{l'}}
		 \atkstate{\alabel}{\atknow' \cup \{n\}}{W'}}
		
		\inferrule*[lab={(A-AtkSer)}, leftskip={22pt}]
    {\timeouts = \mempty \\
		\sample{n} \\ \bid \neq \usr \\ \fnames{p,ck} \subseteq \atknow \\\\
  \alpha = \exsrecv{\bid}{n}{u}{p}{ck}{o}{\batt}{\itop} \\ W \exastep{\einta{\alpha}{\itop}} W'}
		{\atkstate{\ell}{\atknow}{W}
      \exastep{\einta{\alpha}{\itop}}
		 \atkstate{\ell}{\atknow \cup \{n\}}{W'}}
		
		\inferrule*[lab={(A-SerAtk)}]
    {\timeouts = \mempty \\
		n \in \atknow \\ 
      \alpha = \exssend{\bid}{n}{u}{u'}{\lst{v}}{ck}{\page}{s}{\amode}{l} \\\\ 
      W \exastep{\einta{\alpha}{l'}} W' \\
      \atknow' = \atknow \cup \attlower{\fnames{ck, \page, s, \lst{v}}}}
		{\atkstate{\ell}{\atknow}{W}
      \exastep{\einta{\alpha}{l'}}
		 \atkstate{\ell}{\atknow'}{W'}}
		
		\inferrule*[lab={(A-AtkBro)}, leftskip={8pt}]
    {\timeouts = \mempty \\
      \alpha = \exbrecv{\bid}{n}{u}{u'}{\lst{v}}{ck}{\page}{s}{\batt}{\itop} \\
    W \exastep{\einta{\alpha}{l}} W' \\\\
		 I(\ell) \ileq I(\lambda(u)) \\
		 \{ n \} \cup \fnames{ck, \page, s, \lst{v}} \subseteq \atknow \\
     \fvars{s} = \emptyset \todo{add this to main paper?}
   }
		{\atkstate{\ell}{\atknow}{W}
      \exastep{\einta{\alpha}{l}}
		 \atkstate{\ell}{\atknow}{W'}}	
	\end{mathpar}
\end{table*}

\subsubsection{Detailed Explanation of Changes to the Semantics of Web Systems with the attacker}

  For the proof it is required that every rule only performs a single step in a 
  browser. We hence have to split up the rule \irule{A-TimeOut} into two
  separate rules. For this reason we introduce a buffer $\timeouts$ that stores
  the request that requires the timeout-response. As long as this buffer
  contains an element, the only rule that can be taken is
  \irule{A-TimeoutRecv}.

\begin{itemize}
  \item \irule{W-LParallel}, \irule{W-RParallel} 
    and \item\irule{A-Nil}
    simply propagate the annotations.
  \item \irule{A-BrowserServer} ``forwards'' the request with the same
    annotations. We use the sync label of the browser event for the event
    in the websystem and use the integrity label of the browser event as the
    sync label for the server event. 
    This means that in some cases (for example for a load to a URL of low
    integrity) we will require that the browser step is performed in sync,
    while the server step must not be in sync, we just require that the request
    is processed in some form. For example, it is possible to match a server
    receiving a low integrity request with a case where the attacker
    interferes.
  \item \irule{A-ServerBrowser} does the same in the other direction. Again we
    use the browser event's sync integrity label for the websystem event.
    This allows us to synchronize two browsers receiving a
    low integrity a response to a load request with high sync integrity label,
    without synchronizing the server step. For example we can match a server
    responding to the request with the attacker responding to the request. 
  \item \irule{A-TimeoutSend} \irule{A-TimeoutRecv} are two individual
    rules that together equivalent to the rule \irule{A-TimeOut}. 
    In rule \irule{A=TimeoutSend} all relevant information is stored in the
    buffer $\timeouts$ so that rule \irule{A-TimeoutRecv} can send the
    corresponding response.
    Note that the integrity label and the sync integrity label may be different.
  \item \irule{A-BroAtk}``forwards'' the request with the same annotations. 
  \item \irule{A-AtkSer} sends an event labelled with low integrity and
    attacker mode $\batt$ and annotated with low integrity. 
  \item \irule{A-SerAtk} ``forwards'' the request with the same annotations. 
  \item \irule{A-AtkBro} creates a response with low integrity and attacked
    mode $\batt$. The event can have any sync integrity label -- since the
    browser may expect a different label in different situations.
\end{itemize}

We show that the original semantics and the extended semantics are equivalent
for well typed fresh web systems.
Concretely we show that they can produce the same traces.
We use here the notation for well-typed websystems $\systyping{A}$, that is formally introduced in \cref{def:sys-typing}.
\begin{lemma}[Semantic Equivalence]
  \label{lem:sem-equiv}
  Let $A$ be a fresh web system with $\systyping{A}$.
  \begin{enumerate}
    \item if for some $\lst{\alpha}, A'$ we have $A \astepn{*}{\lst{\alpha}} A'$
      then there exists $A''$ such that
      $\trans{A} \exastepn{*}{\lst{\alpha}} A''$,
    \item if for some $\lst{\alpha}, A'$ we have 
      $\trans{A} \astepn{*}{\lst{\alpha}} A'$
      then there exists $A''$ such that
      $A \exastepn{*}{\lst{\alpha}} A''$,
  \end{enumerate}
\end{lemma}
\begin{proof}
  The claim follows directly by induction over the derivation of
    ${\lst{\alpha}}$, using the following observations:
    \begin{itemize}
      \item 
        The integrity label and the attacker state are simply annotations and
        do not prevent or allow additional steps in the semantics. 
      \item The same is true for the type annotations on values, however 
        we must prevent certain joins on credential types, as they are not
        defined. \sketch{Typing ensures that these cases don't occur.}
      \item The command $\resetpc{l}$ is just modifying the integrity label of
        the thread, but is otherwise a no-op \irule{S-Reset}, so adding 
        it in \irule{T-IfTrue} and \irule{T-IfFalse} does not impact the
        behaviour of the program.
      \item The split of \irule{A-Timeout} into two separate rules does not
        impact the semantics as no other rule can be used as long as there is a
        pending timeout response in the buffer $\timeouts$.
    \end{itemize}
\end{proof}

We also present additional or modified typing rules, that allow us to type
situations occurring only at runtime.
As a convention we use $\vDash$ for the extended typing judgements, while we
use $\vdash$ for the original typing judgements.
New rules with the same name as an original rule replace that rule, 
all other original rules also become new rules without modification.
Rules with new names are additional rules.

\begin{table*}
	\caption{Extended Typing Rules}
	\label{tab:ex-typing}

\begin{mathpar}
  \inferrule*[lab={(T-EFresh)}]
    {\stype = (\typebranch = \batt) ~?~ \alabel ~:~ \tcre{\ell}} 
    {\exelabel{\fresh^{\ell}}{\stype}}
%

  \inferrule*[lab=(T-Running)]
  { \bid \neq \usr \Rightarrow \typebranch = \batt \\
    \amode = \bhon \wedge \bid = \usr  \Rightarrow \typebranch = \bhon \\ 
    \amode = \batt \wedge \bid = \usr \Rightarrow \typebranch = \bcsrf \\\\
    \envrg' = (\bid = \usr) ~?~ \envrg ~:~ \subst{\_}{\alabel} \\
    \exstypingpar{(\envu,\envv,\envrg',\envrs,\envt)}{\jlabel{j}}{l}{\typebranch}{c}{\_}}
    {\exsttypingpar{\Gamma}{\alabel}{\exthread{c}{(i,j)}{(n,u,\bid,u)}}}

  \inferrule*[lab={(T-EVal)}]
    {v \not \in \names}
    {\exelabel{v^{\stype}}{\stype}}

  \inferrule*[lab=(T-AuthAtt)]
    {b = \batt}
		{\exstyping{\auth{\lst{se}}{\ell}}} 

		\inferrule*[lab=(T-Halt)]
    {c \in \{\shalt, \sbad\}}
		{\exstyping{c}}

		\inferrule*[lab={(T-Reply)}]
		{\envu(u) = \ulabel, \vec{\stype}, l_r \\ 
		 \pclabel' = \pclabel \ijoin l_r \\
		 \envv' = x_1 \colon \stype_1, \ldots, x_{\len{se}} \colon \stype_{\len{se}} \\
		 \Gamma' = (\envu,\envv',\envrg,\envrs,\envt) \\
		 \forall k \in \intval{1}{\len{se}}. \, \exelabel{se_k}{\stype_k} \wedge \Ct{\stype_k} \ctleq C(\ulabel) \\
		 \forall r \in \domain(ck).\, \exsrtyping{r}{\tref{\stype_r}} \wedge \exelabelpar{\Gamma'}{\seslabel}{ck(r)}{\stype_r}  \wedge \pclabel' \ileq \It{\stype_r} \\
		 \typebranch \neq \batt \Rightarrow \exbstypingpar{\Gamma'}{b}{\pclabel'}{s} \\
		 \isbcsrf \Rightarrow \forall x \in vars(s). \, \isclow{\Ct{\envv'(x)}} \\
		 \isbhon \Rightarrow \pclabel \ileq l_r \wedge 
      \left(\page = \serror \vee \forall v \in \domain(\page).\, \exptyping{\Gamma'}{v}{\pclabel'}{\page(v)}\right) \\ 
		 \isilow{I(\ulabel)} \Rightarrow  \forall k \in \intval{1}{\len{se}}. \, \isclow{\Ct{\stype_k}}}
		{\exstyping{\reply{\page}{s}{ck}{\lst{x} = \lst{se}}}}

		\inferrule*[lab=(T-ReplyErr)]
		{~}
		{\exstyping{\ereply{\serror}{\sskip}{\mempty}}}

  \inferrule*[lab=(T-Reset)]
		{~}
    {\exstypingparpc{\Gamma}{\seslabel}{\pclabel}{\typebranch}{\resetpc{l}}{\seslabel}{l}}

  \inferrule*[lab={(T-BEVal)}]
    {v \not \in \names}
    {\exbelabel{v^{\stype}}{\stype}}

  \inferrule*[lab={(T-BERefFail)}]
    {\typebranch = \batt \\ (\lambda(r)) \not\cleq C(\lambda(u))}
    {\exbelabel{r}{\stype}}

  \inferrule*[lab={(T-BAssignFail)}] 
    {\typebranch = \batt \\ I(\lambda(u)) \not\ileq I(\lambda(r))}
    {\exbstyping{\Gamma}{r := be}}
\end{mathpar}
\end{table*}

\subsubsection{Detailed explanation of changes to Typing Rules}
\begin{itemize}
  \item \irule{T-EFresh} assigns the type $\alabel$ to a $\fresh$ expression if
    it is typed in the attacker's run./
  \item \irule{T-Running} allows us to type running server threads. The typing
    branch is determined based on the browser identity and the attacked mode of
    the thread.
    The typing environment for global variables is determined by the browser
    identity. If it is the honest users' browser, then the original typing
    environment is used (since the cookies come from the honest browser). 
    Otherwise, we use an environment where every type is $\alabel$.
    We then type the code of the thread, inferring the session label
    $\jlabel{j}$ from the session identifier $j$ and using the integrity label
    as $\pclabel$.
  \item \irule{T-EVal} now gives values their annotated type.
  \item \irule{T-AuthAtt} does not perform any checks for authenticated events
    when typing the attackers branch.
  \item \irule{T-Halt} trivially checks the $\shalt$ and $\sbad$ commands
    (which only occur at runtime)
  \item \irule{T-Reply} now only requires the script to be well typed if we are
    not typing the attackers branch (i.e., only if the script is sent to the
    honest user's browser) and additionally passes the URL to the typing
    judgements for scripts.
  \item \irule{T-ReplyErr} trivially checks the response with an error message.
  \item \irule{T-Reset} raises the $\pclabel$ for the continuation to the label
    provided in the reset statement.
  \item \irule{T=BEVal} now gives values their annotated type.
  \item \irule{T-BERefFail} allows us to give type any type $\stype$ 
    to a browser reference if it may not be read by the script.
    This rule (and the next one) is needed to ensure that scripts provided by
    the attacker can be typed (although they will not execute correctly).
  \item \irule{T-BAssignFail} allows us to type any assignment to a browser
    reference, if the script is not allowed to write to it.
\end{itemize}

We now show that typing with the original typing rules implies typing with
the extended rules.

\begin{lemma}[Typing Equivalence]
  \label{lem:typeeq}
  For any fresh server $S = (\mempty,\mempty,t)$,
  whenever we have $\styping{t}$ 
  then we also have $\exstyping{\trans{t}}$.
\end{lemma}
\begin{proof}
  The proof follows by induction on the typing derivation using the
    following observations:
    \begin{itemize}
      \item Every typing rule in the original system is also a typing rule in
        the extended system, with the exception of the modified rules
        \irule{T-EFresh}, \irule{T-EVal}, \irule{T-Reply}, \irule{T-BEVal}.
      \item The changes in rules \irule{T-EVal} and \irule{T-BEval}
        return the type annotations, which are according to the definition of
        $\trans{\cdot}$, $\bot$ for values $v \not \in \names$. 
        Thus the result is the same as in the original typing rule.
      \item The changes in the rule \irule{T-EFresh} and \irule{T-Reply} 
        only affect typing in the typing branch $\typebranch = \batt$, 
        which does not occur in the original type system.
        For $\typebranch \in \{\bhon, \bcsrf\}$ the rules yield the same
        result.
      \item The addition of other rules does not impact the claim
    \end{itemize}
\end{proof}

\subsection{Subject Reduction}
\label{sec:proof-sub-red}
In this section we prove subject reduction for the web system. This is needed 
to ensure that the system is always in a well-typed state, which in turn is
required to prove that our high integrity relation is preserved.

We look at typing of different components of the web system individually.
Concretely we will define typing for requests and responses, browsers, servers 
and websystems as a whole.

We start by defining well-typed requests and responses.
Then we define well-typed browsers and show that typing is preserved when the
browser takes a step, if the browser only receives well-typed responses, and show that the browser only sends out well-typed requests.
We then define well-typed servers and show that typing is preserved whenever the server takes a step, if all requests received by the server are well typed, and that all responses produced by the server are well-typed.
We furthermore show that all requests and responses produced by the attacker are well-typed.
Finally, we define well-typed web-systems and show that typing is preserved whenever the websystem takes a step.


\begin{definition}[Request Typing]
  \label{def:req-ty}
  For a request $\alpha = \exbsend{\bid}{n}{u}{p}{ck}{o}{l}{\amode}$ (resp. 
  $\alpha = \exsrecv{\bid}{n}{u}{p}{ck}{o}{l}{\amode}$)
  with $\envu(u) = \ulabel, \lst{\stype}, l_r$ 
  we have $\reqtyping{\alpha}$ if
  \begin{enumerate}
    \item \label{def:brhp} if $\amode = \bhon $ and $ \bid = \usr$ then 
      \begin{itemize}
        \item for all $k \in \domain(p)$ we have if $p(k)=v_k^{\stype_k'}$ then
          $\stype_k' \tleq \stype_k$ 
        \item $l \ileq I(\ulabel)$
      \end{itemize}
    \item \label{def:brap} if $\amode = \batt$ 
      then 
      \begin{itemize}
        \item for all $k \in \domain(p)$ we have if $p(k)=v_k^{\stype_k'}$ then
          $\stype_k' \tleq \alabel$
        \item $l \tleq I(\alabel)$ 
      \end{itemize}
    \item \label{def:brhc} if $\bid=\usr$ 
      \begin{itemize}
        \item for all $c \in \domain(ck)$
         we have 
         \begin{itemize}
           \item if $ck(c)=v_c^{\stype_c}$ then $\stype_c \tleq \treft{\envrg(c)}$ 
           \item $C(\lambda(r)) \tleq C(\lambda(u))$
         \end{itemize}
      \end{itemize}
    \item \label{def:brac} if $\bid \neq \usr$ then for all $c \in \domain(ck)$
      we have if $ck(c)=v_c^{\stype_c}$ then 
      $\stype_c \tleq \alabel$
    \item \label{def:bro} If $\bid = \usr$,  $u \in \protUrls$ and 
      $o \neq \noorigin$ and $\isihigh{o}$ then $\amode = \bhon$.
  \end{enumerate}
\end{definition}

Intuitively, according to \cref{def:req-ty} a request is well-typed, if
\begin{enumerate}
  \item for all honest requests, all parameter types are respected and the
    integrity label is higher than the integrity label of the URL.
  \item  for all attacked requests, all parameters are of the attacker's type
    and the integrity is low.
  \item For all (attacked and honest) requests from the users browser, all
    cookies respect their type from the environment and their confidentiality
    is as most as high as the one of the URL.
  \item For all requests by the attacker, all cookies have the type of the
    attacker.
  \item Any request with a high integrity origin to a protected URL must be
    honest. 
\end{enumerate}

We now in a similar fashion define well-typed responses.

\begin{definition}[Response Typing] 
  \label{def:res-ty}
  For a response $\alpha = \exssend{\bid}{n}{u}{u'}{\lst{v}}{ck}{page}{s}{l}{\amode}$
  (resp.  $\alpha = \exbrecv{\bid}{n}{u}{u'}{\lst{v}}{ck}{page}{s}{l}{\amode}$)
  with $\envu(u) = \ulabel, \lst{\stype}, l_r$ 
  we have $\reqtyping{\alpha}$ if 
  \begin{enumerate}
    \item \label{def:srhn} For all $v^{\stype} \in \vals{ck,page,s,\lst{v}}$ we have $\Ct{\stype} \cleq C(\ulabel)$ 
    \item \label{def:sra} If $\bid \neq \usr$, then for all $v^{\stype} \in \vals{ck,page,s,\lst{v}}$ we have $\stype \tleq \alabel$
    \item \label{def:srcs} if $\bid = \usr$ and $\amode = \batt$ then
    for all $u''$ with $\isilow{I(\lambda(u'')))}$ we have $\exbstypingparu{\Gamma}{\batt}{\itop}{u''}{s}$ and 
      $\isilow{l}$
    \item \label{def:src} if $\bid = \usr$ then for all 
      $r \in \domain(ck)$ with $ck(r) = v^{\stype}$ we have
      \begin{itemize}
        \item If $\lambda(u) \ileq \lambda(r)$ then $\stype \tleq \treft{\envrg(r)} $ and $l \ileq \It{\treft{\envrg(r)}}$ 
        \item If $\lambda(u) \not\ileq \lambda(r)$ then $\stype \tleq \treft{\envrg(r)}$ or $\stype \tleq \alabel$
      \end{itemize}
    \item \label{def:srhp} if $\bid = \usr$ and $\isihigh{l}$ then
      $page = \serror$ or
        for all $v \in \domain(page)$ with $page(v) = \hform{u''}{\lst{v^\stype}}$ we have 
      \begin{itemize}
        \item $\envu(u'') = \envt(v)$
        \item with $\envu(u'') = \ulabel', \lst{\stype'}, l_r'$,
          \begin{itemize}
            \item for all $i \in \intval{1}{\len{v}}$ we have $\stype_i \tleq \stype_i'$
            \item $l \ileq \ulabel'$
          \end{itemize}
      \end{itemize}
    \item \label{def:srap} if $\bid = \usr$ and 
      $\isilow{l}$ 
      then $page = \serror$ or we have one of the following
      \begin{itemize}
        \item for all  $v \in \domain(page)$ with $page(v) =
          \hform{u''}{\lst{v}}$, for all $i \in \intval{1}{\len{v}}$ we have
          $\stype_i \tleq \alabel$
        \item or $\isihigh{u}$
      \end{itemize}
    \item \label{def:srhs}$\bid = \usr$ and $\amode = \bhon$ then
      $\exbstypingpar{\Gamma}{\bhon}{l_r}{s}$ and
      $l = l_r$ 
    \item \label{def:srru}if $u' \neq \vundef$ and $\bid = \usr$ then with
      $\alpha' = \exbsend{\bid}{n'}{u'}{p}{\mempty}{\noorigin}{l}{\amode}$, for
      any $n'$ and $\forall k \in \intval{1}{\len{v}}: p(k) = v_k$ 
      we have $\reqtyping{\alpha'}$.
  \end{enumerate}
\end{definition}

Intuitively, according to \cref{def:res-ty} a response is well-typed if 
\begin{enumerate}
  \item The confidentiality label of all values contained in the response is 
    at most as high as the confidentiality label of the URL from which the
    response is sent, or the confidentiality is low. 
  \item If the response it not sent to the honest user, then all values must be
    of low confidentiality.
  \item If the response is sent to the honest user and influenced by the
    attacker, then the integrity label is low and the contained script is
    well-typed, using the type branch $\batt$.
  \item For all responses to the honest users, if a cookie may be set by the
    response, then it respects the typing environment (also taking into account
    the integrity label of the response). If the cookie may not be set by the 
    response, then it respects the typing environment or is low.
  \item For all honest responses we have that the page is either the error
    page, or that it is well-typed, i.e., the type of the form name matches the
    type of the URL and all parameters respect the URL type and that the 
    integrity of the current thread is high enough to trigger a request to that
    URL.
  \item For all attacked responses to the honest user,  we have that the page
    is the error page or one of the following holds :
    \begin{itemize}
      \item all parameters contained in the DOM are of type $\alabel$
      \item or the response comes from a high integrity URL (in which case we
          do not make any assumption on the DOM, since the user will not
          interact with it)
    \end{itemize}
  \item For all honest responses, the script is well typed with $\pclabel$ set
    to the expected response integrity of the URL, and the integrity of the
    response must be equal to that label
  \item If the redirect URL is not empty, and the response is sent to the
    honest user's browser, then we know that the request that will result from 
    processing the response at the browser is well typed (using an empty set of
    cookies and an empty origin as placeholders).
\end{enumerate}

\begin{definition}[Browser Typing]
  \label{def:bro-typing}
  Let $B=\exbrowser{N}{M}{P}{T}{Q}{\lst{a}}$ be a browser. We write
  $\browtyping{B}$, if $\bid = \usr$ and 
    \begin{enumerate}
      \item \label{def:btL} $\amode = \batt \Rightarrow \isilow{l}$ 
      \item \label{def:btM} $\forall r \in \domain(M), M(r)= v^{\stype}
        \wedge \stype = \treft{\envrg(r)}$ 
      \item \label{def:btP} For all $tab \in \domain(P)$ with 
         $P(tab) = (u,page,l',\amode')$ and $page \neq \serror$  we have 
         for all $v \in \domain(page)$ with 
         $page(v) = \hform{u'^{\stype_{u'}}}{\lst{v^{\stype}}}$
         \begin{itemize}
           \item $\isclow{\Ct{\stype_{u'}}}$ 
           \item if $\isihigh{l'}$ and $\envu(u') = \ulabel, \lst{\stype'}, l_r$ then
             \begin{itemize}
               \item $\amode = \bhon$
               \item $\envu(u') = \envv(v)$
               \item $ l' \ileq I(\ulabel) $ 
               \item $\forall i \in \intval{1}{\len{v}}. \, \stype_i \tleq \stype_i'$ 
             \end{itemize}
           \item if $\isilow{l'}$ then 
                 one of the following holds
                 \begin{itemize}
                   \item $\forall i \in \intval{1}{\len{v}}.
                     \isclow{\Ct{\stype_i}}$ 
                   \item $\isihigh{I(\lambda(u))}$
                 \end{itemize}
         \end{itemize}
      \item \label{def:btT} If $T = \subst{tab}{s}$ and $P(tab) = (u,page,l',\amode')$ with 
         $\envu(u) = \ulabel, \lst{\stype}, l_r$ then 
         \begin{itemize}
           \item $l = l'$
           \item if $\amode = \bhon$ then $l' \ijoin l \ileq l_r$ and
             $\exbstypingpar{\Gamma}{\bhon}{l_r}{s}$
           \item if $\amode = \batt$ then
             $\exbstypingpar{\Gamma}{\batt}{\itop}{s}$ 

         \end{itemize}
       \item \label{def:btQ} If $Q = \{\einta{\alpha}{l'}\}$, 
         then we have $ \reqtyping{\alpha}$. 
       \item \label{def:btA} For $\lst{a}$ we have
       \begin{itemize}
         \item for the navigation flow
           \begin{itemize}
         \item for every navigation flow $\lst{a}\,'$ in $\lst{a}$, we have
           that $\isilow{I(\lambda(a_j'))}$ implies $
           \isilow{I(\lambda(a_k')})$ for all $j < k \leq \len{a'}$.
         \item If $N = \subst{n}{(tab,u,o)}$ and $T =\mempty$ then
           we have that for all $a \in \nf{\lst{a}}{tab}$ that
           $\isilow{l}$ implies $\isilow{I(\lambda(a))}$.
           Furthermore we have $\isilow{I(\lambda(a_j'))}$ implies $
           \isilow{I(\lambda(a_k')})$ for all $j < k \leq \len{a'}$.
         \item for all $tab \in \domain(P)$ with $P(tab) = (u,page,l',\amode')$
           and $N \neq \subst{n_N}{(tab,u_N,o_N)}$ for all $n_N, u_N, o_N$,
           we have that for all $a \in \nf{\lst{a}}{tab}$ that 
           ($\isilow{I(\lambda(u))}$ or $\amode' = \batt$)
           implies $\isilow{I(\lambda(a))}$.  
           Furthermore we have $\isilow{I(\lambda(a_j'))}$ implies $
           \isilow{I(\lambda(a_k')})$ for all $j < k \leq \len{a'}$.
           \end{itemize}
         \item for all actions $a'$ in $\lst{a}$ we have:
         \begin{itemize}
           \item if $a' = \load{tab}{u}{p}$ and $\envu(u) = \ulabel,
             \lst{\stype}, l_r$ then for all $k \in \domain(p)$ we have that if
             $p(k) = v^{\stype'}$ then $\stype' \tleq \stype_k$;
           \item if $a' = \submit{tab}{u}{v'}{p}$ and $\envt(v') = \ulabel,
             \lst{\stype}, l_r$ then for all $k \in \domain(p)$ we have that if
             $p(k) = v^{\stype'}$ then 
             \begin{itemize}
               \item 
                 $\stype' \tleq \stype_k$.
               \item if $\isilow{\lambda(u)}$ then additionally $\isclow{\Ct{\stype'}}$
             \end{itemize}
           \end{itemize}
       \end{itemize}
     \item \label{def:btI} If $N = \subst{n}{(tab,u,o)}$ and 
       $T = \subst{tab}{s}$ and 
      $\amode = \bhon$ then 
        \begin{itemize}
          \item if $P(tab) = (u',page,l',\amode')$ and $\envu(u) = \ulabel,
            \lst{\stype}, l_r$ and $\envu(u') = \ulabel', \lst{\stype}',
            l_r'$ then $l_r = l_r'$. 
          \item $\isihigh{I(\lambda(u))}$
       \end{itemize}
    \end{enumerate}
\end{definition}

Intuitively, according to \cref{def:bro-typing} a browser is well-typed, if all its components are well-typed.
Concretely, we require that:
\begin{enumerate}
  \item Whenever the state of the browser is directly influenced by the
    attacker, then the integrity of the browser is low.
  \item All values stored in a memory reference have a type annotation that is
    equal to the type of the reference the typing environment.
  \item For any non empty DOM in a tab,
    \begin{itemize}
      \item If the DOM is of high integrity
        \begin{itemize}
          \item The DOM is honest
          \item The type of the form name matches the type of the URL
          \item The integrity label of the DOM is higher than the integrity
            label of the URL 
          \item All parameters have the expected type.
        \end{itemize}
      \item If the DOM is low integrity
        \begin{itemize}
          \item and either
            \begin{itemize}
              \item All parameters have the attacker's type.
              \item or the integrity of the DOM's origin is high
            \end{itemize}
        \end{itemize}
    \end{itemize}
  \item If a script is running in a tab
    \begin{itemize}
      \item the script integrity is equal to the integrity of the DOM in that
        tab.
      \item if the browser is not attacked then the browser's integrity is
        equal the integrity of the expected response type for the URL of the
        DOM in the same tab and the script code is well-typed in the honest
        typing branch using the expected response type as the $\pclabel$.  
      \item if the browser is attacked, then the script is well-typed using
        $\itop$ label as $\pclabel$. 
    \end{itemize}
  \item All requests in the buffer are well typed.
  \item For all user actions we have that
    \begin{itemize}
      \item The user will not submit forms on high integrity pages
        after ``tainting'' the connection, by visiting a low integrity page.
        Concretely the conditions are the following:
        \begin{itemize}
          \item The first condition is exactly the assumption we make on
            well-formed user actions.
          \item The second condition is the same, but taking into account open
            network connections for load or submits. 
          \item The third condition is similarly taking into account pages
            already loaded in browser tabs for the navigation flow. 
            However it is less strict, as it uses the attacked state of the
            page instead of the integrity labels of previously visited pages.
            Concretely, this would allow navigation of high integrity pages
            even after visiting low integrity pages, as long as there has not
            been a direct influence by the attacker.
        \end{itemize}
      \item The user's inputs respect the parameter types and will only input low confidentiality values in forms present in low integrity pages..
    \end{itemize}
  \item Whenever the browser is in an honest state, has a script running in the
    context of URL $u$ and is waiting for the response of a script inclusion
    from URL $u'$, then 
    \begin{itemize}
      \item the two URLs have the same expected response type.
      \item the URL $u'$ is of high integrity.
    \end{itemize}
\end{enumerate}

We now prove that whenever a browser expression containing variables is well
typed in a typing environment, then it is also well-typed if we substitute
the variables with concrete values of the expected type.

\begin{lemma}[Browser Expression Substitution]
  \label{lem:bro-ex-sub}
  Whenever we have $\exbelabel{be}{\stype}$ and we have a substitution $\sigma$ with 
  $\domain(\sigma) = \domain(\envv)$ and 
  $\forall x \in \domain(\envv). \, \sigma(x) = v_x^{\stype_x}$ with 
  $\stype_x \tleq \envv(x)$ then for all
  $\envv'$, we have  
  $\exbelabelpar{(\envu, \envv', \envrg,\envrs,\envt)}{b}{be\sigma}{\stype}$,
\end{lemma}
\begin{proof}
  We perform an induction on the typing derivation of $\exbelabel{be}{\stype}$:
  \begin{itemize}
    \item \irule{T-BEVar}. Then $be = x$ and $be \sigma  = v_x^{\stype_x}$ with
      $\stype_x \tleq \envv(x)$.
      The claim follows directly from rule \irule{T-BVal} and \irule{T-BSub}.
    \item \irule{T-BERef}. Then $be = r =  be\sigma$ and the claim is trivial.
    \item \irule{T-BEVal}. Then $be = v^{\stype_v} =  be\sigma$ and the claim is trivial.
    \item \irule{T-BEUndef}. Then $be = \vundef =  be\sigma$ and the claim is trivial.
    \item \irule{T-BEName}. Then $be = n^{\stype_n} =  be\sigma$ and the claim is trivial.
    \item \irule{T-BEDom}. Then $be = \dom{be_1}{be_2}$ and $be\sigma =
      \dom{be_1\sigma}{be_2\sigma}$ and the claim follows immediately using
      \irule{T-BEDom}.
    \item \irule{T-BEBinOp} Then $be = be_1 \odot be_2$ with
      $\exbelabel{be_1}{\stype_1}$ and $\exbelabel{be_2}{\stype_2}$ and 
      $\stype = \typelabel{\stype_1} \sqcup \typelabel{\stype_2}$.
      We also have $be\sigma = be_1\sigma \odot be_2\sigma$. By induction we
      know that $\exbelabel{be_1\sigma}{\stype_1'}$ and
      $\exbelabel{be_2\sigma}{\stype_2'}$ 
      with $\stype_1' \tleq \stype_1$ and $\stype_2' \tleq \stype_2$.
      We then know that 
      $\typelabel{\stype_1'} \sqcup \typelabel{\stype_2'} \tleq \typelabel{\stype_1} \sqcup \typelabel{\stype_2} = \stype$, and
      the claim follows by \irule{T-BinOp} and \irule{T-BESub}.
    \item \irule{T-BESub} follows by induction and by the transitivity of $\tleq$.
  \end{itemize}
\end{proof}

Next, we prove the same claim on the level of scripts.

\begin{lemma}[Browser Substitution]
  Whenever we have $\exbstyping{\Gamma}{s}$ and we have a substitution $\sigma$
  with 
  $\domain(\sigma) = \domain(\envv)$ and 
  $\forall x \in \domain(\envv). \, \sigma(x) = v_x^{\stype_x}$ with 
  $\stype_x \tleq \envv(x)$ then for all
  $\envv'$, we have 
  $\exbstyping{(\envu, \envv', \envrg,\envrs,\envt)}{s\sigma}$.
\end{lemma}
\begin{proof}
  We do the proof by induction on the typing derivation.
  \begin{itemize}
    \item \irule{T-BSeq}: Then $s = s_1,s_2$. The claim follows by applying the 
      induction hypothesis to $s_1$ and $s_2$ and applying rule \irule{T-BSeq}.
    \item \irule{T-BSkip}: The claim follows trivially.
    \item \irule{T-Bassign}: Then we have $s = r:= be$, with
      \begin{itemize}
        \item $\exbrtyping{r}{\tref{\stype}}$ 
        \item $\exbelabel{be}{\stype}$
        \item $\pclabel \ileq \It{\stype}$
      \end{itemize}
      Using \cref{lem:bro-ex-sub} and \irule{T-BESub}, we get
      $\exbelabel{be\sigma}{\stype}$ and the claim follows immediately.
    \item \irule{T-BSetDom}: Then $s = \setdom{v}{u}{\lst{be}}$. 
      The claim follows by applying of \cref{lem:bro-ex-sub} and
      \irule{T-BESub} for every $be_i$ in $\lst{be}$.
    \item \irule{T-BInclude}: Then $s = \incl{u}{\lst{be}}$,
      The claim follows by applying of \cref{lem:bro-ex-sub} and
      \irule{T-BESub} for every $be_i$ in $\lst{be}$.
  \end{itemize}
\end{proof}

Now, we show that typing is preserved under the evaluation of expressions.

\begin{lemma}[Browser Expression Typing]
  \label{lem:bro-ex-ty}
  Let $B=\exbrowser{N}{M}{P}{T}{Q}{\lst{a}}$ be a browser with $\browtyping{B}$.
  Let $T = \subst{tab}{s}$ and 
  $\subst{tab}{(u,f,l',\amode')} \in P$, $\ell= \lambda(u)$.
  Then for any browser expression $be$, if
  $\exbelabelpar{\Gamma}{\amode}{be}{\stype}$ then 
  $\exbelabelpar{\Gamma}{\amode}{\beval{be}{M}{f}{\ell}}{\stype}$
\end{lemma}
\begin{proof}
  Let $\beval{be}{M}{f}{\ell} = v^{\stype'}$. 
  We show $\stype' \tleq \stype$ and the claim follows using rule
  \irule{T-BESub}.
  We perform the proof by induction over the expression $be$:
  \begin{itemize}
    \item $be = x$: In this case, $\beval{se}{M}{f}{\ell}$ is undefined, so we
      don not have to show anything.
    \item $be = v^{\stype}$ We have $\beval{v^\stype}{M}{f}{\ell} = v^{\stype}$ and the claim is trivial.
    \item $be = be_1 \odot be_2$: By induction we know
      \begin{itemize}
        \item 
      $\exbelabelpar{\Gamma}{\amode}{be_1}{\stype_1}$ and 
      $\beval{be_1}{M}{f}{\ell} = v^{\stype'_1}_1$ and 
      $\stype_1'  \tleq \stype_1$
        \item 
      $\exbelabelpar{\Gamma}{\amode}{be_2}{\stype_2}$ and 
      $\beval{be_2}{M}{f}{\ell} = v^{\stype'_2}_2$ and 
      $\stype_2'  \tleq \stype_2$
      \end{itemize}
      Let now  $v^{\stype'} = v^{\stype'_1}_1 \odot v^{\stype'_2}_2$. 
      Then we know 
      that $\stype' = \typelabel{\stype'_1} \sqcup \typelabel{\stype'_2}$ by rule 
      \irule{BE-BinOp}.
      By rule \irule{T-BEBinOp} we have 
      $\stype = \typelabel{\stype_1} \sqcup \typelabel{\stype_2}$, 
      and the claim follows.
    \item $be = r$: then the claim immediately follows from rule \irule{T-BERef}
      and property \ref{def:btM} of \cref{def:bro-typing}.
    \item $be = \dom{be_1}{be_2}$: 
      We know by property \ref{def:btT} of \cref{def:bro-typing} $l = l'$
      We distinguish two cases:
      \begin{itemize}
        \item If $\isihigh{l'}$  
          then we know  that $\amode = \bhon$ and hence this case
          is impossible, since we do not have a typing rule for the
          expression in the honest type branch.
        \item If $\isilow{l'}$, then we distinguish two cases:
          \begin{itemize}
            \item if $\isihigh{I(\lambda(u))}$ then \sketch{we know that the
                script can also be typed with $\typebranch = \bhon$, and hence 
              this case is impossible.}
            \item if $\isilow{I(\lambda(U))}$ then by rule
          \irule{BE-Dom} the value is either a URL parameter or the URL itself.
          we then know that for all parameters $v^{\stype'}$ of any URL in the
          DOM we have $\isclow{\Ct{\stype'}}$.  For any URL $u^{\stype_u}$ we
          have $\isclow{\Ct{\stype_u}}$ and the claim holds.
          \end{itemize}k
      \end{itemize}
  \end{itemize}
\end{proof}

We now show subject reduction for the browser for internal steps i.e., whenever
a well-typed browser takes a step, it results in another well-typed browser.
We treat browsers sending requests and receiving responses in separate lemmas.

\begin{lemma}[Browser Subject Reduction]
  \label{lem:bro-sub-red}
  Let $B,B'$ be browsers with $\browtyping{B}$ 
  such that $B \bstep{\einta{\blank}{\_}}{B'}$. 
  Then we have $\browtyping{B'}$.
\end{lemma}
\begin{proof}
  Let $B=\exbrowserpar{N}{M}{P}{T}{Q}{\lst{a}}{\amode}{l}$ and
  $B'=\exbrowserpar{N'}{M'}{P'}{T'}{Q'}{\lst{a'}}{\amode'}{l'}$ be browsers as in the lemma.
  We know that $\bid = \usr$ and do a proof by induction on the step taken.
  We show that all properties of \cref{def:bro-typing} hold for $B'$.
  \begin{itemize}
     \item \irule{B-Load}:  
       \begin{itemize}
         \item Property \ref{def:btL} is trivial, since $\amode' = \bhon$.
         \item Property \ref{def:btM} is trivial, since $M = M'$
         \item Property \ref{def:btP} is trivial, since $P = P'$
         \item Property \ref{def:btT} is trivial, since $T = \mempty$.
         \item For property \ref{def:btQ} we have $Q' = \{\alpha\}$ with 
          $\alpha= \exbsend{\bid}{n}{u}{p}{ck}{\noorigin}{I(\lambda(u))}{\bhon}$
           and hence have to show that $\reqtyping{\alpha}$. 
           We show that all the properties of \cref{def:req-ty} are fulfilled.
           \begin{itemize}
             \item Property \ref{def:brhp} follows immediately from property
               \ref{def:btA} of \cref{def:bro-typing} for $B$
             \item Property \ref{def:brap} is trivial since we have $\amode = \bhon$
             \item Property \ref{def:brhc} follows immediately from property
               \ref{def:btM} of \cref{def:bro-typing} for $B$  and the definition of 
               $\getck{\cdot}{\cdot}$.
             \item Property \ref{def:brac} is trivial since $\bid = \usr$
             \item Property \ref{def:bro} is trivial since the origin $o =  \noorigin$.
           \end{itemize}
         \item Property \ref{def:btA} for $B'$ follows directly from property
           \ref{def:btA} of \cref{def:bro-typing} for $B$. 
           \sketch{The navigation flow started by the load action is the same
           as $\nf{\lst{a}}{tab}$ } 
         \item Property \ref{def:btI} is trivial since $T' = \mempty$ 
       \end{itemize}

     \item \irule {B-Include}
           \begin{itemize}
             \item Property \ref{def:btL} is trivial, since $\amode' = \amode$ and $l = l'$
             \item Property \ref{def:btM} is trivial, since $M = M'$
             \item Property \ref{def:btP} is trivial, since $P = P'$
             \item Property \ref{def:btT} is trivial using rule \irule{T-BSkip}, since $T = \subst{tab}{\sskip}$ 
             \item For property \ref{def:btQ} we have $Q' = \{\alpha\}$ with 
              $\alpha = \exbsend{\bid}{n}{u}{p}{ck}{\uorigin{u'}}{l \ijoin I(\lambda(u))}{\amode'}$
               and hence have to show that $\reqtyping{\alpha}$. 
               We show that all the properties of \cref{def:req-ty} are fulfilled.
               \begin{itemize}
                 \item For property \ref{def:brhp} We distinguish two cases: 
                   \begin{enumerate}
                     \item if $\amode = \bhon$ then it  follows from property
                       \ref{def:btT} of \cref{def:bro-typing} for $B$ using
                       rule \irule{T-BInclude} and \cref{lem:bro-ex-ty}
                     \item if $\amode = \batt$ then the claim is trivial
                   \end{enumerate}
                 \item For property \ref{def:brap} We distinguish two cases: 
                   \begin{enumerate}
                     \item if $\amode = \bhon$ then the claim is trivial
                     \item if $\amode = \batt$ then it  follows from property
                       \ref{def:btT} of \cref{def:bro-typing} for $B$ using
                       rule \irule{T-BInclude} and \cref{lem:bro-ex-ty}
                   \end{enumerate}
                 \item Property \ref{def:brhc} follows immediately from property
                   \ref{def:btM} of \cref{def:bro-typing} for $B$ 
                 \item Property \ref{def:brac} is trivial since $\bid = \usr$
                 \item For property \ref{def:bro} we perform a case distinction:
                   \begin{itemize}
                     \item If $u \not\in \protUrls$, $\isilow{\uorigin{u'}}$ or 
                       $\amode' = \bhon$ then the claim is trivial.
                     \item If $u \in \protUrls$, $\isihigh{\uorigin{u'}}$ and 
                       $\amode' = \batt$ then 
                       \sketch{
                         assume that the include statement is contained in the 
                         script $s_{u'}$ served by $u'$. Since $u'$  is of high
                         integrity, we know that the script code can be typed
                         with $\typebranch = \bhon$. This in particular implies
                         that every include statement in the script also has
                         been typed with $\typebranch = \bhon$.  Hence we know
                         by rule \irule{T-BInclude} that $u \not \in \protUrls$
                         and we have a contradiction.
                         If the include statement is not contained in the
                         script $s_{u'}$ served by $u'$, then it must be
                         contained in the script $s_{u''}$ served from some URL
                         $u''$ that is included by the script $s_{u'}$. 
                         Using the same argumentation, we know by rule
                         \irule{T-BInclude} that $\isihigh{I(\lambda(u''))}$
                         and again using the same argumentation we get the
                         contradiction $u \not \in \protUrls$}
                   \end{itemize}
               \end{itemize}
             \item Property \ref{def:btA} of \cref{def:bro-typing} for $B'$
               follows directly from property
               \ref{def:btA} for $B$. 
             \item Property \ref{def:btI} follows from property \ref{def:btT}
               of \cref{def:bro-typing} for $B$ using rule \irule{T-BInclude}
           \end{itemize}

    \item \irule{B-Submit}
        Then we have 
          \begin{itemize}
            \item $a = \submit{tab}{u}{v}{p'}$ 
            \item $ \subst{\tab}{(u, f, l',  \amode')} \in P $
            \item $ \subst{v}{\hform{u'}{\lst{v^\stype}}} \in f $
            \item $ \forall k \in \intval{1}{\len{v}}.\, p(k) = k \in \domain(p') ~?~ p'(k) : v_k^{\stype_k}$
          \end{itemize}
           \begin{itemize}
             \item Property \ref{def:btL} follows from property \ref{def:btP} of \cref{def:bro-typing} for $B$.
             \item Property \ref{def:btM} is trivial, since $M = M'$
             \item Property \ref{def:btP} is trivial, since $P = P'$
             \item Property \ref{def:btT} is trivial, since $T = \mempty$ 
             \item For property \ref{def:btQ} we have $Q' = \{\alpha\}$ with 
               $\alpha = \exbsend{\bid}{n}{u'}{p}{ck}{\uorigin{u}}{l' \ijoin I(\lambda(u'))}{\amode'}$ 
               and hence have to show that $\reqtyping{\alpha}$. 
               We show that all the properties of \cref{def:req-ty} are fulfilled.
               \begin{itemize}
                 \item For property \ref{def:brhp} we distinguish two cases:
                   \begin{enumerate}
                    \item if $\amode' = \bhon$ then it follows from property
                      \ref{def:btP} and \ref{def:btA} of \cref{def:bro-typing}
                      for $B$
                    \item if $\amode' = \batt$ then we distinguish two cases:
                      \begin{itemize}
                        \item if $\isilow{I(\lambda(u))}$ then the claim is trivial
                        \item otherwise, we know from property \ref{def:btA}
                          that $\isilow{\lambda(a)}$. By the definition of
                          $\lambda$ we get $\lambda(a) = \lambda(u)$ which is a
                          contradiction to our assumption. Hence this case
                          cannot happen.
                      \end{itemize}
                   \end{enumerate}
                 \item For property \ref{def:brap} we distinguish two cases:
                   \begin{enumerate}
                    \item if $\amode' = \bhon$ the claim is trivial
                    \item if $\amode' = \batt$ then it follows from property
                      \ref{def:btP} and \ref{def:btA} of \cref{def:bro-typing}
                      for $B$
                   \end{enumerate}
                 \item Property \ref{def:brhc} follows immediately from property
                   \ref{def:btM} of \cref{def:bro-typing} for $B$ and
                   \cref{lem:bro-ex-ty} 
                 \item Property \ref{def:brac} is trivial since $\uid = \usr$
                 \item For property \ref{def:bro} we perform a case distinction:
                   \begin{itemize}
                     \item If $u' \not\in \protUrls$, $\isilow{\uorigin{u}}$ or 
                       $\amode' = \bhon$ then the claim is trivial.
                     \item If $u' \in \protUrls$, $\isihigh{\uorigin{u}}$ and 
                       $\amode' = \batt$ 
                       then we know $\isihigh{\lambda(a)}$.
                       We then get by property \ref{def:btA} of
                       \cref{def:bro-typing} for $B$ that
                       $\isilow{\uorigin{u}}$ or $\amode = \bhon$ and
                       immediately have a contradiction.
                     \end{itemize}
               \end{itemize}
             \item Property \ref{def:btA} of \cref{def:bro-typing} for $B'$
               follows from property \ref{def:btA} for $B$, \sketch{since 
               request from low integrity pages, are also of low integrity and
               since high integrity pages do not include low integrity pages
               (by \irule{T-Form}.}
             \item Property \ref{def:btI} is trivial since $T = \mempty$.
           \end{itemize}
    \item \irule{B-Seq}
      Then $T = \subst{tab}{s}$ 
      with $s = s_1;s_2$ and from \irule{B-BSeq} we know
      $\exbstypingpar{\Gamma}{\amode}{l_r}{s_2}$.
      We apply the induction hypothesis for the browser stepping from script
      $s_1$ to $s_1'$. This immediately gives us all properties from
      \cref{def:bro-typing} except the typing of the script
      $\exbstypingpar{\Gamma}{\amode}{l_r}{s_1' ; s_2}$, but this claim
      follows immediately by applying rule \irule{T-BSeq}.
    \item \irule{B-Skip} 
      Then $T = \subst{tab}{s}$ with $s = \sskip; s'$
      By rule \irule{T-BSeq} we have 
      $\exbstypingpar{\Gamma}{\amode}{l_r}{s'}$. Since nothing besides the
      script changes, the claim follows immediately.
    \item \irule{B-End} 
      \begin{itemize}
        \item Property \ref{def:btL} is trivial since $\amode' = \bhon$
        \item Property \ref{def:btM} is trivial since $M = M'$
        \item Property \ref{def:btP} is trivial since $P = P'$
        \item Property \ref{def:btT} is trivial since $T =\mempty$
        \item Property \ref{def:btQ} is trivial since $Q =\mempty$
        \item Property \ref{def:btA} is trivial since $\lst{a} =\lst{a'}$,
          $P = P'$ and $N = N'$
        \item Property \ref{def:btI} is trivial since $M = \mempty$.
      \end{itemize}
      is trivial, since the only change from $B$ to $B'$ is that 
      $T' = \mempty$, in which case we don't have to show anything for the
      script.
    \item \irule{B-SetReference} 
      Then $T = \subst{tab}{s}$ with $s = r := be$.
      We have $P = P'$ and claim
      \ref{def:btP}  of \cref{def:bro-typing} is trivial and
      since $T' = \subst{tab}{\sskip}$ claim
      \ref{def:btT} follows immediately from rule \irule{T-BSkip}.
  
		  By rule \irule{B-SetReference} we have 
      \begin{itemize}
        \item $\subst{\tab}{(u, f,l',\amode')} \in P$
        \item $\ell = \lambda(u)$
        \item $\exbeval{be}{M}{f}{l'}{\ell} = v^\stype$
        \item $M' = M \subst{r}{v^{\stype_r}}$
          with $\stype_r = \stype \tjoin \treft{\envrg(r)} \tijoin l$
      \end{itemize}

      All properties of \cref{def:bro-typing} except for
      property \ref{def:btM}  are trivial.

      For property \ref{def:btM} it is sufficient to show that 
      $\stype_r \tleq \treft{\envrg(r)}$.

      By rule \irule{T-BAssign} and rule \irule{T-BRef} we get that
      \begin{itemize}
        \item $\exbelabel{be}{\treft{\envrg(r)}}$ 
        \item $l \ileq \It{\treft{\envrg(r)}}$
      \end{itemize}

      By \cref{lem:bro-ex-ty} we get $\stype \tleq \treft{\envrg(r)}$. 
      We hence get $\stype_r = \treft{\envrg(r)}$
      and the claim follows.

    \item \irule{B-SetDom}
      Then $T = \subst{tab}{s}$ 
      with $s = \setdom{be}{u}{\lst{be}}$. 

      All properties of \cref{def:bro-typing} except for
      property \ref{def:btP}  are trivial, so we only show this one.

      We assume the following setting analog to rule \irule{B-SetDom}
      \begin{itemize}
        \item $P = P_0 \uplus \subst{tab}{(u',f,l'',\amode'')}$.
        \item $\ell = \lambda(u')$ 
		    \item $\beval{be'}{M}{f}{\ell} = v'$
        \item $\forall k \in \intval{1}{\len{be}}.\, v_k'^{\stype_k'} =
          \beval{be_k}{M}{f}{\ell} \wedge 
          v_k^\stype = v_k'^{\stype' \tijoin l}$
        \item $\amode''' = (\amode = \batt \vee \amode'' = \batt) ~?~ \batt ~:~ \bhon$
      \end{itemize}

      Then  $P' = P_0 \uplus \subst{tab}{(u',f\subst{v'}{\hform{u^{\pair{\cbot}{l}}}{\lst{v^\stype}}},l'' \ijoin l,\amode''')}$.
      We now do a case analysis:
      \begin{itemize}
        \item $\isihigh{l''}$: Then by property \ref{def:btT} of
          \cref{def:bro-typing}  we know $l = l''$ and hence $l \ijoin l' = l''$
          We now need to show that with $\envu(u) = \ulabel, \lst{\stype_u},
          l_r$  
          \begin{enumerate}
            \item $\envu(u) = \envv(v)$
            \item $l \ijoin l''  \ileq I(\ulabel)$ and 
            \item $\forall i \in \intval{1}{\len{v}}. \, \stype_i \tleq {\stype_u}_i$
            \item $\amode''' = \bhon$
          \end{enumerate}
          $(1)$ follows immediately from rule \irule{T-BSetDom},

          From \cref{def:bro-typing}, we know by property \ref{def:btP} 
          that $l'' \ileq I(\ulabel)$ and by property \ref{def:btT} we know
          with $\envu(u') = \ulabel', \lst{\stype_u'}, l_r'$ that $l'' =l_r$
           and by rule \irule{T-BSetDom}  we know that $l_r' \ileq I(\ulabel)$
           and $(2)$ follows.

          For $(3)$, we get with rule \irule{T-BSetDom} and
          \cref{lem:bro-ex-ty} that 
          $\forall i \in \intval{1}{\len{v}}. \, \stype_i = {\stype_u}_i$ 
          and the claim follows immediately.
  
          $(4)$ is trivial, since with $\isihigh{l}$ and $\isihigh{l''}$
          we also know $\amode = \bhon$ and $\amode'' = \bhon$.

        \item $\isilow{l''}$: 
          Then we need to show that on of the following holds
          \begin{itemize}
            \item $\forall i \in \intval{1}{\len{v}}. \, \stype_i \tleq \alabel$
            \item or $\isihigh{u'}$
          \end{itemize}
          If $\isihigh{u'}$, the claim is trivial, we hence assume
          $\isilow{u'}$. The claim then follows immediately from 
          \sketch{the observation that by rule \irule{T-Reply} scripts of low
            integrity URLs can never contain any values of high
          confidentiality}
      \end{itemize} 
  \end{itemize}
\end{proof}

We now show that Browsers remain well-typed if they send out and request and
that every sent request is well-typed.

\begin{lemma}[Browser Request]
  \label{lem:bro-req}
  Whenever a browser $B \bstep{\alpha} B'$ with
  $\alpha = \exbsend{\bid}{n}{u}{p}{ck}{o}{l}{\amode}$ and $\browtyping{B}$.
  Then $\reqtyping{\alpha}$ and $\browtyping{B'}$
\end{lemma}
\begin{proof}
  Let $B=\exbrowser{N}{M}{P}{T}{Q}{\lst{a}}$ and
  We know that rule \irule{B-Flush} is used. 
  We hence have $Q = \{\einta{\alpha}{l'}\}$ and
    $B'=\exbrowser{N}{M}{P}{T}{\mempty}{\lst{a'}}$.
    $\browtyping{B'}$ then follows immediately from $\browtyping{B}$
  We get $\reqtyping{\alpha}$ by property \ref{def:btQ} of
  \cref{def:bro-typing}.
\end{proof}

The next lemma states that a well-typed browser receiving a well-typed response
is still a well-typed browser. We have the additional assumptions that 
the integrity of the response is at most as high as the integrity of the
browser and that either the attacked mode of the browser and the response are
the same or that the response is attacked and the integrity of the responding
URL is low.

\begin{lemma}[Browser Response]
  \label{lem:bro-res}
  Whenever a for a browser $B=\exbrowser{N}{M}{P}{T}{Q}{\lst{a}}$ we have 
  $B \bstep{\alpha} B'$ with $\browtyping{B}$,
  $\alpha = \exbrecv{\bid}{n}{u}{u'}{ck}{\lst{v}}{page}{s}{l''}{\amode''}$ with 
  $l \ileq l''$ and $\amode = \amode'' \vee \amode'' = \batt
  \wedge \isilow{I(\lambda(u))}$ and 
  $\restyping{\alpha}$ then $\browtyping{B'}$.
\end{lemma}
\begin{proof}
  $B'=\exbrowserpar{N'}{M'}{P'}{T'}{Q'}{\lst{a'}}{l'}{\amode'}$
  We show that $B'$ fulfills the properties of \cref{def:bro-typing}.
  We know that the step $\alpha$ was taken using rule \irule{B-RecvLoad}
  \irule{B-RecvInclude}, or \irule{B-Redirect}. In all cases property
  \ref{def:btM} of \cref{def:bro-typing} follows immediately from property
  \ref{def:src} of \cref{def:res-ty}.
  We now do a case distinction on the rule used 
  \begin{itemize}
    \item \irule{B-RecvLoad} 
      \begin{itemize}
        \item Property \ref{def:btL} follows immediately from property
          \ref{def:srcs} of \cref{def:res-ty}
        \item 
          For property \ref{def:btP} we do a case distinction:
          \begin{itemize}
            \item if $\amode' = \amode'' = \bhon$ then the claim follows from 
              property \ref{def:srhp} of \cref{def:res-ty}. 
            \item if $\amode' = \amode'' = \batt$ then the claim follows from
              properties \ref{def:srap} and \ref{def:srcs} of \cref{def:res-ty} 
          \end{itemize}
        \item Property \ref{def:btT} follows from property \ref{def:srhs} of
          \cref{def:res-ty} for $\amode' = \bhon$ and from \ref{def:srcs} of
          \cref{def:res-ty} if $\amode' = \batt$.
        \item Property \ref{def:btQ} is trivial.
        \item Property \ref{def:btA} follows from the same property for $B$.
          \sketch{the $nf$ on the tab for the page in $B'$ is the same as the
          one for the network connection in $B$}
        \item Property \ref{def:btI} is trivial.
      \end{itemize}
    \item \irule{B-RecvInclude} 
      \begin{itemize}
        \item Property \ref{def:btL} follows immediately from property
          \ref{def:srcs} of \cref{def:res-ty} and property \ref{def:btL} of \cref{def:bro-typing} for $B$.
        \item Property \ref{def:btP} is trivial
        \item For property \ref{def:btT} we do a case distinction:
          \begin{itemize}
            \item If $\amode'' = \bhon$, then $\amode = \amode' = \bhon$ and
              the claim follows from property \ref{def:srhs} of
              \cref{def:res-ty} and property
              \ref{def:btT} of \cref{def:bro-typing} for $B$, using rule
              \irule{T-BSeq} and property
              \ref{def:btI} of \cref{def:bro-typing}
            \item if $\amode'' = \batt$ then $\amode' = \batt \vee \amode = \batt$.
              \sketch{Since we know that 
                $\amode = \batt \Rightarrow \amode' = \batt$ we can conclude that 
              $\amode' = \batt$}
              We distinguish two cases:
              \begin{itemize}
                \item If $\amode = \batt$ the claim follows immediately using
                  property \ref{def:srcs} of \cref{def:res-ty} and property
                  \ref{def:btT} of \cref{def:bro-typing} for $B$, using rule
                  \irule{T-BSeq}
               \item f $\amode = \bhon$ then by the assumption in the lemma we
                 have $\isilow{I(\lambda(u))}$ which is in contradiction to
                 property \ref{def:btI} of \cref{def:bro-typing}, hence this case is impossible.
             \end{itemize}
          \end{itemize}
        \item Property \ref{def:btQ} is trivial.
        \item Property \ref{def:btA} is trivial
        \item Property \ref{def:btI} is trivial.
      \end{itemize}
    \item \irule{B-Redir}
      \begin{itemize}
        \item Property \ref{def:btL} is trivial.
        \item Property \ref{def:btP} is trivial.
        \item Property \ref{def:btT} is trivial.
        \item For property \ref{def:btQ} we know that $Q' = \{ \alpha' \}$ with 
          $\alpha' = \exbsend{\bid}{n'}{u'}{p}{ck'}{o'}{l''}{\amode''}$ 
            where 
            \begin{itemize}
              \item $\forall k \in \intval{1}{\len{v}}: p(k) = v_k$ 
              \item $ck' = \getck{M'}{u'}$
              \item $o' = (o = \uorigin{u}) ~?~ o : \noorigin$
            \end{itemize}
           and hence have to show that $\reqtyping{\alpha'}$. 
           We show that all the properties of \cref{def:req-ty} are fulfilled.
           \begin{itemize}
             \item Property \ref{def:brhp} follows immediately from property
               \ref{def:srru} of \cref{def:req-ty} for $\alpha$
             \item Property \ref{def:brap} follows immediately from property
               \ref{def:srru} of \cref{def:req-ty} for $\alpha$
             \item Property \ref{def:brhc} follows immediately from property
               \ref{def:btM} of \cref{def:bro-typing} for $B$ 
             \item Property \ref{def:brac} is trivial since $\uid = \usr$
             \item For property \ref{def:bro} we perform a case distinction:
                   \begin{itemize}
                     \item If $u' \not\in \protUrls$, $\isilow{o'}$ or 
                       $\amode' = \bhon$ then the claim is trivial.
                     \item If $u \in \protUrls$, $\isihigh{o'}$ and 
                       $\amode' = \batt$ then we know that
                       $\isihigh{\uorigin{u}}$. 
                       \sketch{We then know that the code at endpoint $u$ can
                         be typed with $\typebranch = \bhon$ and we get by 
                         rule \irule{T-Redirect} that $u \not \in \protUrls$.
                         Since the redirect URL must appear as a constant in
                         the code, we apply this result in any case and
                         reach a contradiction.}
                   \end{itemize}
           \end{itemize}
         \item Property \ref{def:btA} is trivial 
           \sketch{since high integrity pages only include high integrity pages}.
        \item For property \ref{def:btI} we distinguish two cases:
          \begin{itemize}
            \item If $T = \mempty$ or $\amode' = \batt$ the claim is trivial
            \item If $T = \mempty$ and $\amode' = \bhon$, then we know that 
              $\amode = \bhon$ and $\amode'' = \bhon$. 
              \sketch{The claim then follows using rule \irule{T-Redir}}
          \end{itemize}
      \end{itemize}
  \end{itemize}
\end{proof}

We have now shown all lemmas for browser steps and move on to the server.
First, we introduce typing for the server:

\begin{definition}[Server Typing]
  \label{def:serv-typing}
  Let $S=\exserver{D}{\phi}{t}$ be a server with $D = (\gmems,\smems)$. We write $\servtyping{S}$, if 
  \begin{enumerate}
    \item \label{def:stG} 
      \begin{itemize}
        \item if $\bid = \usr$ then for all $i \in \domain(\gmems)$, for all
     $r \in \domain(\gmems(i))$ we have if $\gmems(i)(r) = v^{\stype}$ then $\stype \tleq \treft{\envrg(r)}$ 
        \item if $\bid \neq \usr$ then for all $i \in \domain(\gmems)$, for all
     $r \in \domain(\gmems(i))$ we have if $\gmems(i)(r) = v^{\stype}$ then $\stype \tleq \alabel$ 
      \end{itemize}
    \item \label{def:stS} for all $i \in \domain(\smems)$, for all $r \in \domain(\smems(j))$ we have if $\smems(j)(r)=v^{\stype}$ then $\stype = \treft{\envrs(r)} \cremeet \jlabel{j}$ 
    \item \label{def:stP} for all $u \in \surls{S}$, for all $j \in
      \domain(\phi)$we have that $\rho(\phi(j),u) \tleq \jlabel{j}$
    \item \label{def:stT} $\exsttyping{t}$
    \item \label{def:stOI} For all $t \in \threads{S}$ with $t = \exthread{c}{(n,u,\bid,o)}{(i,j)}$ we have if $\bid = \usr$,  $u \in \protUrls$ and 
      $o \neq \noorigin$ and $\isihigh{o}$ then $\amode = \bhon$.
  \end{enumerate}
\end{definition}

Intuitively, according to \cref{def:serv-typing} a server is well typed if
\begin{enumerate}
  \item For the global memories we have that
    \begin{itemize}
      \item for honest users, all values respect the typing environment
      \item for the attacker, all values are of the attackers type $\alabel$
    \end{itemize}
  \item All values in session memories respect the typing environment (taking
    the label of the session identifier into account)
  \item All sessions are protected by session identifiers whose security
    guarantees are stronger than the one of the passwords corresponding to the
    identity stored in the session.
  \item All server threads are well-typed.
  \item For all threads the integrity label is as least as low as the origin 
\end{enumerate}

We now show the same lemmas we showed for the browser on the server side ,
starting with the substitution of variables in server expressions.

\begin{lemma}[Server Expression Substitution]
  \label{lem:serv-ex-sub}
  Whenever we have $\exelabel{se}{\stype}$ and we have a substitution $\sigma$ with 
  $\domain(\sigma) = \domain(\envv)$ and 
  $\forall x \in \domain(\envv). \, \sigma(x) = v_x^{\stype_x}$ with 
  $\stype_x \tleq \envv(x)$ then for all $\envv'$, 
  $\exelabelpar{(\envu, \envv', \envrg,\envrs,\envt)}{b}{se\sigma}{\stype}$.
\end{lemma}
\begin{proof}
  We perform an induction on the typing derivation of $\exelabel{se}{\stype}$:
  \begin{itemize}
    \item \irule{T-EVar}. Then $se = x$ and $se \sigma  = v_x^{\stype_x}$ with
      $\stype_x \tleq \envv(x)$.
      The claim follows directly from rule \irule{T-EVal} and \irule{T-Sub}.
    \item \irule{T-ESesRef}. Then $se =\rg{r} =  se\sigma$ and the claim is trivial.
    \item \irule{T-EGlobRef}. Then $se = \rs{r} =  se\sigma$ and the claim is trivial.
    \item \irule{T-EVal}. Then $se = v^{\stype_v} =  se\sigma$ and the claim is trivial.
    \item \irule{T-EUndef}. Then $se = \vundef =  se\sigma$ and the claim is trivial.
    \item \irule{T-EName}. Then $se = n^{\stype_n} =  se\sigma$ and the claim is trivial.
    \item \irule{T-EFresh}. Then $se = \fresh^{\stype} = se\sigma$ and the claim is trivial.
    \item \irule{T-EBinOp} Then $se = se_1 \odot se_2$ with
      $\exelabel{se_1}{\stype_1}$ and $\exelabel{se_2}{\stype_2}$ and 
      $\stype = \typelabel{\stype_1} \sqcup \typelabel{\stype_2}$.
      We also have $se\sigma = se_1\sigma \odot se_2\sigma$. By induction we
      know that $\exelabel{se_1\sigma}{\stype_1'}$ and
      $\exelabel{se_2\sigma}{\stype_2'}$ 
      with $\stype_1' \tleq \stype_1$ and $\stype_2' \tleq \stype_2$.
       We then know that 
       $\typelabel{\stype_1'} \sqcup \typelabel{\stype_2'} \tleq \typelabel{\stype_1} \sqcup \typelabel{\stype_2} = \stype$, and
      the claim follows by \irule{T-BinOp} and \irule{T-BESub}.
    \item \irule{T-ESub} follows by induction and by the transitivity of $\tleq$.
  \end{itemize}
\end{proof}

To show the substitution lemma for server commands we first need to show
auxiliary lemmas that deal with the program counter.

%
%
%
First, we show that whenever server code can be typed with a $\pclabel$, it can
also be typed with any $\pclabel$ of higher integrity.

\begin{lemma}[Server Program Counter Substitution]
\label{lem:serv-pc-sub} 
Whenever we have $\exstypingparpc{\Gamma}{\seslabel}{\pclabel}{\typebranch}{c}{\seslabel'}{\pclabel'}$ and $\pclabel^* \ileq \pclabel$ then 
$\exstypingparpc{\Gamma}{\seslabel}{\pclabel^*}{\typebranch}{c\sigma}{\seslabel}{\pclabel^{**}}$ with $\pclabel^{**} \ileq \pclabel'$.
\end{lemma}
\begin{proof}
We perform the proof by induction on the typing derivation    
\begin{itemize}
  \item \irule{T-Skip} The claim is trivial
  \item \irule{T-Login} The claim follows from the transitivity of $\ileq$
  \item \irule{T-Start} The claim is trivial
  \item \irule{T-SetGlobal} The claim follows from the transitivity of $\ileq$
  \item \irule{T-SetSession} The claim follows from the transitivity of $\ileq$
  \item \irule{T-Seq} The claim follows by induction on the two subcommands.
  \item \irule{T-If}: Then $c = \ite{se}{c_1}{c_2}$ with 
    \begin{itemize}
      \item $\exelabel{se}{\stype}$.
      \item $\pclabel' = \pclabel \ijoin \It{\stype}$ 
      \item $\exstypingparpc{\Gamma}{\seslabel}{\pclabel''}{\typebranch}{c_1}{\seslabel''}{\pclabel_1}$
      \item $\exstypingparpc{\Gamma}{\seslabel}{\pclabel''}{\typebranch}{c_2}{\seslabel'''}{\pclabel_2}$
      \item $\pclabel'' = ((c \text{ and } c' \text{ do not contain $\mathbf{reply}$, $\mathbf{redir}$, $\mathbf{tokencheck}$ or $\mathbf{origincheck}$}) ~?~ \pclabel ~:~ \pclabel') \ijoin \pclabel_1 \ijoin \pclabel_2$
    \end{itemize}
    Let $\pclabel''' = \pclabel^* \ijoin \It{\stype}$, then $\pclabel''' \ileq
    \pclabel'$ and we can apply the induction hypothesis for $c_1$ and $c_2$ and get 
    \begin{itemize}
      \item $\exstypingparpc{\Gamma}{\seslabel}{\pclabel'''}{\typebranch}{c_1}{\seslabel''}{\pclabel^*_1}$
      \item $\exstypingparpc{\Gamma}{\seslabel}{\pclabel'''}{\typebranch}{c_2}{\seslabel'''}{\pclabel^*_2}$
    \end{itemize}
    with $\pclabel^*_1 \ileq \pclabel_1$ and $\pclabel^*_2 \ileq \pclabel_2$
    and the claim follows by applying \irule{T-If}.

  \item \irule{T-TCheck} The claim is trivial
  \item \irule{T-PruneTCheck} The claim is trivial
  \item \irule{T-OChck} The claim is trivial
  \item \irule{T-PruneOChck} The claim is trivial
  \item \irule{T-Reply}
    With $\envu(u) = \ulabel, \lst{t}, l_r$, we let $\pclabel' = \pclabel \ijoin l_r$
    and $\pclabel'' = \pclabel \ijoin l_r$. 
    We do a case distinction on $\typebranch$:
    \begin{itemize}
      \item If $\typebranch = \bhon$ we get $\pclabel \ileq l_r$, hence
        $\pclabel' = l_r$ and because of of $\pclabel' \ileq \pclabel$ we
        also get $\pclabel'' = l_r$ and the claim follows.
      \item If $\typebranch \neq \bhon$ we have $\pclabel = \itop $
        We hence also have $\pclabel^* = \itop$ and the claim follows.
    \end{itemize}
  \item \irule{T-Redir} The claim follows from the transitivity of $\ileq$
  \item \irule{T-Reset} The claim is trivial.
\end{itemize}
\end{proof}
 
We now show that if server code containing variables is well typed in a typing
environment typing these variables, then the code is also well typed after
instantiating these variables with concrete values of the same type.

\begin{lemma}[Server Substitution]
  \label{lem:serv-sub}
  Whenever we have $\exstypingparpc{\Gamma}{\seslabel}{\pclabel}{\typebranch}{c}{\seslabel'}{\pclabel^*}$ and we have a substitution $\sigma$ with 
  $\domain(\sigma) = \domain(\envv)$ and 
  $\forall x \in \domain(\envv). \, \sigma(x) = v_x^{\stype_x}$ with 
  $\stype_x \tleq \envv(x)$ then for all $\envv'$, we have 
  $\exstypingparpc{\Gamma'}{\seslabel}{\pclabel}{\typebranch}{c\sigma}{\seslabel'}{\pclabel^{**}}$ with
  $\Gamma' = (\envu, \envv', \envrg,\envrs,\envt)$.
\end{lemma}
\begin{proof}
  We do the proof by induction on the typing derivation.
  \begin{itemize}
    \item \irule{T-BSeq}: Then $c = c_1,c_2$. The claim follows by applying the 
      induction hypothesis to $s_1$ and $s_2$ using \cref{lem:serv-pc-sub} and applying rule \irule{T-Seq}
    \item \irule{T-Skip}: The claim follows trivially.
    \item \irule{T-SetSession}: The claim follows from \cref{lem:serv-ex-sub}
      and the transitivity of $\tleq$.
    \item \irule{T-SetGlobal}: The claim follows from \cref{lem:serv-ex-sub}
      and the transitivity of $\tleq$.
    \item \irule{T-Login}: The claim follows from \cref{lem:serv-ex-sub}
      and the transitivity of $\tleq$.
    \item \irule{T-Start}: The claim follows from \cref{lem:serv-ex-sub}
      and the transitivity of $\tleq$.
    \item \irule{T-If}: Then $c = \ite{se}{c_1}{c_2}$ with 
      \begin{itemize}
        \item $\exelabel{se}{\stype}$.
        \item $\pclabel' = \pclabel \ijoin \It{\stype}$ 
        \item $\exstypingparpc{\Gamma}{\seslabel}{\pclabel'}{\typebranch}{c_1}{\seslabel''}{\pclabel_1}$
        \item $\exstypingparpc{\Gamma}{\seslabel}{\pclabel'}{\typebranch}{c_2}{\seslabel'''}{\pclabel_2}$
      \end{itemize}
      By induction we know
      \begin{itemize}
        \item $\exstypingparpc{\Gamma'}{\seslabel}{\pclabel'}{\typebranch}{c_1\sigma}{\seslabel''}{\pclabel^*_1}$
        \item $\exstypingparpc{\Gamma'}{\seslabel}{\pclabel'}{\typebranch}{c_2\sigma}{\seslabel'''}{\pclabel^*_2}$
      \end{itemize}

      By \cref{lem:serv-ex-sub} we know that $\exelabelpar{\Gamma'}{\seslabel}{se\sigma}{\stype}$
      The claim then follows by applying rule \irule{T-If}.
    \item \irule{T-Auth} The claim follows from \cref{lem:serv-ex-sub} and the
      transitivity of $\tleq$.
    \item \irule{T-PruneTCheck} The claim follows from \cref{lem:serv-ex-sub}
      and the fact that there is no subtyping on credentials of high
      confidentiality.
  \item \irule{T-OChckSucc} The claim follows trivially.
  \item \irule{T-OChckFail} The claim follows trivially.
    \item \irule{T-TCheck} The claim follows from \cref{lem:serv-ex-sub}.
    \item \irule{T-Reply} 
      Let variables be assigned as in the rule.
      The claim then follows by applying \cref{lem:serv-ex-sub} for all $se_k$.
     The claim then follows immediately.
  \item \irule{T-Redir}
      Let variables be assigned as in the rule.
      The claim then follows by applying \cref{lem:serv-ex-sub} for all $se_k$.
  \end{itemize}
\end{proof}

We now show that typing of server expressions is preserved under evaluation.

\begin{lemma}[Server Expression Typing]
  \label{lem:serv-ex-ty}
  Let $S=\exserver{D}{\phi}{t}$ be a server with $\servtyping{S}$
  and let $\exthread{c}{R}{i,j} \in \running{S}$.
  Then for any server expression $se$, if
  $\exelabelpar{\Gamma}{\jlabel{j}}{se}{\stype}$ then 
  $\exelabelpar{\Gamma}{\jlabel{j}}{\eval{se}{D}{i,j}}{\stype}$.
\end{lemma}
\begin{proof}
  Proof by induction over the expression $se$.
  \begin{itemize}
    \item $se = v^{\stype_v}$: Then $\eval{se}{D}{i,j}=se$ and the claim is trivial.
    \item $se  = se_1 \odot se_2$: By induction analog to case in in
      \cref{lem:bro-ex-ty}.
    \item $se = \rg{r}$: straightforward from property \ref{def:stG} of
      \cref{def:serv-typing} using \irule{T-EGlobRef} and \irule{T-ESub}
    \item $se = \rs{r}$: straightforward from property \ref{def:stS} of
      \cref{def:serv-typing} using \irule{T-ESesRef} and \irule{T-ESub}
    \item $se = \fresh^{\tau_f}$: straightforward from \irule{SE-Fresh}, \irule{T-Fresh}
      and \irule{T-EName} 
  \end{itemize}
\end{proof}

Next, we show that whenever a server thread is typable with the session label
$\unauth$, then it is also typable with any other session label.

\begin{lemma}[Server Typing with $\seslabel = \unauth$]
  \label{lem:serv-ty-unauth}
  Whenever we have
  $\exstypingpar{\Gamma}{\unauth}{\pclabel}{\typebranch}{c}{\seslabel'}$
  then we also have 
  $\exstypingpar{\Gamma}{\seslabel}{\pclabel}{\typebranch}{c}{\seslabel''}$
  for all $\seslabel$, where $\seslabel' =\unauth$ or $\seslabel' = \seslabel''$.
\end{lemma}
\begin{proof} 
    This is simple by inspecting the typing rules and the observation that
    $\seslabel = \unauth$ implies that the session memory cannot be used.
    Hence the code that is typed with $\seslabel = \unauth$ can be typed with
    any session label.
    Furthermore, if the session label is set to a different label during typing,
    this is unaffected by the old session label.
\end{proof}

We are now ready to show that whenever a well-typed server takes an internal
step, it results in another well-typed server.

\begin{lemma}[Server Subject Reduction]
  \label{lem:serv-sub-red}
  Let $S$ be a server with $\servtypingpar{\Gamma^0}{S}$ and 
  $S \sstep{\alpha}{S'}$, where $\alpha \in \{\blank,\lauth{\lst{v}}{\bid,
  \uid}{\ell}\}$
  Then we have $\servtypingpar{\Gamma^0}{S'}$.
\end{lemma}
\begin{proof}
  Let $S=\exserver{D}{\phi}{t}$ and let 
  $S'=\exserver{D'}{\phi'}{t'}$
  Then there exists  $\exthread{c}{R}{i,j} \in \running{S}$
  with $\exserver{D}{\phi}{\exthread{c}{R}{i,j}} \sstep{\alpha}
  \exserver{D'}{\phi'}{\exthreadpar{c'}{R}{i,j'}{l'}{\amode}}$.

  Because of rules \irule{S-LParallel}, \irule{S-RParallel} and
  \irule{T-Parallel} it is sufficient to show
  $\servtypingpar{\Gamma^0}{\exserver{D}{\phi}{\exthreadpar{c'}{R}{i,j'}{l'}{\amode}}}$, assuming
  $\servtypingpar{\Gamma^0}{\exserver{D}{\phi}{\exthread{c}{R}{i,j}}}$. 

  We chose $\typebranch$, $\seslabel_1$, $\pclabel_1$ and $\Gamma$ as in rule \irule{T-Running}:

  Let $\typebranch = \begin{cases} 
    \bhon &\text{ if } \amode = \bhon \wedge \bid = \usr \\
    \bcsrf &\text{ if } \amode = \batt \wedge \bid = \usr \\
    \batt &\text{ if } \bid \neq \usr
  \end{cases}$  and let $\seslabel_1 = \jlabel{j}$ and let $\pclabel_1 = l$.

  With $\Gamma^0 = (\envu,\envv,\envrg,\envrs,\envt)$ and 
  $\envrg' = (\bid = \usr) ~?~ \envrg ~:~ \subst{\_}{\alabel}$
  we let $\Gamma = (\envu,\envv,\envrg',\envrs,\envt)$.

  We furthermore let $\seslabel_2 = \jlabel{j'}$ and $\pclabel_2 = l'$.

  We now show that $S'$ fulfills all properties of \cref{def:serv-typing}.

  However, for property \ref{def:stT} of \cref{def:serv-typing} we will show the
  following stronger claim:

  Whenever $\exstypingparpc{\Gamma}{\seslabel_1}{\pclabel_1}{\typebranch}{c}{\seslabel_1'}{\pclabel_1'}$
  we have $\exstypingparpc{\Gamma}{\seslabel_2}{\pclabel_2}{\typebranch}{c'}{\seslabel_2'}{\pclabel_2'}$
  where $\pclabel_2' \ileq \pclabel_1'$ and $\seslabel_1' = \unauth$ or $\seslabel_1' = \seslabel_2'$ 

  For all cases property \ref{def:stOI} is trivial.

  We perform the proof by induction the step taken.
  \begin{itemize}
    \item \irule{S-Skip}. This case is trivial.
    \item \irule{S-Seq} Then we know 
      \begin{itemize}
        \item $c = c_1; c_2$ 
        \item $\exserver{D}{\phi}{\exthreadpar{c_1}{R}{i,j}{l}{\amode}} \sstep{\alpha}
          \exserver{D'}{\phi'}{\exthreadpar{c_1'}{R}{i,j'}{l'}{\amode}} $
        \item $c' = c_1' ; c_2$. 
      \end{itemize}
      All properties of \cref{def:serv-typing} except for property
      \ref{def:stT} follow immediately by the induction hypothesis applied to
      $c_1$.

      By rule \irule{T-Seq} we know that for some $\pclabel_1''$ and $\seslabel_1''$ 
      \begin{itemize}
        \item $\exstypingparpctc{\Gamma}{\seslabel_1}{\pclabel_1}{\typebranch}{c_1}{\seslabel_1''}{\pclabel_1''}$ 
        \item $\exstypingparpctc{\Gamma}{\seslabel_1''}{\pclabel_1''}{\typebranch}{c_2}{\seslabel_1'}{\pclabel_1'}$
      \end{itemize}
      
      By induction we know that for some $\pclabel_2$,$\pclabel_2'$,$\seslabel_2$,$\seslabel_2'$
      \begin{itemize}
        \item $\exstypingparpc{\Gamma}{\seslabel_2}{\pclabel_2}{\typebranch}{c_1}{\seslabel_2'}{\pclabel_2'}$
        \item $\pclabel_2' \ileq \pclabel_1''$
        \item $\seslabel_1'' = \unauth$ or $\seslabel_1'' = \seslabel_2'$ 
      \end{itemize}

      Using \cref{lem:serv-pc-sub} we get
      $\exstypingparpctc{\Gamma}{\seslabel_1''}{\pclabel_2'}{\typebranch}{c_2}{\seslabel_1'}{\pclabel_2''}$
      with $\pclabel_2'' \ileq \pclabel_1'$

      Using \cref{lem:serv-ty-unauth} we furthermore get 
      $\exstypingparpctc{\Gamma}{\seslabel_2'}{\pclabel_2'}{\typebranch}{c_2}{\seslabel_2''}{\pclabel_2''}$
      with $\seslabel_1' = \unauth$ or $\seslabel_1' = \seslabel_2''$ 

      Using \irule{T-Seq} we can then conclude
      $\exstypingparpctc{\Gamma}{\seslabel_2}{\pclabel_2}{\typebranch}{c_1';c_2}{\seslabel_2''}{\pclabel_2''}$ and the claim follows.

    \item \irule{S-IfTrue} then
      \begin{itemize}
        \item $c = \ite{se}{c_1}{c_2}$ 
        \item $\eval{se}{D}{i,j}=true^\stype$
        \item $j' = j$ 
        \item $l' = l \ijoin \It{\stype}$
        \item $c' = (\mathbf{reply}, \mathbf{redir}, \mathbf{tokencheck}, \mathbf{origincheck} \in \commands{c}) ~?~ c_1 ~:~  c;\resetpc{l}$
      \end{itemize}

      All properties of \cref{def:serv-typing} except for property
      \ref{def:stT} follow immediately by the induction hypothesis applied to
      $c_1$ using rules \irule{S-Seq} and \irule{S-Reset}.

      By \irule{T-If} we know
      \begin{itemize}
        \item $\exelabel{se}{\stype'}$
        \item $\pclabel' = \pclabel_1 \ijoin \It{\stype}$
        \item $\exstypingparpctc{\Gamma}{\seslabel_1}{\pclabel'}{\typebranch}{c_1}{\seslabel_2'}{\pclabel_2'}$ for some $\seslabel_2', \pclabel_2'$
        \item $\exstypingparpctc{\Gamma}{\seslabel_1}{\pclabel'}{\typebranch}{c_2}{\seslabel_2''}{\pclabel_2''}$ for some $\seslabel_2'', \pclabel_2''$
        \item $\seslabel_1' = \seslabel_2'$ or  $\seslabel_1' =  \unauth$
        \item $\pclabel_1' = \mathbf{reply}, \mathbf{redir}, \mathbf{tokencheck}, \mathbf{origincheck} \in \commands{c} ~?~ \pclabel_2' \ijoin \pclabel_2'' ~:~ \pclabel_1$
      \end{itemize}

      By \cref{lem:serv-ex-ty} we know that $\stype \tleq \stype'$.
      Hence $\pclabel_2 = l' \ileq \pclabel'$.
      We thus have by \cref{lem:serv-pc-sub} that
      $\exstypingparpctc{\Gamma}{\seslabel_2}{\pclabel_2}{\typebranch}{c_1}{\seslabel_2'}{\pclabel_2^*}$
      with $\pclabel_2* \ileq \pclabel_2'$.

      If $\mathbf{reply}, \mathbf{redir}, \mathbf{tokencheck},
      \mathbf{origincheck} \in \commands{c}$, the claim follows immediately.

      Otherwise, we using \irule{T-Seq} and \irule{T-Reset} we observe,
      that
      $\exstypingparpctc{\Gamma}{\seslabel_2}{\pclabel_2}{\typebranch}{c_1;\resetpc{l}}{\seslabel_2'}{l}$.
      With $\pclabel_1' = \pclabel_1 = l = \pclabel_2'$ the claim follows immediately.

    \item \irule{S-IfFalse} then the claim follows analog to the previous one. 

    \item \irule{S-TCTrue} Then
      \begin{itemize}
        \item $c = \tokch{se}{se'}{c'}$
        \item $\eval{se}{D}{i,j}=v_1^{\stype_1}$
        \item $\eval{se'}{D}{i,j}=v_1^{\stype_2}$
        \item $v_1 = v_2$
        \item $l' = l$
        \item $j' = j$
      \end{itemize}

      All properties of \cref{def:serv-typing} except for property
      \ref{def:stT} are trivial.

      We know that typing was done using rule \irule{T-TChk} or
      \irule{T-PruneTChk}. 

      We want to show that \irule{T-TChk} was used. To this end, we assume
      that \irule{T-PruneChk} was used and show a contradiction.

      By rule \irule{T-TChkPrune} we know 
      \begin{itemize}
        \item $se = x$ for some $x$ and $\exelabel{x}{\stype_1'}$
        \item $se' = r$ for some $r$ and $\exelabel{r}{\stype_2'}$ 
        \item $\stype_2 = \tcre{\ell}$
        \item $\stype_1 \neq {\tcre{\ell}}$
        \item $\ischigh{\tcre{\ell}}$
      \end{itemize}

      By \cref{lem:serv-ex-ty} we know that $\stype_1 \tleq \stype_1'$ and
      $\stype_2 \tleq \stype_2'$. By the definition of $\tleq$, we know that 
      $\stype_1 = \stype_1'$. 
      Since the set of credentials at label
      $\typelabel{\stype_1}$ is disjoint from the set of the set of any other
      values, and since $v_1 = v_2$, we know that also $\stype_1 = \stype_2$.
      Using the definition of $\tleq$ we get $\stype_2 = \stype_2'$.
      We hence have $\stype_1' = \stype_2'$ which contradicts the assumption.

      We thus know that \irule{T-TChk}  and we get 
      $\exstypingparpc{\Gamma}{\seslabel_1}{\pclabel_1}{\typebranch}{c'}{\seslabel_1'}{\pclabel_1'}$
      and the claim follows.

    \item \irule{S-TCFalse} Then
      \begin{itemize}
        \item $c = \tokch{se}{se'}{c''}$
        \item $c' = \ereply{\serror}{\sskip}{\mempty}$
        \item $\eval{se}{D}{i,j}=v_1^{\stype_1}$
        \item $\eval{se'}{D}{i,j}=v_1^{\stype_2}$
        \item $v_1 \neq v_2$
        \item $l' = l$
        \item $j' = j$
      \end{itemize}

      All properties of \cref{def:serv-typing} except for property
      \ref{def:stT} are trivial.
    
      By \irule{T-Reply} we immediately get
      $\exstypingparpc{\Gamma}{\seslabel_1}{\pclabel_1}{\typebranch}{\ereply{\serror}{\sskip}{\mempty}}{\seslabel_1}{\pclabel_1}$.

    \item \irule{S-Reset} 
      All properties of \cref{def:serv-typing} 
      are trivial, where property \ref{def:stT} follows immediately from \irule{T-Reset}.

    \item \irule{S-RestoreSession} 
      We have 
      \begin{itemize}
        \item $c = \start{se}$
        \item $c' = \sskip$
        \item $\eval{se}{D}{i,j}=v^{\stype}$
        \item $v \in \domain(\smems)$
        \item $l' = \isclow{\Ct{\stype}} ~?~ \bot ~:~ \typelabel{\stype}$
        \item $j' = v$
      \end{itemize}

      All properties of \cref{def:serv-typing} except for property \ref{def:stT}
      are trivial.

      By \irule{T-Start} we get 
      \begin{itemize}
        \item $\exelabel{se}{\tcre{\ell}}$
        \item $\seslabel_1' = (\isclow{\Ct{\tcre{\ell}}}) ~?~ \low : \ell$
      \end{itemize}

      By \cref{lem:serv-ex-ty} we know that $\stype \tleq \tcre{\ell}'$ 
      We distinguish two cases:
      \begin{itemize}
        \item If $\isclow{\Ct{\tcre{\ell}}}$ then also $\isclow{\Ct{\stype}}$ and 
        we have  $\seslabel_1' = \low = \jlabel{j'} = \seslabel_2'$.
        \item If $\ischigh{\Ct{\tcre{\ell}}}$ we know $\stype = \tcre{\ell}$
          and we have $\seslabel_1' = \ell = \jlabel{j'} = \seslabel_2'$.
      \end{itemize}

    \item \irule{S-NewSession} 
      We immediately get property \ref{def:stG}, \ref{def:stP} of
      \cref{def:serv-typing}.
      Property \ref{def:stS} follows immediately using rule \irule{T-EUndev}
      since the freshly created memory is empty.
      Property \ref{def:stT} follows analog to the previous case.

    \item \irule{S-SetGlobal} 
      We immediately get property \ref{def:stS}, \ref{def:stP}, \ref{def:stT} of
      \cref{def:serv-typing}, using \irule{T-SetGlobal}.

      We have $c = \rg{r} := se$, $\eval{se}{D}{i,j}=v^{\stype}$ and
      $\stype' = \stype \sqcup \treft{\envrg'(r)} \tijoin l$ with
      $ \envrg' = (\bid = \usr) ~?~ \envrg ~:~ \subst{\_}{\alabel}$.

      We know using rule \irule{T-SetGlobal} that
      \begin{itemize}
        \item $\exsrtypingpar{\Gamma'}{\rg{r}}{\tref{\stype'}}$
        \item $\exelabelpar{\Gamma'}{\seslabel_1}{se}{\stype'}$
        \item $\pclabel_1 \ileq \It{\stype'}$
      \end{itemize}

      Using \irule{T-GlobRef} and \irule{T-RefSub} we know that 
      $\stype' \tleq \treft{\envrg'(r)}$.
      Using \cref{lem:serv-ex-ty} we know $\stype \tleq \stype'$.
      
      We hence know that $\stype' = \treft{\envrg'(r)}$ and property
      \ref{def:stG} follows.


    \item \irule{S-SetSession} 
      We immediately get property \ref{def:stG}, \ref{def:stP}, \ref{def:stT}
      of \cref{def:serv-typing}, using rule \irule{T-SetSession}.

      We have $c = \rs{r} := se$, $\eval{se}{D}{i,j}=v^{\stype}$ and
      $\stype' = \stype \tjoin (\treft{\envrs(r)}  \crejoin \jlabel{j}) \tijoin l$.

      We know using rule \irule{T-SetSession} that
      \begin{itemize}
        \item $\exsrtypingpar{\Gamma}{\rs{r}}{\tref{\stype'}}$
        \item $\exelabelpar{\Gamma}{\seslabel_1}{se}{\stype'}$
        \item $\pclabel_1 \ileq \It{\stype'}$
      \end{itemize}

      Using \irule{T-SesRef} and \irule{T-RefSub} we know that 
      $\stype' \tleq (\treft{\envrs(r)} \crejoin \seslabel_1)$.
      Using \cref{lem:serv-ex-ty} we know $\stype \tleq \stype'$.
      
      We hence know that $\stype' = \treft{\envrs(r)} \crejoin \jlabel{j}$ and
      the property \ref{def:stS} follows.

    \item \irule{S-Login}
      We immediately get property \ref{def:stG}, \ref{def:stS}, \ref{def:stT} of
      \cref{def:serv-typing}.
      Property \ref{def:stP} follows from rule \irule{T-Login}.

    \item \irule{S-Auth} All properties are trivial

    \item \irule{S-OChckSucc} 
      Then 
      \begin{itemize}
        \item $c=\och{L}{c'}$ 
        \item $R = n, u, \bid, o$
        \item $o \in L$
      \end{itemize}

      All properties of \cref{def:serv-typing} except for property
      \ref{def:stT} are trivial.

      We know that typing was done using rule \irule{T-OChk} or
      \irule{T-PruneOChk}. 

      We want to show that \irule{T-OChk} was used. To this end, we assume
      that \irule{T-PruneChk} was used and show a contradiction

      By rule \irule{T-OChkPrune} we know 
      \begin{itemize}
        \item $\forall l \in L. \isihigh{l}$
        \item $u \in \protUrls$ 
        \item $\typebranch=\bcsrf$
      \end{itemize}

      We hence have $\isihigh{o}$.

      Then by \ref{def:stOI} of \cref{def:serv-typing}, we know that $\amode =
      \bhon$, which is an immediate contradiction.

    \item \irule{T-OChckFail} This case is analog to the case of rule
      \irule{T-TChkcFail}
  \end{itemize}

\end{proof}

We now show that any expression that is well typed in an honest typing branch 
is also well-typed when typing in the attacker's setting and that all expressions have type $\alabel$ in the attacked setting.

\begin{lemma}[Attacker Server Expression Typability]
  \label{lem:att-serv-ex-ty}
  For all server expressions $se$ we have 
  if 
  \begin{itemize}
    \item $\exelabel{se}{\stype}$
    \item $\forall x \in \lst{x}. \, \envv'(x) = \alabel$  
    \item $\forall r \in \refs. \, \envrg'(r) = \tref{\alabel}$ 
    \item $\seslabel \neq \unauth \Rightarrow \seslabel' = \alabel$
    \item $se = \trans{se'}$ for some $se'$
  \end{itemize} 
  then we have
  $\exelabelpar{(\envu,\envv',\envrg',\envrs,\envt)}{\seslabel'}{se}{\alabel}$
\end{lemma}
\begin{proof}
  We prove the claim by induction over the typing derivation for $\exelabel{se}{\stype}$
  \begin{itemize}
    \item \irule{T-EVal} Since $se = \trans{se'}$ from some $se'$ , we have $se = v^\tbot$. The
      claim then follows since $\tbot \tleq \alabel$.
    \item \irule{T-EFresh} Then we have $se = \fresh^\stype$. The claim is
      trivial because of $\typebranch = \batt$. 
    \item \irule{T-VUndef} Trivial.
    \item \irule{T-EVar} Follows immediately from the definition of $\envv'$.
    \item \irule{T-EGlobRef} Follows immediately from the definition of $\envrg'$.
    \item \irule{T-ESesRef} Then we know that $\seslabel \neq \unauth$ and hence
      $\seslabel' = \alabel$. The claim then follows immediately from
      \irule{T-ESesRef} and \irule{T-ESub} 
    \item \irule{T-EBinOp} Then the claim follows immediately by induction.
    \item \irule{T-ESub} The claim follows immediately by induction.
  \end{itemize}
\end{proof}

Next we show, that any server thread that is well typed in the honest setting
is also well-typed when typing in the attacker's setting.

\begin{lemma}[Attacker Server Typability]
  \label{lem:att-serv-ty}
  Let $t$ be a thread with 
  \begin{itemize}
    \item $t = \listen{u}{\lst{r}}{\lst{x}}{c}$ with $\exsttyping{t}$ 
    \item $\forall x \in \lst{x}. \envv(x) = \alabel$ 
    \item $\forall r \in \refs. \envrg(r) = \alabel$ 
    \item $t = \trans{t'}$ for some $t'$
  \end{itemize}
  we have
  $\exstypingparpc{(\envu^0,\envv,\envrg,\envrs^0,\envt^0)}{\unauth}{\itop}{\batt}{c}{\_}{\_}$
\end{lemma}
\begin{proof}
  By $\exsttyping{t}$ we know by \irule{T-Recv} that with $\envu(u) = \ulabel, \lst{\stype}, l_r$ and 
  $m = \len{x}$ and 
  $\envv^h = x_1 \colon \stype_1, \ldots, x_m \colon \stype_m$ we have 
  \[
    \exstypingparpc{(\envu^0,\envv^h,\envrg^0,\envrs^0,\envt^0)}{\unauth}{I(\ulabel)}{\bhon}{c}{\_}{\_} 
  \]

  We let $\Gamma =(\envu^0,\envv,\envrg,\envrs^0,\envt^0)$ and   now show the
  following stronger claim:
  Whenever 
  \[
    \exstypingparpc{(\envu^0,\envv^h,\envrg^0,\envrs^0,\envt^0)}{\seslabel_h}{\pclabel_h}{\bhon}{c}{\seslabel_h'}{\pclabel_h'} 
  \]
  then
  \[
    \exstypingparpc{\Gamma}{\seslabel}{\itop}{\batt}{c}{\seslabel'}{\itop}
  \]
  where 
  \begin{itemize}
    \item $\seslabel_h = \unauth \Rightarrow \seslabel = \unauth \wedge \seslabel_h \neq \unauth \Rightarrow \seslabel = \alabel$ and 
    \item $\seslabel_h' = \unauth \Rightarrow \seslabel' = \unauth \wedge \seslabel'_h \neq \unauth \Rightarrow \seslabel' = \alabel$ and 
  \end{itemize}

  The proof is by induction on the honest typing derivation for $c$
  \begin{itemize}
    \item \irule{T-Skip} The claim is trivial.
    \item \irule{T-Seq} The claim follows directly from the induction
      hypothesis on the two subcommands.
    \item \irule{T-If} We have $c = \ite{se}{c_t}{c_f}$. With
      $\exelabel{se}{\alabel}$ by \cref{lem:att-serv-ex-ty}.
      We have $\pclabel' = \itop \ijoin I(\alabel) = \itop = \pclabel$ (in rule
      \irule{T-IF}) and
      the claim follows from the induction hypothesis for $c_t$ and $c_f$.
    \item \irule{T-Login} By \cref{lem:att-serv-ty}, we get that all
      expressions are of type $\alabel$. Using rule \irule{T-ESub} we can also
      treat them as expressions of type $\tcre{\low}$. The claim
      then follows immediately using \irule{T-Login}.
    \item \irule{T-Start} By \cref{lem:att-serv-ty} we get that all expressions
      are of type $\alabel$. Using rule \irule{T-ESub} we can also treat them
      as expressions of type $\tcre{\low}$.  The claim then follows
      immediately using \irule{T-Start}
    \item \irule{T-SetGlobal} We have $c = \rg{r} := se$ with 
      $\exelabelpar{\Gamma}{\seslabel'}{se}{\alabel}$ by
      \cref{lem:att-serv-ex-ty} and $\envrg(r) = \tref{\alabel}$.
      Using subtyping we can show 
      $\exelabel{se}{\pair{\cbot}{\itop}}$ and 
      $\exsrtyping{\rg{r}}{\tref{\pair{\cbot}{\itop}}}$ and the claim follows 
      using rule \irule{T-SetGlobal}.
    \item \irule{T-SetSession} We have $c = \rs{r} := se$.
      The claim follows analogous to the previous one, using that
      $\exsrtyping{\rs{r}}{\tref{\alabel}}$ because of $\seslabel = \alabel$.
    \item \irule{T-PruneTCheck}: Impossible since this rule cannot be applied
      for $\typebranch=\bhon$
    \item \irule{T-TokenCheck}: By \cref{lem:att-serv-ty} we get that all
      expressions are of type $\alabel$. Using rule \irule{T-ESub} we can also
      treat them as expressions of type $\tcre{\low}$.  The claim then 
      follows by induction and using \irule{T-TokenCheck}
    \item \irule{T-PruneOChk} Impossible since this rule cannot be applied
      for $\typebranch=\bhon$
    \item \irule{T-OChk} The claim follows immediately by induction.
    \item \irule{T-Auth}: Then the claim follows immediately using rule
      \irule{T-AuthAtt}.
    \item \irule{T-Reply} From \cref{lem:att-serv-ty} and rule \irule{T-ESub}
      we know that for all variables $x$  in the freshly generated environment
      $\envv'$ we have $\envv'(x) = \pair{\cbot}{\itop}$.
      Furthermore, with subtyping we can show
      $\exsrtyping{r}{\tref{\low}}$ for all $r \in \domain(ck)$.
      The claim then follows immediately.
    \item \irule{T-Redir} This case follows analog to the previous case.
  \end{itemize}
\end{proof}

Next we show that whenever a server receives a well typed request, 
the resulting running thread is also well-typed.

\begin{lemma}[Server Request]
  \label{lem:serv-req}
  Whenever a server $S \sstep{\alpha} S'$ with $\servtyping{S}$,
  $\alpha = \exsrecv{\bid}{n}{u}{p}{ck}{o}{l}{\amode}$ and
  $\reqtyping{\alpha}$ then $\servtyping{S'}$
\end{lemma}
\begin{proof}
  Let $S = \exserver{D}{\phi}{t}$ and $S' = \exserver{D'}{\phi'}{t'}$.
  We show that $S'$ fulfills the properties of \cref{def:serv-typing}.
  Property \ref{def:stS} and \ref{def:stP} follow immediately from rule
  \irule{S-Recv} since the session memory and the trust mapping do not change.

  \begin{itemize}
\item 
  For property \ref{def:stG} we perform a case distinction
  \begin{itemize}
    \item if $\bid \neq \usr$ then property \ref{def:stG} follows from property
      \ref{def:brac} of \cref{def:req-ty}.
    \item if $\bid = \usr$ then property \ref{def:stG} follows from property
      \ref{def:brhc} of \cref{def:req-ty}.
  \end{itemize}

\item For property \ref{def:stT} ,because of $S \sstep{\alpha} S'$ we know 
  $\exserver{D}{\phi}{\listen{u}{\lst{r}}{\lst{x}}{c}}
     \sstep{\alpha}
     \exserver{D'}{\phi'}{\para{\exthread{c\sigma}{n, u, \bid}{i,\vundef}}{\listen{u}{\lst{r}}{\lst{x}}{c}}}$.
   It is hence sufficient, because of rule
   \irule{T-Parallel}, to show
   $\exsttyping{\exthread{c\sigma}{(n,u,\bid)}{(i,\vundef)}}$

   We perform a case distinction:
  \begin{itemize}
    \item if $\bid \neq \usr$ 
      then by rule \irule{T-Running} with $\typebranch = \batt$ and $\seslabel
      = \jlabel{\vundef} = \unauth$, $\envrg' = \subst{\_}{\alabel}$ we have to
      show 
    \[
      \exstypingpar{(\envu,\envv,\envrg',\envrs,\envt)}{\seslabel}{l}{\typebranch}{c\sigma}{\seslabel'}
    \]

    Because of \cref{lem:att-serv-ty} we get with $\envv' = x_1 \colon \alabel
    \cdots x_m \colon \alabel $ 

    \[
      \exstypingpar{(\envu,\envv',\envrg',\envrs,\envt)}{\seslabel}{\itop}{\typebranch}{c}{\seslabel'}
    \]

      With property \ref{def:brap} of \cref{def:req-ty} we can use
      \cref{lem:serv-sub} for the substitution $\sigma$ and the claim follows  
      using \cref{lem:serv-pc-sub}.

    \item if $\bid = \usr \wedge \amode = \batt$ 
      then by rule \irule{T-Running} with $\typebranch = \bcsrf$ and $\seslabel
      = \jlabel{\vundef} = \unauth$, we have to show
      $\exstypingpar{\Gamma}{\seslabel}{l}{b}{c\sigma}{\seslabel'}$.

      Since $l \ileq \itop$ 
      using \cref{lem:serv-pc-sub} it is sufficient to show 
    \[
      \exstypingpar{\Gamma}{\seslabel}{\itop}{\typebranch}{c\sigma}{\seslabel'}
    \]

      From rule \irule{T-Recv} we get with
		  $\envv' = x_1 \colon \low, \ldots, x_m \colon \low$ that
      \[
      \exstypingpar{(\envu,\envv',\envrg,\envrs,\envt)}{\unauth}{\ilow}{\bcsrf}{c}{\_}
    \]

      With property \ref{def:brap} of \cref{def:req-ty} we can use \cref{lem:serv-sub}
      for the substitution $\sigma$ and the claim follows.  

    \item if $\bid = \usr \wedge \amode = \bhon$ 
      then by rule \irule{T-Running} with $\typebranch = \bhon$ and$\seslabel
      = \jlabel{\vundef} = \unauth$, we have to show
      $\exstypingpar{\Gamma}{\seslabel}{l}{b}{c\sigma}{\seslabel'}$.
    
      Since we know $l \ileq I(\ulabel)$ by property \ref{def:brhp} of
      \cref{def:req-ty}, using \cref{lem:serv-pc-sub} it is sufficient to show 
      \[
      \exstypingpar{\Gamma}{\seslabel}{I(\ulabel)}{\typebranch}{c\sigma}{\seslabel'}
    \]

      From rule \irule{T-Recv} we get with $\envu(u) = \ulabel, \lst{t}, lr$ and 
		  $\envv' = \envv^0, x_1 \colon t_1, \ldots, x_m \colon t_m$
      \[
      \exstypingpar{(\envu,\envv',\envrg,\envrs,\envt)}{\unauth}{\ilow}{\bhon}{c}{\_}
    \]

      With property \ref{def:brhp} of \cref{def:req-ty} we can use \cref{lem:serv-sub}
      for the substitution $\sigma$ and the claim follows.  
  \end{itemize}

\item Property \ref{def:stOI} follows immediately from property \ref{def:bro}
  of \cref{def:req-ty}.
  \end{itemize}
\end{proof}

We now show that all responses by the server fulfill these conditions.

\begin{lemma}[Server Response]
  \label{lem:serv-res}
  Whenever a server $S \sstep{\alpha} S'$ with $\servtyping{S}$,
  $\alpha = \exssend{\bid}{n}{u}{u'}{\lst{v}}{ck}{page}{s}{l}{\amode}$ then
  $\restyping{\alpha}$ and $\servtyping{S'}$
\end{lemma}
\begin{proof}
  \ptodo{4}{Formalize this more}
  $\servtyping{S'}$ is trivial in all cases.
  We show that $\alpha$ fulfills all properties of \cref{def:res-ty}.
  We perform a case distinction on the rule used to type the reply. 
  \begin{itemize}
    \item 
  \irule{T-Redir}:
  Property \ref{def:srhn} follows directly from \irule{T-Redir}
  \ptodo{1}{Not if there are non-constant or non-variable expressions in the
    reply}
  We perform a case distinction: 
  \begin{itemize}
    \item If $\bid \neq \usr$ Then we need to show property \ref{def:sra}
      which follows immediately from the typing rule, using \cref{lem:att-serv-ex-ty} \ptodo{3}{does not apply, since we don't have ``static'' precondition. But holds, since all constants are at most $\alabel$}
    \item If $\bid = \usr$ and $\amode = \bhon$,
      then property \ref{def:src} follows from \irule{T-Redir}.
      Properties \ref{def:srhp} and \ref{def:srhs} are trivial.
      Property \ref{def:srru} follows from \irule{T-Redir}.  \item If $\bid =
      \usr$ and $\amode = \batt$ then properties \ref{def:srcs} and
      \ref{def:srap} are  trivial.
      Property \ref{def:src} follows from \irule{T-Redir}.
      Property \ref{def:srru} follows from \irule{T-Redir}.
  \end{itemize}
    \item 
  \irule{T-Reply}:
  Property \ref{def:srhn} follows directly from \irule{T-Reply} and property \ref{def:srru} is trivial.
  We perform a case distinction: 
  \begin{itemize}
    \item If $\bid \neq \usr$ Then we need to show property \ref{def:sra}
      which follows immediately from the typing rule, using \cref{lem:att-serv-ex-ty}
    \item If $\bid = \usr$ and $\amode = \bhon$,
      then properties \ref{def:srhp}, \ref{def:srhs} and \ref{def:src} of
      \cref{def:res-ty} follow from \irule{T-Reply} and \irule{T-Form}
    \item If $\bid = \usr$ and $\amode = \batt$
      then properties \ref{def:srap} and \ref{def:src} of
      \cref{def:res-ty} follow from \irule{T-Reply}.
      Property \ref{def:srcs} follows immediately from \irule{T-Reply} and
      from the observation that rule \irule{T-BERefFail} and
      \irule{T-BAssignFail} are not used for typing the script, as the script
      can also be typed in the honest typing branch $\typebranch = \bhon$.
  \end{itemize}
  \item \irule{T-ReplyErr}: All claims are trivial.
  \end{itemize}
\end{proof}

We can now define the typing of websystems, which simply states that all 
browsers and servers contained in the system are well typed.

\begin{definition}[System Typing]
  \label{def:sys-typing}
  Let $W$ be a websystem. We write $\systyping{\exatkstate{\alabel}{\atknow}{W}}$, if 
  \begin{enumerate}
    \item \label{def:systB} for all $S \in \servers{W}$ we have  $\servtyping{S}$ 
    \item \label{def:systS} for all $B \in \browsers{W}$ we have $\browtyping{B}$ 
    \item \label{def:systL} for all $B \in \browsers{W}$ with $B =
      \exbrowser{N}{M}{P}{T}{Q}{\lst{a}}$ and $N = \subst{n}{u}$ we have one of the following:
      \begin{itemize}
        \item there exists $S \in \servers{W}$ with $t \in \running{S}$, 
          $t = \exthreadpar{c}{(n,u,\bid)}{(i,j)}{l'}{\amode}$ and 
          $l \ileq \tlowestint{t}$ 
          for some $c, l', i, j$,
        \item or $\isilow{\lambda(u)}$
        \item or $\timeouts = \{(\bid, n, u, l', \amode)\}$ for some $l'$
      \end{itemize}
    \item \label{def:systA} for all $v^{\stype} \in \atknow$ we have $\stype \tleq \alabel$
  \end{enumerate}
\end{definition}

%


Next, we show that any script created by the attacker, that is served over a
low integrity network connection is well-typed in the users browser.

\begin{lemma}[Attacker Script Typability]
  \label{lem:att-bro-ty}
  For all scripts $s$ and well formed environments $\Gamma$, URLs $u$ with
  $\isilow{I(\lambda(u))}$, $\forall n^\stype \in \vals{s}. \, \stype \tleq
  \alabel$, $\fvars{s} = \emptyset$ we have
  $\exbstypingpar{\Gamma}{I(\alabel)}{\bcsrf}{s}$.
\end{lemma}
\begin{proof}
  We first show that for all browser expressions $be$ we have 
  $\exbelabelpar{\Gamma}{\bcsrf}{be}{\alabel}$.
  We show the claim by induction over $be$.
  \begin{itemize}
    \item $be = r$. 
      \begin{itemize}
        \item if $C(\lambda(r)) \not\cleq C(\lambda(u))$ the claim follows
          immediately using rule \irule{T-BERefFail}
        \item if $C(\lambda(r)) \cleq C(\lambda(u))$ we know by the
          well-formedness of $\Gamma$ that $\isclow{\Ct{\treft{\envrg(r)}}}$ 
          and hence also $\isilow{\It{\treft{\envrg(r)}}}$
          and the claim follows using \irule{T-BERef} and \irule{T-BESub}
      \end{itemize}
    \item $be = v^\stype$: The claim follows from the assumption $\stype \tleq
      \alabel$ and rule \irule{T-BEVal} 
    \item $be = \dom{be'}{be''}$: Immediately by rule \irule{T-BEDom}.
    \item $be = be_1 \odot be_2$: By induction and rule \irule{T-BEBinOp}
  \end{itemize}

  We now show the main claim by induction over $s$.
  \begin{itemize} \item $s = s_1;s_2$: the claim follows from the induction hypothesis for
      $s_1$ and $s_2$ and \irule{T-Bseq}
    \item $ s = \sskip$ : trivial with \irule{T-BSkip}
    \item $ s = r := be$  We distinguish two cases
      \begin{itemize}
        \item if $I(\lambda(u)) \not\ileq I(\lambda(r))$ then the claim is
          trivial with rule \irule{T-BAssignFail}.
        \item if $I(\lambda(u)) \ileq I(\lambda(r))$ then we know 
          because of $\isilow{\lambda(u)}$ that also
          $\isilow{\lambda(r)}$.
          Therefore, we know by well-formedness of $\Gamma$ that 
          if $\envrg(r) = \stype$ with $\stype = \tcre{\cdot}$ then 
          $\isclow{\Ct{\stype}}$.
          We can hence show $\exbrtyping{r}{\tref{\alabel}}$.
          Since we know that $\exbelabelpar{\Gamma}{\typebranch}{be}{\alabel}$
          the claim follows.
      \end{itemize}
    \item $ s =\setdom{v}{u}{\lst{be}}$: The claim follows from rule
      \irule{T-BSetDom}, using our observation about expression types.
    \item $ s =\incl{u}{\lst{be}}$: The claim follows from rule
      \irule{T-BInclude}, using our observation about expression types.
  \end{itemize}

\end{proof}

Next, we show that the attacker can only learn low confidentiality values from 
the network.

\begin{lemma}{Attacker Knowledge for low confidentiality requests}
  \label{lem:atknow-low-conf-req}
  Whenever we have $\alpha = \exbsend{\usr}{n}{u}{p}{ck}{o}{l}{\amode}$ with 
  $\reqtyping{\alpha}$ and $\isclow{\lambda(u)}$ then for all $n^\stype \in
  \fnames{p,ck}$ we have $\isclow{\Ct{\stype}}$.
\end{lemma}
\begin{proof}
  For $\amode = \bhon$ the claim for $\fnames{p}$ follows immediately from
  the well-formedness of URLs and property \ref{def:brhp} of
  \cref{def:req-ty}, 
  otherwise the claim follows directly from 
  property \ref{def:brap} of \cref{def:req-ty}, 

  The claim for $\fnames{ck}$ follows immediately from property 
  \ref{def:brhc} of \cref{def:req-ty}.
\end{proof}

The next two lemmas show that requests and responses crafted by the attacker
are well-typed.

\begin{lemma}[Attacker Request]
  \label{lem:att-req}
   Let $\alpha = \exbsend{\bid}{n}{u}{p}{ck}{o}{\ibot}{\batt}$ with
   $\bid \neq \usr$ and 
   for all $v^\stype \in \vals{p,ck}$ we have 
   $\stype \tleq \alabel$.
   Then $\reqtyping{\alpha}$.
\end{lemma}
\begin{proof}
      Since $\bid \neq \usr$ and $\amode = \batt$, we have to show properties
      \ref{def:brap} and \ref{def:brac} of \cref{def:req-ty}.
      Both claims follow immediately since 
      $\forall n^\stype \in \fnames{p,ck}, \stype \tleq \alabel$.
\end{proof}

\begin{lemma}[Attacker Response]
  \label{lem:att-res}
  Let $\alpha = \exssend{\usr}{n}{u}{u'}{\lst{v}}{ck}{page}{s}{\itop}{\batt}$
  with $\isilow{\lambda(u)}$ and  
   for all $v^\stype \in \vals{p,ck}$ we have 
   $\stype \tleq \alabel$.
  Then $\restyping{\alpha}$
\end{lemma}
\begin{proof}
  We show that $\alpha$ fulfills the properties of \cref{def:res-ty}.

  We have to show properties \ref{def:srhn}, \ref{def:srap}, \ref{def:srcs}, 
  \ref{def:src} and \ref{def:srru} of \cref{def:res-ty}.

  With $\forall n^\stype \in \fnames{\lst{v},ck,page,s}, \stype \tleq \alabel$.
  Properties \ref{def:srhn} and \ref{def:srap} are trivial. 
  Property \ref{def:srcs} follows immediately from \cref{lem:att-bro-ty}

  For Property \ref{def:src} we look at all $r \in ck$ and perform a case
  distinction:
  \begin{itemize}
    \item If $\lambda(u) \ileq \lambda(r)$ then by transitivity of $\ileq$
      we know $\isilow{\lambda(r)}$ and hence by well-formedness of
      $\Gamma$ we know that if $\envrg(r) = \tcre{\cdot}$ then
      $\isclow{\Ct{\envrg(r)}}$. We hence know that $\alabel \tleq \treft{\envrg(r)}$.
    \item If $\lambda(u) \not\ileq \lambda(r)$ then the property is
      trivially true.
  \end{itemize}
   
  For property \ref{def:srru}, we have to show property \ref{def:brap} of
  \cref{def:req-ty}, which follows immediately.
\end{proof}

Finally, we show that whenever a well-formed system takes a step, it produces
another well-typed system.

\begin{lemma}[System Subject Reduction]
  \label{lem:sys-sub-red}
  Let $W$ be a websystem with $\systyping{\atkstate{\alabel}{\atknow}{W}}$ and 
  $\atkstate{\alabel}{\atknow}{W}\astep{\alpha} 
    \atkstate{\alabel}{\atknow'}{W'}$. 
  Then we have $\systyping{\atkstate{\alabel}{\atknow'}{W'}}$
\end{lemma}
\begin{proof}
  We do a proof by a case analysis over the derivation of $\astep{\alpha}$
  \begin{itemize}
    \item \irule{A-Nil} If the step was taken using rule \irule{A-Nil} then we
      perform an induction on the internal step.
      If the step is taken through rule \irule{W-LParallel} or
      \irule{W-RParallel} the claim follows by induction. 
      If it is taken locally in one browser or
      server the claim follows from \cref{lem:bro-sub-red} or
      \cref{lem:serv-sub-red} and the fact that $\atknow = \atknow'$.
      \sketch{Property \ref{def:systL} of \cref{def:sys-typing} follows
      from the observation that raising the server integrity label can only
      happen in rule \irule{S-Reset}.}
    \item \irule{A-BroSer} Follows immediately from \cref{lem:bro-req} and
      \cref{lem:serv-req}. Property \ref{def:systL} follows from the semantics
      rules for browsers and servers.
      Property \ref{def:systA} follows from \cref{lem:atknow-low-conf-req}.
    \item \irule{A-SerBro} Follows immediately from \cref{lem:serv-res} and
      \cref{lem:bro-res}. We can apply \cref{lem:bro-res} because of 
      property \ref{def:systL} of \cref{def:sys-typing}. 
    \item \irule{A-TimeOutSend}
      Follows immediately from \cref{lem:bro-req}.
    \item \irule{A-TimeOutRecv} 
      Let $\alpha'$ be the response sent in the rule. Then we trivially have 
      $\restyping{\alpha'}$ and the claim follows 
      using \cref{lem:bro-res} and property \cref{def:systL} of
      \cref{def:sys-typing}.
    \item \irule{A-BroAtk} We then have $W \astep{\alpha} W'$ with 
      $\alpha = \exbsend{\usr}{n}{u}{p}{ck}{o}{l}{\amode}$ and
      $\isilow{\lambda(u)}$.
      The typing of the browser follows immediately from \cref{lem:bro-req}.
      Property \ref{def:systA} follows from \cref{lem:atknow-low-conf-req}.

    \item \irule{A-AtkSer} We then have $W \astep{\alpha} W'$ with 
      $\alpha = \exsrecv{\bid}{n}{u}{p}{ck}{o}{I(\alabel)}{\batt}$, where
      $\fnames{p,ck} \subset \atknow$. 

      We hence get by \cref{lem:att-req} that $\reqtyping{\alpha}$ and the
      claim follows from \cref{lem:serv-req}.

    \item \irule{A-SerAtk}
      We then have $W \astep{\alpha} W'$  with
      $\alpha = \exssend{\bid}{n}{u}{u'}{\lst{v}}{ck}{page}{s}{l}{\amode}$, 
      where $n \in \atknow$.

      From \cref{lem:serv-res} we get that the resulting server state is
      well-typed and that $\alpha$ is a well-typed response.
      We now have to show that for all $v^{\stype} \in \atknow'$ we have
      $\stype \tleq \alabel$. 
      Since $n \in \atknow$
      \sketch{and the only point where the attacker can
        learn $n$ is in rule \irule{A-AtkSer}} 
      we know that 
      $\bid \neq \usr$ and $\amode = \batt$.
      The claim follows directly from property 
      \ref{def:sra} 
      of \cref{def:res-ty}.

    \item \irule{A-AtkBro}
      We then have $W \astep{\alpha} W'$  with
      $\alpha = \exbrecv{\bid}{n}{u}{u'}{\lst{v}}{ck}{page}{s}{l}{\amode}$ where
      $\bid =\usr$, $l = \itop$, $\amode = \batt$ and $\isilow{\lambda(u)}$. 
      By \cref{lem:att-res} we get that $\alpha$ is a well-typed response, then
      the claim follows from \cref{lem:bro-res} 
      (which we can apply, since $\amode = \batt$ and $l = \itop$).
  \end{itemize}
  
\end{proof}

\subsection{Relation}
\label{sec:proof-rel}

We now define a notion of \emph{High Equality} between different components,
that will be used to relate two websystems.

The general intuition is, that everything that is of high integrity must be
equal, while values of low integrity can be arbitrarily different.

\begin{definition}[High Equality]
  \label{def:heq}
  We define high equality in different contexts:
  
  \begin{enumerate}

    \item
  For two (browser or server) expressions $e$, $e'$ we inductively define $e \heqe e'$
  by the following rules.

  \begin{mathpar}
    \inferrule*{~}{e \heqe e}

    \inferrule*{\isilow{\It{\stype} \imeet \It{\stype'}  }}{v^{\stype} \heqe v'^{\stype'}}

    \inferrule*{\isilow{\It{\stype} \imeet \It{\stype'} }}{\fresh^{\stype} \heqe \fresh^{\stype'}}

    \inferrule*{e_1 \heqe e_2 \\ e_2 \heqe e_2'}{e_1 \odot e_1' \heqe e_2 \odot e_2'}

    \inferrule*{e_1 \heqe e_1' \\ e_2 \heqe e_2'}{\dom{e_1}{e_2} \heqe \dom{e_1'}{e_2'}}

    \inferrule*{\len{e} = \len{e'} \\ \forall i \in \intval{1}{\len{e}}.  e_i \heqe e_i'}{\lst{e} \heqe \lst{e'}}
  \end{mathpar}

    \item
  For two pages $page$, $page'$ we define $page \heqe page'$ as 

  \begin{mathpar}
    \inferrule*
    {
      \domain(page) = \domain(page') \\
      \forall v \in \domain(page). \, 
      page(v) = \exhform{u_i}{\lst{v_i}}{\amode} \wedge 
      page'(v) = \exhform{u_i'}{\lst{v'_i}}{\amode'} \wedge 
       u_i \heqe u'_i \wedge 
       v_i \heqe v_i'
    }
    {page \heqe page'}
  \end{mathpar}

    \item
  For two scripts $s$, $s'$ we define $s \heqe s'$ as 
  \begin{mathpar}
    \inferrule*{~}{\sskip \heqe \sskip}

    \inferrule*{s_1 \heqe s_2 \\ s_2 \heqe s_2'}{s_1 ; s_1' \heqe s_2 ; s_2'}

    \inferrule*{be \heqe be'}{r := be \heqe r := be'}

    \inferrule*{\lst{be} \heqe \lst{be'}}
    {\incl{u}{\lst{be}} \heqe \incl{u}{\lst{be'}}}

    \inferrule*{ be \heqe be' \\
      \lst{be} \heqe \lst{be'}}
    {\setdom{be}{u}{\lst{be}} \heqe \setdom{be'}{u}{\lst{be'}}}
  \end{mathpar}

    \item
  For two commands $c$, $c'$ we define $c \heqe c'$ as 
  \begin{mathpar}
    \inferrule*{~}{\sskip \heqe \sskip}

    \inferrule*{~}{\halt \heqe \halt}

    \inferrule*{c_1 \heqe c_2 \\ c_2 \heqe c_2'}{c_1 ; c_1' \heqe c_2 ; c_2'}

    \inferrule*{se \heqe se' \\ c_1 \heqe c_1' \\ c_2 \heqe c_2'}{\ite{se}{c_1}{c_2} \heqe \ite{se'}{c_1'}{c_2'}}

    \inferrule*{se_1 \heqe se_1' \\ se_2 \heqe se_2' \\ se_3\heqe se_3'}{\login{se_1}{se_2}{se_3} \heqe \ite{se_1'}{se_2'}{se_3'}}

    \inferrule*{se \heqe se'}{\start{se} \heqe \start{se'}}

    \inferrule*{ \forall i \in \intval{1}{\len{se}}. \, se_i \heqe se'_i}
    {\auth{\lst{se}}{l} \heqe \auth{\lst{se'}}{l}}

    \inferrule*{\forall i \in \intval{1}{\len{se}}. \, se_i \heqe se'_i}
    {\reply{page}{s}{ck}{\lst{x} = \lst{se}} \heqe \reply{page}{s}{ck}{\lst{x} = \lst{se'}}}

    \inferrule*{\forall i \in \intval{1}{\len{se}}. \, se_i \heqe se'_i}
    {\redr{u}{\lst{z}}{ck}{\lst{x} = \lst{se}} \heqe \redr{u}{\lst{z}}{ck}{\lst{x} =\lst{se'}}}

    \inferrule*{\forall i \in \intval{1}{\len{se}}. \, se_i \heqe se'_i}
    {\redr{u}{\lst{z}}{ck}{\lst{se}} \heqe \redr{u}{\lst{z}}{ck}{\lst{se'}}}
  \end{mathpar}

    \item 
  For two memories $M$, $M'$ we write 
  $M \heqm M'$ , if 
  \begin{itemize}
    \item for all $r \in \domain(M) \cup \domain(M')$ 
      we have $M(r) \heqe M'(r)$ 
    \item for all $r \in \domain(M) \setminus \domain(M')$ with
      $M(r) = v^{\stype}$ we have $\isilow{\stype}$
    \item for all $r \in \domain(M') \setminus \domain(M)$ with
      $M'(r) = v^{\stype}$ we have $\isilow{\stype}$
  \end{itemize}  


  \item
    \label{def:low-int-req}
    For two requests $\alpha = \exbsend{\bid}{n}{u}{p}{ck}{o}{l}{\amode}$ (resp. 
    $\alpha = \exsrecv{\bid}{n}{u}{p}{ck}{o}{l}{\amode}$)
    and $\beta = \exbsend{\bid'}{n'}{u'}{p'}{ck'}{o'}{l'}{\amode'}$ (resp. 
    $\beta = \exsrecv{\bid'}{n'}{u'}{p'}{ck'}{o'}{l'}{\amode'}$)
    we let 
    \begin{align*}
      \alpha \heqe \beta \iff
        & \isihigh{l \ijoin l'} \Rightarrow  \\
        & \bid = \bid' \wedge n = n' \wedge u = u' \wedge \\ 
        & \domain(p) = \domain(p') \wedge \forall x \in \domain(p). \, p(x) \heqe p'(x)  \\
        & ck \heqm ck' \\
        & \wedge l = l' \wedge \amode = \amode'
    \end{align*}

  \item
    \label{def:low-int-res}
    For two responses $\alpha = \exssend{\bid}{n}{u}{u_r}{\lst{v}}{ck}{page}{s}{l}{\amode}$
    (resp.  $\alpha = \exbrecv{\bid}{n}{u}{u_r}{\lst{v}}{ck}{page}{s}{l}{\amode}$) and 
    $\beta = \exssend{\bid'}{n'}{u'}{u_r'}{\lst{v'}}{ck'}{page'}{s'}{'l}{\amode'}$
    (resp.  $\beta = \exbrecv{\bid'}{n'}{u'}{u_r'}{\lst{v'}}{ck'}{page'}{s'}{'l}{\amode'}$) 
    we let
    \begin{align*}
      \alpha \heqe  \beta \iff 
      & ck \heqmpar{\Gamma} ck' \\
      & \isihigh{l \ijoin l'} \Rightarrow  \\
    &\bid = \bid' \wedge n = n' \wedge u = u' \wedge u_r = u_r' \wedge \lst{v} \heqe \lst{v'}\\   
      & page \heqe page' \wedge s \heqe s' \wedge  \\
      & l = l' \wedge \amode = \amode'
    \end{align*}

  \item For two authentication events 
    $\alpha = \lauth{\vec{v}}{\bid,\uid}{\ell}$ and 
    $\alpha' = \lauth{\vec{v'}}{\bid',\uid'}{\ell'}$ we let 
    $\alpha \heqe \alpha'$ if 
    \begin{itemize}
      \item $\isilow{I(\ell)}$ and $\isilow{I(\ell')}$ or
      \item $\alpha = \alpha'$
    \end{itemize}

%
%
%
%
%
%
  \end{enumerate}

\end{definition}

We introduce a predicate $\rbad{\cdot}$ which we use to denote that the system
has entered a state in which the browser will perform no more actions because
  it received an error message from the server.

\begin{definition}[Bad State]
  \label{def:bad}
  \begin{itemize}
    \item A browser $B=\exebrowser{N}{K}{P}{T}{Q}{\lst{a}}{\usr}{l}{\amode}$
      is in a bad state and we write $\rbad{B}$ if $\halt \in \lst{a}$.

    \item A server $S = \exserver{D}{\phi}{t}$ is in a bad state and we write
      $\rbad{S}$ if there is a $t \in  \running{S}$ with 
      $t = \exthreadpar{c}{E}{R}{l}{\amode}$ and 
      \begin{itemize}
        \item $\bid = \usr$
        \item $c = \ereply{\serror}{\sskip}{\mempty}$ or $c = \sbad$
      \end{itemize}

    \item A web system $\atkstate{\alabel}{\atknow}{W}$ is in a bad state and
      we write $\rbad{A}$ if 
      \begin{itemize}
        \item with $\{B\} = \browsers{W}$, we have $\rbad{B}$
        \item for any $S \in \servers{W}$, we have $\rbad{S}$
      \end{itemize}
  \end{itemize}

  The following properties are straightforward by inspecting the semantic
  rules:
  \begin{itemize}
    \item If $\rbad{A}$ and $A \astepn{*}{\lst{\alpha}} A'$, then $\rbad{A'}$.
    \item If $\rbad{A}$ and $A \astepn{*}{\lst{\alpha}} A'$, then there does
      not exist $\lauth{\lst{v}}{\bid, \uid}{\ell} \in \lst{\alpha}$ with
      $\isihigh{\It{\ell}}$. 
      \ptodo{1}{Not so straight forward, make this a lemma}
  \end{itemize}
\end{definition}

We now define a relation between two browsers.

\begin{definition}[Browser Relation]
  \label{def:brel}
  Let $B = \exebrowser{N}{K}{P}{T}{Q}{\lst{a}}{\usr}{l}{\amode}$ and  
  $B' = \exebrowser{N'}{K'}{P'}{T'}{Q'}{\lst{a'}}{\usr}{l'}{\amode'}$ be browsers.
  Then we write  $B \bprel B'$ if
      the following conditions hold 
    \begin{enumerate}
      \item \label{def:brelTy} $\browtyping{B}$ and $\browtyping{B'}$
      \item \label{def:brelL} $\isilow{l} \iff \isilow{l'}$ and $\isihigh{l} \Rightarrow l = l'$
      \item \label{def:brelN} If $\isihigh{l}$ then $N = N'$
      \item \label{def:brelK} $K \heqmpar{\envrg} K'$,
      \item \label{def:brelP} 
        For all $ t \in \domain(P)$
        if $P(t)= (u_1,page_1,l_1,\amode_1)$ then $\isilow{l_1}$ or 
        $t \in \domain(P')$ with $P'(t)= (u_2,page_2,l_2,\amode_2)$
        and $u_1 = u_2$, $page_1 \heqe page_2$
        and vice versa
      \item \label{def:brelT} If $\isihigh{l}$ then $\domain(T) = \domain(T')$
        and if $T = \subst{t}{s}$ and $T'= \subst{t}{s'}$ then $s \heqe s'$
      \item \label{def:brelA} $\lst{a} = \lst{a'}$ 
      \item \label{def:brelQ} If $\isihigh{l}$ and $Q = \{\alpha \}$ then $Q' = \{ \alpha' \}$ with $\alpha \heqe \alpha'$.
    \end{enumerate}
  
  We let $B \brel B'$ if 
  \begin{itemize}
    \item $\rbad{B}$ 
    \item or $B \bprel B'$
  \end{itemize}

\end{definition}

Intuitively, two browsers are related by the relation $\bprel$
if 
\begin{enumerate}
  \item Both browsers are well typed
  \item Either both have low or high integrity. If the integrity is high, it must be the same.
  \item If the integrity is high, then the network connections are equal
  \item The cookie jars fulfill high equality
  \item For any high integrity page in a tab of one browser, there exists a
    page  in the same tab of the other browser, with same URL, integrity and
    attacked mode, and a DOM that fulfills high equality.
  \item For high integrity browsers the scripts fulfill high equality
  \item The list of user actions is equal
  \item If the browsers are in high integrity states, then the events in the
    output buffer must fulfill high equality.
\end{enumerate}
We then define the relation $\brel$, which holds
if the left browser is in a bad state, or the browsers are in the relation
$\bprel$.

We then show that the relation $\bprel$ is symmetric and transitive. Note that this does not hold for $\brel$.

\begin{lemma}[$\bprel$ is symmetric and transitive]
  \label{lem:brel-trans}
  The relation $\bprel$ is symmetric and transitive.
\end{lemma}
\begin{proof}
  Trivial, by checking the individual properties of \cref{def:brel}
\end{proof}

Next, we show that high equality on browser expressions is preserved under
evaluation in the browser.

\begin{lemma}[Preservation of $\heqe$ under browser evaluation]
  \label{lem:pres-heq-beval}
  Let $be$ and $be'$ be browser expressions with $be \heqe be'$, let $M,M'$ be
  memories with $M \heqmpar{\envrg} M'$, let $u$ be a URL and let $page=f$ and
  $page'=f'$ be pages with $page \heqe page'$. 
  Then $\beval{be}{M}{f}{\lambda(u)} \heqe \beval{be'}{M'}{f'}{\lambda(u)}$.
\end{lemma}
\begin{proof}
  Let $v^{\stype} = \beval{be}{M}{f}{\lambda(u)}$ and 
  $v'^{\stype'} = \beval{be'}{M'}{f'}{\lambda(u)}$.
  If $\isilow{I(\stype) \imeet I(\stype')}$ the claim is trivial. We hence now
  assume $\isihigh{I(\stype) \imeet I(\stype')}$, i.e., $\isihigh{\It{\stype}} \vee \isihigh{\It{\stype'}}$
  \begin{itemize}
    \item $be = x$ : Impossible, since evaluation is not defined on variables.
    \item $be = v^{\stype}$ : Trivial, since evaluation on values is the identity (\irule{BE-Val}).
    \item $be = be_1 \odot be_2$: Then $be' = be_1' \odot be_2'$ with $be_1
      \heqe be_1'$ and $be_2 \heqe be_2'$.
      Let $v_1^{\stype_1} = \beval{be_1}{M}{f}{\lambda(u)}$,
      let $v_2^{\stype_2} = \beval{be_2}{M}{f}{\lambda(u)}$,
      let $v_1'^{\stype_1'} = \beval{be_1'}{M'}{f'}{\lambda(u)}$ and
      let $v_2'^{\stype_2'} = \beval{be_2'}{M'}{f'}{\lambda(u)}$.
      By induction we get 
      $v_1^{\stype_1} \heqe v_1'^{\stype_1'}$ and $v_2^{\stype_2} \heqe v_2'^{\stype_2'}$.
      By rule \irule{BE-BinOp} we know that
      $\It{\stype} = \It{\stype_1} \ijoin \It{\stype_2}$ and 
      $\It{\stype'} = \It{\stype_1'} \ijoin \It{\stype_2}'$.

      We know that $\isihigh{\It{\stype}}$ or $\isihigh{\It{\stype'}}$.
      We perform a case distinction:
      \begin{itemize}
        \item If $\isihigh{\It{\stype}}$ then we know that 
          $\isihigh{\It{\stype_1}}$ and $\isihigh{It{\stype_2}}$.
          By the definition of $\heqe$ we then know that 
          $v_1^{\stype_1} =  v_1'^{\stype_1'}$ and
          $v_2^{\stype_2} =  v_2'^{\stype_2'}$ and we get that 
          $v^{\stype} = v'^{\stype'}$.
        \item  If $\isihigh{\It{\stype'}}$ the claim follows analog.
      \end{itemize}

    \item $be = r$: Then $be' = r$. 
      By rule \irule{Be-BE-Reference} we have $v^{\stype} = M(r)$ and
      $v'^{\stype'} = M'(r)$ and the claim immediately follows because of
      $M \heqe M'$.  
    \item $be = \dom{be_1}{be_2}$: 
      Then $be' = \dom{be_1'}{be_2'}$
      with $be_1 \heqe be_1'$ and $be_2 \heqe be_2'$.
      Let $v_1^{\stype_1} = \beval{be_1}{M}{f}{\lambda(u)}$,
      let $v_2^{\stype_2} = \beval{be_2}{M}{f}{\lambda(u)}$,
      let $v_1'^{\stype_1'} = \beval{be_1'}{M'}{f'}{\lambda(u)}$ and
      let $v_2'^{\stype_2'} = \beval{be_2'}{M'}{f'}{\lambda(u)}$.
      By induction we get 
      $v_1^{\stype_1} \heqe v_1'^{\stype_1'}$ and $v_2^{\stype_2} \heqe v_2'^{\stype_2'}$.

      We distinguish the following cases:
      \begin{itemize}
        \item If $\isilow{\It{\stype_1} \ijoin \It{\stype_1'}}$ then by the
          definition of $\heqe$ we also know that 
          $\isilow{\It{\stype_2} \ijoin \It{\stype_2'}}$.
          Then the claim is trivial, since then $\isilow{\It{\stype} \ijoin
          \It{\stype'}}$ by \irule{BE-Dom}.
        \item If $\isilow{\It{\stype_2} \ijoin \It{\stype_2'}}$ the claim
          follows analog to the previous one.
        \item If $\isihigh{\It{\stype_1}}$, $\isihigh{\It{\stype_1'}}$,
          $\isihigh{\It{\stype_2}}$ and $\isihigh{\It{\stype_2'}}$ 
          Then we know by that $v_1 = v_1'$ and $v_2 = v_2'$.
          The claim then follows from $page \heqe page'$ and rule
         \irule{BE-Dom}.  \end{itemize}
  \end{itemize}
\end{proof}

Now we introduce the notion of deterministic termination. This property states
that a system terminates and can only produce a single trace. This is a
property that holds in the honest run, as the assumptions on user behaviour
allow only terminating runs and without the attacker there is no point of
non-determinism.

\begin{definition}[Deterministic Termination]
  \label{def:termination}
  We say that a websystem $W$ is deterministically terminating for a user
  $\usr$ if there exists exactly one unattacked trace $\gamma$ such that
  $\atkstate{\alabel}{\atknow}{W} \astepn{*}{\gamma}
  \atkstate{\alabel}{\atknow'}{W'}$ where 
  $W ' = {\para{\sbrowser{M'}{P'}{\lempty}{\usr}}{W'}}$ 
  \sketch{and $\browsers{W'} = \emptyset$}
  for some $\atknow', W', M', P'$.

  We say that a server thread $t = \exthreadpar{c}{E}{R}{l}{\amode}$ is
  deterministically terminating if there exists exactly one $\lst{\alpha}$
  with $t \sstepn{*}{\alpha} t'$ for some 
  $t' = \exthreadpar{c'}{E'}{R'}{l'}{\amode'}$ with
  \begin{itemize}
    \item $c' = \reply{page}{\cdot}{\cdot}{\lst{x} = \cdot}$, where $page \neq \serror$ 
    \item or $c' = \redr{\cdot}{\cdot}{\cdot}{\lst{x} = \cdot}$
  \end{itemize}

  We say that server $S$ is deterministically terminating if 
  all $t \in \running{S}$ are deterministically terminating.

  Note that it immediately follows that all servers in a deterministically
  terminating web system are also deterministically terminating.
\end{definition}

Next we define a relation between two servers:

\begin{definition}[Server Relation]
  \label{def:srel}
  Let $S = \exserver{D}{\phi}{t}$ and  $S = \exserver{D'}{\phi'}{t'}$
  Then we write $S \sprel S'$ if 
    \item or 
    the following conditions hold
    \begin{enumerate}
      \item \label{def:srelTy} $\servtyping{S}$ and $\servtyping{S'}$
      \item 
        Let $t_1^{H} := \{t_1 | t_1 \in \running{t} \wedge \isihigh{\thighestint{t_1}}\}$
        and   
        $t_2^{H} := \{t_2 | t_2 \in \running{t} \wedge \isihigh{\thighestint{t_2}}\}$.
        Then there is a bijection $c : t_1^{H} \rightarrow t_2^{H}$  such that for all
        $t_1 \in t_1^{H}$ and  $t_2 = c(t_1) \in t_2^{H}$, if we let 
        $t_1 = \exthreadpar{c_1}{E_1}{R_1}{l_1}{\amode_1}$ and $t_2 = \exthreadpar{c_2}{E_2}{R_2}{l_2}{\amode_2}$
        then we have 
        \begin{enumerate}
          \item \label{def:srelE} $R_1 = R_1$ and with $E_1 = i_1,j_1$ and $E_2 = i_2,j_2$ we
            have $i_1 = i_2$ and $j_1 \heqe j_2$.
          \item \label{def:srelG}
            With $E_1 = (i_1,j_1)$ we have
            ${\gmems}_1(i_1) \heq{\Gamma} {\gmems}_2(i_1)$ 
          \item \label{def:srelL} the following holds:
            \begin{enumerate}
              \item $\isilow{l_1} \iff \isilow{l_2}$ and $\isihigh{l_1} \Rightarrow l_1 = l_2$
              \item if $\isihigh{l_1}$ then $c_1 \heqe c_2$
              \item if $\isilow{l_1}$ and there exists an $l$ with $\isihigh{l}$
                and $c_1'$ and $c_1''$ such that $c_1  = c_1';\resetpc{l};c_1''$ 
                then there exist 
                $c_2'$ and $c_2''$ such that $c_2  = c_2';\resetpc{l};c_2''$ 
                with $c_1'' \heqe c_2''$
                and vice versa.
            \end{enumerate}
        \end{enumerate}
      \item \label{def:srelS} We have 
        \begin{itemize}
          \item for all $j^\stype \in \domain(\smems) \cap \domain(\smems')$ with $\ischigh{C(\stype)}$  that
            $\smems(j) \heq{\Gamma} \smems'(j)$ 
          \item for all $j^\stype \in (\domain(\smems) \setminus \domain(\smems'))$ with $\ischigh{C(\stype)}$ 
            that for all $r \in \refs$ with $\isihigh{\envrg(r)}$ we have $\smems(j)(r) = \vundef$.
          \item for all $j^\stype \in (\domain(\smems') \setminus \domain(\smems))$  with $\ischigh{C(\stype)}$ 
            that for all $r \in \refs$ with $\isihigh{\envrg(r)}$ we have $\smems'(j)(r) = \vundef$.
        \end{itemize}
      \item  \label{def:srelP} For all $j^\stype$ with
        $\isihigh{\stype}$ we have that  $\phi(j) = \phi'(j)$.
    \end{enumerate}
  
    We let $S \srel S'$ if 
    \begin{itemize}
      \item $\rbad{S}$ 
      \item or $S \sprel S'$ and $S'$ is deterministically terminating.
    \end{itemize}

\end{definition}
 
Intuitively, two servers are in the relation $\sprel$ if
\begin{enumerate}
  \item Both servers are well-typed. 
  \item There is a bijection between high integrity running threads on the two servers. 
    For each pair we have that 
    \begin{enumerate}
      \item They have the same request context and global memory index. For high
        integrity threads they also have the same session memory index.
      \item High equality holds between the two global memories. 
      \item  
        \begin{enumerate}
          \item Either both threads have high or both have low integrity. If it
            is high it has to be equal.
          \item For high integrity threads, the two codes have to be high equal 
          \item If the integrity of one thread is low, but it can be raised
            to high using a reset command, then there also has to be a reset with
            the same high integrity label in the other thread.
        \end{enumerate}
    \end{enumerate}
  \item 
    \begin{itemize}
      \item For all session identifiers appearing in both threads that are
        secret, the session memories indexed by the identifiers are high equal 
      \item For all session identifiers present in only one thread, that are
        secret, all high integrity references are unset.
    \end{itemize}
  \item For all high integrity session identifiers, the user information
    ($\phi$) is equal.
\end{enumerate}
We then define the relation $\srel$ which holds if the left server is in a bad
state, or the servers are in the relation $\srel$ and the right server is
deterministically terminating.

We then show that the relation $\sprel$ is symmetric and transitive. Note that
this does not hold for $\srel$.

\begin{lemma}[$\sprel$ is symmetric and transitive]
  \label{lem:srel-trans}
  The relation $\sprel$ is symmetric and transitive.
\end{lemma}
\begin{proof}
  Trivial, by checking the individual properties of \cref{def:brel}
\end{proof}

Next, we show that high equality for server expressions is preserved under
evaluation.

\begin{lemma}[Preservation of $\heqe$ under server evaluation]
  \label{lem:pres-heq-eval}
  Let $se$ and $se'$ be server expressions with $se \heqe se'$,
  let $E = i,j$ and $E' = i, j'$ with $j \heqe j'$,
  let $D,D'$ be databases and $\envrg, \envrs$ typing environments with 
  $\gmems(i) \heqmpar{\envrg} \gmems(i')$ and 
  if $j \neq \vundef$ then $\smems(j) \heqmpar{\envrs'} \smems'(j')$
  with 
  $ \forall r . \, \envrs'(r) = \envrs(r) \cremeet \jlabel{j} $.
  Then $\eval{se}{D}{E} \heqe \eval{se'}{D'}{E'}$.
\end{lemma}
\begin{proof}
  Let $v^{\stype} = \eval{se}{D}{E}$ and 
  $v'^{\stype'} = \eval{se'}{D'}{E'}$.
  If $\isilow{I(\stype) \imeet I(\stype')}$ the claim is trivial. We hence now
  assume $\isihigh{I(\stype) \imeet I(\stype')}$, i.e., $\isihigh{\It{\stype}} \vee \isihigh{\It{\stype'}}$
  \begin{itemize}
    \item $be = x$ : Impossible, since evaluation is  not defined on variables.
    \item $be = v^{\stype}$ : Trivial, since evaluation on values is the identity  \irule{E-Val}
    \item $be = \fresh^{\stype}$ : Then $be' = \fresh^{\stype}$.
      By rule \irule{SE-Fresh} we know that $v, v' \in \names_{\stype}$.
      \sketch{For simplicity, we assume that $v =v'$ and that the sampled names
      are fresh (i.e., have not been sampled before and will not be sampled
    again).}
    \item $se = se_1 \odot se_2$: Then $se' = se_1' \odot se_2'$ with $se_1
      \heqe se_1'$ and $se_2 \heqe se_2'$ and the claim follows by induction
      analog to \cref{lem:pres-heq-beval}.
    \item $se = \rg{r}$: Then $se' = \rg{r}$. The claim then follows from
      $\gmems(i) \heqmpar{\envrg} \gmems'(i')'$. 
    \item $se = \rs{r}$: Then $se' = \rs{r}$ and the claim then follows from
      $\smems(j) \heqmpar{\envrs} \smems'(j')$. 
  \end{itemize}
\end{proof}

Now we introduce a relation between websystems:
 
\begin{definition}[Integrity Relation]
  \label{def:relation}
  Given a typing environment $\Gamma$, we consider two websystems
  $A = \atkstate{\alabel}{\atknow}{W}$ and $A' = \atkstate{\alabel}{\atknow'}W'$ to be
  in the integrity relation $\rel$ if
  \begin{enumerate}
    \item \label{def:relT} $\systyping{W}$ and $\systyping{W'}$
    \item \label{def:relSC} For each server $S \in \servers{W}$ there exists exactly one server
      $S' \in \servers{W}'$ such that $\surls{S} = \surls{S'}$ and vice versa,
      i.e. the available URLs and the code associated to them are the same in
      both web systems. We will call these servers $S$ and $S'$
      \emph{corresponding} servers. \sketch{Formally, the correspondence is a
      bijection between the sets $\servers{W}$ and $\servers{W'}$.}
    \item \label{def:relSR} For all servers
      $S$ in $W$ and the corresponding servers $S'$ in $W'$ we have that 
      $S \sprel S'$ 
    \item \label{def:relBR} $W$ contains exactly one browser 
      \
        $B = \exebrowser{N}{K}{P}{T}{Q}{\lst{a}}{\usr}{l}{\amode}$,
        and $W'$ contains exactly one browser 
        $B' = \exebrowser{N'}{K'}{P'}{T'}{Q'}{\lst{a'}}{\usr}{l'}{\amode'}$. 
        and we have $B \bprel B'$.
  \end{enumerate}

  We furthermore let $A \rel A'$ if
  \begin{enumerate}
    \item $\rbad{A}$
    \item or $A \prel A'$ and $A'$ is deterministically terminating.
  \end{enumerate}
\end{definition}

Intuitively, we require that 
\begin{enumerate}
  \item Both websystems are well-typed 
  \item All servers have a matching server in the other websystem that contains
    the same URLS and commands (i.e., statically the websystems are equal)
  \item All corresponding servers are in the relation $\sprel$.
  \item Both websystems contain exactly one browser, and they are in the
    relation $\bprel$
\end{enumerate}
We then define the relation $\rel$, which holds
if the left system is in a bad state, or the systems are in the relation $\prel$
and the right system is deterministically terminating.

Next, we show that the relation $\prel$ is transitive. This property is helpful
for proofs of upcoming lemmas, where we consider the case where only one of the
system does a step. Then it is enough to show that the system before and after
taking the step are in the relation.  

\begin{lemma}[Transitivity of $\rel$]
  \label{lem:rel-trans}
  The relation $\rel$ is transitive.
\end{lemma}
\begin{proof}
  Trivial, by inspecting the single conditions.
\end{proof}

Now we show that whenever a browser processes an event with low sync integrity
for an internal step, then the state before and after taking the step are in
the relation.

\begin{lemma}[Low Sync Integrity Browser Steps]
  \label{lem:low-int-bro-step}
  Let $B = \exebrowser{N}{K}{P}{T}{Q}{\lst{a}}{\usr}{l}{\amode}$ and  
  $B' = \exebrowser{N'}{K'}{P'}{T'}{Q'}{\lst{a'}}{\usr}{l'}{\amode'}$ be 
  browsers with $B \bstep{\einta{\blank}{l''}} B'$ and $\isilow{l''}$ and 
  $\browtyping{B}$.
  Then $B \bprel B'$
\end{lemma}
\begin{proof}
  We show that all the properties of \cref{def:brel} are fulfilled.
  In all cases property \ref{def:brelTy} follows immediately from 
  \cref{lem:bro-sub-red}.
  Proof by induction over the derivation of the step $\alpha$
  \begin{itemize}
    \item \irule{B-Seq} follows from induction.
    \item \irule{B-Skip}
      Properties \ref{def:brelL}, \ref{def:brelP}, \ref{def:brelK}, \ref{def:brelN}, \ref{def:brelA} and \ref{def:brelQ} are
      trivial, since $l = l'$, $P = P'$, $K = K'$, $N = N'$, $\lst{a} =
      \lst{a'}$ and $Q = Q'$.
      Property \ref{def:brelT} is trivial, since 
      because of  $\isilow{\eint{\alpha}}$ we know $\isilow{l }$.
    \item \irule{B-End} Impossible, since $\isilow{\eint{\alpha}}$.n
    \item \irule{B-SetReference}
      Properties \ref{def:brelL}, \ref{def:brelP}, \ref{def:brelN},
      \ref{def:brelA} and \ref{def:brelQ} are trivial, since $l = l'$, $P = P'$,
      $N = N'$, $\lst{a} = \lst{a'}$ and $Q = Q'$.
      Property \ref{def:brelT} is trivial, since 
      because $\isilow{\eint{\alpha}}$ we know $\isilow{l}$.

      We know that $T = \subst{t}{r := be}$.

      For property \ref{def:brelK}, 
      by rule \irule{T-BAssign} we then know that
      $\isilow{\treft{\envrg(r)}}$
      and by property \ref{def:btM} of \cref{def:bro-typing} we then know that
      with $K(r) = v^\stype$ we have $\stype = \treft{\envrg(r)}$.
      With $K'(r) = v'^{\stype'}$ we also have $\stype' = \treft{\envrg(r)}$.
      Because of $\isilow{\treft{\envrg(r)}}$ the claim follows immediately.
    \item \irule{B-SetDom}
      Properties \ref{def:brelL}, \ref{def:brelK}, \ref{def:brelN},
      \ref{def:brelA} and \ref{def:brelQ} are trivial, since $l = l'$, $K =
      K'$, $N = N'$, $\lst{a} = \lst{a'}$ and $Q = Q'$.
      Property  \ref{def:brelT} is trivial, since 
      because $\isilow{\eint{\alpha}}$ we know $\isilow{l}$.

      We know that $T = \subst{\tab}{\setdom{v}{u}{\lst{be}}}$ and 
      $ P(\tab)  = (u', f, l_P, \amode_P)$

      Since by \prop{def:btT} we know $l_P = l$ we have $\isilow{l_P}$ and the
      claim is trivial.


    \item \irule{B-Load}: Impossible since $\isilow{l}$
    \item \irule{B-Submit}: Impossible since $\isilow{l}$
    \item \irule{B-Include}: 
      Properties \ref{def:brelL}, \ref{def:brelP}, \ref{def:brelK} and
      \ref{def:brelA} are trivial, since 
      $l = l'$, $P = P'$, $K = K'$ and $\lst{a} = \lst{a'}$.
      Properties \ref{def:brelN},  \ref{def:brelT} and 
      \ref{def:brelQ} are trivial, since 
      because $\isilow{\eint{\alpha}}$ we know $\isilow{l \ijoin l'}$.
  \end{itemize}
\end{proof}

We show the same for internal steps on the server side with low sync integrity.

\begin{lemma}[Low Sync Integrity Server Steps]
  \label{lem:low-int-serv-step}
  Let $S = \exserver{D}{\phi}{t}$ and  $S = \exserver{D'}{\phi'}{t'}$
  with $S \sstep{\alpha} S'$ and 
  $\alpha \in \{\blank, \lauth{\cdot}{\cdot}{\cdot}, \eerror\}$ 
  and $\isilow{\eint{\alpha}}$ and 
  $\servtyping{S}$.
  Then $S \sprel S'$
\end{lemma}
\begin{proof}
  We show that all the properties of \cref{def:srel} are fulfilled.
  In all cases property \ref{def:srelTy} follows immediately from 
  \cref{lem:serv-sub-red}.

  Let $t_1 \in \running{S}$ and 
  let $t_1 = \exthreadpar{c}{E}{R}{l}{\amode}$.
  Then there exists 
  $t_1' \in \running{S'}$ with 
  $t_1' = \exthreadpar{c'}{E'}{R}{l'}{\amode'}$ and 
  $\exserver{D}{\phi}{t_1} \sstep{\alpha} \exserver{D'}{\phi'}{t_1'}$

  We prove the claim by induction over the derivation of step $\alpha$.
  \begin{itemize}
    \item \irule{S-Seq} The claim follows from the induction hypothesis 
    \item \irule{S-IfTrue}
      Properties \ref{def:srelE}, \ref{def:srelG} \ref{def:srelS} and
      \ref{def:srelP} are trivial since $\isilow{l}$, $E = E'$, $D=D'$ and $\phi =
      \phi$.
      Property \ref{def:srelL} is trivial since $\isilow{l}$ and $\isilow{l'}$
    \item \irule{S-TCTrue} All properties are trivial since
      since $\isilow{l}$, $E = E'$, $D=D'$, $l = l'$ and $\phi = \phi$.
    \item \irule{S-Skip} 
      All properties are trivial since
      $\isilow{l}$, $E = E'$, $D=D'$, $l = l'$ and $\phi = \phi$.
    \item \irule{S-Reset} Then $c = \resetpc{l''}$
      Properties \ref{def:srelE}, \ref{def:srelG} \ref{def:srelS} and
      \ref{def:srelP} are trivial since $E = E'$, $D=D'$, $l = l'$ and $\phi =
      \phi$.
      Property \ref{def:srelL} is trivial since $\isilow{l}$ \sketch{(because the reset command will never lower the integrity label)} and $\isilow{l'}$.
    \item \irule{S-IfFalse} Analog to rule \irule{S-IfTrue}
    \item \irule{S-TCFalse} All properties are trivial since
      since $\isilow{l}$, $E = E'$, $D=D'$, $l = l'$ and $\phi = \phi$.
      \sketch{$Property (iii)$ of \ref{def:srelL} does hold since we know that a token check is never followed by a reset -- this is enforced in \irule{S-IfFalse} and \irule{S-IfTrue}}

    \item \irule{S-RestoreSession} 
      Properties \ref{def:srelG}, \ref{def:srelL}
      \ref{def:srelS} and \ref{def:srelP} are trivial since
      $D=D'$, $l = l'$ and $\phi = \phi$.
      Let $E = i,j$ and $E' = i, j'$.
      If for $t_1$ we do not have $\isihigh{\thighestint{t_1}}$,
      property \ref{def:srelE} is trivial. 
      Otherwise, we know that $\amode = \bhon$.
      Then we know by rule \irule{T-Running} and \irule{T-Start} that
      because of $\isilow{l}$ we have 
      $\isilow{\It{\jlabel{j}}}$ and $\isilow{\It{\jlabel{j'}}}$. Property
      \ref{def:srelE} immediately follows.

    \item \irule{S-NewSession}
      Properties \ref{def:srelG}, \ref{def:srelL}
      and \ref{def:srelP} are trivial since
      with $D=(\gmems,\smems)$, $D'=(\gmems',\smems')$ we have
      $\gmems=\gmems'$, $l = l'$ and $\phi = \phi$.
      If for $t_1$ we do not have $\isihigh{\thighestint{t_1}}$,
      property \ref{def:srelE} is trivial. 
      Otherwise, we know that $\amode = \bhon$.
      Then we know by rule \irule{T-Running} and \irule{T-Start} that
      because of $\isilow{l}$ we have 
      $\isilow{\It{\jlabel{j}}}$ and $\isilow{\It{\jlabel{j'}}}$. Property
      \ref{def:srelE} immediately follows.

      Let $E' = i, j'$
      For property \ref{def:srelS} we know that 
      $j' \in \smems' \setminus \smems$. 
      Since for all $r$, $(\smems'(j')(r) = \vundef$ ($\smems'(j')$ is a fresh
        memory) property \ref{def:srelS} immediately follows.

    \item \irule{S-SetGlobal} 
      Then we have $c = \rg{r} := se$.

      Properties \ref{def:srelE},  \ref{def:srelL}
      \ref{def:srelS} and \ref{def:srelP} are trivial since
      with $D=(\gmems,\smems)$, $D'=(\gmems',\smems')$ we have
      $E = E'$, $\smems=\smems'$, $l = l'$ and $\phi = \phi$.

      Let $E = i,j$ and let $v^\stype = \gmems(i)(r)$ and
      $v'^{\stype'} = \gmems'(i)(r)$.
      By rule \irule{T-SetGlobal} we know that $\isilow{\It{\treft{\envrg(r)}}}$.
      By property \ref{def:stG} we know that 
      $\stype = \treft{\envrg(r)} = \stype'$.
      We hence have $\isilow{\It{\stype}}$ and $\isilow{\It{\stype'}}$ and the
      claim follows immediately.

    \item \irule{S-SetSession} 
      Then we have $c = \rs{r} := se$.

      Properties \ref{def:srelE},  \ref{def:srelL}
      \ref{def:srelG} and \ref{def:srelP} are trivial since
      with $D=(\gmems,\smems)$, $D'=(\gmems',\smems')$ we have
      $E = E'$, $\gmems=\gmems'$, $l = l'$ and $\phi = \phi$.

      Let $E = i,j$ and let $v^\stype = \smems(j)(r)$ and
      $v'^{\stype'} = \smems'(j)(r)$.
      By rule \irule{T-SetSession} we know that 
      $\isilow{\It{\treft{\envrs(r)} \tijoin \jlabel{j}}}$.
      By property \ref{def:stG} we know that 
      $\stype = \treft{\envrg(r)} \tijoin \jlabel{j} = \stype'$.
      We hence have $\isilow{\It{\stype}}$ and $\isilow{\It{\stype'}}$ and the
      claim follows immediately.

    \item \irule{S-Login} 
      $t_1 = \exthread{\login{se_{usr}}{se_{pw}}{se_{side}}}{E}{R}$.

      Properties \ref{def:srelE}, \ref{def:srelG} \ref{def:srelL} and
      \ref{def:srelS}  are trivial since
      $E = E'$, $D=D'$ and $l = l'$.

      By rule \irule{T-Login} with $\exelabel{se_{sid}}{\stype}$ we get that
      $\isilow{\stype}$. 
      Property \ref{def:srelP} then follows immediately using
      \cref{lem:bro-ex-ty}.

    \item \irule{S-Auth}
      Properties \ref{def:srelE}, \ref{def:srelG} \ref{def:srelL}
      \ref{def:srelS} and \ref{def:srelP} are trivial since
      $E = E'$, $D=D'$, $l = l'$ and $\phi = \phi$.

    \item \irule{S-OChkSucc}
        All properties are trivial since since $\isilow{l}$, $E = E'$, $D=D'$,
        $l = l'$ and $\phi = \phi$.  
    \item \irule{S-OChkFail} 
        All properties are trivial since since $\isilow{l}$, $E = E'$, $D=D'$,
        $l = l'$ and $\phi = \phi$.  

    \item \irule{S-LParallel} The claim follows from induction hypothesis 
    \item \irule{S-RParallel} The claim follows from induction hypothesis 
  \end{itemize}
\end{proof}

Now, we show the same for browsers issuing a request with low sync integrity.

\begin{lemma}[Low Sync Integrity Browser Request]
  \label{lem:low-int-bro-req}
  Let $B = \exebrowser{N}{K}{P}{T}{Q}{\lst{a}}{\usr}{l}{\amode}$ and  
  $B' = \exebrowser{N'}{K'}{P'}{T'}{Q'}{\lst{a'}}{\usr}{l'}{\amode'}$ be 
  browsers with $B \bstep{\alpha} B'$ and $\isilow{\eint{\alpha}}$ and 
  $\browtyping{B}$ and $\alpha = \exbsend{\bid}{n}{u}{p}{o}{ck}{l''}{\amode''}$
  Then $B \bprel B'$ and $\eint{\alpha} \ileq l''$.
\end{lemma}
\begin{proof}
  We show that all the properties of \cref{def:brel} are fulfilled.
  We know that $\alpha$ has been produces using rule \irule{B-Flush}
  Property \ref{def:brelTy} follows immediately from \cref{lem:bro-req}.

  Properties \ref{def:brelL}, \ref{def:brelP}, \ref{def:brelK},
  \ref{def:brelN}, \ref{def:brelT} and
  \ref{def:brelA} are trivial, since 
  $l = l'$, $P = P'$, $K = K'$, $N = N'$, $T = T'$ and $\lst{a} = \lst{a'}$.
  Properties \ref{def:brelQ} is trivial, since 
  because $\isilow{\eint{\alpha}}$ we know $\isilow{l \ijoin l'}$.

  The claim $\eint{\alpha} \ileq l''$ follows immediately, by inspecting the
  rules \irule{B-Load}, \irule{B-Include}, \irule{B-Submit} and
  \irule{B-Redirect}.
\end{proof}

Next, we show the same for a browser receiving a response of low sync integrity .

\begin{lemma}[Low Sync Integrity Browser Response]
  \label{lem:low-int-bro-res}
  Let $B = \exebrowser{N}{K}{P}{T}{Q}{\lst{a}}{\usr}{l}{\amode}$ and  
  $B' = \exebrowser{N'}{K'}{P'}{T'}{Q'}{\lst{a'}}{\usr}{l'}{\amode'}$ be 
  browsers with $B \bstep{\alpha} B'$ and $\isilow{\eint{\alpha}}$ 
  and $\alpha = \exbrecv{\bid}{n}{u}{u'}{\lst{v}}{ck}{page}{s}{l''}{\amode''}$ and
  $\browtyping{B}$ and $\restyping{\alpha}$.
  Then $B \bprel B'$ and $\eint{\alpha} \ileq l''$.
\end{lemma}
\begin{proof}
  We show that all the properties of \cref{def:brel} are fulfilled.
  We perform a case distinction on the rule used to derive $\alpha$.

  In all cases for property \ref{def:brelK} we get from property \ref{def:src}
  of \ref{def:res-ty} that for all updated references $r$, we have
  $\isilow{\It{\treft{\envrg}}}$. The claim then follows from property
  \ref{def:btM} for $B$ and $B'$
\begin{itemize}
  \item \irule{B-RecvLoad}: 
    \sketch{The claim $\eint{\alpha} \ileq l''$ follows from the observation
      that the integrity can only be lowered between the request and the
    response}

    Property \ref{def:brelTy} follows immediately from \cref{lem:bro-res}.

    Property \ref{def:brelA} is trivial, since we have $\lst{a} = \lst{a'}$.

    Properties \ref{def:brelN}, \ref{def:brelL} \ref{def:brelT}, \ref{def:brelQ} are trivial,
    since because $\isilow{\eint{\alpha}}$ we know $\isilow{l}$ and
    $\isilow{l'}$.

    Property \ref{def:brelP} follows immediately from $\isilow{l''}$, which 
    we get from $\eint{\alpha} \ileq l''$.

  \item \irule{B-RecvInclude}
    The claim $\eint{\alpha} \ileq l''$ is trivial.

    Property \ref{def:brelTy} follows immediately from \cref{lem:bro-res}.

    Properties \ref{def:brelP} and \ref{def:brelA} are trivial, since we have 
    $P = P'$ and $\lst{a} = \lst{a'}$.

    Properties \ref{def:brelN}, \ref{def:brelL} \ref{def:brelT}, \ref{def:brelQ} are trivial,
    since because $\isilow{\eint{\alpha}}$ we know $\isilow{l}$ and
    $\isilow{l'}$.

  \item \irule{B-Redir} 
    The claim $\eint{\alpha} \ileq l''$ is trivial.

    Property \ref{def:brelTy} follows immediately from \cref{lem:bro-res}.

    Properties \ref{def:brelP} and \ref{def:brelA} are trivial, since we have 
    $P = P'$ and $\lst{a} = \lst{a'}$.

    Properties \ref{def:brelL}, \ref{def:brelN}, \ref{def:brelT} and \ref{def:brelQ} are
    trivial, since because $\isilow{\eint{\alpha}}$ we know $\isilow{l'}$ and
    $\isilow{l}$.

 \end{itemize}
\end{proof}

Now we show the same for servers receiving a request of low  sync integrity.

\begin{lemma}[Low Sync Integrity Server Request]
  \label{lem:low-int-serv-req}
  Let $S = \exserver{D}{\phi}{t}$ and  $S = \exserver{D'}{\phi'}{t'}$
  with $S \sstep{\alpha} S'$ and $\isilow{\eint{\alpha}}$ and 
  $\alpha = \exsrecv{\bid}{n}{u}{p}{ck}{o}{l''}{\amode''}$
  $\servtyping{S}$ and $\restyping{\alpha}$.
  Then $S \sprel S'$
\end{lemma}
\begin{proof}
  We show that all the properties of \cref{def:srel} are fulfilled.
  Properties \ref{def:srelE}, \ref{def:srelG} \ref{def:srelL} are trivial since
  $\isilow{\eint{\alpha}}$. 
  \ref{def:srelS} and \ref{def:srelP} are trivial since
  with $D=(\gmems,\smems)$, $D'=(\gmems',\smems')$ we have
  $\smems=\smems'$, and $\phi = \phi$.
  Property \ref{def:srelTy} follows immediately from \cref{lem:serv-req}.
\end{proof}

Now we show the same for servers sending a response of low sync integrity.

\begin{lemma}[Low Sync Integrity Server Response]
  \label{lem:low-int-serv-res}
  Let $S = \exserver{D}{\phi}{t}$ and  $S = \exserver{D'}{\phi'}{t'}$
  with $S \sstep{\alpha} S'$ and $\isilow{\eint{\alpha}}$ 
  and $\alpha = \exssend{\bid}{n}{u}{u'}{\lst{v}}{ck}{page}{s}{l''}{\amode''}$ and
  $\servtyping{S}$.
  Then $S \sprel S'$
\end{lemma}
\begin{proof}
  Then the event $\alpha$ was produced using rule \irule{S-Reply} or
  \irule{S-Redir}. 
  In both cases
  properties \ref{def:srelE}, \ref{def:srelG} \ref{def:srelL}
  \ref{def:srelS} and \ref{def:srelP} are trivial since
  $E = E'$, $D=D'$, $l = l'$ and $\phi = \phi$.
  Property \ref{def:srelTy} follows immediately from \cref{lem:serv-res}.
\end{proof}

Finally, we use the previous lemmas to show that if a websystem takes a 
step of low sync integrity, then the state before and after the step are in the relation.

\begin{lemma}[Low Sync Integrity Steps]
  \label{lem:low-int-step}
  Let $A,A'$ be web systems with 
  $A \astep{\alpha} A'$ for some $\alpha$ with
  $\isilow{\eint{\alpha}}$.
  Then $A \prel A'$.
\end{lemma}
\begin{proof}
  We perform an induction on the rule used to derive the step $\alpha$. 
  \begin{itemize}
    \item \irule{A-Nil} Then, if the step is derived through rule
      \irule{W-LParallel} or \irule{W-RParallel} the claim follows by induction.
      The claim for internal browser steps follows from 
      \cref{lem:low-int-bro-step}.
      The claim for internal server steps follows from 
      \cref{lem:low-int-serv-step}.
    \item \irule{A-BrowserServer} Then we know by
      \cref{lem:low-int-bro-req} that the browser relation is preserved
      and that the server step is of low integrity.
      By \cref{lem:bro-req} we know that the request is well typed.
      The claim for the server relation then follows from 
      \cref{lem:low-int-serv-req}.
    \item \irule{A-ServerBrowser} 
      Then by \cref{lem:serv-res} we get that the response is well-typed.
      By \cref{lem:low-int-bro-req} we hence know that the browser relation is
      preserved and that the server step is of low integrity.
      Then we know by 
      \cref{lem:low-int-serv-res} that the server relation is preserved.
    \item \irule{A-TimeoutSend} Then the claim follows from 
      \cref{lem:low-int-bro-req}.
    \item \irule{A-TimeoutRecv} Then the claim follows from 
      \cref{lem:low-int-bro-res}.
    \item \irule{A-BroAtk} Then the claim follows from 
      \cref{lem:low-int-bro-req} for the browser step.
    \item \irule{A-AtkSer} Then the claim follows from 
      \cref{lem:low-int-serv-req} for the server step, using \cref{lem:att-req}
    \item \irule{A-SerAtk} Then the claim follows from 
      \cref{lem:low-int-serv-res} for the server step.
    \item \irule{A-AtkBro} Then the claim follows from 
      \cref{lem:low-int-bro-res} for the browser step, using \cref{lem:att-res}
  \end{itemize}
\end{proof}

We now define the \emph{next high integrity state} of a deterministically terminating
websystem as the state that is just before processing the next event with high sync integrity. This state can be reached by processing a number of events with low sync integrity. We furthermore show that 
\begin{enumerate}
  \item this state is unique 
  \item The websystem before and after taking the steps with low sync integrity are in
    the relation.
  \item The websystem in the newly reached state is still deterministically
    terminating
  \item The websystem has a special form (one of the few specified in the lemma)
\end{enumerate}

\begin{lemma}[Low Integrity Catch Up]
  \label{lem:lint-catchup}
  Let $A = \atkstate{\atknow}{\alabel}{W}$ be a deterministically terminating websystem. We say that it is in a 
  low integrity state if:
  \begin{itemize}
    \item $\{B\} = \browsers{W}$, 
      $B = \exebrowser{N}{M}{P}{T}{Q}{\lst{a}}{\usr}{l}{\amode}$ 
      and $\isilow{l}$
    \item or there is a 
      $t \in \{ t' ~|~ S \in \servers{W} \wedge t' \in \running{S} \}$ with 
      $t = \exthreadpar{c}{R}{E}{l}{\amode}$ with 
      \begin{itemize}
        \item $\shalt \not \in \commands{c}$
        \item $\isilow{l}$
        \item $R = n, u, \bid, o \wedge \bid = \usr$
      \end{itemize}
  \end{itemize}

  We then let $\nexth{A}$ be the websystem $A' = \atkstate{\atknow'}{\alabel}{W'}$ such that 
  \begin{itemize}
    \item $A \astepn{*}{\beta} A'$ with $\isilow{\eint{\beta}}$ for all $\beta
      \in \lst{\beta}$.
    \item for all $A''$, $\alpha$ with $A' \astep{\alpha} A''$ $\isihigh{\eint{\alpha}}$
  \end{itemize}
  We then know that 
  \begin{enumerate}
    \item There exists such a unique $A'$ 
    \item $A \prel A'$ 
    \item $A'$ is deterministically terminating
    \item Let $B = \exebrowser{N}{M}{P}{T}{Q}{\lst{a}}{\usr}{l}{\amode}$
      be the honest browser in $W$ and let 
      $B'$ be the honest browser in $W'$.
      Then exactly one of the following claims about $W'$ holds 
      \begin{enumerate}
        \item $B' = \exebrowser{\mempty}{M'}{P'}{\subst{tab}{\sskip}}{\mempty}{\lst{a'}}{\usr}{l'}{\amode'}$ 
        \item 
          there exists a server $S$ in $W'$ with $t \in \running{S}$ and 
          $t = \exthreadpar{\resetpc{l''};c}{R}{E}{l'}{\amode'}$ and $\isihigh{l''}$
        \item $B' = \exebrowser{N'}{M'}{P'}{T'}{\mempty}{\lst{a'}}{\usr}{l'}{\amode'}$ 
          with $N = \subst{n}{\_}$, $\isihigh{l'}$ and 
          there exists a server $S$ in $W'$ with $t \in \running{S}$ and 
          $t = \exthreadpar{\reply{page}{s}{ck}{\lst{se}}}{R}{E}{l'}{\amode'}$
          and $R = n, \_, \_, \_$.
        \item $B' = \exebrowser{N'}{M'}{P'}{T'}{\mempty}{\lst{a'}}{\usr}{l'}{\amode'}$ 
          with $N = \subst{n}{\_}$, $\isihigh{l'}$ and 
          there exists a server $S$ in $W'$ with $t \in \running{S}$ and 
          $t = \exthreadpar{\redr{u}{p}{ck}{\lst{se}}}{R}{E}{l'}{\amode'}$
          and $R = n, \_, \_, \_$.
        \item $B' = \exebrowser{N'}{M'}{P'}{T'}{\mempty}{\lst{a'}}{\usr}{l'}{\amode'}$ 
          with $N = \subst{n}{\_}$, $\isihigh{l'}$ and 
            $\timeouts' = \{ (\_, n, \_, \_, \_) \}$
      \end{enumerate}
  \end{enumerate}
\end{lemma}
\begin{proof}
  We show that the different claims hold:
  \begin{enumerate}
    \item The existence and uniqueness follow immediately from the fact the $W$
      is deterministically terminating, using \cref{def:termination}
    \item $A \prel A'$ follows from repeated application of
      \cref{lem:low-int-step} and \cref{lem:rel-trans}.
    \item Deterministic termination for $W'$follows immediately from
      deterministic termination of $W$, using \cref{def:termination}
    \item The form of $W'$ follows from the observation that these five points
      are the only ones in the semantic rules, where the integrity is raised.
  \end{enumerate}
\end{proof}

%

Next, we show that if two high integrity browsers are in the relation and the
left browser takes an internal step of high sync integrity, then also the right browser
can take the same step and the resulting browsers are still in the relation.

\begin{lemma}[High Sync Integrity Browser Steps]
  \label{lem:hint-bro-step}
  Let $B_1 = \exebrowser{N}{M}{P}{T}{Q}{\lst{a}}{\usr}{l}{\amode}$
  and $B_2 = \exebrowser{N'}{M'}{P'}{T'}{Q'}{\lst{a'}}{\usr}{l'}{\amode'}$
  be browsers with $B_1 \bprel B_2$ and 
  $\isihigh{l}$ and let $B_1 \bstep{\einta{\blank}{l_s}}
  B_1'$ with $\isihigh{l_s}$. Then there exist $B_2'$ 
  such that
  $B_2 \bstep{\einta{\blank}{l_s}} B_2'$ 
  and $B_1' \bprel B_2'$.
\end{lemma}
\begin{proof}
  We show that all properties of \cref{def:brel} are fulfilled.
  In all cases property \ref{def:brelT} follows immediately from 
  \cref{lem:bro-sub-red}.
  Because of $B_1 \bprel B_2$ we know 
  \begin{itemize}
    \item $\browtyping{B}$ and $\browtyping{B'}$
    \item $l = l'$
    \item $N = N'$
    \item $K \heqmpar{\envrg} K'$
    \item $\domain(T) = \domain(T')$ and if $T = \subst{t}{s}$ and 
      $T' = \subst{t}{s'}$ then $s \heqe s'$
    \item $\lst{a} = \lst{a'}$
  \end{itemize}
  By property \ref{def:btL} of \cref{def:bro-typing} and because of
  $\isihigh{l}$ we know that $\amode = \bhon$.
  
  We perform an induction on the derivation of the step $\alpha$.
  \begin{itemize}
    \item \irule{B-Seq} The claim follows by induction.
    \item \irule{B-Skip} Trivial because $s \heqe s'$
    \item \irule{B-End} Trivial because $s \heqe s'$. 
    \item \irule{B-SetReference} 
      Then because of $N = N'$, $s \heqt s'$, we can also apply
      \irule{B-SetReference} for $B_2$. 
      We have that $s = r := be$ and $s' = r := be'$
      where $be \heqe be'$ 
      and the claim follows immediately using \cref{lem:pres-heq-beval}.
    \item \irule{B-SetDom} Then because of $N = N'$, $s \heqt s'$, we can also
      apply \irule{B-SetDom} for $B_2$.  We have that 
      $s = {\setdom{be}{u}{\lst{be}}}$ and 
      $s' = {\setdom{be'}{u}{\lst{be'}}}$ 
      where $be \heqe be'$ and 
      $\forall k \in \intval{1}{\len{be}}. \, be_k \heqe be'_k$. 
      By rule \irule{T-BsetDom} we known that 
      $be = v^\stype$ and $be' = v'^{\stype'}$ are primitive values with
      $\It{\stype} = \It{\stype'} = \ibot$ \todo{by translation}. 
      Hence we know $v = v'$ by the definition of $\heqe$.
      Using \cref{lem:pres-heq-beval} for all expressions in $\lst{be}$, we get
      $page \heqe page'$ and the claim follows.
    \item \irule{B-Load} 
      Because of $N = N'$, $\domain(T) = \domain(T')$, and $\lst{a} =
      \lst{a'}$ we can also apply rule \irule{B-Load} in $B_2$
      
      All properties except for property \ref{def:brelN} and \ref{def:brelQ}
      are trivial.

      \sketch{For simplicity we assume that the names $n$ and $n'$ sampled in the two
        browser are the same, i.e., we have $n = n'$, and property
        \ref{def:brelN} follows immediately,}


      \sketch{
      For property \ref{def:brelQ}
      the only non-trivial condition is the claim on the cookies of the
      produced event.
      This however follows immediately from  $K \heqmpar{\envrg} K'$}
    \item \irule{B-Include} Because of $N = N'$, 
      $s \heqt s'$, we can also apply \irule{B-Include} for $B_2$.
      \sketch{For simplicity we assume that the names $n$ and $n'$ sampled in the two
      browser are the same, i.e., we have $n = n'$}, and property
        \ref{def:brelN} follows immediately using \prop{def:btT} to get that
        the DOM is of high integrity and hence the origins of the two requests
        are the same.
      \sketch{
      For property \ref{def:brelQ}
      the only non-trivial conditions are the claim on the parameters and the
      cookies of the produced event.
      These however follow immediately from  $s \heqe s'$ using
    \cref{lem:pres-heq-beval} and $K \heqmpar{\envrg} K'$.}
    \item \irule{B-Submit} 
      Because of 
      $N = N'$, $\domain(T) = \domain(T')$, and $\lst{a} = \lst{a'}$. 
      Hence we can also apply \irule{B-Submit} in $B_2$ and all properties
      except for property \ref{def:brelN} and  \ref{def:brelQ} follow
      immediately.
      \sketch{
      Let $l_D$ be the integrity label of the DOM
      We distinguish two cases:
      \begin{itemize}
        \item $\isihigh{l_D}$
          Then property \ref{def:brelN} follows immediately.
          For property \ref{def:brelQ}
          the only non-trivial conditions are the claim on the parameters and
          the cookies of the produced event.
          The claim on the parameters follows directly from $\heqe$ on the DOM
          and $\lst{a} = \lst{a'}$ and the claim of the cookies follows
          immediately from $K \heqm K'$.
          %
        \item $\isilow{l_\alpha}$ the claim is trivial.
      \end{itemize}}
  \end{itemize}
\end{proof}

Next, we show the same property for browsers sending out a request of  high sync integrity.

\begin{lemma}[High Sync Integrity Browser Request]
  \label{lem:hint-bro-req}
  Let $B_1 = \exebrowser{N}{M}{P}{T}{Q}{\lst{a}}{\usr}{l}{\amode}$
  and $B_2 = \exebrowser{N'}{M'}{P'}{T'}{Q'}{\lst{a'}}{\usr}{l'}{\amode'}$
  be browsers with $B_1 \bprel B_2$ and 
  let $B_1 \bstep{\alpha}
  B_1'$ with $\isihigh{\eint{\alpha}}$ 
  and $\alpha = \exbsend{\bid}{n}{u}{p}{ck}{o}{l_\alpha}{\amode_\alpha}$
  Then there exist $B_2'$ and $\alpha'$ such that
  $B_2 \bstep{\alpha'} B_2'$ and $\alpha \heqe \alpha'$
  and $B_1' \bprel B_2'$.
\end{lemma}
\begin{proof}
  We know that rule \irule{B-Flush} was used and we know that $Q = \{\alpha\}$.

  We then know by $B_1 \bprel B_2$ that if $Q' = \{\alpha'\}$ with $\alpha \heqe
  \alpha'$.

  We can thus also apply rule \irule{B-Flush} in $B_2$ and all claims follows
  immediately.
\end{proof}


Next we show the same property for high integrity browsers receiving a response
of high sync integrity.

\begin{lemma}[High Sync Integrity Browser Response]
  \label{lem:hint-bro-res}
  Let $B_1, B_2$ be browsers with $B_1 \bprel B_2$ 
  and let $B_1 \bstep{\alpha}
  B_1'$ with $\isihigh{\eint{\alpha}}$
  and $\alpha =
  \exbrecv{\bid}{n}{u}{u'}{\lst{v}}{ck}{page}{s}{l_\alpha}{\amode_\alpha}$ with 
  $\isihigh{l_\alpha}$.
  Let  $\alpha' =
  \exbrecv{\bid}{n}{u}{u'}{\lst{v'}}{ck'}{page'}{s'}{l_\alpha}{\amode_\alpha}$
  with $\alpha' \heqe \alpha$, $\restyping{\alpha}$ and $\restyping{\alpha'}$.
  Then there exist $B_2'$ and such that
  $B_2 \bstep{\alpha'} B_2'$ and $B_1 \bprel B_2'$
\end{lemma}
\begin{proof}
  We show that all properties of \cref{def:brel} are fulfilled.

  In all cases property \ref{def:brelT} follows immediately from 
  \cref{lem:bro-res}.

  Let $B_1 = \exebrowser{N}{M}{P}{T}{Q}{\lst{a}}{\usr}{l}{\amode}$
  and $B_2 = \exebrowser{N'}{M'}{P'}{T'}{Q'}{\lst{a'}}{\usr}{l'}{\amode'}$

  Because of $\isihigh{l_\alpha}$ we then know  
  $\isihigh{l}$ ,\sketch{since the integrity label can not be
  raised between the request and the response},
  and $\isihigh{l'}$ by inspection of the possible rules.

  We perform a case distinction between the three possible rules.
  \begin{itemize}
    \item \irule{B-RecvLoad} 
      Then because of $N = N'$, $\domain(T) = \domain(T')$, we can
      also apply \irule{B-RecvLoad} for $B_2$.  
      \sketch{We get $B_1' \bprel B_2'$ from $\alpha \heqe \alpha'$}.
    \item \irule{B-RecvInclude} 
      Let $T = \subst{tab}{s}$ and $T' = \subst{tab}{s'}$
      Then because of $N = N'$, $s \heqt s'$, we can
      also apply \irule{B-RecvInclude} for $B_2$. 
      \sketch{We get $B_1' \bprel B_2'$ from $\alpha \heqe \alpha'$.}
    \item \irule{B-Redirect}
      Then because of $N = N'$, , we can
      also apply \irule{B-Redirect} for $B_2$. 
      \sketch{We get $B_1' \bprel B_2'$ from $\alpha \heqe \alpha'$ and $K \heqm K$.}
  \end{itemize}
\end{proof}

The next lemma treats the case, where a browser receives a response
that is of high sync integrity, but of low integrity,

\begin{lemma}[High Sync Integrity Browser Response of Low Integrity]
  \label{lem:lhint-bro-res}
  Let $B_1, B_2$ be browsers with $B_1 \bprel B_2$ 
  and let $B_1 \bstep{\alpha}
  B_1'$ with $\isihigh{\eint{\alpha}}$
  and $\alpha =
  \exbrecv{\bid}{n}{u}{u'}{\lst{v}}{ck}{page}{s}{l_\alpha}{\amode_\alpha}$ with 
  $\isilow{l_\alpha}$.
  Let $\alpha' =
  \exbrecv{\bid}{n}{u}{u'}{\lst{v'}}{ck'}{page'}{s'}{l_\alpha}{\amode_\alpha}$
  with $\alpha \heqe \alpha'$, $\restyping{\alpha}$ and $\restyping{\alpha'}$.
  Then there exist $B_2'$ and such that
  $B_2 \bstep{\alpha'} B_2'$ and $B_1 \bprel B_2'$
\end{lemma}
\begin{proof}
  We show that all properties of \cref{def:brel} are fulfilled.

  In all cases property \ref{def:brelT} follows immediately from 
  \cref{lem:bro-res}.

  Let $B_1 = \exebrowser{N}{M}{P}{T}{Q}{\lst{a}}{\usr}{l}{\amode}$
  and $B_2 = \exebrowser{N'}{M'}{P'}{T'}{Q'}{\lst{a'}}{\usr}{l'}{\amode'}$.
  
  We perform a case distinction between the three possible rules.
  \begin{itemize}
    \item \irule{B-RecvLoad} 
      Then we know that $\isihigh{l}$
      Then because of $B_1 \bprel B_2$ we get that  $N = N'$, $\domain(T) =
      \domain(T')$ and we can also apply \irule{B-RecvLoad} or \irule {B-Redirect}
      for $B_2$.  Because of $\isilow{l_alpha}$ we immediately get $B_1' \bprel
      B_2'$.
    \item \irule{B-RecvInclude} 
      Then we know that $\isihigh{l}$.
      \sketch{As a high integrity script cannot receive a low integrity
      response, this case is impossible}
      \ptodo{1}
      {This case should not occur (High integrity script receiving a low
      integrity response to include)} 
    \item \irule{B-Redirect}
      Then we know that $\isihigh{l}$
      Then because of $B_1 \bprel B_2$ we get that  $N = N'$, $\domain(T) =
      \domain(T')$ and we can also apply \irule{B-RecvLoad} or \irule {B-Redirect}
      for $B_2$.  Because of $\isilow{l_alpha}$ we immediately get $B_1' \bprel
      B_2'$.
      Then because of $N = N'$, we can also apply \irule{B-Redirect} or
      \irule{B-Load} for $B_2$. 
  \end{itemize}

\end{proof}


Next we show the same property for servers taking an internal step of high sync integrity.

\begin{lemma}[High Sync Integrity Server Steps]
  \label{lem:hint-serv-step}
  Let $S_1, S_2$ be servers with $S_1 \sprel S_2$ and 
  let $S_2$ be deterministically terminating.
  Let $S_1 \sstep{\alpha}
  S_1'$ with $\isihigh{\eint{\alpha}}$ and $\alpha \in \{\blank,
  \lauth{\cdot}{\cdot}{\cdot}\}$. 
  Then $\rbad{S_1'}$ or there exist $S_2'$ and $\alpha', \beta$ such that
  $S_2 \sstepn{*}{\beta \cdot \alpha'} S_2'$ with $\isilow{\eint{\beta_k}}$ for all $\beta_k \in \beta$ and  $\alpha \heqe \alpha'$
  and $S_1' \sprel S_2'$.
\end{lemma}
\begin{proof}
  Let $S_1 = \exserver{D_1}{\phi_1}{t^0_1}$,
  $S_1' = \exserver{D_1'}{\phi_1'}{t^{0'}_1}$,
  $S_2 = \exserver{D_2}{\phi_2}{t^0_2}$,
  $S_2' = \exserver{D_2'}{\phi_2'}{t^{0'}_2}$.
  Then there is $t_1 \in \running{S_1}$ with $\exserver{D_1}{\phi_1}{t_1} \sstep{\alpha} \exserver{D_1'}{\phi_1'}{t_1'}$
  and $t_1' \in \running{S_1'}$.
  Let $t_1 = \exthreadpar{c_1}{E_1}{R_1}{l_1}{\amode_1}$ and $t_1' =
  \exthreadpar{c_1'}{E_1'}{R_1}{l_1'}{\amode_1}$.

  Because of $\isihigh{\alpha}$ we know that $\isihigh{\thighestint{t}}$ and by
  the definition of $\sprel$ we know that there exists a corresponding thread
  $c(t_1) = t_2 \in \running{S_2}$ 
  with $t_2 = \exthreadpar{c_2}{E_2}{R_2}{l_2}{\amode_2}$.  

  We now show that there are $\alpha,\beta$ and $t_2' =
  \exthreadpar{c_2'}{E_2'}{R_2}{l_2'}{\amode_2}$,  $t_2'' =
  \exthreadpar{c_2''}{E_2''}{R_2}{l_2''}{\amode_2}$
  with 
  $\exserver{D_2}{\phi_2}{t_2} \sstepn{*}{\beta} \exserver{D_2''}{\phi_2''}{t_2''} \sstep{\alpha} \exserver{D_2'}{\phi_2'}{t_2'}$.

  Let $S_2'' = \exserver{D_2''}{\phi_2''}{t^{0''}_2}$. \ptodo{3}{Correctly introduce this}

  We perform the proof by induction over the derivation of the step $\alpha$.

  For all cases except \irule{S-Reset} we let $\beta = \epsilon$ and $S_2'' =
  S_2$.
  \begin{itemize}
    \item \irule{S-Seq} Claim follows by induction.
    \item \irule{S-IfTrue} 
     Then $c_1= \ite{se}{c_{11}}{c_{12}}$ and $c_2=\ite{se'}{c_{21}}{c_{22}}$
     with $se \heqe se'$. 
     Let $v^{\stype} = \eval{se}{D_1}{E_1}$ and 
     $v'^{\stype'} = \eval{se'}{D_2}{E_2}$. 
     Then by \cref{lem:pres-heq-eval} we get $v^\stype \heqe v'^{\stype'}$.
     We distinguish two cases:
      \begin{itemize}
        \item If $\mathsf{reply}, \mathsf{redir}, \mathsf{tokencheck}, \mathsf{origincheck} \in \commands{c_{11}} \cup \commands{c_{12}}$.
          We distinguish to cases
          \begin{itemize}
            \item If $\isihigh{\It{\stype}}$ then we also have
              $\isihigh{\It{\stype'}}$ and we have $v = v'$. 
              Hence the continuations are $c_1 = c_{11}$
              and $c_2 = c_{21}$ and the claim follows immediately.
            \item If $\isilow{\It{\stype}}$ then we also have
              $\isilow{\It{\stype'}}$.
              We hence have $\isilow{l_1'}$ and $\isilow{l_2'}$ and the claim
              follows. 
          \end{itemize}
          
        \item If $\mathsf{reply}, \mathsf{redir}, \mathsf{tokencheck},
          \mathsf{origincheck} \not\in \commands{c_{11}} \cup \commands{c_{12}}$.
          We distinguish to cases
        \begin{itemize}
          \item If $\isihigh{\It{\stype}}$ then we also have $\isihigh{\It{\stype'}}$
            and we have $v = v'$. Hence the continuations are $c_{11};\resetpc{l}$
            and $c_{21};\resetpc{l}$ and the claim follows immediately.
          \item If $\isilow{\It{\stype}}$ then we also have $\isilow{\stype'}$.
            Then $t_1' =  \exthreadpar{c_{12};\resetpc{l_1}}{E_1}{R_1}{l_1 \ijoin \It{\stype}}{\amode_1}$
            and $t_2' =  \exthreadpar{c_2';\resetpc{l_2}}{E_2}{R_2}{l_2 \ijoin \It{\stype'}}{\amode_2}$
            where $c_2' \in \{c_{11}', c_{12}\}$. 
            The claim then follows immediately.
        \end{itemize}
      \end{itemize}
    \item \irule{S-False} This case is analog to the case of rule \irule{T-True}.
    \item \irule{S-TokenCheckTrue}, 
      Then $c_1 = \tokch{se_{11}}{se_{12}}{c_1''}$
      and $c_2 = \tokch{se_{21}}{se_{22}}{c_2''}$.

      Let $v_{11} = \eval{se_{11}}{D_1}{E_1}, 
      v_{12} = \eval{se_{12}}{D_1}{E_1}, 
      v_{21} = \eval{se_{21}}{D_2}{E_2}, 
      v_{22} = \eval{se_{22}}{D_2}{E_2}$

      We then know that $v_{11} = v_{12}$.

      We distinguish two cases:
      \begin{itemize}
        \item if $v_{21} = v_{22}$ then $c_1 = c_1''$ and
          $c_2 = c_2''$ and the claim is trivial.
        \item if $v_{21} \neq v_{22}$ then we have 
          $c_2' = \ereply{\serror}{\sskip}{\mempty}$.
          This however is a contradiction to the
          assumption of the deterministic termination 
      \end{itemize}
    \item \irule{S-TokenCheckFalse} 
        Then $t_1' =
        \exthreadpar{\ereply{\serror}{\sskip}{\mempty}}{E_1}{R_1}{l_1}{\amode_1}$
        and the claim is trivial, since we have $\rbad{S_1'}$ 
    \item \irule{S-Skip} Trivial
    \item \irule{S-Reset} 
      We distinguish two cases:
      \begin{itemize}
        \item If $\isihigh{l_1}$ then we have $l_1 = l_2$ and the claim is trivial.
        \item Otherwise we know that $c_1 = \resetpc{l_r}$ where
          $\isihigh{l_r}$. Then we know by property \ref{def:srelL}
          of \cref{def:srel} that $c_2 = {c_2}_r; \resetpc{l_1}; {c_2}_r'$
          for some ${c_2}_r, {c_2}_r'$, with $c_1' \heqe {c_2}_r'$.
          
          We then let $c'' = \resetpc{l_1} ; {t_2}_r'$ and 
          $c' = {t_2}_r'$ and show that they fulfill the claim.

          Since we know that 
          $\mathbf{reply}, \mathbf{redir}, \mathbf{tokencheck},
          \mathbf{origincheck} \not \in \commands{c_2}_r$, we know by 
          deterministic termination of $t_2$ that 
          $t_2 \sstepn{*}{\beta} \exthreadpar{c''}{R''}{E''}{l''}{\amode}$.

          By repeated application of \cref{lem:low-int-serv-step} and
          \cref{lem:srel-trans} we get
          $S_1 \sprel S_2''$. 

          Now we need to show $S_1' \sprel S_2'$. All claims from \cref{def:srel}
          except for property \ref{def:srelL} are trivial. 
          For property \ref{def:srelL}  $l_1' = l_2'$
          follows immediately from rule \irule{S-Reset} and $c_1' \heqe c_2'$
          follows immediately from $c_1' \heqe {c_2}_r'$.
      \end{itemize}
    \item \irule{S-RestoreSession} Then $c_1 = \start{se}$ and $c_2 = \start{se'}$, 
      with $se \heqe se'$. Then for $S_2$ we can apply rule
      \irule{S-RestoreSession} or \irule{S-NewSession} and the claim follows
      because using \cref{lem:pres-heq-eval} we immediately get $j_1' \heqe
      j_2'$.
    \item \irule{S-NewSession} Analog to previous case.
    \item \irule{S-SetGlobal} We have $c_1 = r := se$ and $c_2 = r := se'$
      with $se \heqe se'$. The claim then follows immediately using
      \cref{lem:pres-heq-eval}.
    \item \irule{S-SetSession} This case follows analog to the previous one.
    \item \irule{S-Login} We have $c_1 = \login{se_1}{se_2}{se_3}$
      and $c_2 = \login{se_1'}{se_2'}{se_3'}$ with $se_1 \heqe se_1'$,
      $se_2 \heqe se_2'$ and  $se_3 \heqe se_3'$.
      Let $j_1^{\stype} = \eval{se_3}{D_1}{E_1}$ and 
      let $j_2^{\stype'} = \eval{se_3'}{D_2}{E_2'}$.
      We distinguish two cases: 
      \begin{itemize}
        \item If $\isilow{\It{\stype}}$ then also $\isilow{\It{\stype'}}$ and the claim
          follows immediately.
        \item If $\isihigh{\It{\stype}}$ then $j_1^{\stype} = j_1'^{\stype'}$.
          By rule \irule{T-Login} and \cref{lem:bro-ex-ty} we know that $\stype
          = \tcre{\ell}$.
        Let $v_1^{\stype_1} = \eval{se_1}{D_1}{E_1}$, 
        let $v_1'^{\stype_1'} = \eval{se_1'}{D_1}{E_1}$, 
        let $v_2^{\stype_2} = \eval{se_2}{D_1}{E_1}$ and 
        let $v_2'^{\stype_2'} = \eval{se_2'}{D_2}{E_2}$. Then 
        by rule \irule{T-Login} and \cref{lem:bro-ex-ty} we know that
        and $\isihigh{\stype_1}$ and  $\isihigh{\stype_2}$ and hence by $se_1
        \heqe se_1'$ and $se_2 \heqe se_2'$ we get 
        $v_1^{\stype_1} = v_1'^{\stype_1'}$ and 
        $v_2^{\stype_2} = v_2'^{\stype_2'}$. 
        With $\bid = \eval{se_1}{D_1}{E_1}$ and 
        let $\bid' = \eval{se_1'}{D_2}{E_2}$ we get using the properties of
        $\rho$ 
      \end{itemize}
    \item \irule{T-Auth} 
      Then we have $c_1 = \auth{\lst{se_1}}{\ell}$ and 
      $c_2 = \auth{\lst{se_2}}{\ell}$ with $\lst{se_1} \heqe \lst{se_2}$.
      If $\isilow{\ell}$, then the claim is trivial. 
      We hence assume $\isihigh{\ell}$.

      We then know by rule \irule{T-Auth} that $\isihigh{l}$.

      Let $v_{1,i}^{\stype_{1,i}} = \eval{se_{1,i}}{D_1}{E_1}$ and 
      Let $v_{2,i}^{\stype_{2,i}} = \eval{se_{2,i}}{D_2}{E_2}$.
      Let $R_1 = R_2 = n,u,\bid,o$, let $E_1 = i_1, j_1$ and $E_2 = i_2, j_2$,
      and let ${\sid}_1 = \phi(j_1)$ and ${\sid}_2 = \phi(j_2)$.

      We then know $\lst{se_1} \heqe \lst{se_2}$ and $j_1 \heqe j_2$

      We have $\alpha = \lauth{\lst{v_1}}{\bid, {\sid}_1}{\ell}$
      and $\alpha' = \lauth{\lst{v_2}}{\bid, {\sid}_2}{\ell}$.
    
      By rule \irule{T-Auth} we know that 
      $\isihigh{\It{\stype_{1,i}}}$ and 
      $\isihigh{\It{\stype_{2,i}}}$, we thus have 
      $\lst{v_1^{\stype_1}} = \lst{v_2^{\stype_2}}$ by \cref{lem:pres-heq-eval}.
      
      By rule \irule{T-Auth} we also know that that $\isihigh{I(\jlabel{j_1})}$
      and $\isihigh{I(\jlabel{j_2})}$. We thus get by property \ref{def:srelP}
      of \cref{def:srel} that ${\sid}_1 = {\sid}_2$.

      We thus have $\alpha = \alpha$ and the claim follows.

    \item \irule{S-OCheckSucc}
      We then have $c_1 = \och{O}{c_1'}$ and $c_2 = \och{O}{c_2'}$.
      With $R_1 = R_2 = n,u,\bid,o$ we know that we can also apply rule
      \irule{S-OCheckSucc} in $t_2$ and the claim follows immediately.

    \item \irule{S-OCheckFail}
      We then have $c_1 = \och{O}{c_1'}$ and $c_2 = \och{O}{c_2'}$.
      With $R_1 = R_2 = n,u,\bid,o$ we know that we can also apply rule
      \irule{S-OCheckFail} in $t_2$ which contradicts our assumption about the 
      termination of $t_2$. This case is thus impossible
  \end{itemize}
\end{proof}

Next, we show the same property for servers receiving a request of high sync integrity.

\begin{lemma}[High Sync Integrity Server Request]
  \label{lem:hint-serv-req}
  Let $S_1, S_2$ be servers with $S_1 \sprel S_2$ and 
  let $S_2$ be the corresponding server of $S_1$ as defined in
  \cref{def:relation}.
  Let $S_1 \sstep{\alpha}
  S_1'$ with $\isihigh{\alpha}$ 
  and $\alpha = \exsrecv{\bid}{n}{u}{p}{ck}{o}{l}{\amode}$.
  Let $\alpha'$ with $\alpha \heqe \alpha'$ and $\reqtyping{\alpha}$,
  $\reqtyping{\alpha'}$.
  Then there exist $S_2'$ such that
  $S_2 \bstep{\alpha'} S_2'$ 
  and $S_1' \sprel S_2'$.
\end{lemma}
\begin{proof}
  Then the step is taken using rule \irule{S-Recv}, We thus have 
  $t = \listen{u}{\lst{r}}{\lst{x}}{c} \in \threads{S_1}$.
  Because $S_2$ is the corresponding server of $S_1$ we know that 
  $t \in \threads{S_2}$.

  We can thus apply rule \irule{S-Recv} in and take the step 
  $S_2 \sstep{\alpha'} S_2'$.

  We now show $S_1' \sprel S_2'$ by showing the properties of \cref{def:srel}.
  \begin{itemize}
    \item Property \ref{def:srelTy} follows immediately from \cref{lem:serv-req}
    \item Property \ref{def:srelE} is trivial \sketch{(For simplicity we assume
      that sampling returns the same result on both servers)}
    \item Property \ref{def:srelG} follows immediately from the claim on
      cookies in $\alpha \heqe \alpha$.
    \item Property \ref{def:srelL} follows from the claim on the parameters in 
      $\alpha \heqe \alpha$.
    \item Properties \ref{def:srelS} and \ref{def:srelP} are trivial since the
      session memory and trust mapping are not modified in rule \irule{S-Recv}.
  \end{itemize}

  Let $t$ and $t'$ be the freshly generated running threads. 
  Then $t \heqe t'$ follows from the claim on $p$ in $\alpha \heqe \alpha'$ . 
\end{proof}

Next, we show the same property for servers sending a response of high sync integrity.

\begin{lemma}[High Sync Integrity Server Response]
  \label{lem:hint-serv-res}
  Let $S_1, S_2$ be servers with $S_1 \sprel S_2$ and 
  let $S_1 \bstep{\alpha}
  S_1'$ with $\isihigh{\eint{\alpha}}$ 
  and $\alpha = \exssend{\bid}{n}{u}{u'}{\lst{v}}{ck}{page}{s}{l}{\amode}$
  Then there exist $S_2'$ and $\alpha'$ such that
  $S_2 \bstep{\alpha'} S_2'$ 
  and $S_1' \sprel S_2'$ and 
  $\alpha \heqe \alpha'$.
\end{lemma}
\begin{proof}
  We distinguish two cases for the rule applied to take the step:
  \begin{itemize}
    \item \irule{S-Reply}
      We have $c= \reply{page}{s}{ck}{\lst{x} = \lst{se}}$
      and $c'= \reply{page}{s}{ck}{\lst{x} = \lst{se'}}$ with
      $\forall i \in \intval{1}{\len{se}} se_i \heqe se_i'$.

      Let $v_i = \eval{se_i}{D}{E}$ and $v_i' = \eval{se_i'}{D'}{E'}$.
      Then by \cref{lem:pres-heq-eval} we get $v_i \heqe v_i'$.

      With $\sigma = \{x_1 \mapsto v_1 \cdots x_m \mapsto v_m\}$ 
      and  $\sigma' = \{x_1 \mapsto v_1' \cdots x_m \mapsto v_m'\}$ 

      We immediately get $s\sigma \heqe s\sigma'$, 
      $page\sigma \heqe page\sigma'$ and $ck \sigma \heqm ck'$.

      The claim then follows using \cref{lem:serv-res}.
    \item \irule{S-Redir} 
      We have $c= \redr{u}{\lst{z}}{ck}{\lst{x} = \lst{se}}$
      and $c'= \redr{u}{\lst{z}}{ck}{\lst{x} = \lst{se'}}$ with
      $\forall i \in \intval{1}{\len{se}} se_i \heqe se_i'$.

      Let $v_i = \eval{se_i}{D}{E}$ and $v_i' = \eval{se_i'}{D'}{E'}$.
      Then by \cref{lem:pres-heq-eval} we get $v_i \heqe v_i'$.

      With $\sigma = \{x_1 \mapsto v_1 \cdots x_m \mapsto v_m\}$ 
      and  $\sigma' = \{x_1 \mapsto v_1' \cdots x_m \mapsto v_m'\}$ 

      We immediately get $\lst{z}\sigma \heqe \lst{z'}\sigma'$ and 
      $ck\sigma \heqe ck'\sigma$.

      The claim then follows using \cref{lem:serv-res}.

\end{itemize}
\end{proof}

Finally, we show the same property on websystem level.

\begin{lemma}[High Sync Integrity Steps]
  \label{lem:high-int-step}
  Let $A_1 = \atkstate{\atknow_1}{\alabel}{W_1}$ and
  $A_2 \atkstate{\atknow_2}{\alabel}{W_2}$ be web systems with $A_1 \prel A_2$ and
  let $A_2$ be deterministically terminating.
  Then whenever $A_1 \astep{\alpha} A_1'$ with $\isihigh{\eint{\alpha}}$
  then $\rbad{A_1'}$ or there exist $\lst{\beta}$, $\alpha'$ and $A_2'$ such that 
  $A_2 \astep{\lst{\beta} \cdot \alpha'} A_2'$ with $\alpha \heqe \alpha'$
  and for all $\beta \in \lst{\beta}$ we have
  $\isilow{\eint{\beta}}$ and $A_1' \prel A_2'$.
\end{lemma}
\begin{proof}
    If $A_2$ is not in a low integrity state as defined in \cref{lem:lint-catchup},
    then let $A_2'' = A_2$.  \ptodo{2}{However, the reset state is already covered}
  Otherwise, let $A_2'' = \nexth{A_2}$ as in \cref{lem:lint-catchup}.
  We then know that $A_2 \astep{\lst{\beta}} A_2''$ where for
  $\beta \in \lst{\beta}$ we have $\isilow{\eint{\beta}}$ and $A_2 \prel A_2''$.
  By transitivity we hence get $A_1 \prel A_2''$. We furthermore know that
  $A_2''$ is in one of the five states described in \cref{lem:lint-catchup}.
   
  We now show that $A_2'' \astep{\alpha'} A_2'$ with $\alpha \heqe \alpha'$ 
  and $A_2 \prel A_2'$.
  We prove the claim by induction over the derivation of the step $\alpha$.
  \begin{itemize}
    \item \irule{A-Nil} Then, if the step is derived through rule
      \irule{W-LParallel} or \irule{W-RParallel} the claim follows by induction.
      The claim for internal server steps follows from 
      \cref{lem:hint-serv-step}.
      For internal browser steps, we perform a case distinction:
      Let $\browsers{W} \ni B = \exebrowser{N}{K}{P}{T}{Q}{\lst{a}}{\usr}{l}{\amode}$ 
      \begin{itemize}
        \item if $\isihigh{l}$ then the claim follows immediately from
          \cref{lem:hint-bro-step}.
        \item if $\isilow{l}$ then we know that rule \irule{B-End} is used.
            We know that $A_2'' = \nexth{A_2}$ is in one of the five states
            described in \cref{lem:lint-catchup}. Since we already excluded one
            possible state, and three other states require $\isihigh{l}$, 
            we know that the browser $B_2''$ in $A_2''$ is in a state where 
            rule \irule{B-End} can be used. The claim then follows immediately.
      \end{itemize}
    \item \irule{A-BrowserServer} Then 
      we can also apply \irule{A-BrowserServer} for $A_2''$ and 
      the claim follows from
      \cref{lem:hint-bro-req} 
      for the browser step 
      and \cref{lem:hint-serv-req} for the server in case of a high integrity
      request or \cref{lem:low-int-serv-req} for the server in case of a low
      integrity request.
    \item \irule{A-ServerBrowser}
      Then we distinguish two cases
      \begin{itemize}
        \item Integrity of the response is high:
          Then we can also apply rule \irule{A-ServerBrowser} in $A_2'' = A_2$
          and the claim follows from \cref{lem:hint-serv-res} for the server 
          and \cref{lem:hint-bro-res} for the browser step.
        \item Integrity of the response is low:
          \sketch{Then we know by \cref{lem:lint-catchup} that 
            $A_2'' = \nexth{A_2}$ is in a state where the browser can receive a
            request and the server can send a reply or a redirect or a timeout
          response is ready to be sent.}
          We immediately get $\alpha \heqe \alpha'$.
          Then we can also apply rule \irule{A-ServerBrowser} and the claim
          follows from \cref{lem:low-int-serv-step} for the server step and
          from \cref{lem:lhint-bro-res} for the browser step.
      \end{itemize}
    \item \irule{A-TimeoutSend}
      Then we can also apply \irule{A-TimeoutSend} in $A_2''$ and 
      the claim follows from \cref{lem:hint-bro-req} for the browser step
      in case of a high integrity browser state 
      or from \cref{lem:low-int-bro-req}
      in case of a low integrity browser state.
    \item \irule{A-TimeoutRecv}
      Then we can also apply \irule{A-TimeoutRecv} in $A_2$'' and 
      the claim follows from \cref{lem:hint-bro-res} 
      or \cref{lem:low-int-bro-req} for the browser step.
    \item \irule{A-BroAtk} Then we distinguish two cases
      \begin{itemize}
        \item $W_2$ can perform a step  using \irule{A-BrowserServer}:
          Then the claim follows from
          \cref{lem:hint-bro-req} 
          or \cref{lem:low-int-bro-req} 
          for the browser step 
         and \cref{lem:low-int-serv-step} for the server.
        \item $W_2$ can perform a step  using \irule{A-TimeoutSend}
          Then the claim follows from
          \cref{lem:hint-bro-req} or \cref{lem:low-int-bro-req} 
          for the browser step.
      \end{itemize}
    \item \irule{A-AtkSer} Cannot happen, event is of high integrity
    \item \irule{A-SerAtk} Cannot happen, event is of high integrity 
    \item \irule{A-AtkBro} Then we distinguish two cases:
      \begin{itemize}
        \item $W_2$ can perform a step  using \irule{A-ServerBrowser}:
          Then the claim follows from
          \cref{lem:hint-bro-res} for the browser step.
        \item $W_2$ can perform a step  using \irule{A-TimeoutRecv}
          Then the claim follows from
          \cref{lem:hint-bro-res} for the browser step.
      \end{itemize}
  \end{itemize}
\end{proof}

Using the previous lemmas, we can conclude that the relation $\rel$ fulfills
core properties, that will allow us to prove the main theorem.

\begin{lemma}
  \label{lem:rel-step-rel}
  Let $A_1$ and $A_2$ be web systems with $A_1 \rel A_2$. Then
  \begin{enumerate}
    \item $\rbad{A_1}$ 
    \item or the following properties hold:
      \begin{enumerate}
        \item if $A_1 \exastep{\alpha} A_1'$ and $\isihigh{\eint{\alpha}}$ 
          then there exists $\alpha', \lst{\beta}$ and $A_2'$ such that 
          \begin{itemize}
            \item for all $\beta \in \lst{\beta}$ we have $\isilow{\eint{\beta}}$
            \item $A_2 \exastepn{*}{\vec{\beta} \cdot \alpha'} A_2'$ \
            \item $\alpha \heqe \alpha'$
            \item $A_1' \rel A_2'$
          \end{itemize} 
        \item if $A_1 \exastep{\alpha} A_1'$ for some $\alpha$ with
        $\isilow{\eint{\alpha}}$
        then $A_1' \rel A_2$.
      \end{enumerate}
  \end{enumerate}
\end{lemma}
\begin{proof}
  If $\rbad{A_1}$ then the claim is trivial.
  We hence assume $\neg \rbad{A_1}$, which then immediately gives us 
  $A_1 \prel A_2$.
  The claim for the low integrity step then follows immediately from
  \cref{lem:low-int-step} and the transitivity of $\prel$ (\cref{lem:rel-trans})
  and the claim for the
  high integrity step follows from \cref{lem:high-int-step}.
\end{proof}

Intuitively,the relation fulfills the following properties:
Either the first websystem is in a bad state, or
\begin{enumerate}
  \item  Whenever the first system takes a step of high sync integrity, then the second
    system can take a number of steps of low sync integrity, followed by the same step
    of high sync integrity, and the resulting websystems are in the relation.
  \item If the first system takes a step of low sync integrity, then it remains in
    relation with the second system (which didn't take a step).
\end{enumerate}

\subsection{Main Theorem}
\label{sec:proof-main}
In this section we bring together the results from the previous sections in
order to show our main theorem.

First, we show that whenever an attacked and an unattacked websystem are in 
the relation $\rel$, and the attacked system generates a trace, then the
unattacked websystem can generate a trace that has the same events with high
sync integrity,

\begin{lemma}[High Integrity Trace Equality]
  \label{lem:high-trace-eq}
  Let $high(\gamma)$ be the trace containing only the events $\alpha \neq
  \blank$ with $\isihigh{\eint{\alpha}}$.

  Let $A_1$ be an attacked and $A_2$ and unattacked websystem with $A_1 \rel
  A_2$. Then if $A_1$ generates the trace $\gamma_1$, then $A_2$
  can generate a trace $\gamma_2$ such that
  $high(\gamma_1) = high(\gamma_2)$.

\end{lemma}
\begin{proof}
  We prove the claim by induction over the generated trace $\gamma$, using the
  properties of $\rel$ from \cref{lem:rel-step-rel}
  \begin{enumerate}
    \item If {$\gamma_1 = \epsilon$} then the claim is trivially fulfilled.
    \item If {$\gamma_1 = \alpha \cdot \gamma_1':$} then
      $A_1$ takes the step $\alpha$ to reach state $A_1'$ and produces the trace
      $\alpha \cdot \gamma_1'$.
      If $\rbad{A_1}$ then we know that $high(\alpha \cdot \gamma) =
      \epsilon$, since according to \cref{def:bad} 
      we either have $\alpha = \blank$ or  $\isilow{int(\alpha)}$ and the
      claim is trivial.
      We hence assume $\neg \rbad{A_1}$ and hence know $A_1 \prel A_2$.

      We distinguish two cases
      \begin{enumerate}
        \item If {$\isihigh{\eint{\alpha}}$} then
          by \cref{lem:rel-step-rel} $A_2$ can take the steps $\vec{\beta} \cdot \alpha$,
          where $\isilow{\eint{\beta}}$ for all $\beta \in \vec{\beta}$ and hence
          produces the trace $\vec{\beta} \cdot \alpha \cdot \gamma_2'$. We
          hence have $high(\vec{\beta} \cdot \alpha) = \alpha$.
          Since $A_1' \prel A_2'$, we can apply the induction hypothesis and
          get
          $high(\gamma_1') = high(\gamma_2')$, hence we also have
          $high (\gamma_1) = high(\alpha \cdot \gamma_1') = 
          high(\alpha \cdot \gamma_2') = high(\gamma_2)$.
        \item If {$\isilow{\eint{\alpha}}$}
          then we have $high(\gamma_1') = high(\gamma_1)$ and since
          by \cref{lem:rel-step-rel} we know
          $A_1' \prel A_2$, we can apply the induction hypothesis and get 
          $high(\gamma_2) = high(\gamma_1') = high(\gamma_1)$.
      \end{enumerate}
  \end{enumerate}
\end{proof}

Next, we show that whenever a well-typed websystem produces a high integrity
authenticated event then it also has high sync integrity.

\begin{lemma}[High Integrity Auth Events]
  \label{lem:high_int_auth}
  Let $\usr$ be the honest user and for all $u$ with $\rho(\usr,u) = n^\stype$
  we have $\ischigh{C(\stype)}$.
  Let $\alabel$ be an attacker.
  For any $A$ with $\systyping{A}$ and $A \astepn{*}{\lst{\alpha}} A'$, if 
  for $\beta = \lauth{\lst{v}}{\bid, \uid}{\ell}$ we have 
  $\beta \in \lst{\alpha}$ and $\isihigh{I(\ell)}$ and $\bid = \usr$ or $\uid =
  \usr$ then we have 
      $\isihigh{\eint{\beta}}$, 
\end{lemma}
\begin{proof}
  Let $l = \eint{\beta}$.

  There exist $A_1, A_2$ such that $A \astepn{*}{\lst{\alpha_1}} A_1 \astep{\beta} A_2 \astepn{*}{\lst{\alpha_2}} A'$ with $\systyping{A_1}$ by \cref{lem:sys-sub-red}.

  We also know that $S_1 \in \servers{A_1}, S_2 \in \servers{A_2}$ with
  $S_1 \astep{\beta} S_2$, with $\servtyping{S_1}$ by \cref{def:sys-typing}
   
  Let $S_1 = \exserver{D}{\phi}{t}$. Then there is $\exthread{c}{R}{E} \in \running{S_1}$ with 
  $
    \exserver{D}{\phi}{\exthread{c}{R}{E}}
    \astep{\beta}
    \exserver{D}{\phi}{\exthread{c'}{R}{E}}
    $
    for some $c, c'$.

  We know that the event $\beta$ is produces using rule \irule{S-Auth}, hence
  have 
  \[
    \exserver{D}{\phi}{\exthread{\auth{\lst{se}}{\ell}}{R}{E}}
    \astep{\beta}
    \exserver{D}{\phi}{\exthread{\sskip}{R}{E}}
  \]
  with $R = n, u, \bid, o$ and $E = i,j$ with $\phi(j) = \uid$.

  Let
  \begin{align*}
    \typebranch = 
      \begin{cases}
        \batt & \text{ if } \bid \neq \usr \\
        \bhon & \text{ if } \amode = \bhon \wedge \bid = \usr \\
        \bcsrf & \text{ if } \amode = \batt \wedge \bid = \usr 
      \end{cases}
      \\
    \envrg' = 
      \begin{cases}
        \envrg & \text{ if } (\bid = \usr) \\
        \subst{\_}{\alabel} & \text{ if } (\bid \neq \usr) 
      \end{cases} \\
    \Gamma' = (\envu,\envv,\envrg',\envrs,\envt)
  \end{align*}


  We then get by rule \irule{T-Running}
  \[
    \stypingpar{\Gamma'}{\jlabel{j}}{l}{\typebranch}{\auth{\lst{se}}{\ell}}{\_}
  \]

  We distinguish two cases:
  \begin{itemize}
    \item If $\typebranch \neq \batt$ then 
    by typing we know from rule \irule{T-Auth} that 
      we have
      $\isihigh{l}$ and the claim follows immediately.
  

    \item If $\typebranch = \batt$ then we know that $\bid \neq \usr$.
      Hence we must have $\phi(j) = \uid = \usr$.
      We now show that this case can also not happen.
      Because of  $\ischigh{\Ct{\rho(\uid)}}$ and property \ref{def:stP} of
      \cref{def:serv-typing} we know that $\Ct{\rho(\uid)} \cleq
      \Ct{\jlabel{j}}$.
      Since an attacker can never have a session with a high confidentiality session label, 
      we know $\isclow{\Ct{\jlabel{j}}}$ and we immediately have a
      contradiction.
  \end{itemize}
\end{proof}


%
%

We define a well-formed attacker to be an attacker whose knowledge is limited
by his label.

\begin{definition}[Well-formed attacker]
  An attacker $(\alabel,\atknow)$ is \emph{well-formed} if  
  $\forall n^{\stype} \in \atknow$ we have  $\stype \tleq \alabel$.
\end{definition}

This lemma shows that the initial state is in the relation $\rel$ with itself.

\begin{lemma}[The initial state is in $\rel$]
  \label{lem:init-rel}
  Assume a well-formed server cluster $W_0$, an honest browser of the user
  $\usr$ ${B_{\usr}(\mempty,\lst{a})}$ with well formed user actions $\lst{a}$,
  a well-formed attacker ($\alabel$,$\atknow$) and let
  $A=\atkstate{\atknow}{\alabel}{\para{B_{\usr}(\mempty,\lst{a})}{W_0}}$.
  If for all servers $S=\exserver{D}{\phi}{t}$ of $W_0$, we have $\sttyping{t}$, then 
  $\trans{A} \rel \trans{A}$.
\end{lemma}
\begin{proof}
  We fist show $\trans{A} \prel \trans{A}$ by showing the different properties
  of \cref{def:relation}.
  \begin{itemize}
    \item For \prop{def:relT}, $\systyping{A}$ we show that the properties of
      \cref{def:sys-typing} are fulfilled:
      \begin{itemize}
        \item We get property \ref{def:systB}, $\browtyping{\trans{B_{\usr}(\mempty, \lst{a})}}$ by 
          checking that all the properties of \cref{def:bro-typing} hold.
          Property \ref{def:btA} follows from the well-formedness of $\lst{a}$.
          All other properties are trivial,
        \item We get \ref{def:systS}, $\servtyping{\trans{S}}$ for all servers 
          $S = \exserver{D}{\phi}{t} \in \servers{W_0}$ by checking that all properties of
          \cref{def:serv-typing}.
          Property \ref{def:stT} follows from $\sttyping{t}$, using
          \cref{lem:typeeq}. All other properties are trivial for fresh
          servers (as defined in \cref{def:cluster}).
        \item Property \ref{def:systL} is trivial since there are no network
          connections in the browser.
        \item Property \ref{def:systA} follows immediately from the
          well-formedness of the attacker.
      \end{itemize}
    \item \prop{def:relSC} is trivial
    \item For \prop{def:relSR}, $\trans{S} \sprel \trans{S}$ we show that the properties of
      \cref{def:srel} hold:
      \begin{itemize}
        \item \prop{def:srelTy} follows immediately from $\systyping{\trans{A}}$, which
          we have already shown.
        \item All other properties are trivial for fresh servers.
      \end{itemize}
    \item For \prop{def:relSR}, ${\trans{B_{\usr}(\mempty, \lst{a})}} \bprel
      {\trans{B_{\usr}(\mempty, \lst{a})}}$ we show that the properties of
      \cref{def:brel} hold:
      \begin{itemize}
        \item \prop{def:brelTy} follows immediately from $\systyping{\trans{A}}$, which
          we have already shown.
        \item All other properties are trivial for fresh browsers.
      \end{itemize}
  \end{itemize}
\end{proof}

Our main theorem states that typing ensures web session integrity -- if we
consider all ingredients to be well-formed.

\begin{theorem}[Typing implies Web Session Integrity]
  \label{lem:main-theorem}
  Let $W$ be a fresh cluster, ($\alabel$,$\atknow$) a well-formed attacker, 
  $\Gamma^0$ a
  typing environment with $\envtyping{\Gamma^0}$ and let $\lst{a}$ be a list of
  well-formed user actions for $\usr$ in $W$ with respect to $\Gamma^0$ and
  $\alabel$. 
  Assume that for all $u$ with $\rho(\usr,u) = n^\stype$ we have
  $\ischigh{C(\stype)}$
  and that we have 
  $\sttyping{t}$ for all servers $S=\exserver{\mempty}{\mempty}{t}$ in
  $W$.
  Then $W$ preserves \emph{session integrity} against ($\alabel$,$\atknow$) 
  for the honest user $\usr$ performing the list of actions $\lst{a}$.
\end{theorem}
\begin{proof}
  Let $W'=\para{B_{\usr}(\mempty,\lst{a})}{W}$ and let 
  $A = \atkstate{\alabel}{\atknow}{W'}$.

  We have to show that for any attacked trace $\gamma$ generated by the
  attacked system $A$ there
  exists a corresponding unattacked trace $\gamma'$ generated by $A$ such that
  \[
    \forall I(\ell) \not\ileq I(\ell'): \proj{\gamma}{(\usr, \ell')} =
     \proj{\gamma'}{(\usr, \ell')}
  \]

  By lemma \cref{lem:init-rel}, we know $\trans{A} \rel \trans{A}$.



  By lemma \cref{lem:sem-equiv} we know that also $\trans{A}$ can produce the
  trace $\alpha$. 

  Applying \cref{lem:high-trace-eq} , we know that there exists an unattacked
  trace $\gamma'$ produced by $\trans{A}$, such that 
  $high(\gamma) = high(\gamma')$. 

  Since for all  $\alpha = \lauth{\lst{v}}{\sid}{\ell'}$ 
  with $\ell' \not\sqsubseteq \ell$ we have $int(\alpha) = \ihigh$
  by \cref{lem:high_int_auth}, we know that 

  \[
    \forall I(\ell) \not\ileq I(\ell'): \proj{\gamma}{(\usr, \ell')} =
     \proj{\gamma'}{(\usr, \ell')}
  \]

  By lemma \cref{lem:sem-equiv} we know that this trace $\gamma'$ can also be
  produced by $A$.


\end{proof}

\end{document}